# Large deviations principles of Non-Freidlin-Wentzell type.

Jaykov Foukzon
Center for Mathematical Sciences, Israel Institution of Technology, Haifa, Israel.
E-mail: jaykovfoukzon@list.ru
E-mail: jaikovfoukzon@gmail.com
https://orcid.org/0000-0002-6837-4004

**Abstract**: Generalized Large deviation principles was developed for weakly time inhomogeneous diffusions prove to be key tools for a treatment of the problem of diffusion exit from a domain and thus for the approach of stochastic resonance via transition probabilities between meta-stable states. We expand the classical theory of large deviations for randomly perturbed dynamical systems developed by Freidlin and Wentzell.

**Content**

# Introduction.

# I.Large deviations of Freidlin-Wentzell type for diffusion proces.

Let us consider dynamical systems driven by slowly time dependent vector fields, perturbed by Gaussian noise of small intensity. We shall be interested in their large deviation behavior. Due to the slow time inhomogeneity, the task we face is not covered by the classical theory presented in Freidlin,Wentzell [4] and Dembo, Zeitouni [2]. For this reason we shall have to extend the theory of large deviations for randomly perturbed dynamical systems developed by Freidlin, Wentzell [4] to drift terms depending in a weak form to be made precise below on the time parameter. Before doing so in the second subsection, we shall recall the classical results (suections **I.1**-**I.2**) and main non-classical results (suection**I.3**.Theorem ) on time homogeneous diffusions in the following brief overview.The main general result Theorem  is stated in subsection **I.4**.

# I.1 The time homogeneous case: classical results.

For a more detailed account of the following well known theory see [2] or [4].We consider the family of $\mathbb{R}^d$-valued processes $X^\varepsilon, \varepsilon > 0,$ defined by Ito type equation

$$dX_t^\varepsilon = b(X_t^\varepsilon)dt + \sqrt{\varepsilon}\,\mathbf{W}_t, \quad X_0^\varepsilon = x_0, \tag{1.1}$$

on a fixed time interval $[0,T],$ where $b$ is Lipschitz continuous and $\mathbf{W}_t$ is a $d$-dimensional Brownian motion on a fixed time interval $[0,\mathbf{T}],$ where $b(x)$ is Lipschitz continuous This family of diffusion processes satisfies in the small noise limit, i.e. as $\varepsilon \to 0,$ a large deviations principle ($\mathbf{LDP}$) in the space $C([0,\mathbf{T}];\mathbb{R}^d)$ equipped with the topology of uniform convergence induced by the metric $\rho_{0\mathbf{T}}(\varphi,\psi)$ :

$$\rho_{0\mathbf{T}}(\varphi,\psi) = \sup_{0 \leqslant t \leqslant \mathbf{T}} \|\varphi_t - \psi_t\|, \varphi,\psi \in C([0,\mathbf{T}];\mathbb{R}^d).$$

The Freidlin-Wentzell, rate function (**FW** rate function) or Freidlin-Wentzell action functional (**FW** action functional) $\mathbf{I}_{0\mathbf{T}}^{x_0}(\varphi) : C([0,\mathbf{T}];\mathbb{R}^d) \to [0,\infty]$ is given via formula

$$\mathbf{I}_{0\mathbf{T}}^{x_0}(\varphi) = \begin{cases} \frac{1}{2}\int_0^{\mathbf{T}} \|\dot{\varphi}_t - b(\varphi)\|^2 dt & \text{if } \varphi \text{ is absolutely continuous and } \varphi_0 = x_0, \\ +\infty & \text{otherwise.} \end{cases} \tag{1.2}$$

Moreover, $\mathbf{I}_{0\mathbf{T}}^{x_0}(\varphi)$ is a good rate function, i.e. it has compact level sets. The classical **LDP** for this family of processes is mainly obtained as an application of the contraction principle to the **LDP** for the processes $\sqrt{\varepsilon}\,\mathbf{W}_t, \varepsilon > 0.$ More precisely, in the language of Freidlin and Wentzell, the functional $\mathbf{I}_{0\mathbf{T}}^{x_0}(\varphi)$ is the normalized action functional corresponding to the normalizing coefficient $1/\varepsilon$.In the sequel we will not considerscalings other than this one. We have $\mathbf{I}_{0\mathbf{T}}^{x_0}(\varphi) < \infty$ if and only if $\varphi$ belongs to the Cameron-Martin space of absolutely continuous functions with square integrable derivatives starting at $x_0,$ i.e.

$$\varphi \in \mathbf{H}_{0\mathbf{T}}^{x_0} \Leftrightarrow \left\{ \varphi : [0,T] \to \mathbb{R}^d |\ \varphi(t) = x_0 + \int_0^T g(s)ds \text{ for some } g \in \mathbf{L}^2([0,T]) \right\}.$$

We omit the superscript $x_0$ whenever there is no confusion about the initial condition we are referring $t_0$. Observe that $\mathbf{I}_{0\mathbf{T}}^{x_0}(\varphi)$ means that $\varphi$ (up to time $T$) is a solution of the deterministic equation

$$\dot{\xi}_t = b(\xi_t), \tag{1.3}$$

so $\mathbf{I}_{0T}^{x_0}(\varphi)$ is essentially the $\mathbf{L}^2$-deviation of $\varphi$ from the deterministic solution $\xi$. The cost function $V$ of $X_t^\varepsilon$, defined by

$$V(x,y,t) = \inf_\varphi \{\mathbf{I}_{0\mathbf{T}}^x(\varphi) : \varphi \in C_{0t}, \varphi_0 = x, \varphi_t = y\}$$

takes into account all continuous paths connecting $x,y \in \mathbb{R}^d$ in a fixed time interval of length $t$, and the quasi-potential

$$V(x,y) = \inf_{t>0} V(x,y,t)$$

describes the cost of $X_t^\varepsilon$ going from $x$ to $y$ eventually. In the potential case, $V$ agrees up to a constant with the potential energy to spend in order to pass from $x$ to $y$ in the potential landscape, hence the term quasi-potential.

As far as we know, the **LDP** for the process $X_t^\varepsilon$ is only proven in the case of the usual *global Lipschitz and linear growth conditions* from the standard existence and uniqueness results for SDE. In our setting the coefficients will not be globally Lipschitz. Though the extension is immediate, we therefore state it for completeness in the following proposition.

**Proposition 1.1**. Assume that the equation (1.1) has a unique strong solution that never explodes and that the drift is locally Lipschitz. Then $X_t^\varepsilon$ satisfies on any time interval $[0,\mathbf{T}]$ a weak **LDP** (**WLDP**) with rate function $\mathbf{I}_{0\mathbf{T}}^x(\varphi)$ .More precisely, for any compact $\mathbf{F} \subset C_{0\mathbf{T}}$ we have:

$$\limsup_{\varepsilon\to 0} \varepsilon \log \mathbf{P}_x((X_t^\varepsilon)_{0\leqslant t\leqslant \mathbf{T}} \in \mathbf{F}) \leq -\inf_F \mathbf{I}_{0\mathbf{T}}^x(\varphi), \quad \forall \mathbf{F} \subset C_{0\mathbf{T}}, \tag{1.4}$$

and for any open $\mathbf{G} \subset C_{0\mathbf{T}}$ we have:

$$\liminf_{\varepsilon\to 0} \varepsilon \log P_x((X_t^\varepsilon)_{0\leqslant t\leqslant \mathbf{T}} \in \mathbf{G}) \geq -\inf_G \mathbf{I}_{0\mathbf{T}}^x(\varphi), \quad \forall \mathbf{G} \subset C_{0\mathbf{T}}, \tag{1.5}$$

**Proof**. For $R > 0$ let $b_R(x)$ be a continuous function with $b_R(x) = b(x)$ for $x \in B_R(x_0)$ and $b_R(x) = 0$ for $x \notin B_{2R}(x_0)$, and let $\tilde{X}_t^\varepsilon$ be the solution of (1.1) with $b$ replaced by $b_R$ with the same initial condition $x_0$. We denote by $B_R(x_0)$ the ball of radius $R$ in $C_{0\mathbf{T}}$ for the uniform topology. Then there exists $R > 0$ such that $K \subset B_R(x_0)$. Hence $\mathbf{P}(X_t^\varepsilon \in K) = \mathbf{P}(\tilde{X}_t^\varepsilon \in K)$. Since the drift of $\tilde{X}_t^\varepsilon$ is globally Lipschitz it satisfies a large deviations principle with some good rate function $\mathbf{I}_{0\mathbf{T}}^R$. Applying this large deviations principle we obtain

$$\limsup_{\varepsilon\to 0} \varepsilon \log \mathbf{P}_x((X_t^\varepsilon)_{0\leqslant t\leqslant \mathbf{T}} \in \mathbf{F}) \leq -\inf_F \mathbf{I}_{0\mathbf{T}}^R(\varphi) = -\inf_F \mathbf{I}_{0\mathbf{T}}(\varphi),$$

which is the claimed upper bound.For the lower bound, due to its local nature (see,for instance, Theorem 3.3 in [4]), it is sufficient to show that for all $\delta > 0, \varphi \in C_{0\mathbf{T}}$

$$\liminf_{\varepsilon\to 0} \varepsilon \log P_x((X_t^\varepsilon)_{0\leqslant t\leqslant \mathbf{T}} \in B_\delta(\varphi)) \geq -\inf_G \mathbf{I}_{0\mathbf{T}}^x(\varphi).$$

This is obvious due to the W**LDP** for $\tilde{X}_t^\varepsilon$ and since $\mathbf{I}_{0\mathbf{T}}^R(\varphi) = \mathbf{I}_{0\mathbf{T}}(\varphi)$ for $R$ large enough.

**Remark 1.2**.(**i**) A sufficient condition for the existence of a non-exploding and unique strong solution is a locally Lipschitz drift term $b$ which satisfies

$$\langle x, b(x)\rangle \leq \gamma\left(1 + \|x\|^2\right) \text{ for all } x \in \mathbb{R}^d \tag{1.6}$$

for some constant $\gamma > 0$ (see [14], Theorem 10.2.2). This still rather weak condition is obviously satisfied if $\langle x, b(x)\rangle \leq 0$ for large enough $x$, which means that $b$ contains a component that pulls $X$ back to the origin. In the gradient case $b(x) = -\nabla U(x)$, (1.6) means that the potential may not grow stronger than linearly in the same direction as $x$.

(**ii**) A strengthening of condition (1.6) ensuring superlinear growth will be used in subsequent sections. In that case, the law of $X_t^\varepsilon$ is exponentially tight, and so $X_t^\varepsilon$ satisfies not only a weak but the *strong* **LDP** (**SLDP**) (i.e. the upper bound (1.4) holds for all closed sets), and $\mathbf{I}_{0\mathbf{T}}$ is a good rate function. Recall that the laws of $X_t^\varepsilon$ are exponentially tight if there exist some $R_0 > 0$ and a positive function $\varphi$ satisfying $\lim\limits_{x\to\infty} \varphi(x) = +\infty$ such that

$$\limsup_{\varepsilon\to 0} \varepsilon \log \mathbf{P}_x(\tau_R^\varepsilon \leqslant \mathbf{T}) \leq -\inf_F \mathbf{I}_{0\mathbf{T}}^R(\varphi) \text{ for all } R \geqslant R_0. \tag{1.7}$$

Here $\tau_R^\varepsilon$ denotes the first time that $X_t^\varepsilon$ exits from $B_R(0)$.

We will also make use of the following strengthening of (1.4) and (1.5) which expresses the fact that the convergence statements in the asymptotic results of Proposition 1.1 are uniform on compact sets of the state space. Let us denote by $\mathbf{P}_y((X_t^\varepsilon)_{t\geqslant 0} \in \cdot)$ the law of the diffusion $X_t^\varepsilon$ starting in $y \in \mathbb{R}^d$. For the proof see [2], Corollary 5.6.15.

**Corollary 1.3** (Uniformity of **WLDP** w.r.t. initial conditions). Assume the conditions of Proposition 1.1 and that $(X_t^\varepsilon)_{t\geqslant 0}$ is exponentially tight. Let $K \subset \mathbb{R}^d$ be compact.

(i) For any closed set $\mathbf{F} \subset C_{0\mathbf{T}}$

$$\limsup_{\varepsilon\to 0} \varepsilon \log \sup_{y\in K} \mathbf{P}_y((X_t^\varepsilon)_{0\leqslant t\leqslant \mathbf{T}} \in \mathbf{F}) \leq -\inf_{y\in K} \left(\inf_{\varphi\in \mathbf{F}} \mathbf{I}_{0\mathbf{T}}^y(\varphi)\right). \tag{1.8}$$

(ii) For any open set $\mathbf{G} \subset C_{0\mathbf{T}}$

$$\limsup_{\varepsilon\to 0} \varepsilon \log \inf_{y\in K} \mathbf{P}_y((X_t^\varepsilon)_{0\leqslant t\leqslant \mathbf{T}} \in G) \geqslant -\sup_{y\in K} \left(\inf_{\varphi\in G} \mathbf{I}_{0\mathbf{T}}^y(\varphi)\right). \tag{1.9}$$

# I.2. General classical results on weakly time inhomogeneous diffusions.

Let us now come to inhomogeneous diffusions with slowly time dependent drift coefficients. For our understanding of stochastic resonance effects of dynamical systems with slow time dependence, we have to adopt the large deviations results of the previous subsection to diffusions moving in potential landscapes with different valleys slowly and periodically changing their depths and positions. In this subsection we shall extend the large deviations results of Freidlin and Wentzell to time inhomogeneous diffusions which are almost homogeneous in the small noise limit, so that in fact we are able to compare to the large deviation principle for time homogeneous diffusions. The result we present in this subsection is not strong enough for the treatment of stochastic resonance (one needs uniformity in some of the system parameters), but it most clearly exhibits the idea of the approach, which is why we state it here. Consider the family $X_t^{\varepsilon}, \varepsilon > 0,$ of solutions of the SDE

$$dX_t^{\varepsilon} = b^{\varepsilon}(t, X_t^{\varepsilon})dt + \sqrt{\varepsilon}\, d\mathbf{W}_t, \quad X_0^{\varepsilon} = x_0 \in \mathbb{R}^d. \tag{1.10}$$

We assume that (1.10) has a global strong solution for all $\varepsilon > 0.$ Our main large deviations result for diffusions for which time inhomogeneity fades out in the small noise limit is summarized in the following

**Proposition**.**1**.**4** (**Large deviations principle**).Assume that the drift of the SDE (1.10) satisfies

$$\lim_{\varepsilon \to 0} b^{\varepsilon}(t, x) = b(x) \tag{1.11}$$

for all $t \geqslant 0,$ uniformly w.r.t. $x$ on compact subsets of $\mathbb{R}^d,$ for some locally Lipschitz function $b : \mathbb{R}^d \to \mathbb{R}^d.$ If the laws of $(X_t^{\varepsilon})$ are exponentially tight then $(X_t^{\varepsilon})$ satisfies a large deviations principle on any finite time interval $[0, T]$ with good rate function $\mathbf{I}_{0\mathbf{T}}$ given by (1.2). More precisely, for any closed $\mathbf{F} \subset C_{0\mathbf{T}}$ we have

$$\lim\sup_{\varepsilon \to 0} \varepsilon \log \mathbf{P}_x((X_t^{\varepsilon})_{0 \leqslant t \leqslant \mathbf{T}} \in \mathbf{F}) \leq -\inf_F \mathbf{I}_{0\mathbf{T}}^x(\varphi), \quad \forall \mathbf{F} \subset C_{0\mathbf{T}}, \tag{1.12}$$

and for any open $\mathbf{G} \subset C_{0\mathbf{T}}$ we have

$$\lim\inf_{\varepsilon \to 0} \varepsilon \log P_x((X_t^{\varepsilon})_{0 \leqslant t \leqslant \mathbf{T}} \in \mathbf{G}) \geq -\inf_G \mathbf{I}_{0\mathbf{T}}^x(\varphi), \quad \forall \mathbf{G} \subset C_{0\mathbf{T}}. \tag{1.13}$$

It is easy to see that **Corollary1**.**3** also holds for the weakly inhomogeneous process $X_t^{\varepsilon}$ of this proposition. One only has to carry over Proposition 5.6.14 in [2], which is easily done using some Gronwall argument. Then the proof of the Corollary is the same as in the homogeneous case (see [2], Corollary 5.6.15). We omit the details.

# II.Strong large deviations principle. The time inhomogeneous case. Main nonclassical results.

## II.1. Colombeau generalized stochastic processes.

Generalized random processes first introduced by Gel'fand and Vilenkin as elements of $L(V, \mathcal{L}_2(\Theta))$ i.e. as linear continuous mappings of a test space $V$ into the space of random variables with finite second moments. (In this paper we use notation $\Theta$ or $\Omega$ for the probability space and $\Delta$ for an open set of $\mathbb{R}^n$.)

Stochastic process with paths in algebras of Colombeau generalized functions first considered by Oberguggenberger, Russo and their coauthors in [20]-[25].The algebra of generalized functions considered in these papers is the Colombeau algebra of generalized functions denoted by $G(\Delta)$,where $\Delta \subseteq \mathbb{R}^n$. This algebra contains the space of Schwartz distributions $D'(\Delta)$ as a subspace. Stochastic processes are defined as mappings $\Theta \to G(\Delta)$ and called Colombeau generalized stochastic processes.Oberguggenberger and Russo have shown that distribution-valued stochastic processes i.e. weakly measurable mappings $X : \Theta \to D'(\Delta)$, where $X : \omega \to \langle X(\omega), \varphi \rangle$ is measurable for every $\varphi \in D(\Delta)$,can be embedded into Colombeau generalized stochastic processes. We introduce Colombeau generalized stochastic processes as done in [20]-[25].We will deal with the the algebra of generalized functions first constructed by Colombeau. Omitting the general construction [26], we recall only the definition of the algebra $G(\mathbb{R}^n)$ on $\mathbb{R}^n$.

**Definition 2.1.1.** Set $E(\mathbb{R}^n) \triangleq (C^\infty(\mathbb{R}^n))^I, I = (0,1]$.

$$E_M(\mathbb{R}^n) =$$

$$\left\{ (u_\varepsilon)_\varepsilon \in E(\mathbb{R}^n) | (\forall K \subseteq \mathbb{R}^n)(\forall \alpha \in \mathbb{N}_0^\infty)(\exists p \in \mathbb{N}) \left( \sup_{x \in K} |\partial^\alpha u_\varepsilon(x)| = O(\varepsilon^{-p}) \right) \right\},$$

$$N(\mathbb{R}^n) = \tag{2.1.1}$$

$$\left\{ (u_\varepsilon)_\varepsilon \in E(\mathbb{R}^n) | (\forall K \subseteq \mathbb{R}^n)(\forall \alpha \in \mathbb{N}_0^\infty)(\forall q \in \mathbb{N}) \left( \sup_{x \in K} |\partial^\alpha u_\varepsilon(x)| = O(\varepsilon^{q}) \right) \right\},$$

$$G(\mathbb{R}^n) = E_M(\mathbb{R}^n)/N(\mathbb{R}^n).$$

Elements of $E_M(\mathbb{R}^n)$ and $N(\mathbb{R}^n)$ are called moderate,resp.negligible functions.Landau symbol $b_\varepsilon = O(a_\varepsilon)$ having the following meaning: $(\exists C > 0)(\exists \varepsilon_0 \in I)(\forall \varepsilon \in (0, \varepsilon_0))(a_\varepsilon \leq C b_\varepsilon)$. Note that $E_M(\mathbb{R}^n)$ is a differential algebra with pointwise operations. It is the largest differential subalgebra of $E_M(\mathbb{R}^n)$ in which $N(\mathbb{R}^n)$ is a differential ideal. Thus, $G(\mathbb{R}^n)$ is an associative, commutative differential algebra. If $(u_\varepsilon)_\varepsilon \in E_M(\mathbb{R}^n)$ is a representative of $u \in G(\mathbb{R}^n)$, we write $u = [(u_\varepsilon)_\varepsilon]$ or $u = \mathbf{cl}[(u_\varepsilon)_\varepsilon]$.

Let $(\Theta,\mathcal{F},\mathbf{P})$ denotes a probability space.

**Definition 2.1.2.**Let $E(\mathbb{R}^n,\Theta)$ be the set of nets $(u_\varepsilon(\omega,x))_\varepsilon,\omega \in \Theta,x \in \mathbb{R}^n,\varepsilon \in I,$ such that, for almost every (a.e) $\omega \in \Theta,$ it holds $(u_\varepsilon(\omega,\cdot))_\varepsilon \in E(\Omega),$ and for every $x \in \mathbb{R}^n$ : $(u_\varepsilon(\cdot,x))_\varepsilon$ is the net of measurable functions on $\Theta.$ Set

$$E_M(\mathbb{R}^n,\Theta) =$$

$$\left\{(u_\varepsilon)_\varepsilon \in E(\mathbb{R}^n,\Theta) | \left(\text{for a.e. } \omega \in \Theta\right)(\forall K \subseteq \mathbb{R}^n)(\forall \alpha \in \mathbb{N}_0^\infty)\right.$$

$$\left.(\exists p \in \mathbb{N})\left(\sup_{x\in K} |\partial^\alpha u_\varepsilon(\omega,x)| = O(\varepsilon^{-p})\right)\right\},$$

$$N(\mathbb{R}^n,\Theta) = \tag{2.1.2}$$

$$\left\{(u_\varepsilon)_\varepsilon \in E(\mathbb{R}^n,\Theta) | \left(\text{for a.e. } \omega \in \Theta\right)(\forall K \subseteq \mathbb{R}^n)(\forall \alpha \in \mathbb{N}_0^\infty)\right.$$

$$\left.(\forall q \in \mathbb{N})\left(\sup_{x\in K} |\partial^\alpha u_\varepsilon(\omega,x)| = O(\varepsilon^{q})\right)\right\},$$

$$G(\mathbb{R}^n,\Theta) = E_M(\mathbb{R}^n,\Theta)/N(\mathbb{R}^n,\Theta).$$

Elements of $E_M(\mathbb{R}^n,\Theta)$ and $N(\mathbb{R}^n,\Theta)$ are called moderate and negligible functions respectively.

**Definition 2.1.3.**Denote:$E_M(\mathbb{R}^n,\mathcal{L}_p(\Theta)) \triangleq (C^\infty(\mathbb{R}^n,,\mathcal{L}_p(\Theta)))^I.$ Then:

$$E_M(\mathbb{R}^n, \mathcal{L}_p(\Theta)) =$$

$$\{(u_\varepsilon)_\varepsilon \in E(\mathbb{R}^n, \mathcal{L}_p(\Theta)) | (\forall K \subseteq \mathbb{R}^n)(\forall \alpha \in \mathbb{N}_0^\infty)$$

$$(\exists p \in \mathbb{N})\Big(\|\partial^\alpha u_\varepsilon(\omega, x)\|_{\mathcal{L}_p} = O(\varepsilon^{-p})\Big)\Big\},$$

$$N(\mathbb{R}^n, \mathcal{L}_p(\Theta)) = \tag{2.1.3}$$

$$\{(u_\varepsilon)_\varepsilon \in E(\mathbb{R}^n, \mathcal{L}_p(\Theta)) | (\forall K \subseteq \mathbb{R}^n)(\forall \alpha \in \mathbb{N}_0^\infty)$$

$$(\forall q \in \mathbb{N})\left(\sup_{x \in K} |\partial^\alpha u_\varepsilon(\omega, x)| = O(\varepsilon^q)\right)\Big\},$$

$$G(\mathbb{R}^n, \mathcal{L}_p(\Theta)) = E_M(\mathbb{R}^n, \mathcal{L}_p(\Theta))/N(\mathbb{R}^n, \mathcal{L}_p(\Theta)).$$

Schwartz

**Definition 2.2.4**.Let $(\Theta, \mathcal{F}, \mathbf{P})$ denotes a probability space.Weakly measurable mapping $X : \Theta \to D'(\mathbb{R}^d)$ is called canonical generalized stochastic process on $\mathbb{R}^d$.

**Remark 2.1.1**.Note that for each fixed test function $\varphi \in D(\mathbb{R}^d)$,the mapping $\Theta \to \mathbb{R}$ defined by $\omega \to \langle X(\omega), \varphi \rangle$ is random variable.The space of generalized stochastic processes will be denoted by $D'(\Theta, \mathbb{R}^d)$.
**Definition 2.1.5**.The characteristic functional $C_X(\varphi), \varphi \in D(\mathbb{R}^d)$ of process $X(\omega)$ is

$$C_X(\varphi) = \int \exp[i\langle X(\omega), \varphi \rangle] d\mathbf{P}(\omega). \tag{2.1.4}$$

Construction of the white noise $\dot{W}(x)$ on $\mathbb{R}^d$ given as follows.The probability space will be the space of tempered distributions $\Theta = S'(\mathbb{R}^d)$ and $\mathcal{F}$ will be the Borel $\sigma$-algebra generated by the weak topology.There is a unique probability measure $\mu$ on $(\Theta, \mathcal{F})$ such that for any $\varphi \in S'(\mathbb{R}^d)$ :

$$\int \exp[i\langle X(\omega), \varphi \rangle] d\mu(\omega) = \exp\left(-\frac{1}{2}\|\varphi\|^2_{L_2(\mathbb{R}^d)}\right). \tag{2.1.5}$$

**Definition 2.1.6**.We define the white noise $\dot{W}(\omega) : \Theta \to D'(\mathbb{R}^d)$ as the identity mapping:

$$\langle \dot{W}(\omega), \varphi \rangle = \langle \omega, \varphi \rangle \tag{2.1.6}$$

for $\varphi \in D(\mathbb{R}^d)$.

Note that (2.1.5) determines its characteristic functional.Thus $\dot{W}(\varphi)$ is a generalized Gaussian process with mean zero and variance

$$\mathbf{D}[\dot{W}(\varphi)] = \mathbf{E}[\dot{W}^2(\varphi)] = \|\varphi\|^2_{L_2(\mathbb{R}^d)}, \tag{2.1.7}$$

where $\mathbf{E}$ denotes mathematical expectation. Its covariance is

$$\mathbf{E}[\dot{W}(\varphi)\dot{W}(\phi)] = \int_{\mathbb{R}^d} \varphi(x)\phi(x)dx. \tag{2.1.8}$$

**Definition 2.1.7**.For $x \in \mathbb{R}^d$ we define its signed indicator function as

$$v(x,y) = \prod_{j=1}^{d} sign(x_j)\kappa(x,y), \tag{2.1.9}$$

where $\kappa(x,y)$ is the indicator function of the $d$-dimensional interval from the origin to the point $x$ as extremal corner. We define the Wiener process on $\mathbb{R}^d$ as follows

$$B(x) = \lim_{\varepsilon \to 0} \langle \dot{W}, v(x,y) * \varphi_\varepsilon(y) \rangle, \tag{2.1.10}$$

where $\varphi_\varepsilon$ are the molifiers of the form

$$\varphi_\varepsilon(y) = \frac{1}{\varepsilon^d}\varphi\left(\frac{y}{\varepsilon}\right),$$

$$\varphi \in D(\mathbb{R}^d), \tag{2.1.11}$$

$$\int\limits_{\mathbb{R}^d} \varphi(y)dy = 1.$$

Note that the limit on the right-hand side (1.3.1.10) exists in $L_2(\Omega)$. Mapping $(x,\omega) \to B(x,\omega)$ has a version with almost surely continuous paths and it is a Wiener process on $\mathbb{R}^d$. It follows from the construction that

$$\dot{W}(x,\omega) = \partial_{x_1}\ldots\partial_{x_d}B(x,\omega) \tag{2.1.12}$$

almost surely in $D'(\mathbb{R}^d)$.

Smoothed white noise process on $\mathbb{R}_+$ is defined as

$$\dot{W}_\varepsilon(t,\omega) = \langle \dot{W}(t,\omega), \varphi_\varepsilon(s-t)\rangle, \tag{2.1.13}$$

where $\dot{W}(t,\omega)$ is the white noise process on $\mathbb{R}_+$ and $\varphi_\varepsilon$ is a model delta net.

**Definition 2.1.8**.Let $(\Omega,\Sigma,\mathbf{P})$ be a probability space.Denote by $E_\Omega(n,[0,\infty))$ the space of all nets $(X_\varepsilon(t))_{\varepsilon\in(0,1]}$ of stochastic processes $X_\varepsilon(t)$ with almost surely continuous paths, i.e.,the space of nets of processes $X_\varepsilon : (0,1] \times [0,\infty) \times \Omega \to \mathbb{R}^n$ such that:
(1) mapping $(t,\omega) \to X_\varepsilon(t,\omega)$ is jointly measurable, for all $\varepsilon \in (0,1]$
(2) $t \to X_\varepsilon(t,\omega)$ belongs to $C^\infty([0,\infty))$,for all $\varepsilon \in (0,1]$ and almost all $\omega \in \Omega$.
**Definition 2.1.9**.Let $(\Omega,\Sigma,\mathbf{P})$ be a probability space. Denote by $E_M^\Omega(n,[0,\infty))$ the space of nets of processes $(X_\varepsilon(t))_\varepsilon \in E_\Omega(n,[0,\infty))$,with the property that for almost all $\omega \in \Omega$,for all $T > 0$ and $\alpha \in \mathbb{N}_0$, there exist constants $N,C > 0$ and $\varepsilon_0 \in (0,1]$ such that

$$\forall\varepsilon(\varepsilon \leq \varepsilon_0)\left[\sup_{t\in[0,T]} \|\partial^\alpha X_\varepsilon(t,\omega)\| \leq C\varepsilon^{-N}\right]. \tag{2.1.14}$$

Denote by $N^\Omega(n,[0,\infty))$ the space of nets of processes $(X_\varepsilon(t))_\varepsilon \in E_\Omega(n,[0,\infty))$,with the property that for almost all $\omega \in \Omega$, for all $T > 0$ and $\alpha \in \mathbb{N}_0$, and all $v \in \mathbb{R}_+$ there exist constants $C > 0$ and $\varepsilon_0 \in (0,1]$ such that

$$\forall\varepsilon(\varepsilon \le \varepsilon_0)\left[\sup_{t\in[0,T]} \|\partial^\alpha X_\varepsilon(t,\omega)\| \le C\varepsilon^v\right]. \qquad (2.1.15)$$

We say that $(X_\varepsilon(t))_\varepsilon \in N^\Omega(n,[0,\infty))$ is negligible.We define now

$$G_n^\Omega = G_n^\Omega(n,[0,\infty)) = E_M^\Omega(n,[0,\infty))/N^\Omega(n,[0,\infty)). \qquad (2.1.16)$$

Then $G_n^\Omega$ is a differential algebra (differentiation with respect to $t$ and pointwise multiplication) called algebra of Colombeau generalized stochastic processes. The elements of $G_n^\Omega$ will be denoted by $[X_\varepsilon]$ or $\mathbf{cl}(X_\varepsilon)$, where $(X_\varepsilon(t))_\varepsilon$ is a representative of the class.

**Remark 2.1.1**. Both Brownian motion and white noise process can be viewed as Colombeau generalized stochastic processes. It follows from the usual imbedding arguments of Colombeau theory. For instance, the Colombeau generalized white noise process has the representative given by Eq.(2.1.13).

**Definition 2.1.10**.Let $(\Omega,\Sigma,\mathbf{P})$ be a probability space. Denote by $E^\Omega_{M,C^k}(n,[0,\infty))$ the space of nets of continuous processes $(X_\varepsilon(t))_\varepsilon$, with the property that for almost all $\omega \in \Omega$, for all $T > 0$ and $\alpha \in \mathbb{N}_0$, there exist constants $N, C > 0$ and $\varepsilon_0 \in (0,1]$ such that

$$\forall\varepsilon(\varepsilon \le \varepsilon_0)\forall m(0 \le m \le k)\left[\sup_{t\in[0,T]} \|\partial^m X_\varepsilon(t,\omega)\| \le C\varepsilon^{-N}\right]. \qquad (2.1.17)$$

Denote by $N^\Omega_{C^k}(n,[0,\infty))$ the space of nets of processes $(X_\varepsilon(t))_\varepsilon \in E^\Omega_{M,C^k}(n,[0,\infty))$, with the property that for almost all $\omega \in \Omega$, for all $T > 0$ and all $v \in \mathbb{R}_+$ there exist constants $C > 0$ and $\varepsilon_0 \in (0,1]$ such that

$$\forall\varepsilon(\varepsilon \le \varepsilon_0)\forall m(0 \le m \le k)\left[\sup_{t\in[0,T]} \|\partial^m X_\varepsilon(t,\omega)\| \le C\varepsilon^{v}\right]. \qquad (2.1.18)$$

We say that $(X_\varepsilon(t))_\varepsilon \in N^\Omega_{C^k}(n,[0,\infty))$ is negligible.We define now

$$G^{\Omega}_{n,C^k} = G^{\Omega}_{n,C^k}(n,[0,\infty)) = E^{\Omega}_{M,C^k}(n,[0,\infty))/N^{\Omega}_{C^k}(n,[0,\infty)). \qquad (2.1.19)$$

Then $G^{\Omega}_{n,C^k}$ is a differential algebra (differentiation with respect to $t$ and pointwise multiplication) called algebra of $C^k$-Colombeau generalized stochastic processes. The elements of $G^{\Omega}_{n,C^k}$ will be denoted by $[X_\varepsilon]$ or $\mathbf{cl}(X_\varepsilon)$, where $(X_\varepsilon(t))_\varepsilon$ is a representative of the class.

**Definition 2.1.11**.Let $(\Omega,\Sigma,\mathbf{P})$ be a probability space.Denote by $E^{\Omega}_{C^{k,l}}([0,\infty)\times\mathbb{R}^n)$ the space of all nets $(X_\varepsilon(t,\omega;x))_{\varepsilon\in(0,1]}$ of stochastic processes $X_\varepsilon(t,\omega;x)\in\mathbb{R}^n, x\in\mathbb{R}^n$ with almost surely continuous paths,i.e.,the space of nets of processes
$X_\varepsilon : (0,1]\times[0,\infty)\times\Omega\times\mathbb{R}^n\to\mathbb{R}^n$ such that:

(1) $X_\varepsilon(0,\omega;x) = x\in\mathbb{R}^n$ for almost all $\omega\in\Omega$,

(2) mapping $(t,\omega;x)\to X_\varepsilon(t,\omega;x)$ is jointly measurable, for all $\varepsilon\in(0,1]$ and all $x\in\mathbb{R}^n$,

(3) $t\to X_\varepsilon(t,\omega;\bullet)$ belongs to $C^k([0,\infty))$, for all $\varepsilon\in(0,1]$ and almost all $\omega\in\Omega$,

(4) $x\to X_\varepsilon(\bullet,\omega;x)$ belongs to $C^l(\mathbb{R}^n)$, for all $\varepsilon\in(0,1]$ and almost all $\omega\in\Omega$.

**Definition 2.1.12**.Let $(\Omega,\Sigma,\mathbf{P})$ be a probability space. Denote by $E^{\Omega}_{C^{k,l}}([0,\infty)\times\mathbb{R}^n)$ the space of nets of processes $(X_\varepsilon(t))_\varepsilon\in E^{\Omega}_{C^{k,l}}([0,\infty)\times\mathbb{R}^n)$, with the properties that for almost all $\omega\in\Omega$, for all $T>0$, there exist constants $N_1,C_1,N_2,C_2>0$ and $\varepsilon_0\in(0,1]$ such that

$$\forall\varepsilon(\varepsilon\leq\varepsilon_0)\forall m(0\leq m\leq k)\left[\sup_{(t,x)\in[0,T]\times\mathbb{R}^n}\|\partial_t^m X_\varepsilon(t,\omega;x)\|\leq C_1\varepsilon^{-N_1}\right]. \qquad (2.1.20)$$

and

$$\forall\varepsilon(\varepsilon\leq\varepsilon_0)\forall p(0\leq p\leq l)\left[\sup_{(t,x)\in[0,T]\times\mathbb{R}^n}\|\partial_x^p X_\varepsilon(t,\omega;x)\|\leq C_2\varepsilon^{-N_2}\right]. \qquad (2.1.21)$$

Denote by $N^{\Omega}_{C^{k,l}}([0,\infty)\times\mathbb{R}^n)$ the space of nets of processes $(X_\varepsilon(t,\omega;x))_\varepsilon\in E^{\Omega}_{M,C^{k,l}}([0,\infty)\times\mathbb{R}^n)$, with the properties that for almost all $\omega\in\Omega$, for all $T>0$ and all $\nu\in\mathbb{R}_+$ there exist constants $C_1,C_2>0$ and $\varepsilon_0\in(0,1]$ such that

$$\forall\varepsilon(\varepsilon\leq\varepsilon_0)\forall m(0\leq m\leq k)\left[\sup_{(t,x)\in[0,T]\times\mathbb{R}^n}\|\partial_t^m X_\varepsilon(t,\omega)\|\leq C_1\varepsilon^{\nu}\right]. \qquad (2.1.22)$$

and

$$\forall\varepsilon(\varepsilon \leq \varepsilon_0)\forall p(0 \leq p \leq l)\left[\sup_{(t,x)\in[0,T]\times\mathbb{R}^n} \|\partial_x^p X_\varepsilon(t,\omega)\| \leq C_1\varepsilon^\nu\right]. \quad (2.1.23)$$

We say that $(X_\varepsilon(t))_\varepsilon \in N^\Omega_{C^{k,l}}([0,\infty)\times\mathbb{R}^n)$ is negligible.We define now

$$G^\Omega_{C^{k,l}} = G^\Omega_{C^{k,l}}([0,\infty)\times\mathbb{R}^n) = E^\Omega_{M,C^{k,l}}([0,\infty))/N^\Omega_{C^{k,l}}([0,\infty)\times\mathbb{R}^n). \quad (2.1.24)$$

Then $G^\Omega_{C^{k,l}}$ is a differential algebra (differentiation with respect to $(t,x)$ and pointwise multiplication) called algebra of $C^{k,l}$-Colombeau generalized stochastic processes. The elements of $G^\Omega_{C^{k,l}}$ will be denoted by $[X_\varepsilon]$ or $\mathbf{cl}(X_\varepsilon)$,where $(X_\varepsilon(t))_\varepsilon$ is a representative of the class.

## II.2.Strong large deviations principle for Colombeau generalized stochastic processes .

.Let us consider the family of the Ito SDE:

$$d\mathbf{X}_t^\delta = \mathbf{b}(\mathbf{X}_t^\delta,t)dt + \sqrt{\delta}\,d\mathbf{W}_t,\quad \mathbf{X}_0^\delta = x_0 \in \mathbb{R}^d. \quad (2.2.1)$$

Here $b(\bullet,t):\mathbb{R}^d \to \mathbb{R}^d$ is a polynomial transform i.e.,

$$b_i(x,t) = \sum_{\|\alpha\|\leqslant n} b^i_\alpha(t)x^\alpha, \alpha = (i_1,\ldots,i_k), \|\alpha\| = \sum\nolimits_r i_r, i = 1,\ldots,d. \quad (2.2.2)$$

Let us introduce regularized SDE (2.2.1) in the following way:

$$d\mathbf{X}_{t,\varepsilon}^\delta = \mathbf{b}_\varepsilon(\mathbf{X}_{t,\varepsilon}^\delta,t)dt + \sqrt{\delta}\,d\mathbf{W}_t,\quad \mathbf{X}_{0,\varepsilon}^\delta = x_0 \in \mathbb{R}^d, \quad (2.2.3)$$

where

$$b_{i,\varepsilon}(x,t) = \sum_{\|\alpha\|\leqslant n} b_{\alpha}^{i}(t)(x_{\varepsilon})^{\alpha}, \alpha = (i_1,\ldots,i_k), \|\alpha\| = \sum_r i_r, i = 1,\ldots,d,$$

$$x_{\varepsilon} = \frac{x}{1+\varepsilon^{m}(x^{r})}, m \geq 1, r \geq 2. \tag{2.2.4}$$

Let us consider now the family $(X_{t,\varepsilon}^{\delta})_{\varepsilon\in(0,1]}, \delta > 0,$ of Colombeau generalized stochastic proceses which is a solutions of the Ito-Colombeau SDE

$$(d\mathbf{X}_{t,\varepsilon}^{\delta})_{\varepsilon\in(0,1]} = (\mathbf{b}_{\varepsilon}(\mathbf{X}_{t,\varepsilon}^{\delta},t)dt)_{\varepsilon\in(0,1]} + \sqrt{\delta}\, d\mathbf{W}_t =$$

$$(\mathbf{b}(\mathbf{X}_{t,\varepsilon}^{\delta},t)dt)_{\varepsilon\in(0,1]} + \sqrt{\delta}\, d\mathbf{W}_t, \tag{2.2.5}$$

$$\mathbf{X}_0^{\delta} = x_0 \in \mathbb{R}^d,$$

where

$$(\mathbf{X}_{t,\varepsilon}^{\delta})_{\varepsilon\in(0,1]} = \frac{(\mathbf{X}_{t,\varepsilon}^{\delta})_{\varepsilon\in(0,1]}}{1+\varepsilon^{m}((\mathbf{X}_{t,\varepsilon}^{\delta})^{r})_{\varepsilon\in(0,1]}},$$

$$m \geq 1, r \geq 2. \tag{2.2.6}$$

**Remark 2.2.1.** We note that Eq.(2.2.5) has a global strong solution for all $\varepsilon \in (0,1]$ and $\mathbf{cl}\left[(\mathbf{X}_{t,\varepsilon}^{\delta})_{\varepsilon}\right] \in G_{C^{0,\infty}}^{\Omega}$.

Our main large deviations result for diffusions for which time inhomogeneity no fades out in the small noise limit is summarized in the following:

**Theorem 2.2.1.** ( **Strong large deviations principle**). For all solutions $(\mathbf{X}_{t,\varepsilon}^{\delta}(\omega))_{\varepsilon}$ of the Ito-Colombeau SDE (2.2.5) and any $\mathbb{R}$ valued parameters $\lambda_1,\lambda_2,\ldots,\lambda_d \in \mathbb{R}$

$$\liminf_{\delta\to 0,\varepsilon\to 0,\varepsilon/\delta\to 0} (\mathbf{M}\|\mathbf{X}^{\delta}_{t,\varepsilon}(\omega)-\boldsymbol{\lambda}\|) \leqslant \|\mathbf{u}(t,\boldsymbol{\lambda})\|,$$

$$t\in\mathbb{R}_+, \mathbf{X}^{\delta}_{0,\varepsilon}(\omega) = x_0, \tag{2.2.7}$$

$$\boldsymbol{\lambda} = (\lambda_1,\lambda_2,\ldots,\lambda_d).$$

Here $\mathbf{u}(t,\boldsymbol{\lambda})$ is the solution of the linear *differential master equation*

$$\frac{d\mathbf{u}(t)}{dt} = \mathbf{J}[\mathbf{b}(\boldsymbol{\lambda},t)]\mathbf{u}(t)+\mathbf{b}(\boldsymbol{\lambda},t), \mathbf{u}(0) = x_0-\boldsymbol{\lambda}, \tag{2.2.8}$$

where $\mathbf{J}[\mathbf{b}(\lambda,t)]$ is the Jacobian

$$\mathbf{J}[\mathbf{b}(\boldsymbol{\lambda},t)] = \mathbf{J}[\mathbf{b}(x,t)]_{x=\boldsymbol{\lambda}} =$$

$$\begin{bmatrix} \frac{db_1(x,t)}{dx_1} & \cdots & \frac{db_1(x,t)}{dx_d} \\ \cdot & \cdot & \cdot \\ \cdot & \cdot & \cdot \\ \cdot & \cdot & \cdot \\ \frac{db_d(x,t)}{dx_1} & \cdots & \frac{db_d(x,t)}{dx_d} \end{bmatrix}_{x=\boldsymbol{\lambda}}. \tag{2.2.9}$$

**Theorem 2.2.2**. Assume the conditions of Theorem 1.1. Then for any $\boldsymbol{\lambda}\in\mathbb{R}^d, t\in\mathbb{R}_+$ sutch that $\|\mathbf{u}(t,\lambda)\| \equiv 0,$ we have

$$\liminf_{\varepsilon\to 0} (\mathbf{M}\|\mathbf{X}^{\varepsilon}_t(\omega)-\boldsymbol{\lambda}(t)\|) = 0. \tag{2.2.10}$$

More precisely, for any $t\in\mathbb{R}_+$ and $\boldsymbol{\lambda}(t) = (\lambda_1(t),\lambda_2(t),\ldots,\lambda_d(t))$ sutch that

$$
\begin{aligned}
u_1(t, \lambda_1(t), \lambda_2(t), \ldots, \lambda_d(t)) &= 0, \\
\ldots\ldots\ldots\ldots\ldots\ldots\ldots\ldots \\
u_d(t, \lambda_1(t), \lambda_2(t), \ldots, \lambda_d(t)) &= 0
\end{aligned} \tag{2.2.11}
$$

and for some infinite sequences $\delta_n, \varepsilon_m, n, m \in \mathbb{N}$ we have

$$
\lim_{n\to\infty,\ m\to\infty, \varepsilon_m/\delta_n \to 0} (\mathbf{M} \| \mathbf{X}^{\delta_n}_{t,\varepsilon_m}(\omega) - \boldsymbol{\lambda}(t) \|) \equiv 0. \tag{2.2.12}
$$

## II.3.Strong large deviations principle. Colombeau-Ito SDE with a random coefficients inhomogeneous case. Main nonclassical result.

We introduce Colombeau-Ito SDE with a random coefficients in the following way. Let us consider the family of the Ito SDE with a random coefficients:

$$
d\mathbf{X}^{\delta}_t(\omega, \varpi) = \mathbf{b}(\omega, \mathbf{X}^{\delta}_t(\omega, \varpi), t)dt + \mathbf{C}(\omega, \mathbf{X}^{\delta}_t(\omega, \varpi), t)\dot{\mathbf{W}}_t(\omega)dt + \sqrt{\delta}\, d\mathbf{W}_t(\varpi),
$$

$$
\mathbf{X}^{\delta}_0 = x_0 \in \mathbb{R}^d, \tag{2.3.1}
$$

$$
(\omega, \varpi) \in \Omega \times \widetilde{\Omega}, \Omega \cap \widetilde{\Omega} = \varnothing.
$$

Here:

(**1**) $\mathbf{W}_t(\omega)$ and $\mathbf{W}_t(\varpi)$ is a $d$-dimensional Brownian motions on a $\Omega \times [0, T]$, and $\widetilde{\Omega} \times [0, T]$ accordingly,

(**2**) $\mathbf{b}(\omega, \cdot, t) : \mathbb{R}^d \to \mathbb{R}^d$ is a random polinomial transform,i.e.

$$b_i(\omega,x,t) = \sum_{\|\alpha\|\leqslant n} b_\alpha^i(\omega,t)x^\alpha, \alpha = (i_1,\ldots,i_k),$$

$$\|\alpha\| = \sum_r i_r, i = 1,\ldots,d \tag{2.3.2}$$

(**3**) $\mathbf{C}(\omega,\bullet,t) : \mathbb{R}^d \to \mathbb{R}^d$ is a random polinomial transform,i.e.

$$C_{i,j}(\omega,x,t) = \sum_{\|\alpha\|\leqslant n} C_\alpha^{i,j}(\omega,t)x^\alpha, \alpha = (i_1,\ldots,i_k),$$

$$\|\alpha\| = \sum_r i_r, i = 1,\ldots,d. \tag{2.3.3}$$

Let us introduce regularized SDE (2.3.1) in the following way:

$$d\mathbf{X}_{t,\varepsilon}^\delta(\omega,\varpi) = \mathbf{b}_\varepsilon(\omega,\mathbf{X}_{t,\varepsilon}^\delta(\omega,\varpi),t)dt + \mathbf{C}_\varepsilon(\omega,\mathbf{X}_{t,\varepsilon}^\delta(\omega,\varpi),t)\dot{\mathbf{W}}_{t,\varepsilon}(\omega)dt + \sqrt{\delta}\, d\mathbf{W}_t(\varpi),$$

$$\mathbf{X}_{0,\varepsilon}^\delta = x_0 \in \mathbb{R}^d, \tag{2.3.4}$$

$$(\omega,\varpi) \in \Omega \times \widetilde{\Omega}, \Omega \cap \widetilde{\Omega} = \varnothing.$$

Here:

(**1**) $\dot{\mathbf{W}}_{t,\varepsilon}(\omega)$ is a smoothed $d$-dimensional white noise process on a $\Omega \times [0,T]$, given via Eq.(2.3.13) and $\mathbf{W}_t(\varpi)$ is a $d$-dimensional Brownian motion on a $\widetilde{\Omega} \times [0,T]$,
(**2**) $\mathbf{b}_\varepsilon(\omega,\cdot,t) : \mathbb{R}^d \to \mathbb{R}^d$ is a random transform such that

$$b_{i,\varepsilon}(\omega,x,t) = \sum_{\|\alpha\|\leqslant n} b_\alpha^i(\omega,t)(x_\varepsilon)^\alpha, \alpha = (i_1,\ldots,i_k),$$

$$\|\alpha\| = \sum_r i_r, i = 1,\ldots,d, \tag{2.3.5}$$

$$x_\varepsilon = \frac{x}{1+\varepsilon^m(x^r)}, m \geq 1, r \geq 2,$$

**(3)** $\mathbf{C}(\omega,\bullet,t):\mathbb{R}^d\to\mathbb{R}^d$ is a random transform such that

$$C_{i,j}(\omega,x,t)=\sum_{\|\alpha\|\leqslant n}C_{\alpha}^{i,j}(\omega,t)x^{\alpha},\alpha=(i_1,\ldots,i_k),$$

$$\|\alpha\|=\sum_r i_r, i=1,\ldots,d, \tag{2.3.6}$$

$$x_{\varepsilon}=\frac{x}{1+\varepsilon^m(x^r)},m\geq 1,r\geq 2.$$

We assume that Eq.(2.3.4) has a global strong solution for all $\varepsilon\in(0,1]$ and $\delta>0$.
Let us consider a family $(X_t^{\delta}(\omega,\varpi))_{\varepsilon\in(0,1]}$ ,$\varepsilon\in(0,1],\delta>0,$ of Colombeau generalized stochastic processes (where pair $(\omega,\varpi)\in\Omega\times\tilde{\Omega}$ and $\Omega\cap\tilde{\Omega}=\varnothing$) which is a solution of the Ito-Colombeau SDE with a random coefficients:

$$(d\mathbf{X}_{t,\varepsilon}^{\delta}(\omega,\varpi))_{\varepsilon}=(\mathbf{b}_{\varepsilon}(\omega,\mathbf{X}_{t,\varepsilon}^{\delta}(\omega,\varpi),t)dt)_{\varepsilon}+$$

$$(\mathbf{C}_{\varepsilon}(\omega,\mathbf{X}_{t,\varepsilon}^{\delta}(\omega,\varpi),t)\mathbf{W}_{t,\varepsilon}(\omega)dt)_{\varepsilon}+\sqrt{\delta}\,d\mathbf{W}_t(\varpi), \tag{2.3.7}$$

$$\mathbf{X}_{0,\varepsilon}^{\delta}=x_0\in\mathbb{R}^d.$$

Our main large deviations result for generalyzed diffusions is summarized in the following:

**Theorem 2.3.1**. (**Strong large deviations principle**). For all solutions $(\mathbf{X}_{t,\varepsilon}^{\delta}(\omega,\varpi))_{\varepsilon}$ of the equation $(2.3.7)$ and $\mathbb{R}$ valued parameters $\lambda_1,\lambda_2,\ldots,\lambda_d\in\mathbb{R}$

$$\mathbf{P}_{\varpi}\left\{\underline{\lim}_{\varepsilon\to 0,\delta\to 0,\delta/\varepsilon\to 0}(\mathbf{M}_{\varpi}\|\mathbf{X}_{t,\varepsilon}^{\delta}(\omega,\varpi)-\boldsymbol{\lambda}\|)\leqslant\|\mathbf{u}(\omega,t,\boldsymbol{\lambda})\|\right\}=1, \tag{2.3.8}$$

$$t\in\mathbb{R}_+,\boldsymbol{\lambda}=(\lambda_1,\lambda_2,\ldots,\lambda_d),$$

where $\mathbf{u}(\omega,t,\lambda)$ the solution of the linear *stochastic differential master equation*

$$d\mathbf{u}(\omega,t) = (\mathbf{J}[\mathbf{b}(\omega,\lambda,t)]\mathbf{u}(\omega,t) + \mathbf{b}(\omega,\lambda))dt +$$

$$\sum_{k=1}^{d}\left\langle \hat{\mathbf{J}}_k[\mathbf{C}(\omega,\boldsymbol{\lambda},t)]\mathbf{u}(\omega,t), d\mathbf{W}_t(\omega)\right\rangle + \mathbf{C}(\omega,\boldsymbol{\lambda},t)d\mathbf{W}_t(\omega), \qquad (2.3.9)$$

$$\mathbf{u}(\omega,0) = x_0 - \boldsymbol{\lambda}.$$

Here:

(**1**) $\mathbf{J}[\mathbf{b}(\omega,\lambda,t)]$ the Jacobi random matrix,i.e. random Jacobian:

$$\mathbf{J}[\mathbf{b}(\omega,\boldsymbol{\lambda},t)] = \mathbf{J}[\mathbf{b}(\omega,x,t)]_{x=\boldsymbol{\lambda}} =$$

$$\begin{bmatrix} \frac{db_1(\omega,x,t)}{dx_1} & \cdots & \frac{db_1(\omega,x,t)}{dx_d} \\ \cdot & \cdot & \cdot \\ \cdot & \cdot & \cdot \\ \cdot & \cdot & \cdot \\ \frac{db_d(\omega,x,t)}{dx_1} & \cdots & \frac{db_d(\omega,x,t)}{dx_d} \end{bmatrix}_{x=\boldsymbol{\lambda}}, \qquad (2.3.10)$$

(**2**) $\hat{\mathbf{J}}_k[\mathbf{C}(\omega,\boldsymbol{\lambda},t)]$ the generalyzed Jacobian random matrix

$$\hat{\mathbf{J}}_k[\mathbf{C}(\omega,\boldsymbol{\lambda},t)] = \hat{\mathbf{J}}_k[\mathbf{C}(\omega,x,t)]_{x=\boldsymbol{\lambda}} =$$

$$\begin{bmatrix} \frac{dC_{1,k}(\omega,x,t)}{dx_1} & \cdots & \frac{dC_{1,k}(\omega,x,t)}{dx_d} \\ \cdot & \cdot & \cdot \\ \cdot & \cdot & \cdot \\ \cdot & \cdot & \cdot \\ \frac{dC_{d,k}(\omega,x,t)}{dx_1} & \cdots & \frac{dC_{d,k}(\omega,x,t)}{dx_d} \end{bmatrix}_{x=\boldsymbol{\lambda}}, \qquad (2.3.11)$$

**(3)** $\langle x,y\rangle = \sum_{i=1}^{d} x_i y_i,$ where $x,y \in \mathbb{R}^d$.

**Theorem 2.3.2**. Assume the conditions of the theorem 2.3.1.Then for any $\lambda = \boldsymbol{\lambda}(\omega,t) \in \mathbb{R}^d, t \in \mathbb{R}_+,\ \omega \in \Omega$ sutch that $\|\mathbf{u}(\omega,t,\boldsymbol{\lambda})\| \equiv 0,$ we have

$$\underline{\lim}_{\delta\to 0,\varepsilon\to 0,\delta/\varepsilon\to 0} (\mathbf{M}_{\varpi}\|\mathbf{X}^{\delta}_{t,\varepsilon}(\omega) - \boldsymbol{\lambda}\|) \equiv 0. \qquad (2.3.12)$$

More precisely, for any $t \in \mathbb{R}_+, \lambda = \lambda(\omega,t) = (\lambda_1(\omega,t),\lambda_2(\omega,t),\ldots,\lambda_d(\omega,t)), \omega \in \Omega$ sutch that

$$\begin{gathered} u_1(\omega,t,\lambda_1(\omega,t),\lambda_2(\omega,t),\ldots,\lambda_d(\omega,t)) = 0, \\ \cdots\cdots\cdots\cdots\cdots\cdots \\ u_d(\omega,t,\lambda_1(\omega,t),\lambda_2(\omega,t),\ldots,\lambda_d(\omega,t)) = 0, \end{gathered} \qquad (2.3.13)$$

and for some infinite sequences $\delta_n, \varepsilon_m, n,m \in \mathbb{N}$ we have

$$\mathbf{P}_{\omega}\left\{ \underline{\lim}_{n,m\to\infty,\delta_n/\varepsilon_m\to 0} (\mathbf{M}_{\varpi}\|\mathbf{X}^{\delta_n}_{t,\varepsilon_m}(\omega) - \lambda(\omega,t)\|) \equiv 0 \right\} = 1. \qquad (2.3.14)$$

**Theorem 2.3.3**. Assume the conditions of the Theorem 2.3.1 with $\delta = 0$. Then for any $\lambda = \boldsymbol{\lambda}(\omega,t) \in \mathbb{R}^d, t \in \mathbb{R}_+,\ \omega \in \Omega$ sutch that $\|\mathbf{u}(\omega,t,\boldsymbol{\lambda})\| \equiv 0,$ we have

$$\underline{\lim}_{\varepsilon\to 0} \left(\mathbf{M}_{\varpi}\|\mathbf{X}_{t,\varepsilon}(\omega) - \boldsymbol{\lambda}\|\right) \equiv 0. \quad (2.3.15)$$

# II.4. Numerical examples.

The stochastic dynamics (2.3.1) we take now of the form

$$\dot{x}(t) = \mathbf{F}(x(t),t) + \sqrt{\varepsilon}\ \xi(t), \mathbf{F}(x,t) = \mathbf{b}(x) + \mathbf{f}\,(t). \quad (2.4.1)$$

The force field $\mathbf{F}(x,t)$ in (2.4.1) is assumed to derive from a metastable potential which undergoes an arbitrary periodic modulation in time with period $\tau$:

$$\mathbf{F}(x,t+\mathcal{T}) = \mathbf{F}(x,t)\ . \quad (2.4.2)$$

An examples is a static potential $V(x)$, supplemented by an additive sinusoidal and more general driving. The time-dependent force field $\mathbf{F}(x,t)$ takes the following form:

$$\mathbf{F}(x,t) = -V'(x) + A\sin(\Omega t) + B\cos(\Theta t), \Omega = 2\pi/\mathcal{T}. \quad (2.4.3)$$

### 2.4.1. Cubic potential.

As a first example we consider a force field (2.4.3) with a cubic metastable potential $V(x)$ as cartooned in Fig.1,

$$V(x) = -\frac{a}{3}x^3 + \frac{b}{2}x^2\ ,\ \ a,b > 0\ . \quad (2.4.4)$$

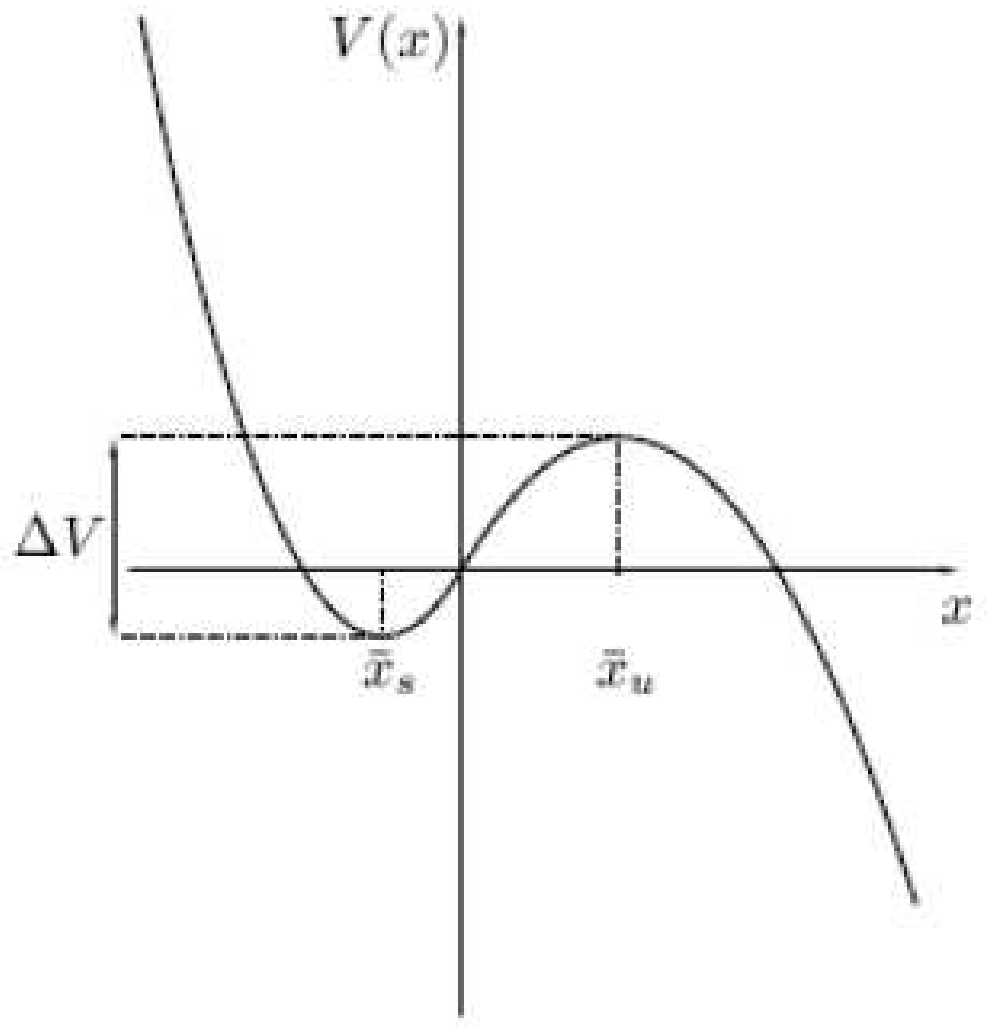

**Fig. 1.** Cubic metastable potential.

The time-dependent force field (2.5.3) takes the following form:

$$F(x,t) = ax^2 - bx + A\sin(\Omega t) . \qquad (2.5.5)$$

The stochastic dynamics (2.5.1) takes the following form:

$$\dot{x}(t) = ax^2 - bx + A\sin(\Omega t) + \sqrt{\varepsilon}\ \xi(t), x(0) = x_0 . \qquad (2.4.6)$$

From master equation $(2.2.8)$ we have the next differential linear master equation

$$\dot{u} = (2a\lambda - b)u + a\lambda^2 - b\lambda + A\sin(\Omega t), u(0) = x_0 - \lambda . \qquad (2.4.7)$$

From Theorem 2.2.2 we have the next transcendental master equation

$$(x_0 - \lambda(t))\exp[(2a\lambda(t) - b)t] +$$

$$(a\lambda^2(t) - b\lambda(t))\int_0^t \exp[(2a\lambda(t) - b)(t-\tau)]d\tau + \qquad (2.4.8)$$

$$A\int_0^t \sin(\Omega\tau)\exp[(2a\lambda(t) - b)(t-\tau)]d\tau = 0.$$

**Comparison of the classical and non-perturbative quasiclassical dynamics**.

We have compared by $\delta(t) = |x(t) - \lambda(t)|$ the above analytical predictions for the limit (2.2.7) with very accurate numerical results for classical $x(t)$ and limiting stochastic dynamics.

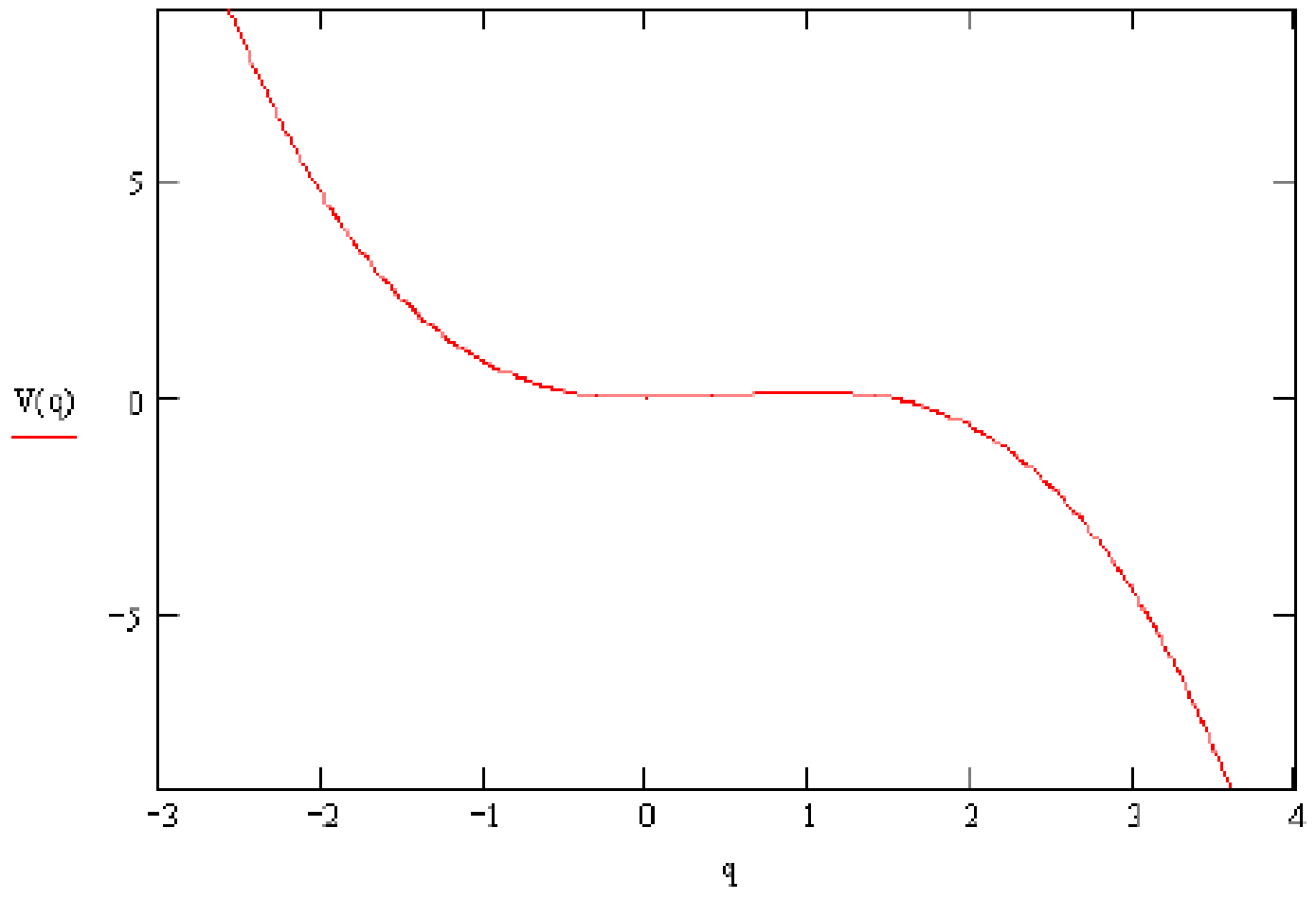

**Fig. 2.** Cubic metastable potential.

$a = 1, b = 1.$

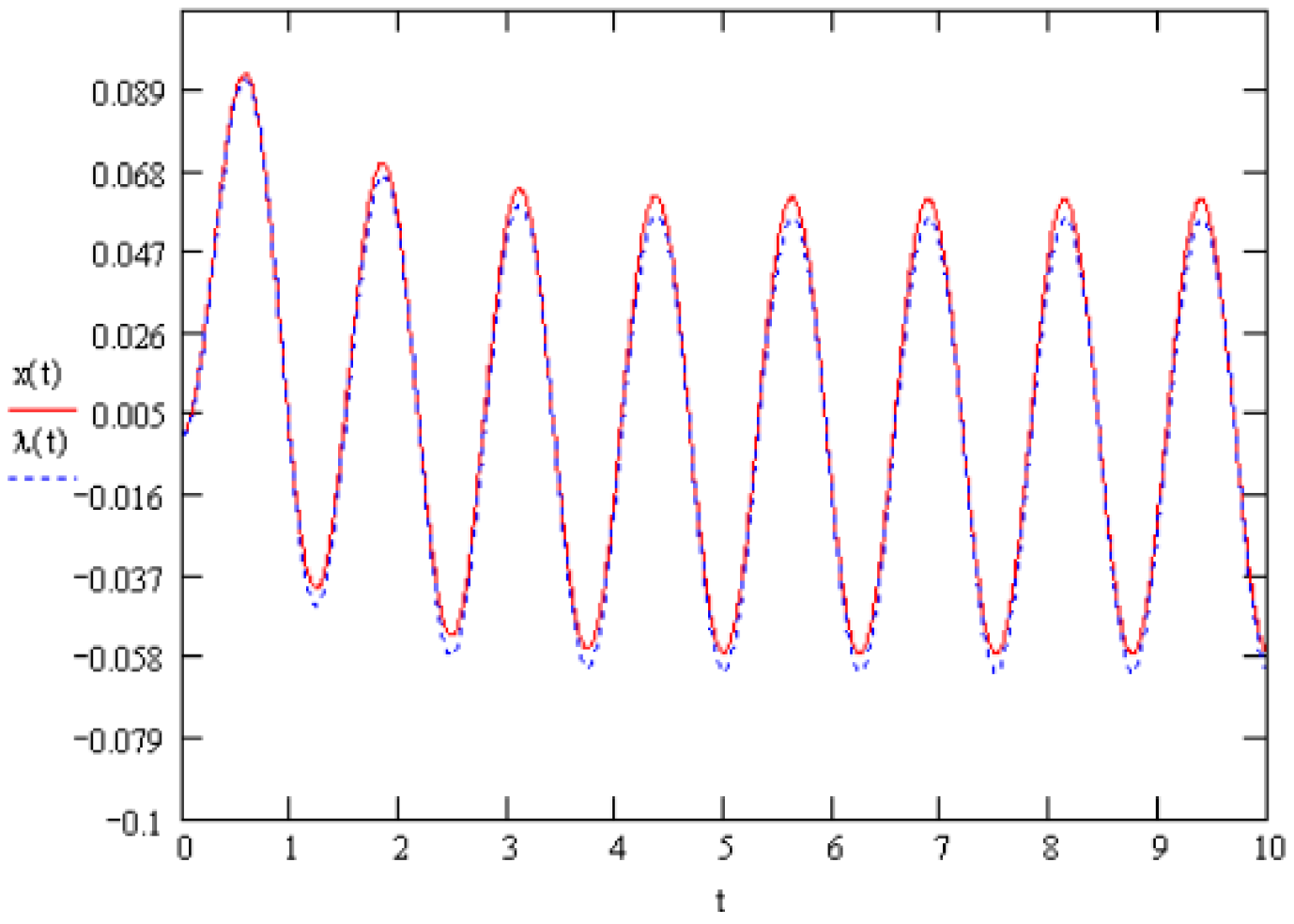

**Fig. 3**. Comparison of classical (red curve) and quasiclassical (blue curve) dynamics calculated from **SLDP**. $a = 1, b = 1, A = 0,3, \Omega = 5, x_0 = 0.$

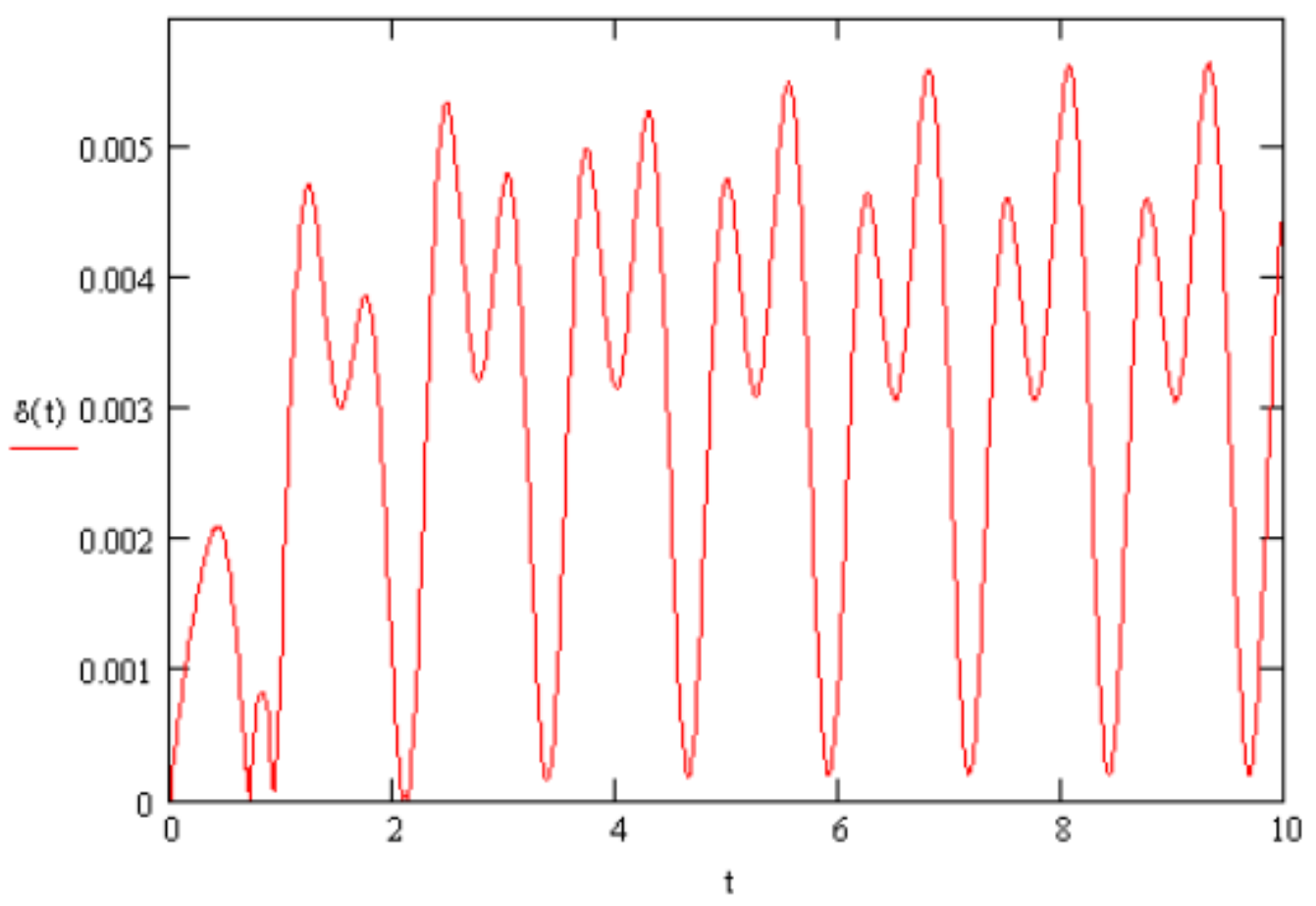

**Fig. 4**. Comparison of classical (red curve) and quasiclassical (blue curve) dynamics calculated from **SLDP**, by norm $\delta(t)$.
$a = 1, b = 1, A = 0,3, \Omega = 5,\ x_0 = 0.$

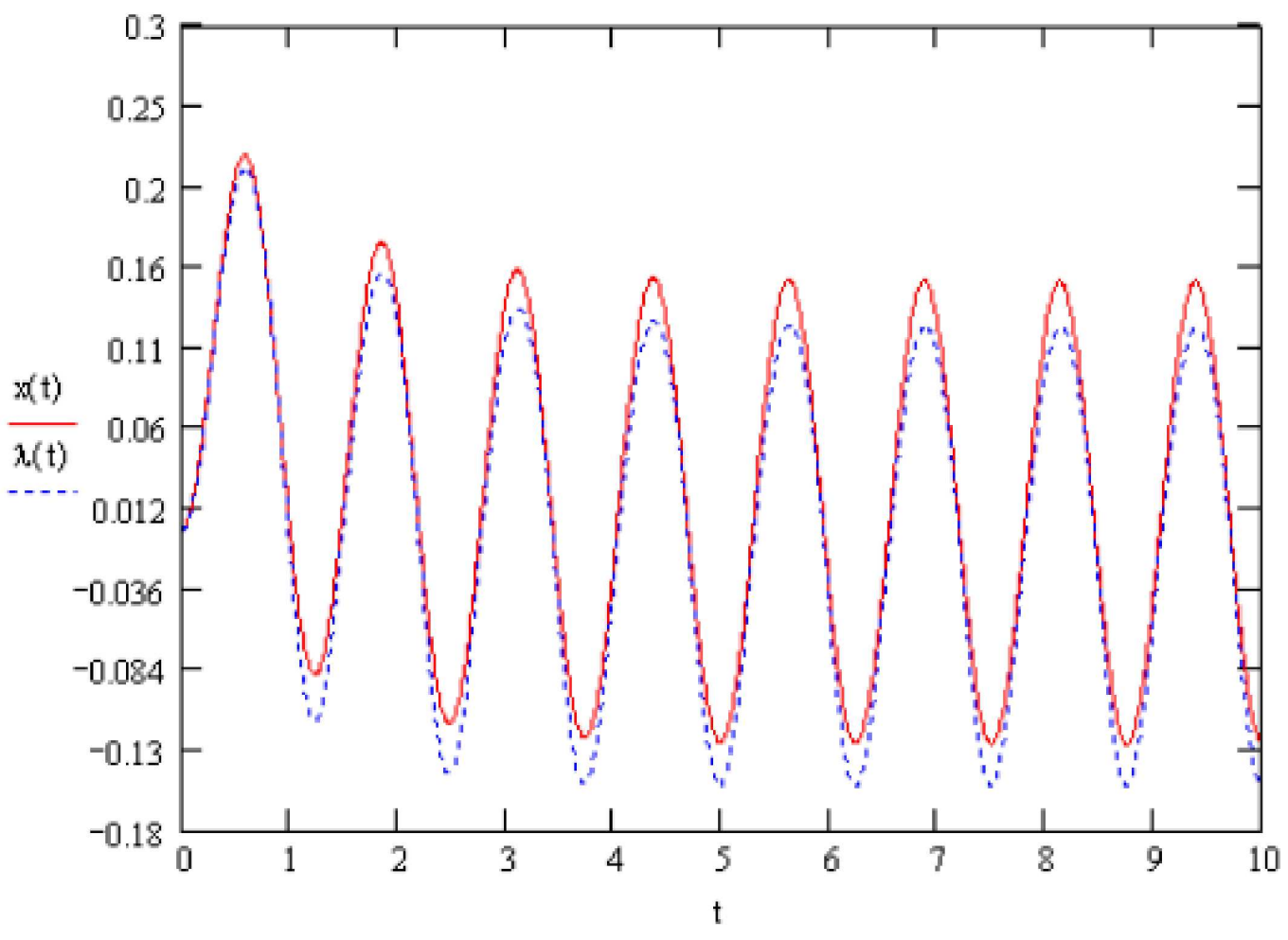

**Fig. 5**. Comparison of classical (red curve) and quasiclassical (blue curve) dynamics calculated from **SLDP**. $a = 1, b = 1, A = 0.7, \Omega = 5,\ x_0 = 0.$

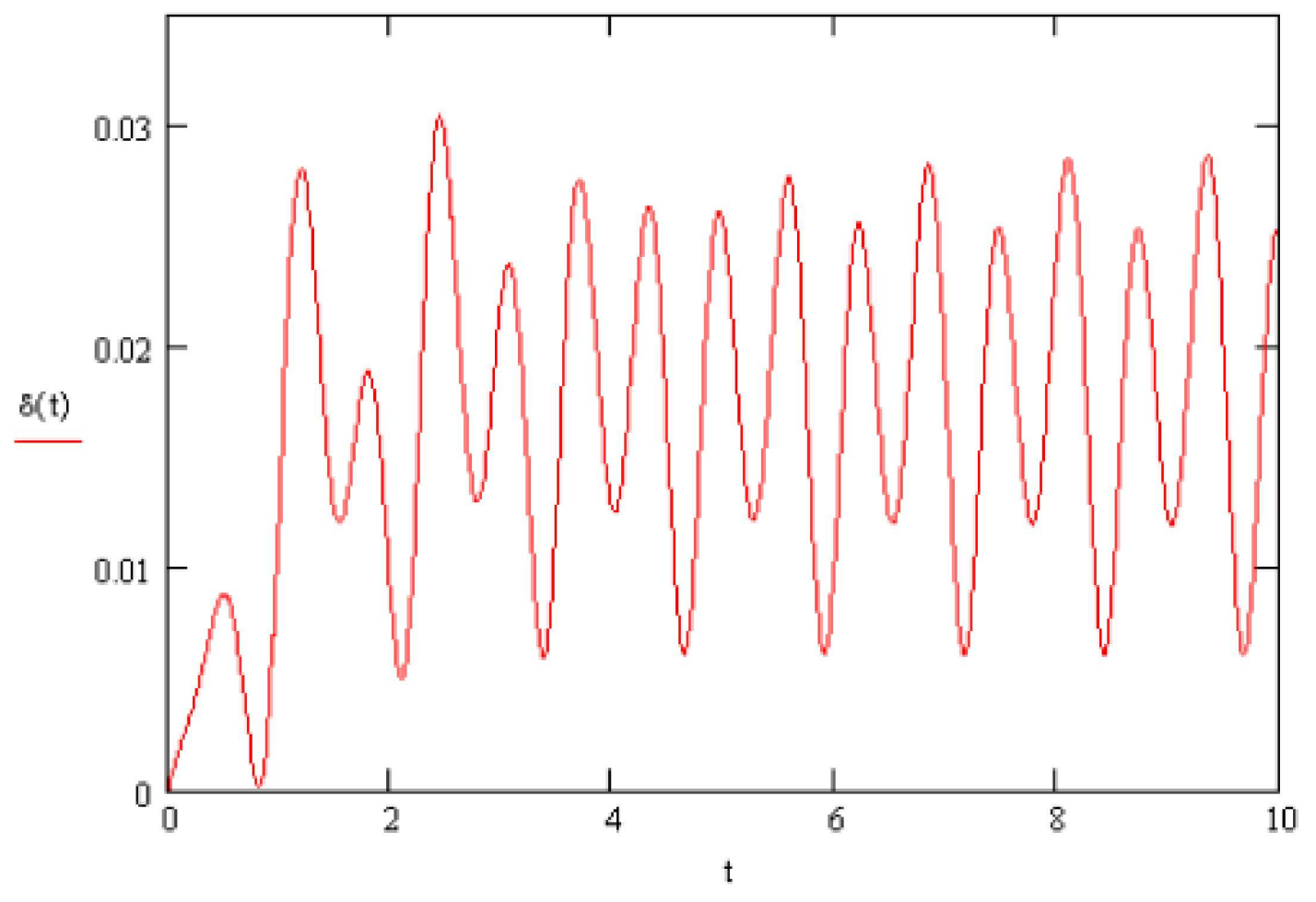

**Fig. 6**. $a = 1, b = 1, A = 0.7, \Omega = 5, x_0 = 0.$

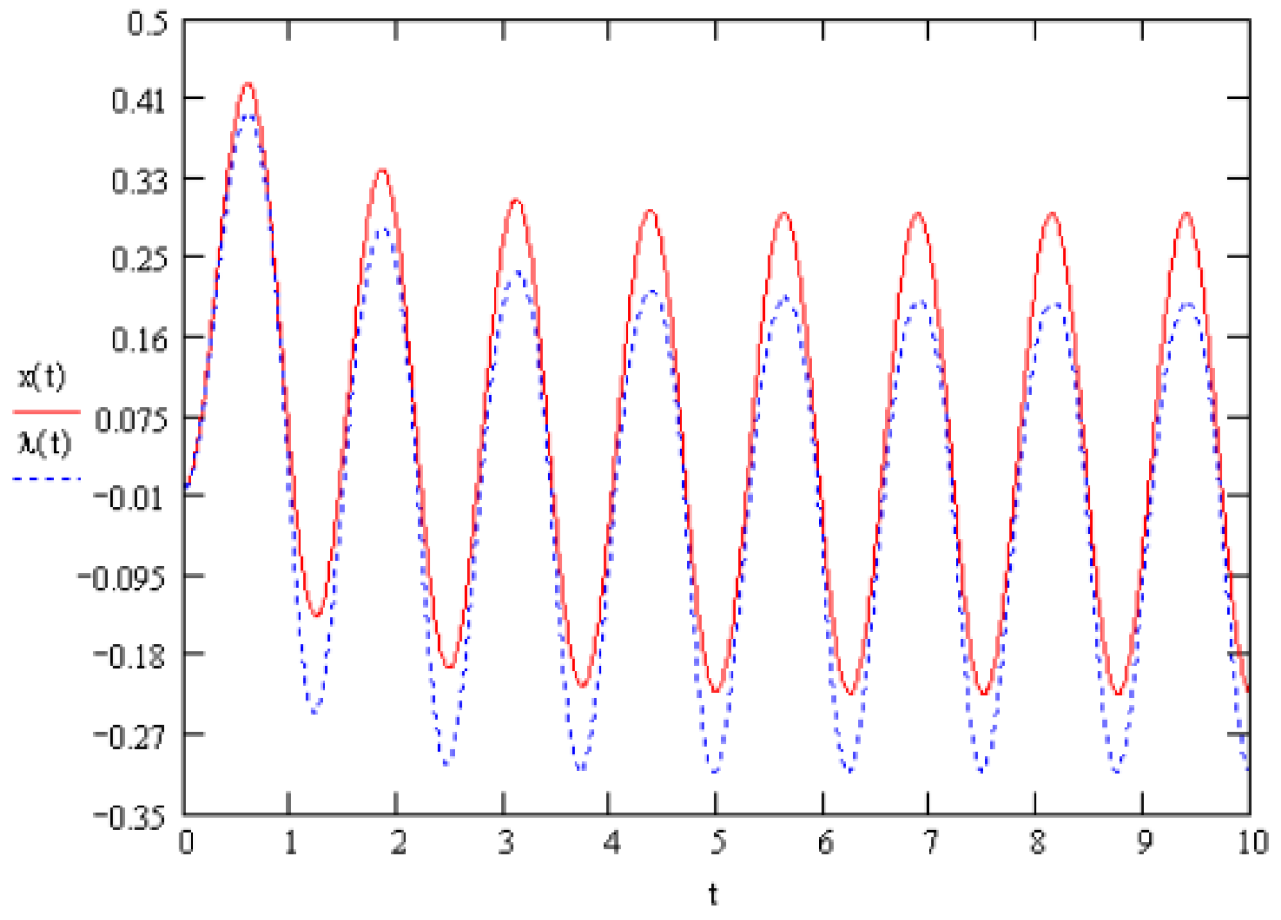

**Fig. 6**. Comparison of classical (red curve) and quasiclassical (blue curve) dynamics calculated from **SLDP**. $a = 1, b = 1, A = 1.3, \Omega = 5, x_0 = 0.$

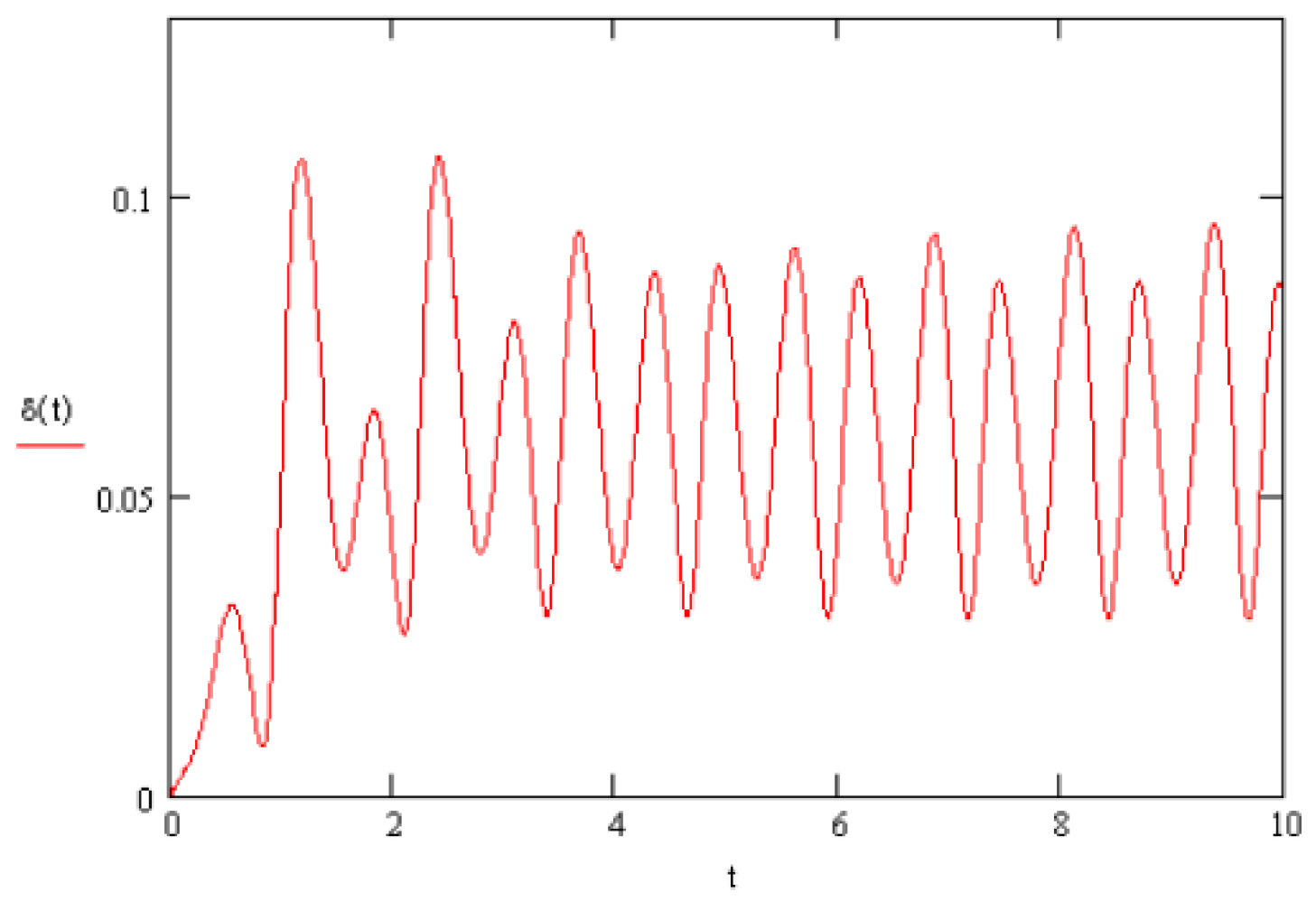

**Fig. 6.** $a = 1, b = 1, A = 1.3, \Omega = 5, x_0 = 0.$

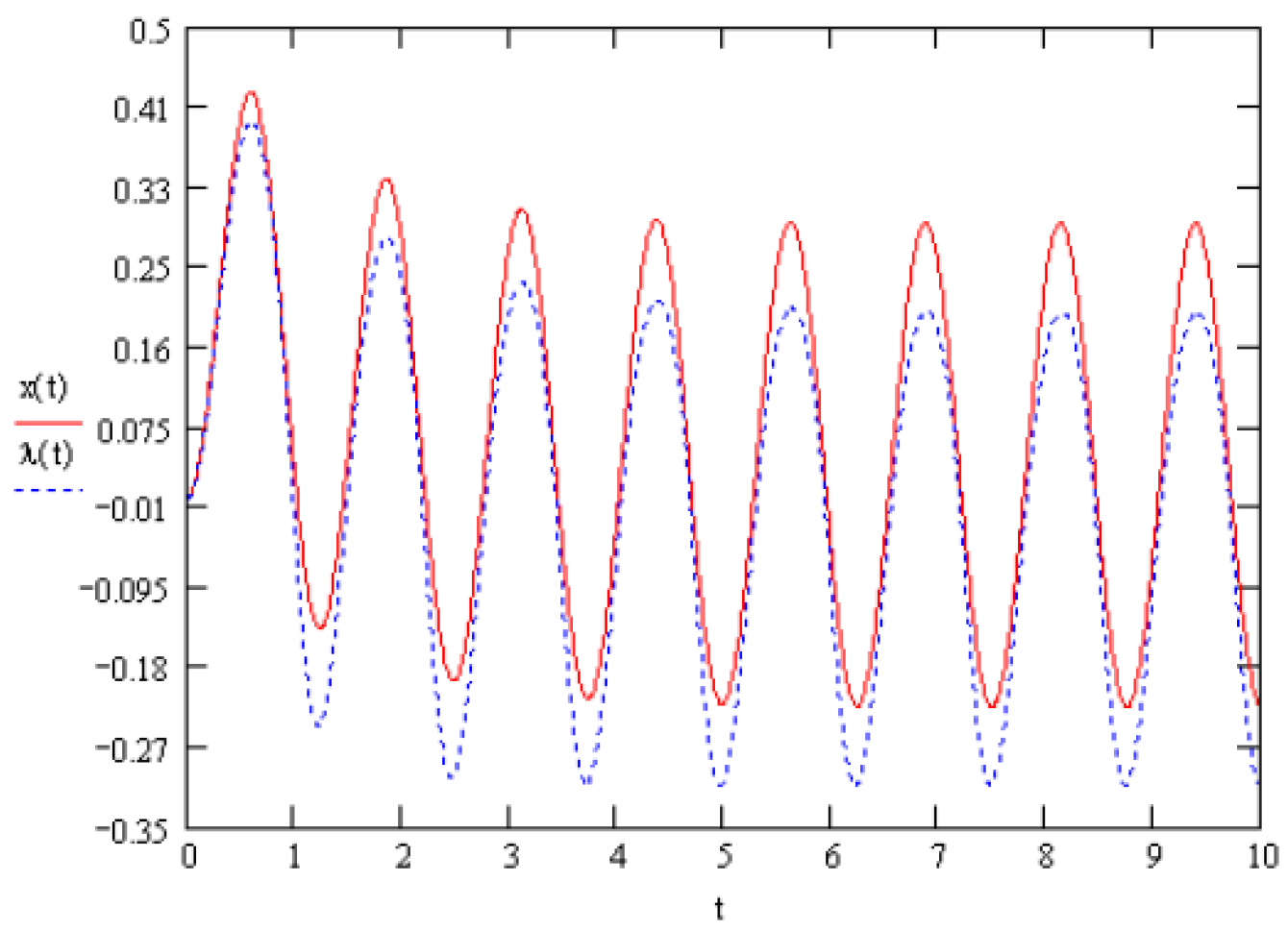

**Fig. 7.** Comparison of classical (red curve) and quasiclassical (blue curve) dynamics calculated from **SLDP**. $a = 1, b = 1, A = 2, \Omega = 5, x_0 = 0.$

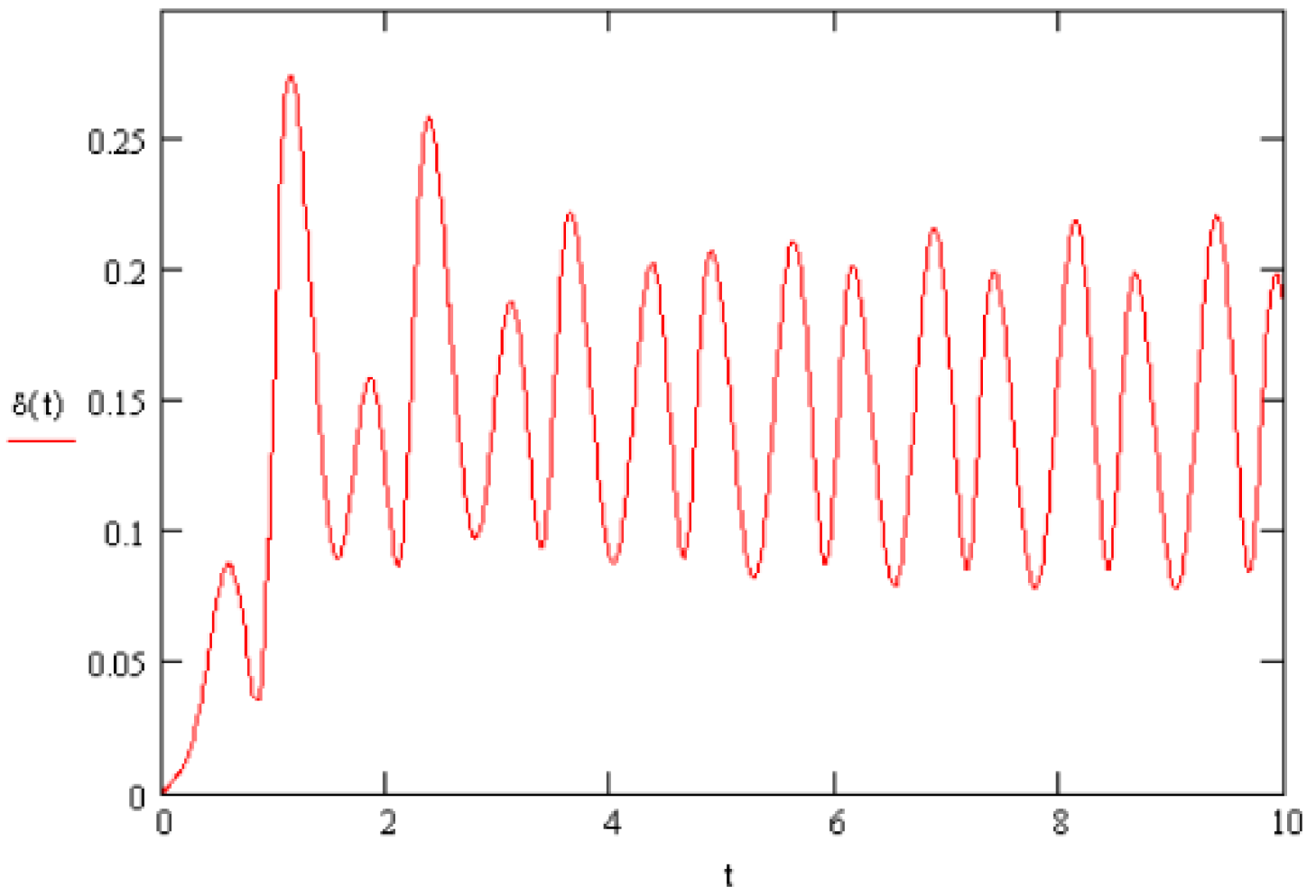

**Fig. 8.** $a = 1, b = 1, A = 2, \Omega = 5, x_0 = 0.$

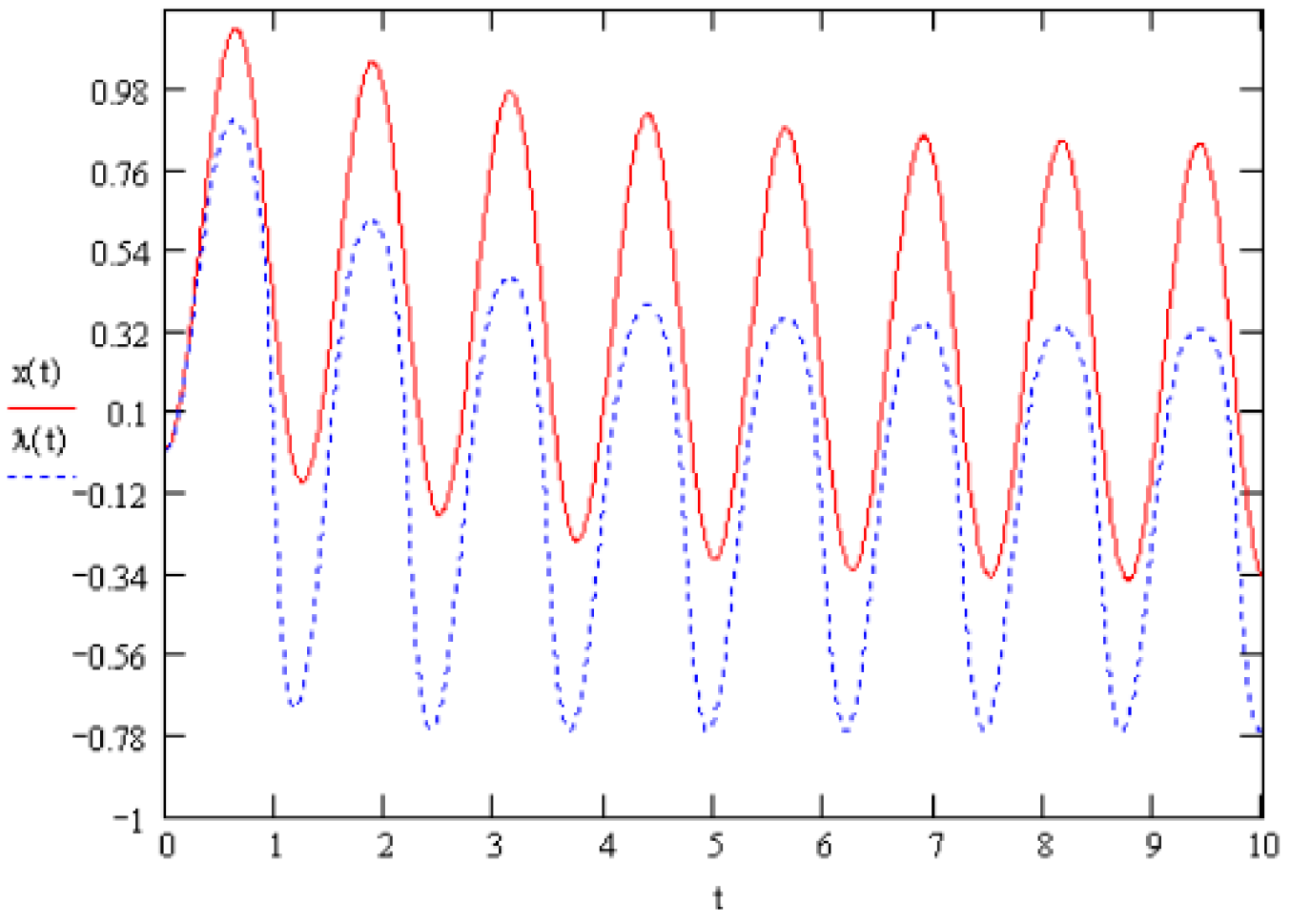

**Fig. 9.** Comparison of classical (red curve) and quasiclassical (blue curve) dynamics calculated from **SLDP**. $a = 1, b = 1, A = 3, \Omega = 5, x_0 = 0.$

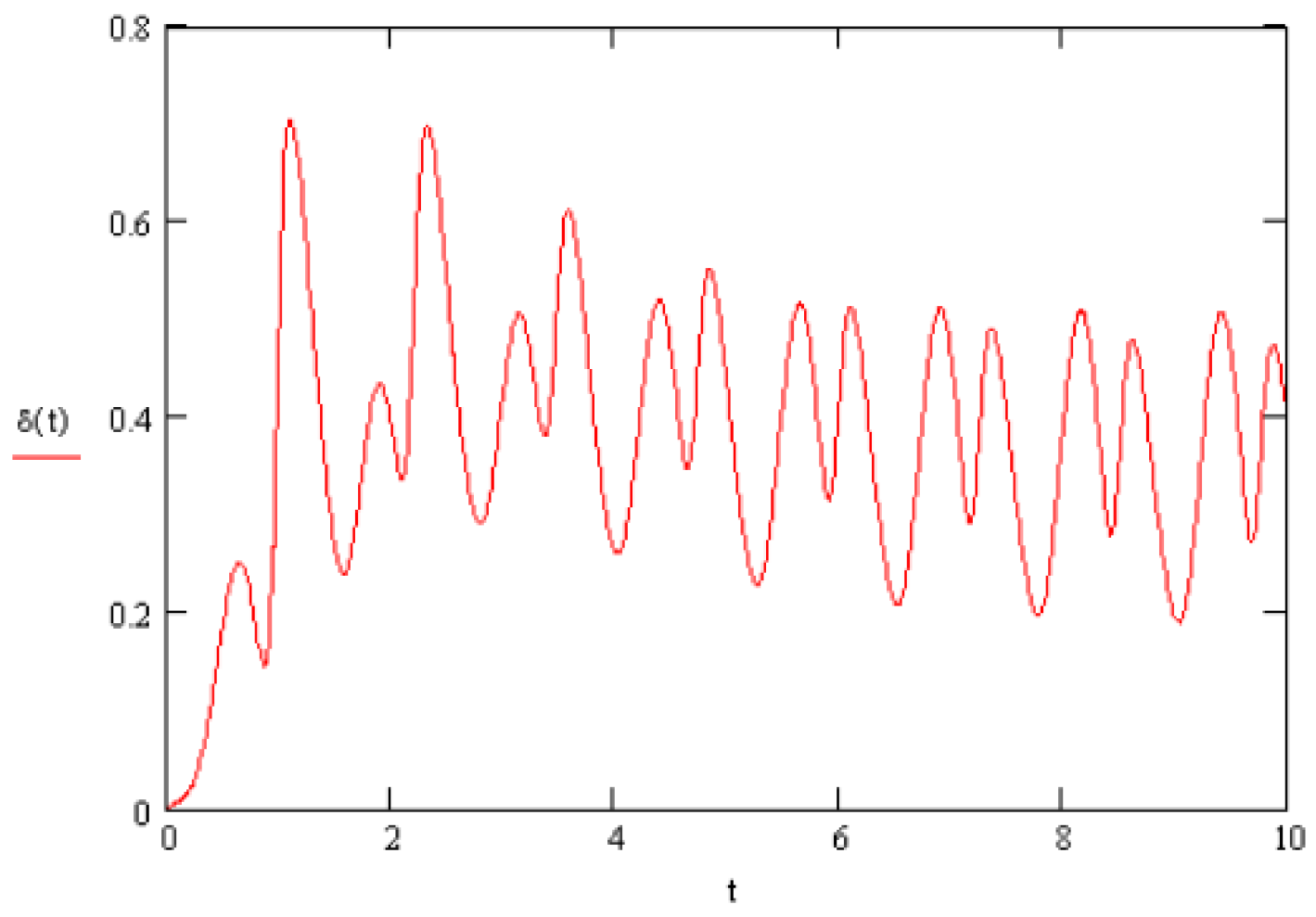

**Fig. 10**. Comparison of classical (red curve) and quasiclassical (blue curve) dynamics calculated from **SLDP**. $a = 1, b = 1, A = 3, \Omega = 5, x_0 = 0$.

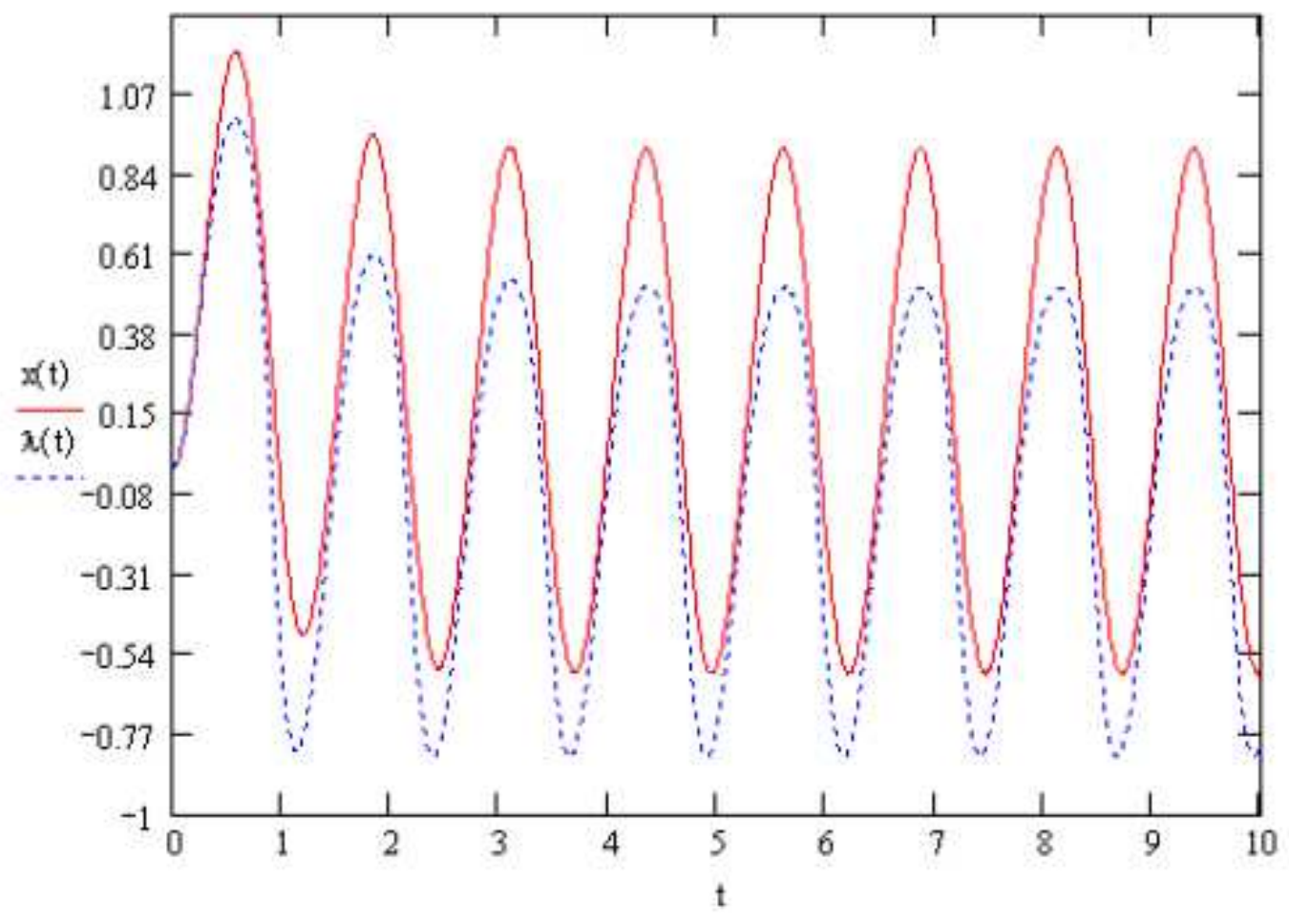

**Fig. 11.** Comparison of classical (red curve) and quasiclassical (blue curve) dynamics calculated from **SLDP**. $a = 1, b = 2, A = 4, \Omega = 5, x_0 = 0.$

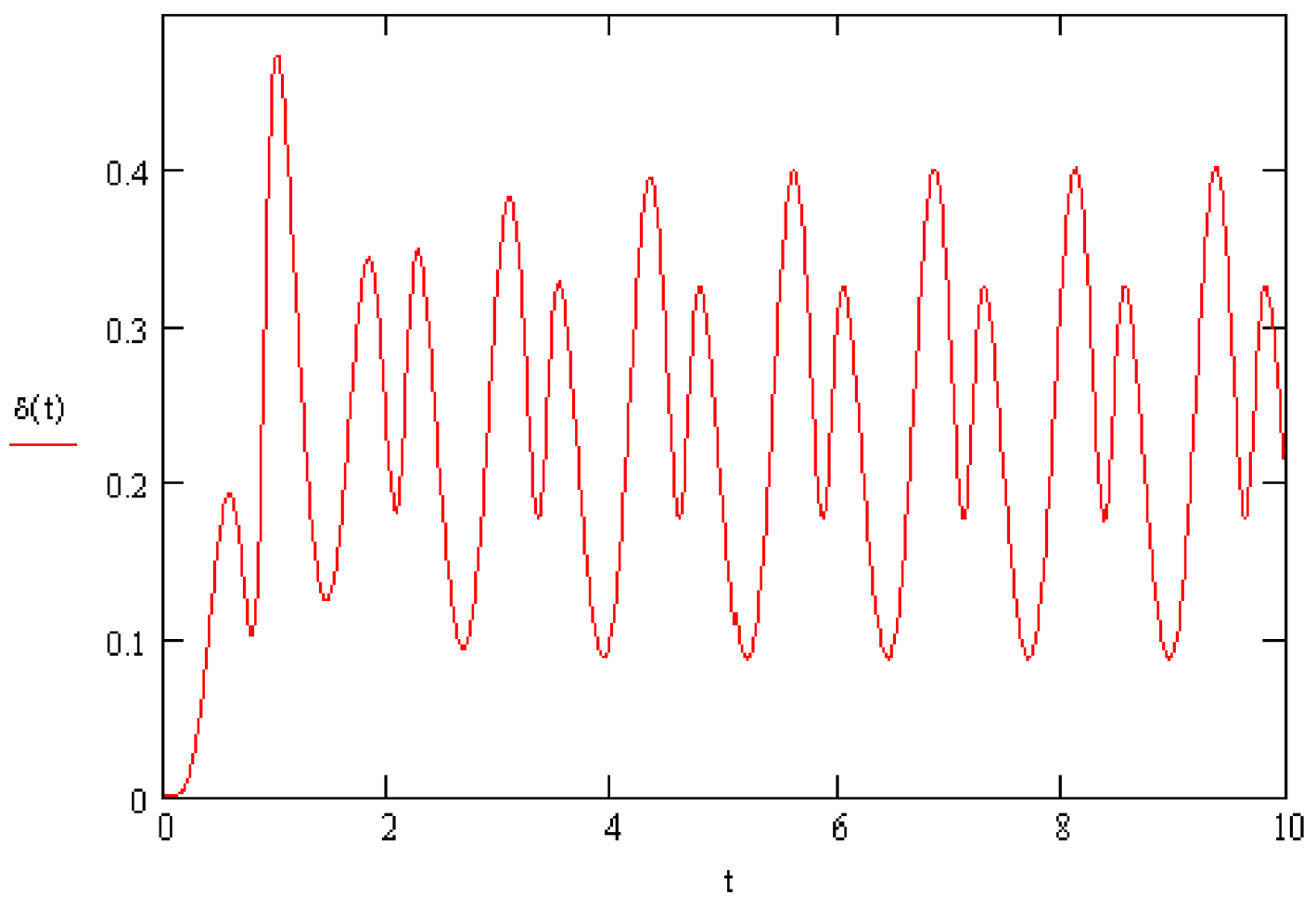

**Fig. 12.** Comparison of classical (red curve) and quasiclassical (blue curve) dynamics calculated from **SLDP**. $a = 1, b = 2, A = 4, \Omega = 5, x_0 = 0.$

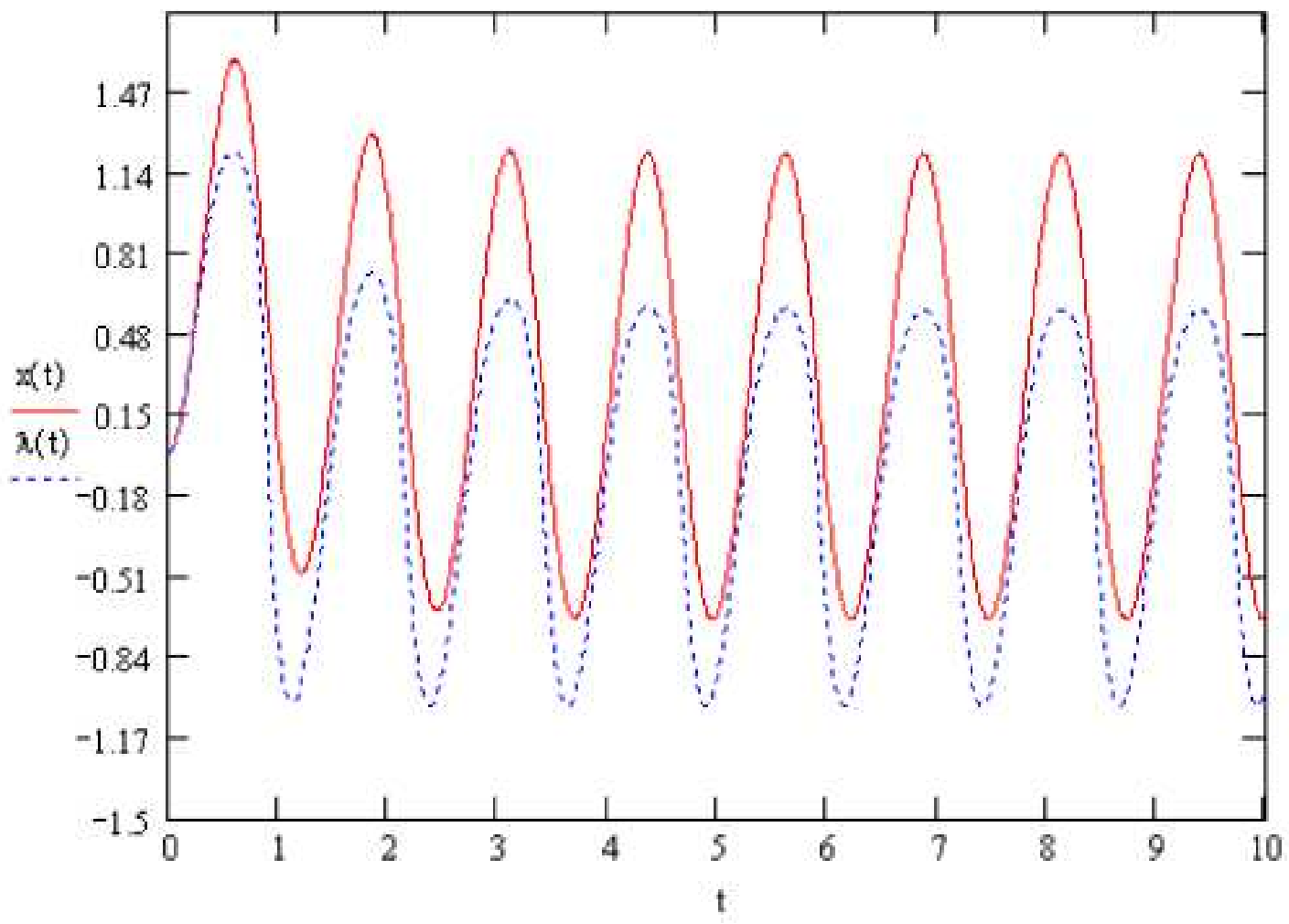

**Fig. 13.** Comparison of classical (red curve) and quasiclassical (blue curve) dynamics calculated from **SLDP**. $a = 1, b = 2, A = 5, \Omega = 5, x_0 = 0.$

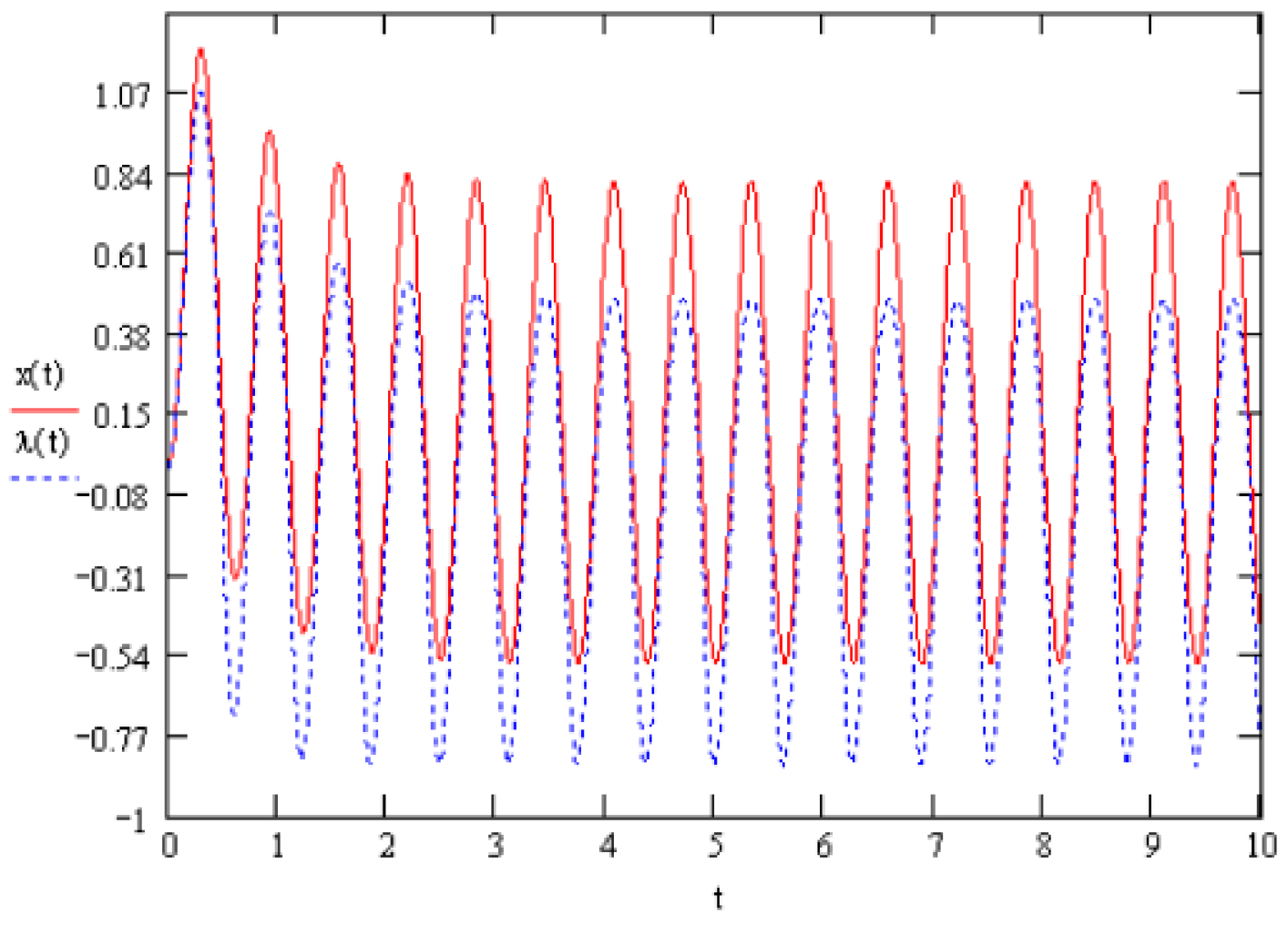

**Fig. 14.** Comparison of classical (red curve) and quasiclassical (blue curve) dynamics calculated from **SLDP**. $a = 1, b = 2, A = 7, \Omega = 10, x_0 = 0.$

**2.4.2. Duble well potential.**

As a a second example we consider a force field (2.4.3) with a duble well potential

$$V(x) = \frac{a}{4}x^4 - \frac{b}{2}x^2 - cx\,,\quad a,b > 0\,. \tag{2.4.9}$$

The time-dependent force field (2.4.3) takes the following form:

$$F(x,t) = -ax^3 + bx + A\sin(\Omega t) + B\cos(\Theta t) + c\,. \tag{2.4.10}$$

The stochastic dynamics (2.4.1) takes the following form:

$$\dot{x}(t) = -ax^3 + bx + A\sin(\Omega t) + \sqrt{\varepsilon}\;\xi(t)\,, x(0) = x_0. \tag{2.4.11}$$

From master equation (2.2.8) we have the next differential linear master equation

$$\dot{u}(t) = -(3a\lambda^2 - b\lambda)u(t) - a\lambda^3 + b\lambda + A\sin(\Omega t)\,, u(0) = x_0 - \lambda. \qquad (2.4.12)$$

From Theorem 2.2.2 we have the next transcendental master equation

$$(x_0 - \lambda(t))\exp[-(3a\lambda^2(t) - b)t] -$$
$$-(a\lambda^3(t) - b\lambda(t))\int_0^t \exp[-(3a\lambda^2(t) - b)(t-\tau)]d\tau + \qquad (2.4.13)$$
$$A\int_0^t \sin(\Omega\tau)\exp[-(3a\lambda^2(t) - b)(t-\tau)]d\tau = 0.$$

**Comparison of classical and non-perturbative quasiclassical dynamics**.

We have compared the above analytical predictions for the limit (2.2.10) with very accurate numerical results for classical dynamics

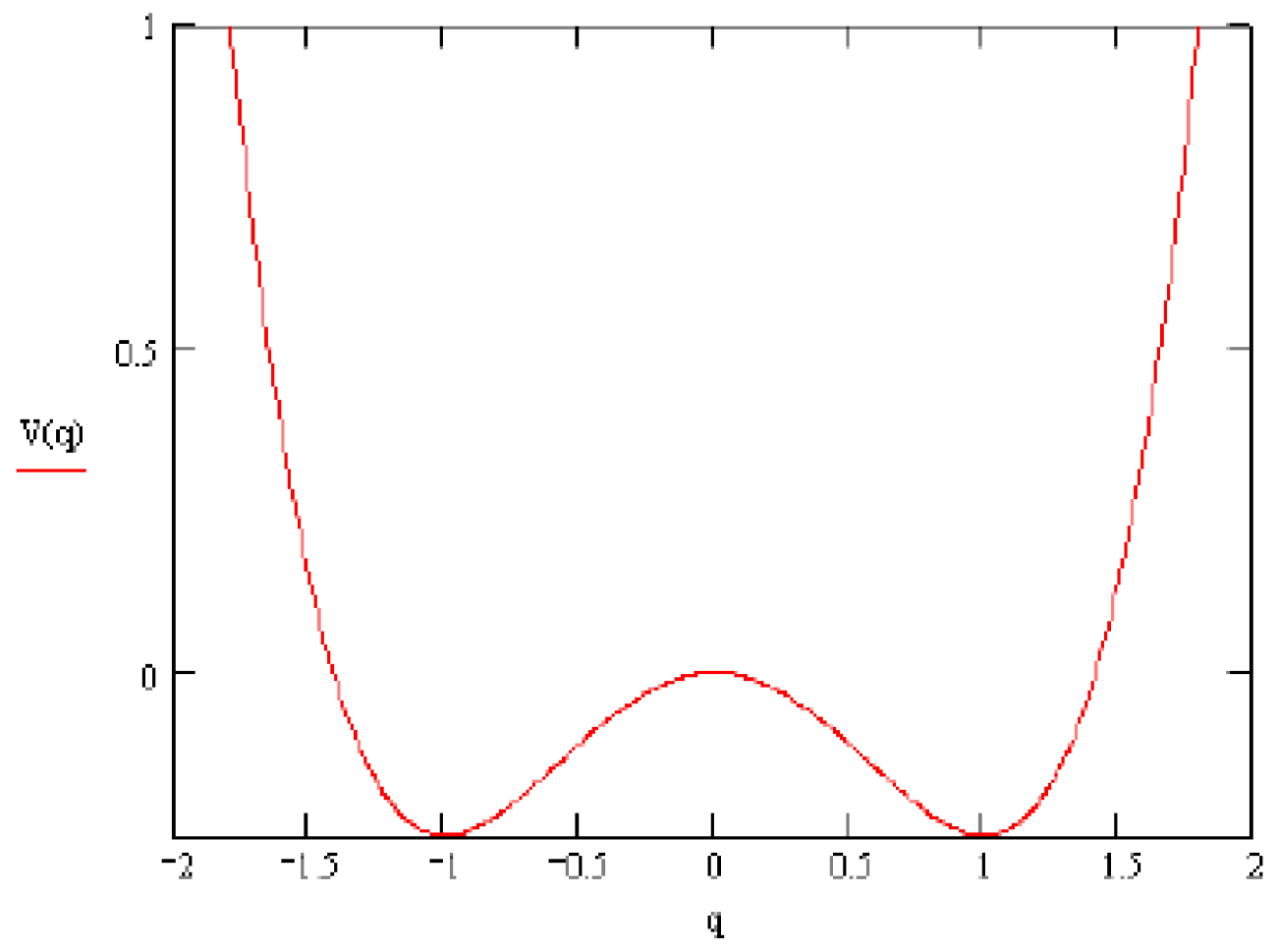

**Fig. 15.** Double Well Potential $a = 1, b = 1, c = 0.$

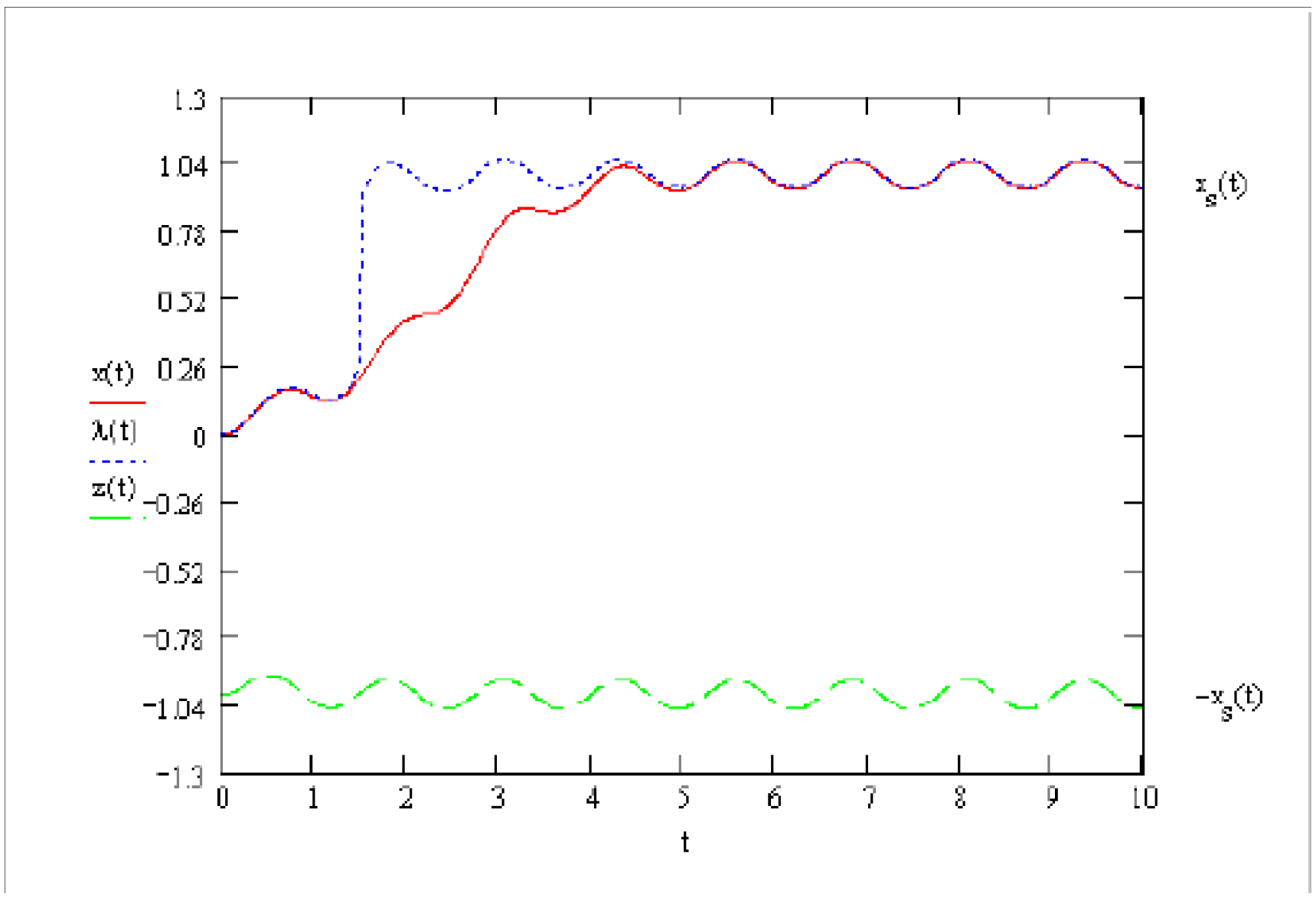

**Fig. 16.** Comparison of classical (red curve) and stochastic (blue curve) dynamics calculated by using **SLDP**. $a = 1, b = 1, c = 0, A = 0.3, B = 0, \Omega = 5, x_0 = 0.$

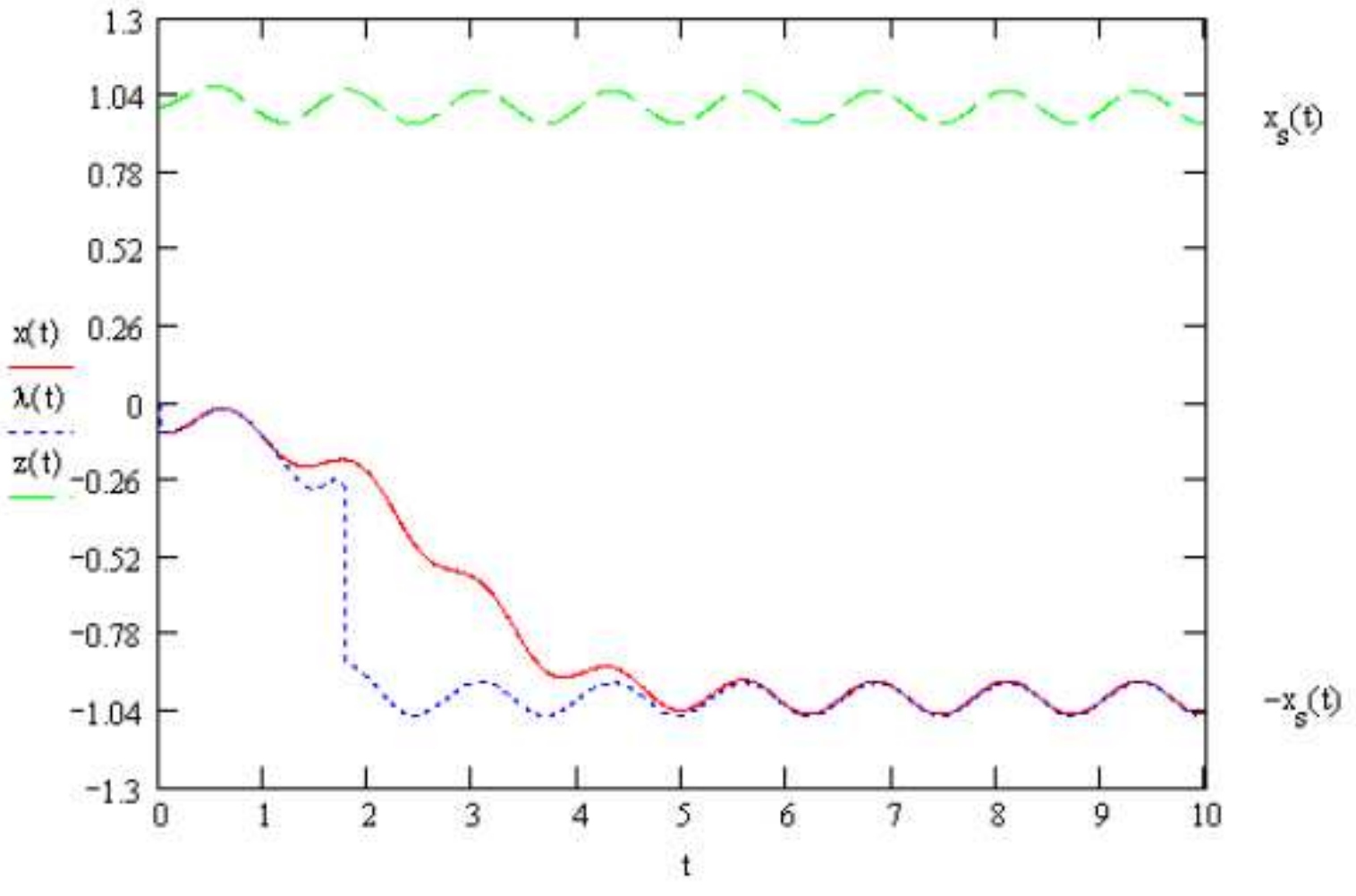

**Fig. 17.** Comparison of classical (red curve) and stochastic (blue curve) dynamics calculated by using **SLDP**. $a = 1, b = 1, c = 0, A = 0.3, B = 0, \Omega = 5, x_0 = -0.1.$

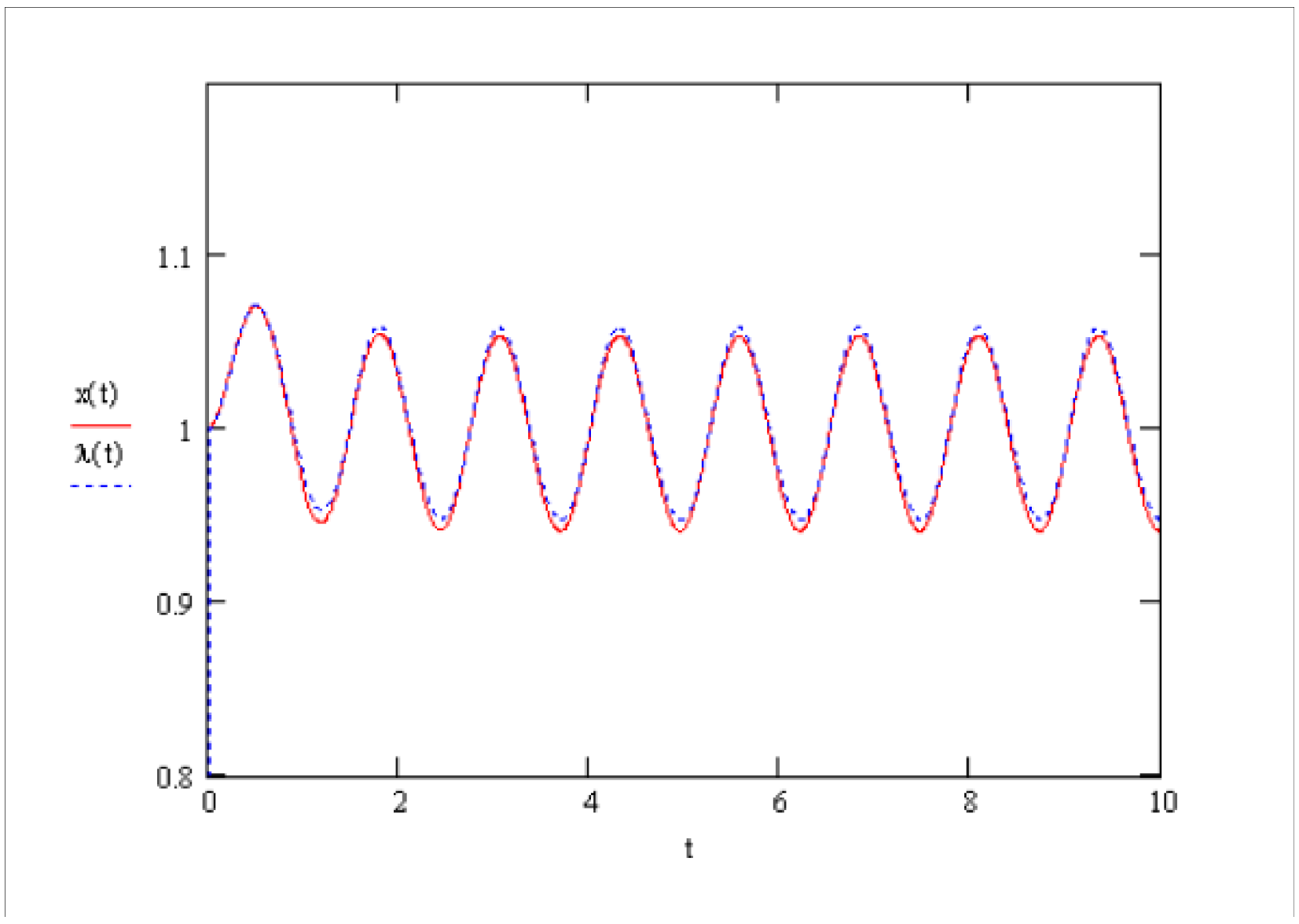

**Fig. 18.** Comparison of classical (red curve) and stochastic (blue curve) dynamics calculated by using **SLDP**.

$a = 1, b = 1, c = 0, A = 0.3, B = 0, \Omega = 5, x_0 = 0.$

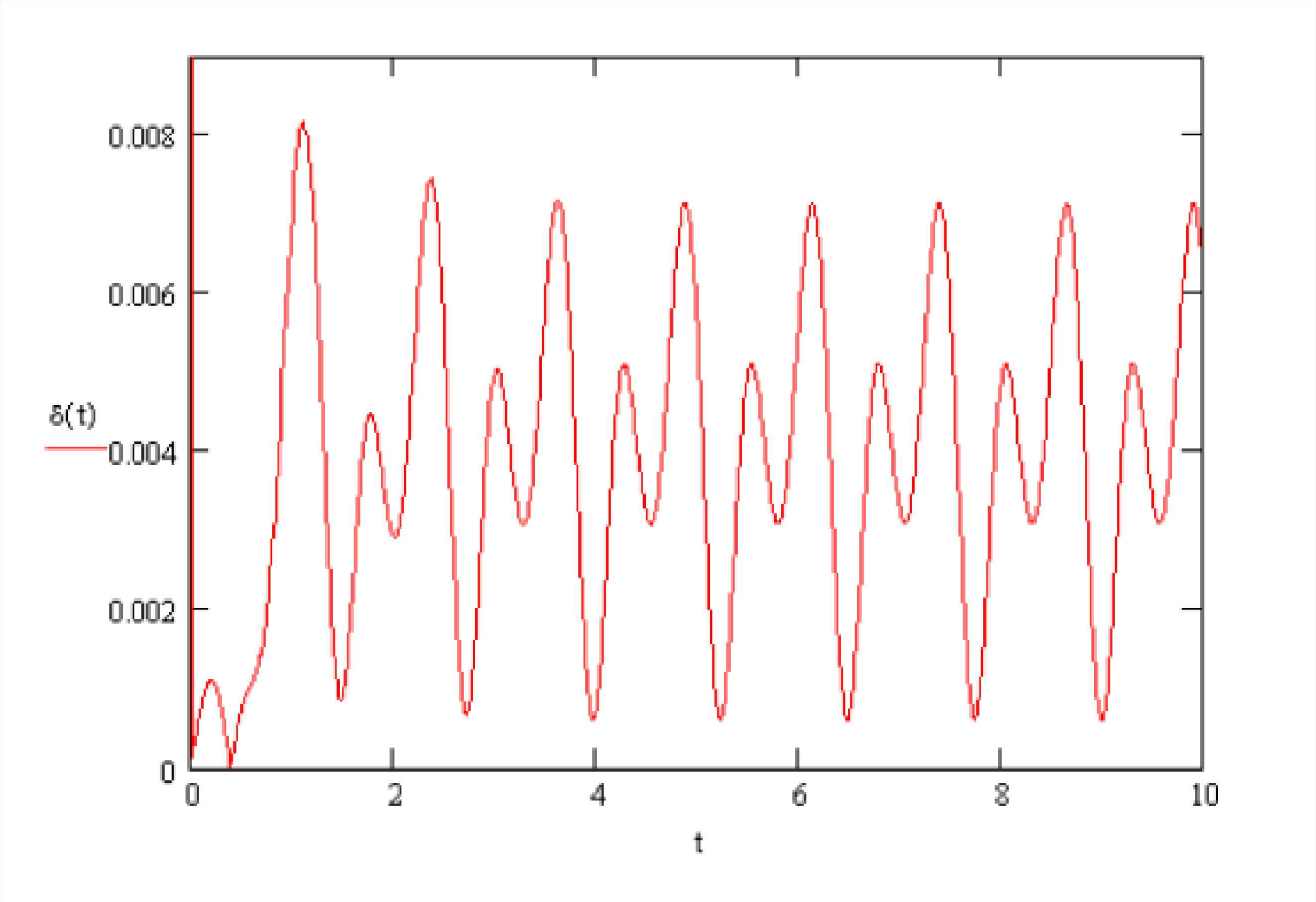

**Fig. 19.** $a = 1, b = 1, c = 0, A = 0.3, B = 0, \Omega = 5, x_0 = 0.$

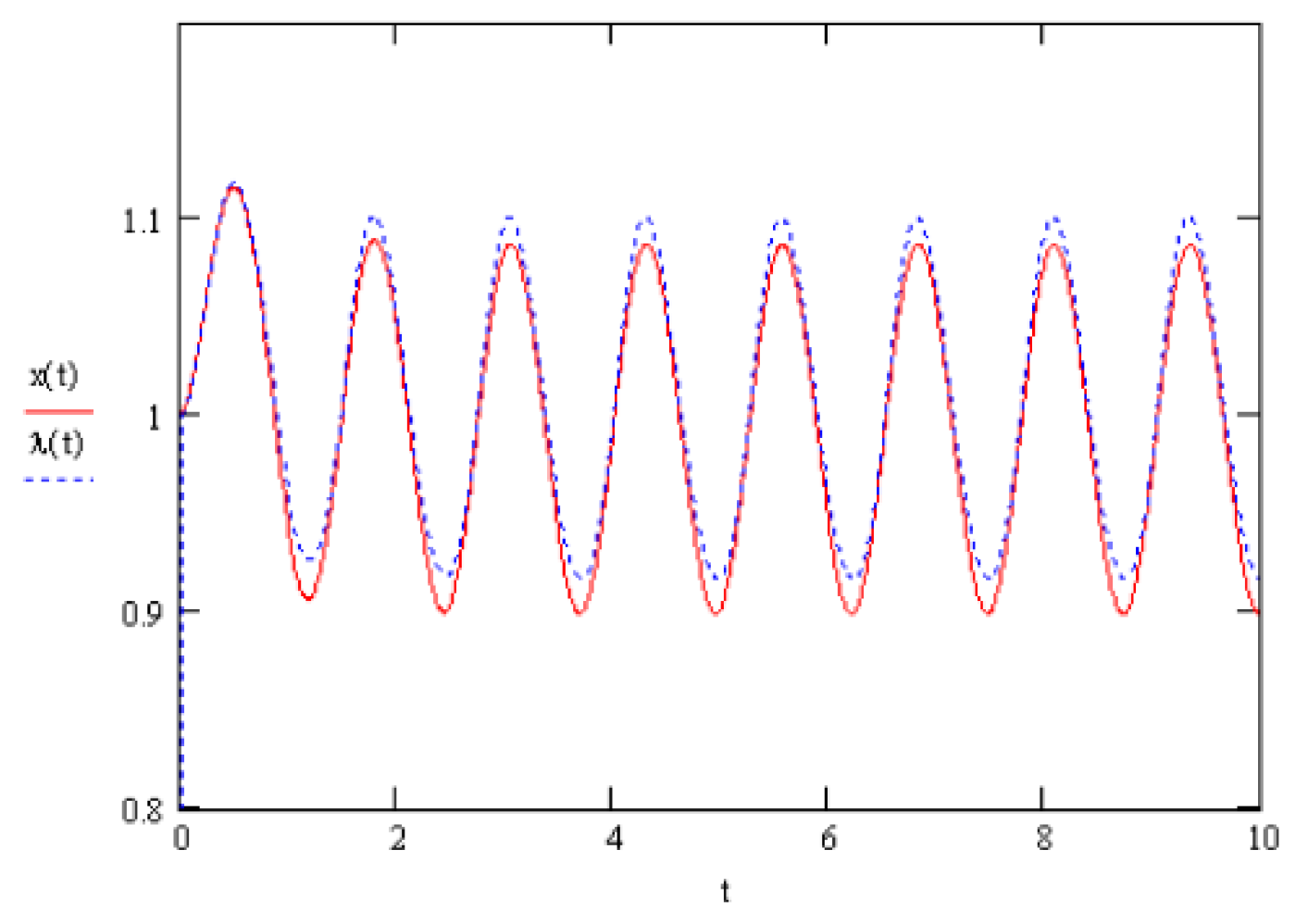

**Fig. 20.** Comparison of classical (red curve) and stochastic (blue curve) dynamics calculated by using **SLDP**.
$a = 1, b = 1, c = 0, A = 0.5, B = 0, \Omega = 5, x_0 = 0.$

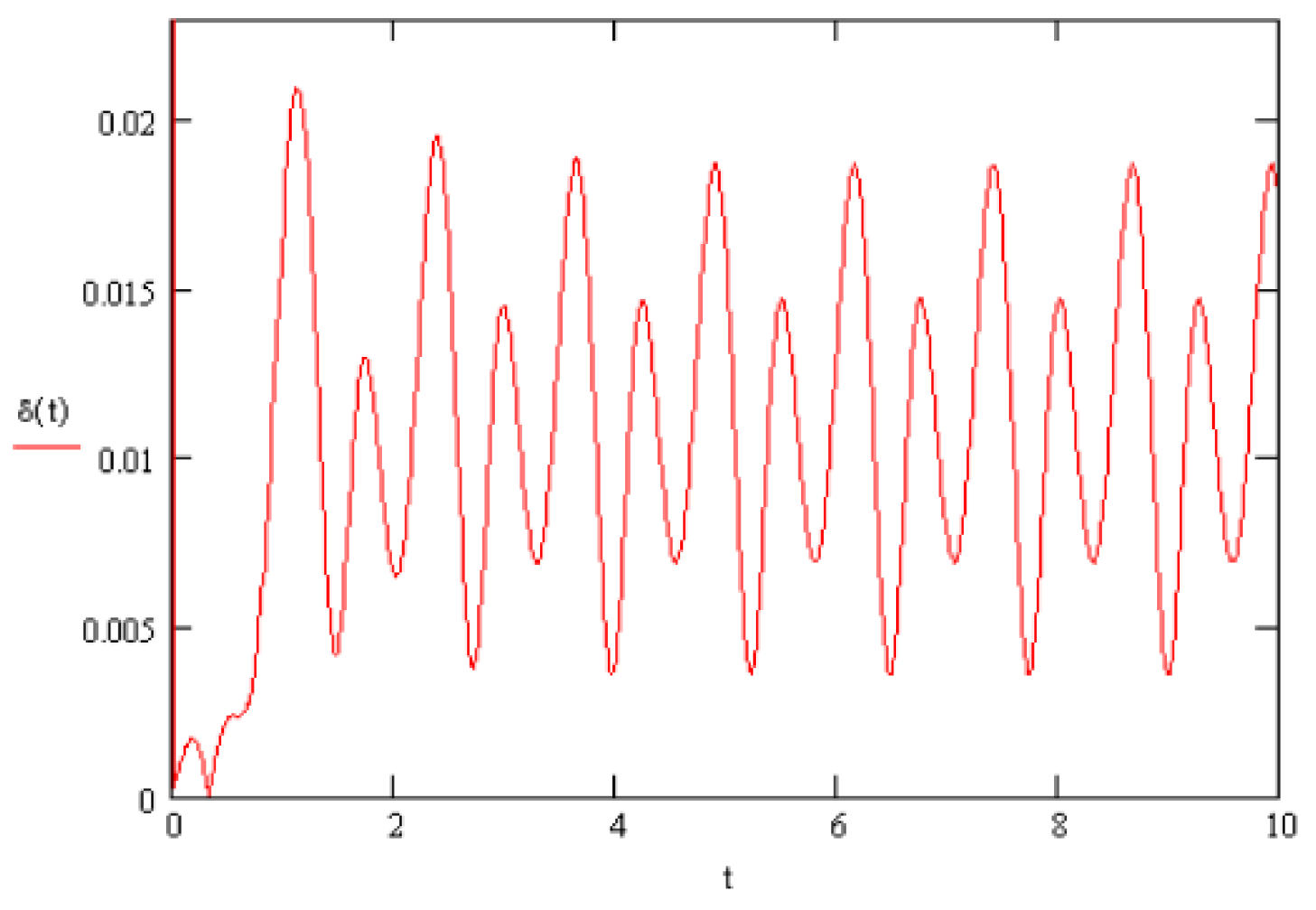

**Fig. 21.** $a = 1, b = 1, c = 0, A = 0.5, B = 0, , \Omega = 5,$
$x_0 = 0.$

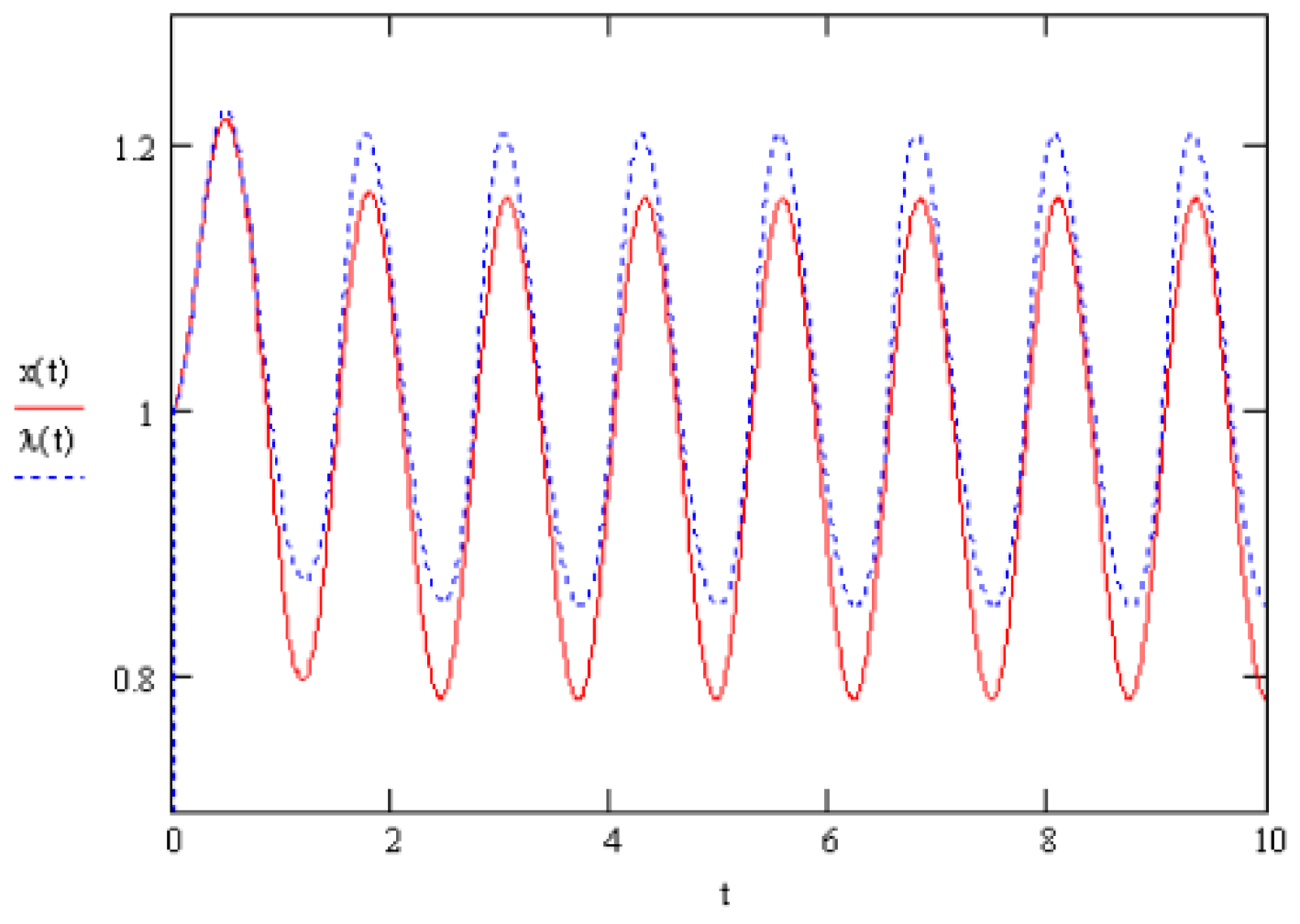

**Fig. 22.** Comparison of classical (red curve) and stochastic (blue curve) dynamics calculated by using **SLDP**.

$a = 1, b = 1, c = 0, A = 1, B = 0, \Omega = 5, x_0 = 0.$

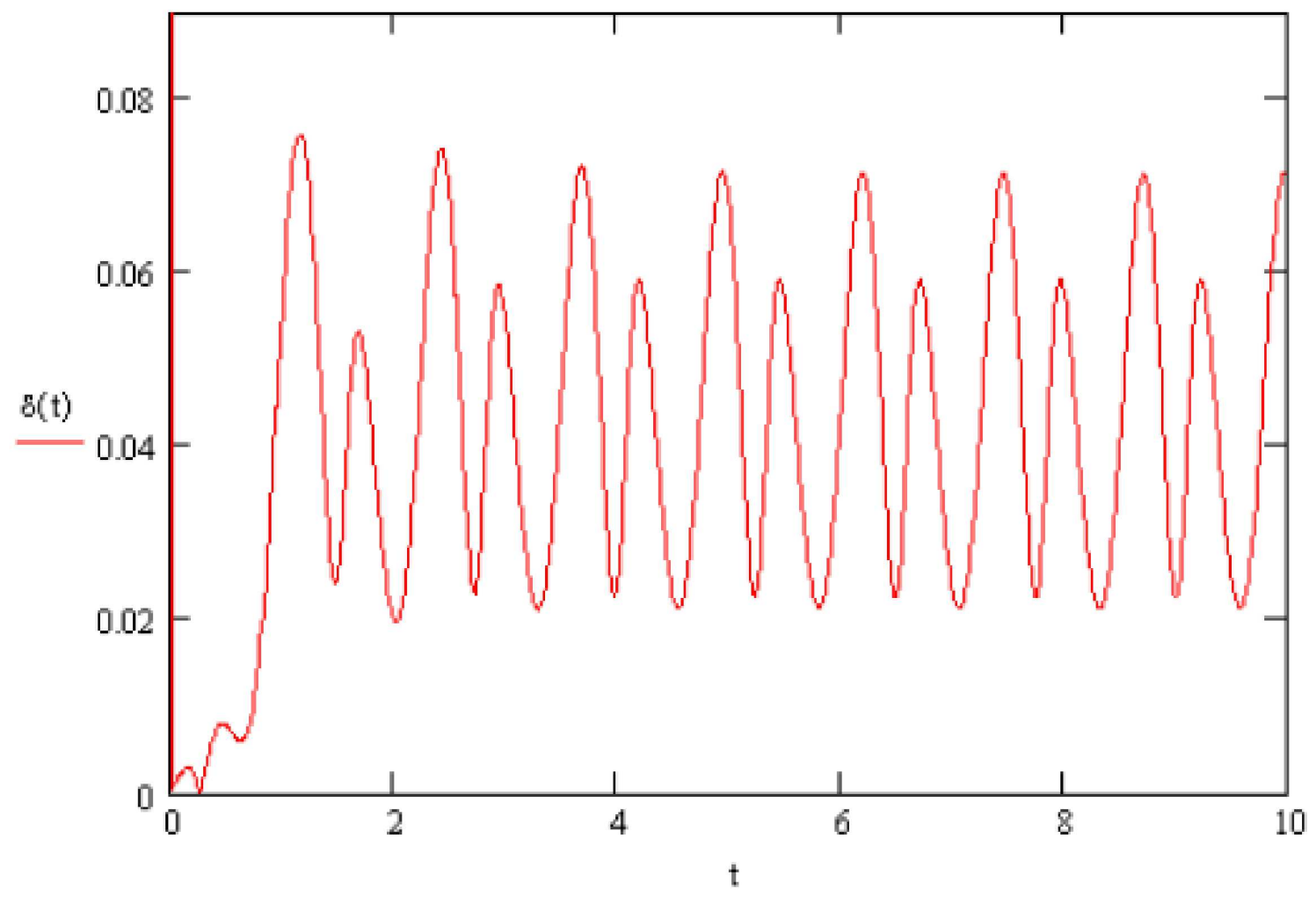

**Fig. 23.** $a = 1, b = 1, c = 0, A = 1, B = 0, \Omega = 5, x_0 = 0.$

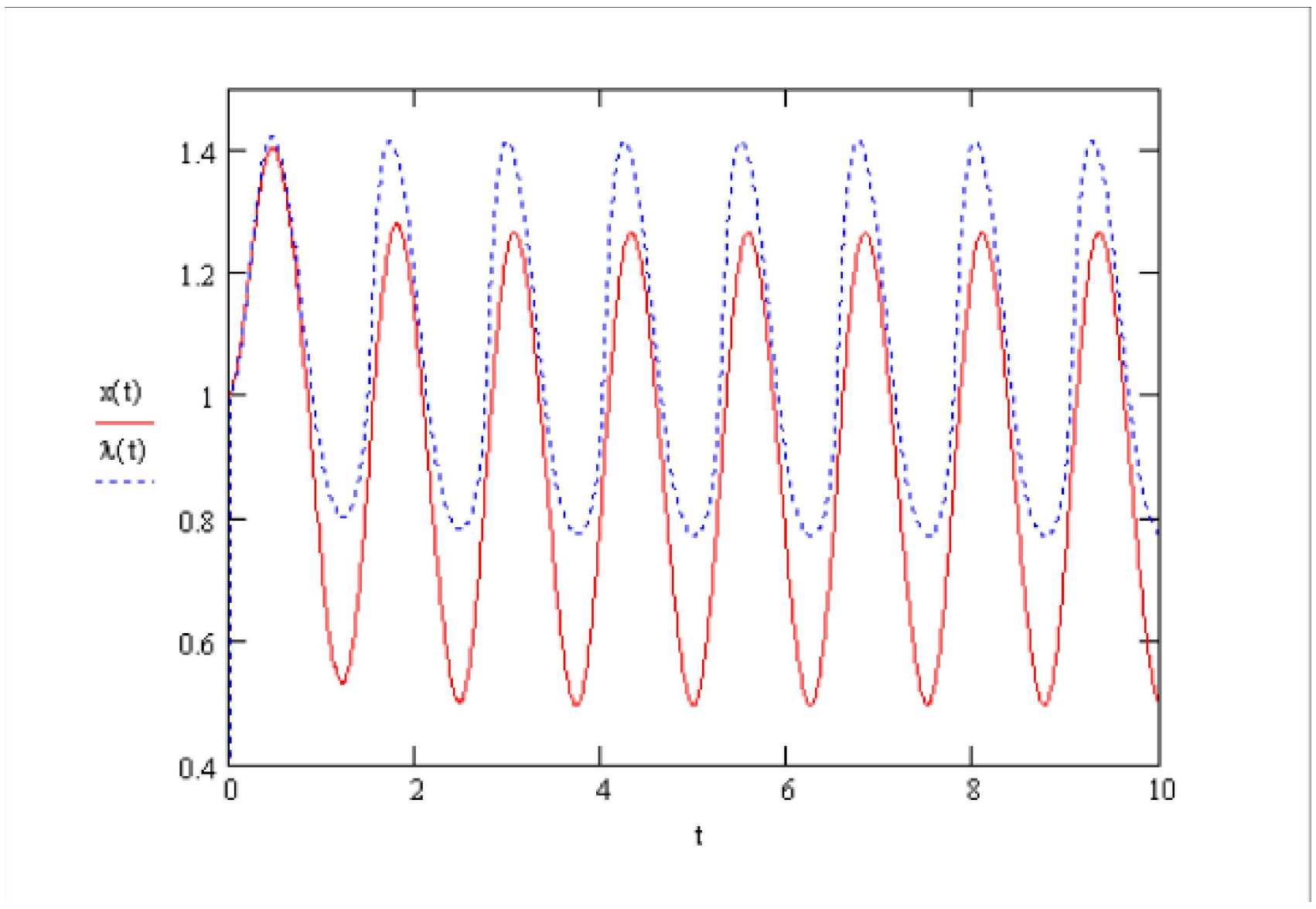

**Fig. 24.** Comparison of classical (red curve) and stochastic (blue curve) dynamics calculated by using **SLDP**.

$a = 1, b = 1, c = 0, A = 2, B = 0, \Omega = 5, x_0 = 0.$

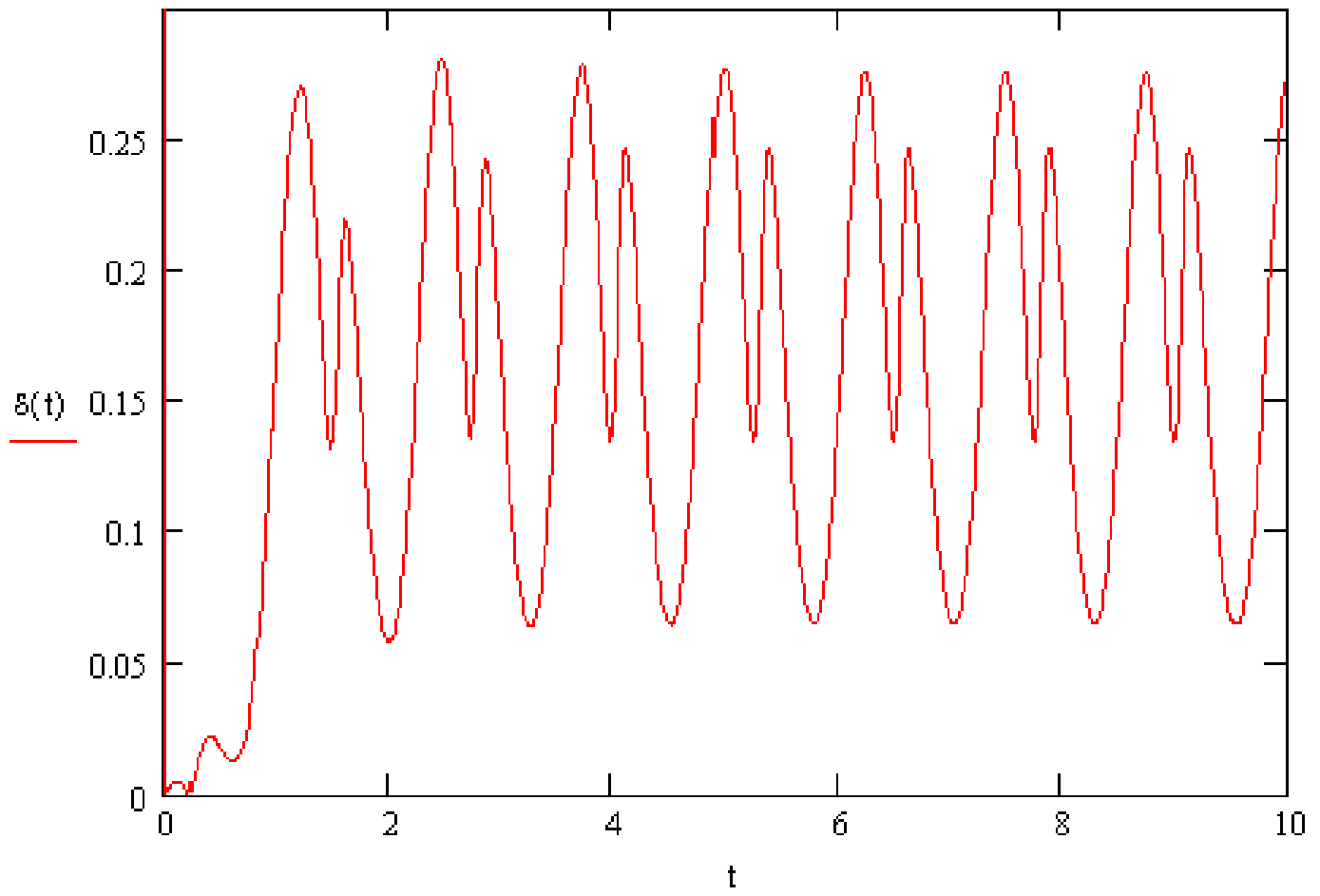

**Fig. 25.** $a = 1, b = 1, c = 0, A = 2, B = 0, \Omega = 5, x_0 = 0.$

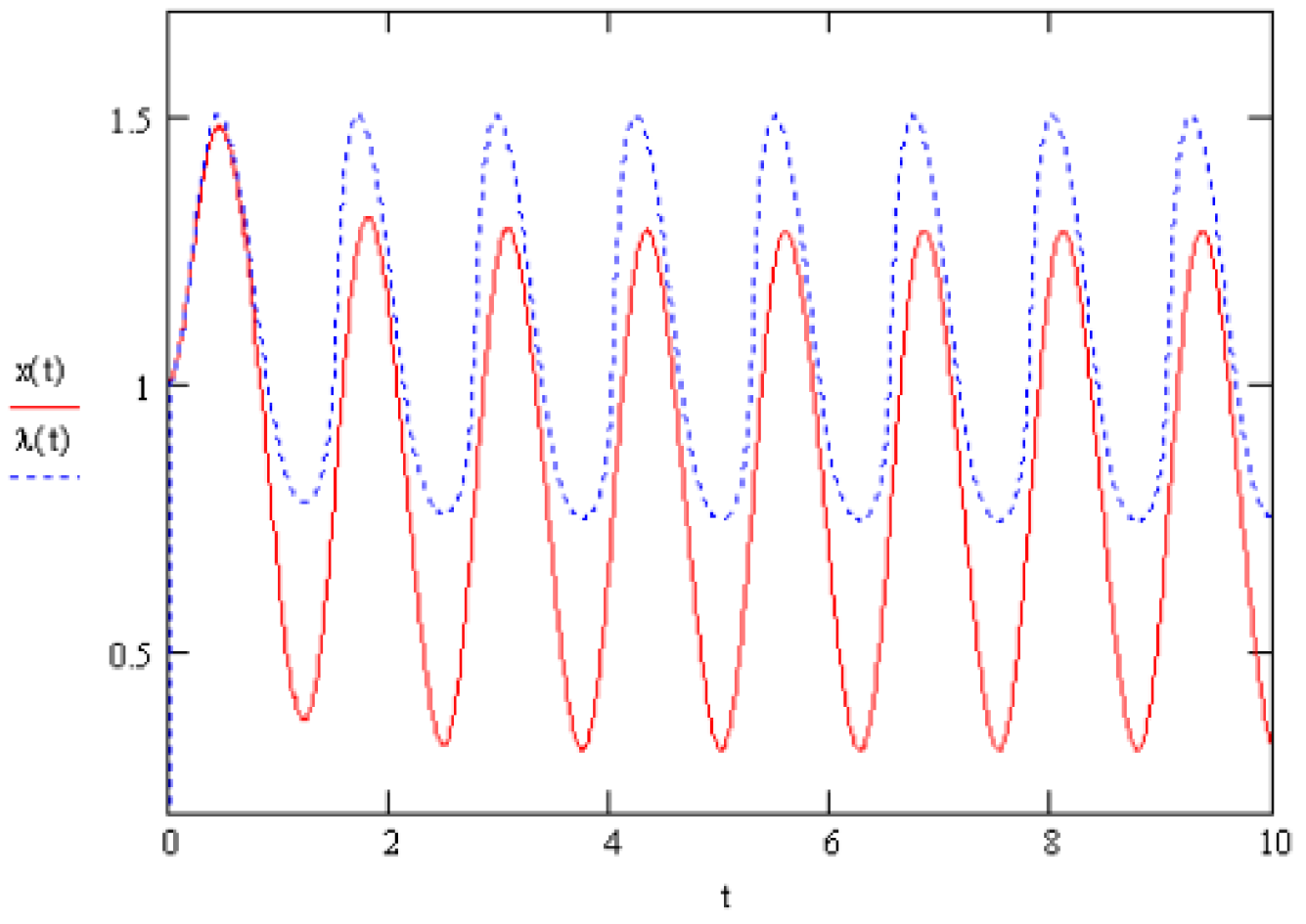

**Fig. 26.** Comparison of classical (red curve) and stochastic (blue curve) dynamics calculated by using **SLDP**.
$a = 1, b = 1, c = 0, A = 2.5, B = 0, \Omega = 5, x_0 = 0.$

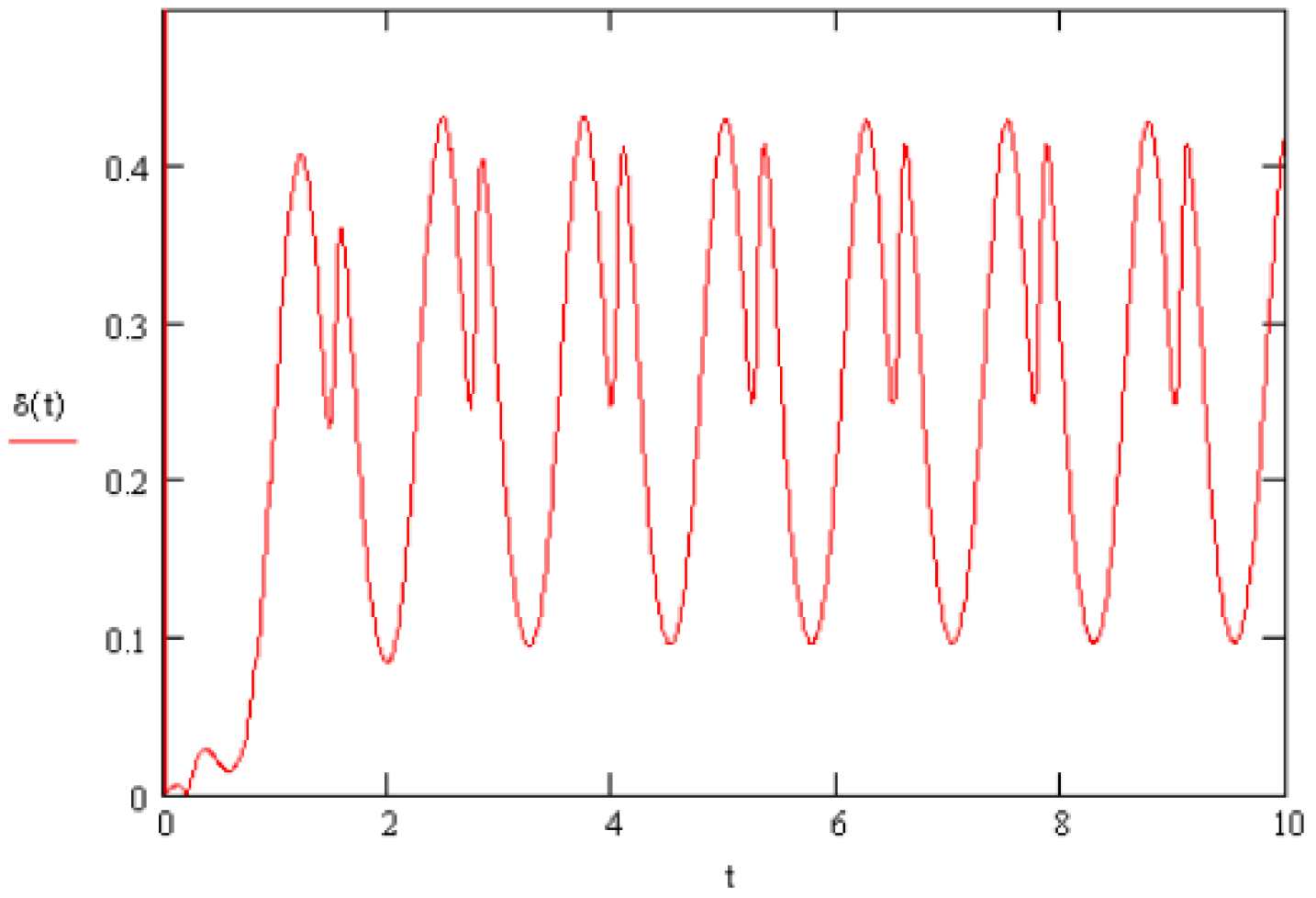

**Fig. 27.** $a = 1, b = 1, c = 0, A = 2.5, B = 0, \Omega = 5, x_0 = 0.$

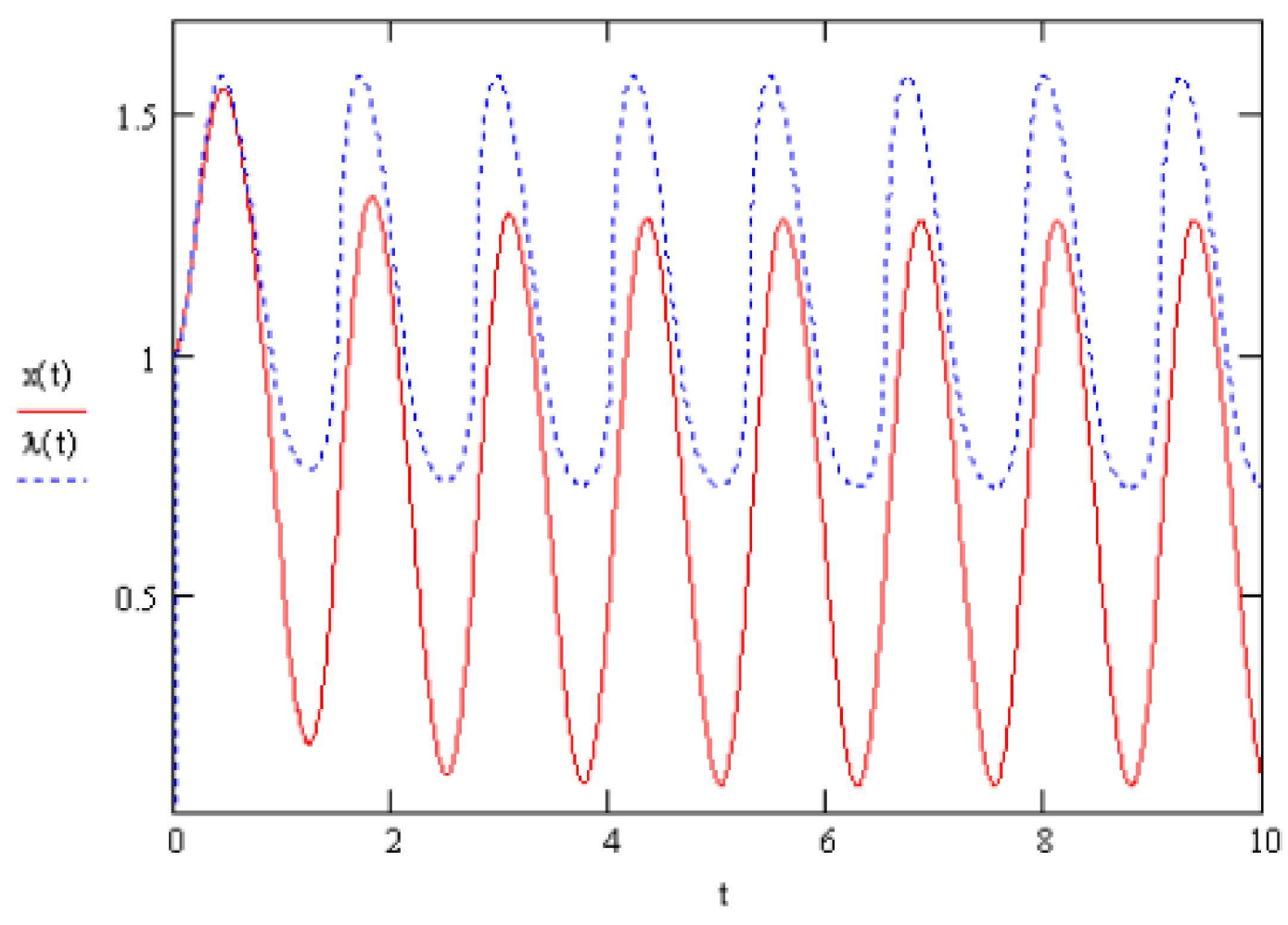

**Fig. 28.** Comparison of classical (red curve) and stochastic (blue curve) dynamics calculated by using **SLDP**.

$a = 1, b = 1, c = 0, A = 3, B = 0, \Omega = 5, x_0 = 0.$

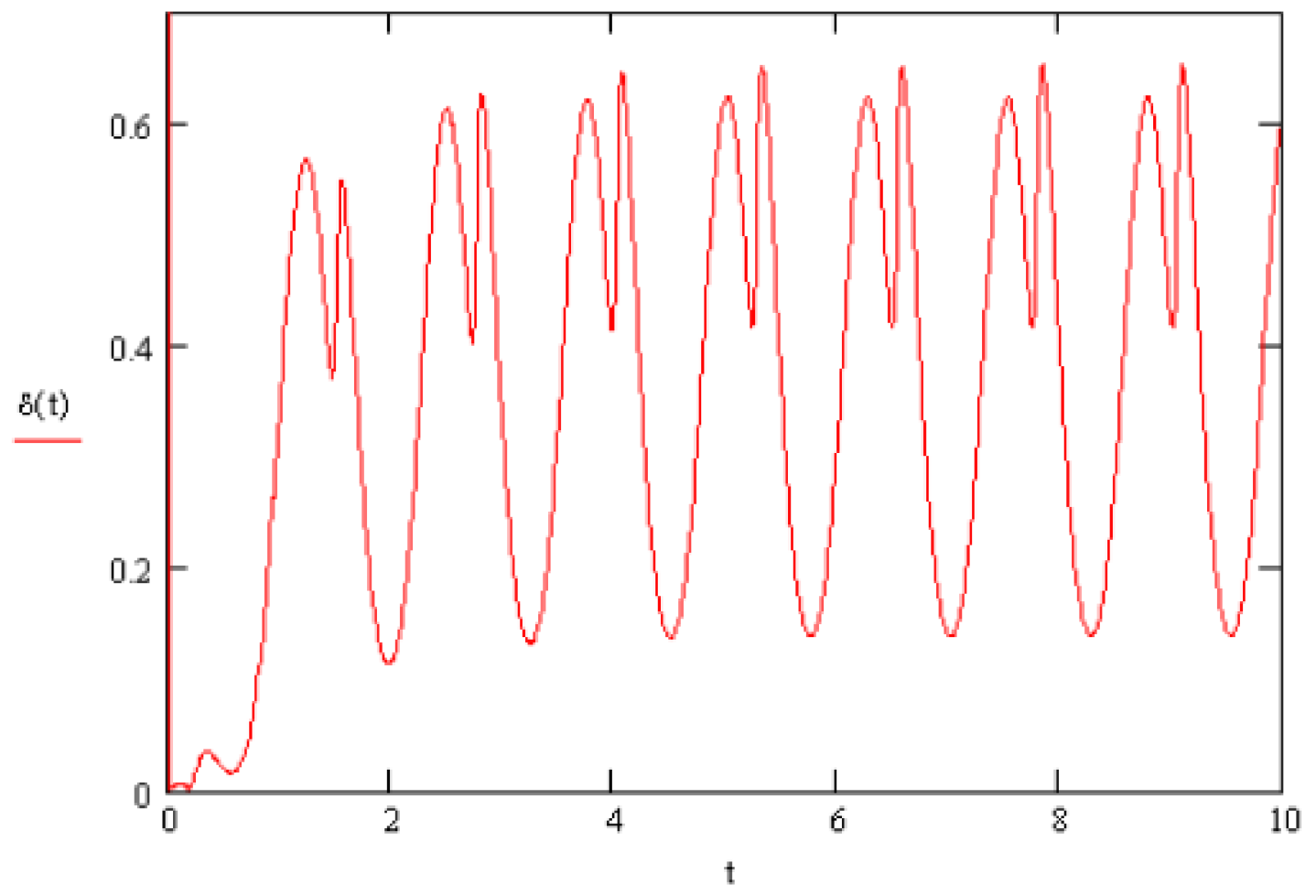

**Fig. 29.** $a = 1, b = 1, c = 0, A = 3, B = 0, \Omega = 5.$

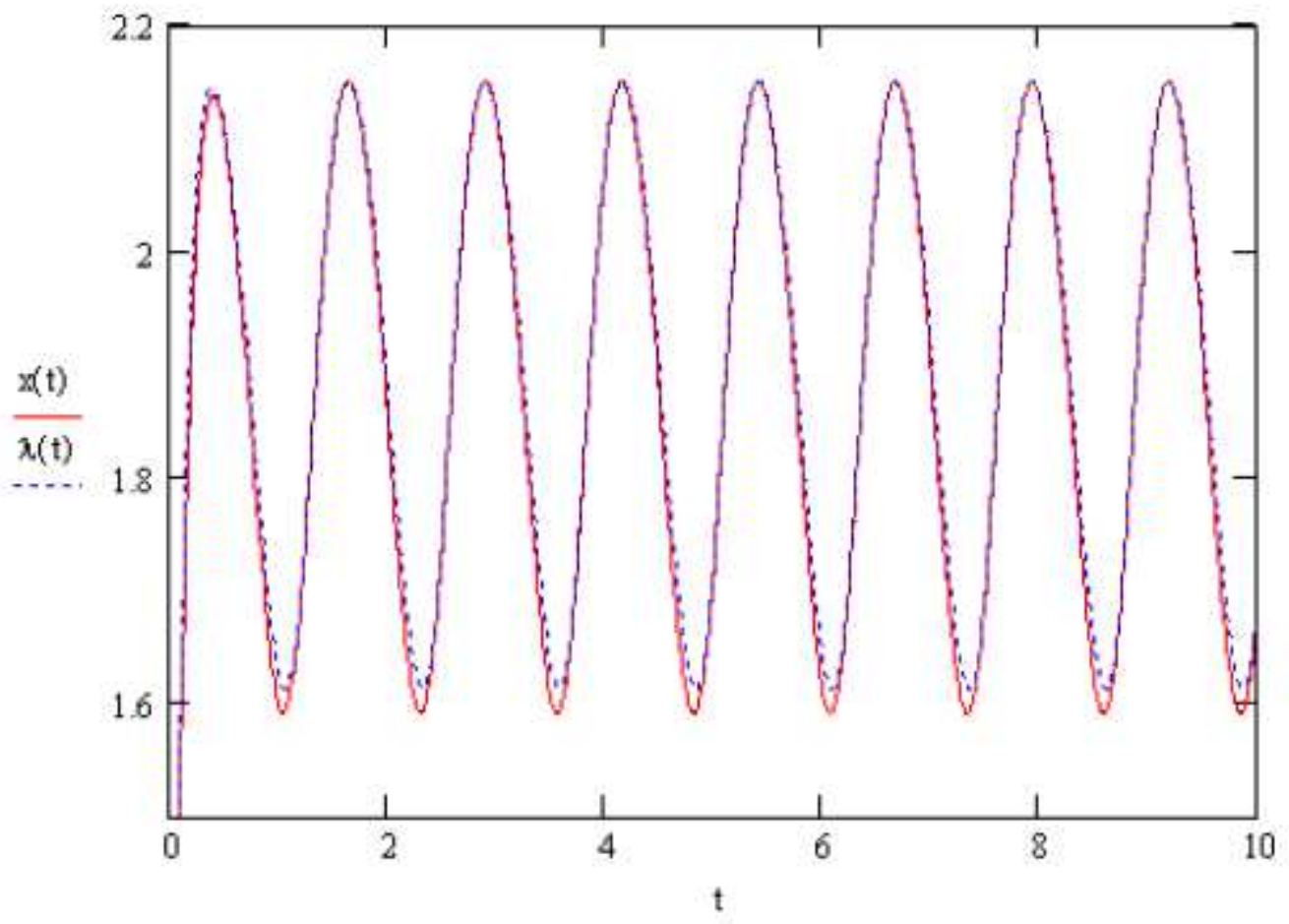

**Fig. 30.** Comparison of classical (red curve) and limiting stochastic (blue curve) dynamics calculated by using **SLDP** : $a = 1, b = 1, c = 5, A = 3, B = 0, \Omega = 5, x_0 = 0.$

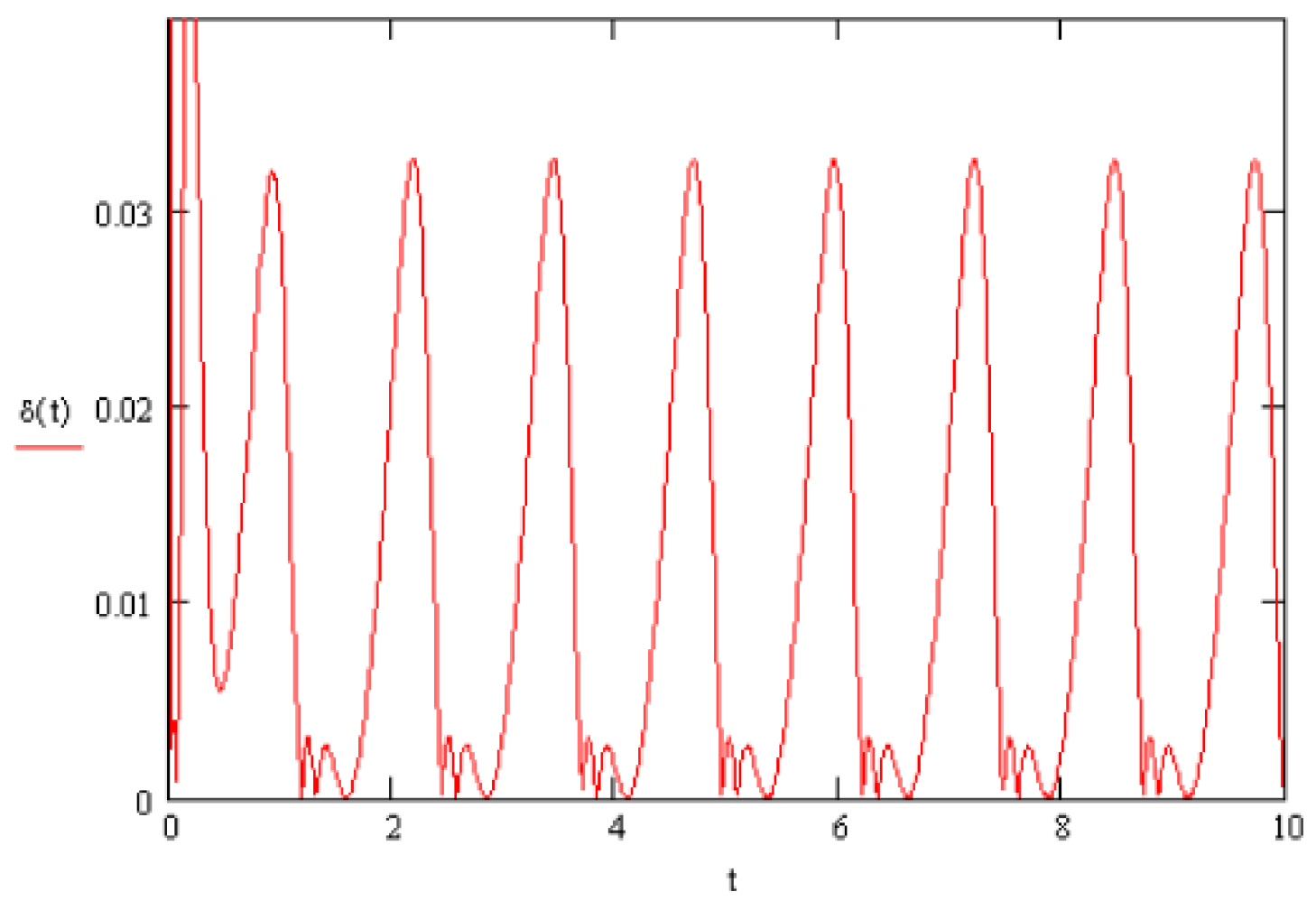

**Fig. 31.** $a = 1, b = 1, c = 5, A = 3, B = 0, \Omega = 5.$

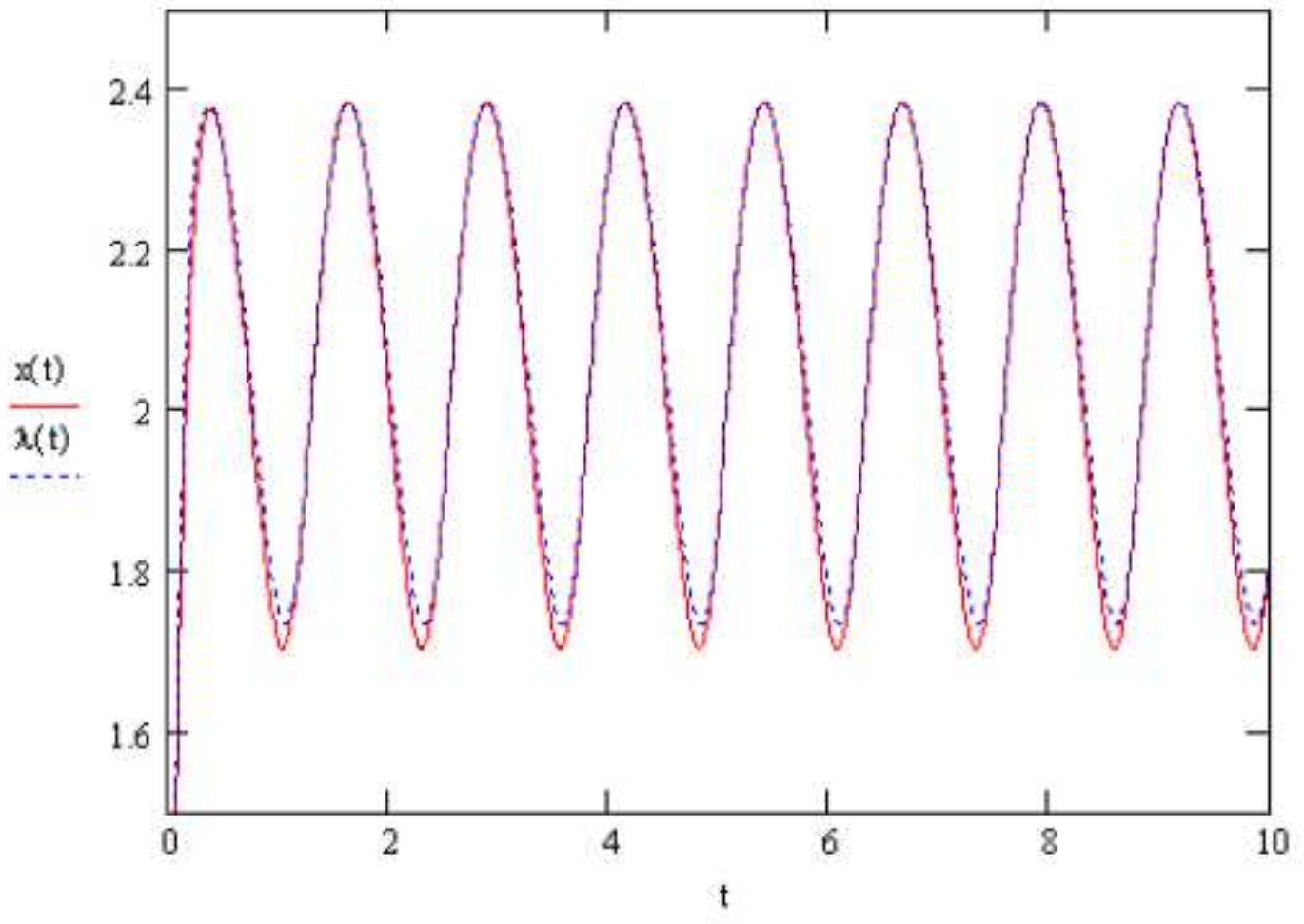

**Fig. 32.** Comparison of classical (red curve) and limiting stochastic (blue curve) dynamics calculated by using **SLDP** : $a = 1, b = 2, c = 5, A = 4, B = 0, \Omega = 5, x_0 = 0.$

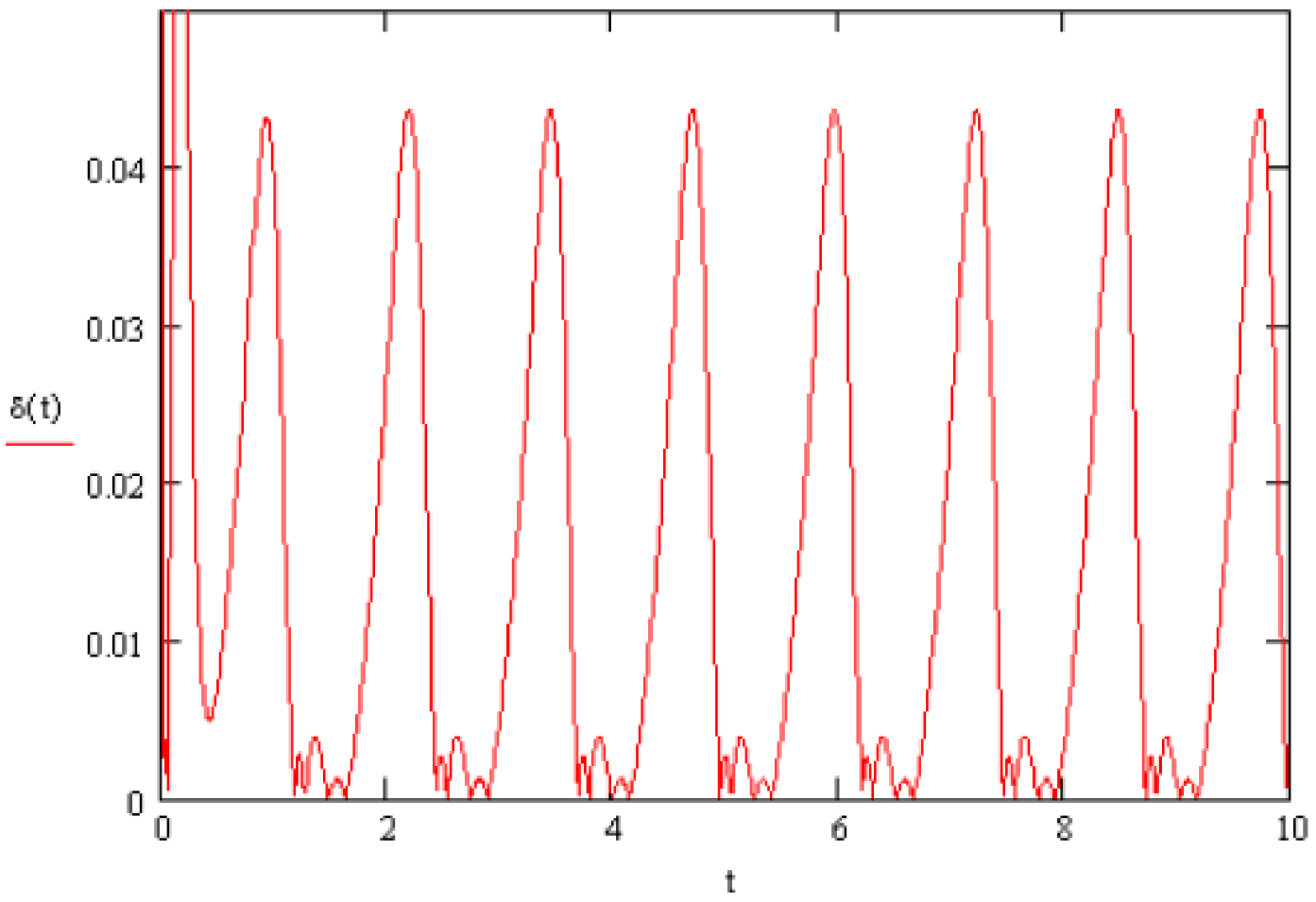

**Fig. 33.** $a = 1, b = 2, c = 5, A = 4, B = 0, \Omega = 5.$

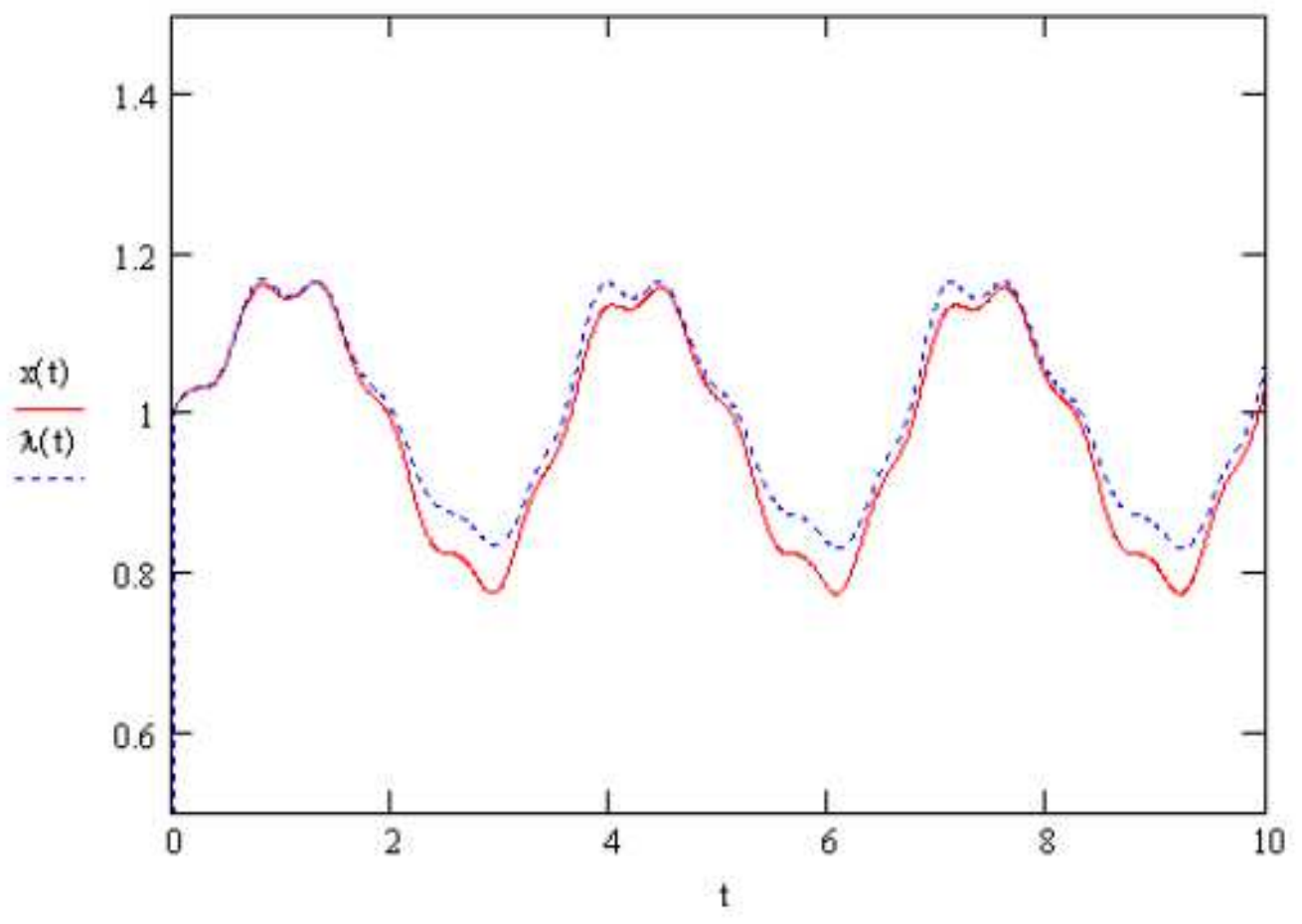

**Fig. 34.** Comparison of classical (red curve) and limiting stochastic (blue curve) dynamics calculated by using **SLDP** : $a = 1, b = 1, c = 0, A = 0.5, B = 0.2, \Omega = 2,$ $\Theta = 10,\ x_0 = 0.$

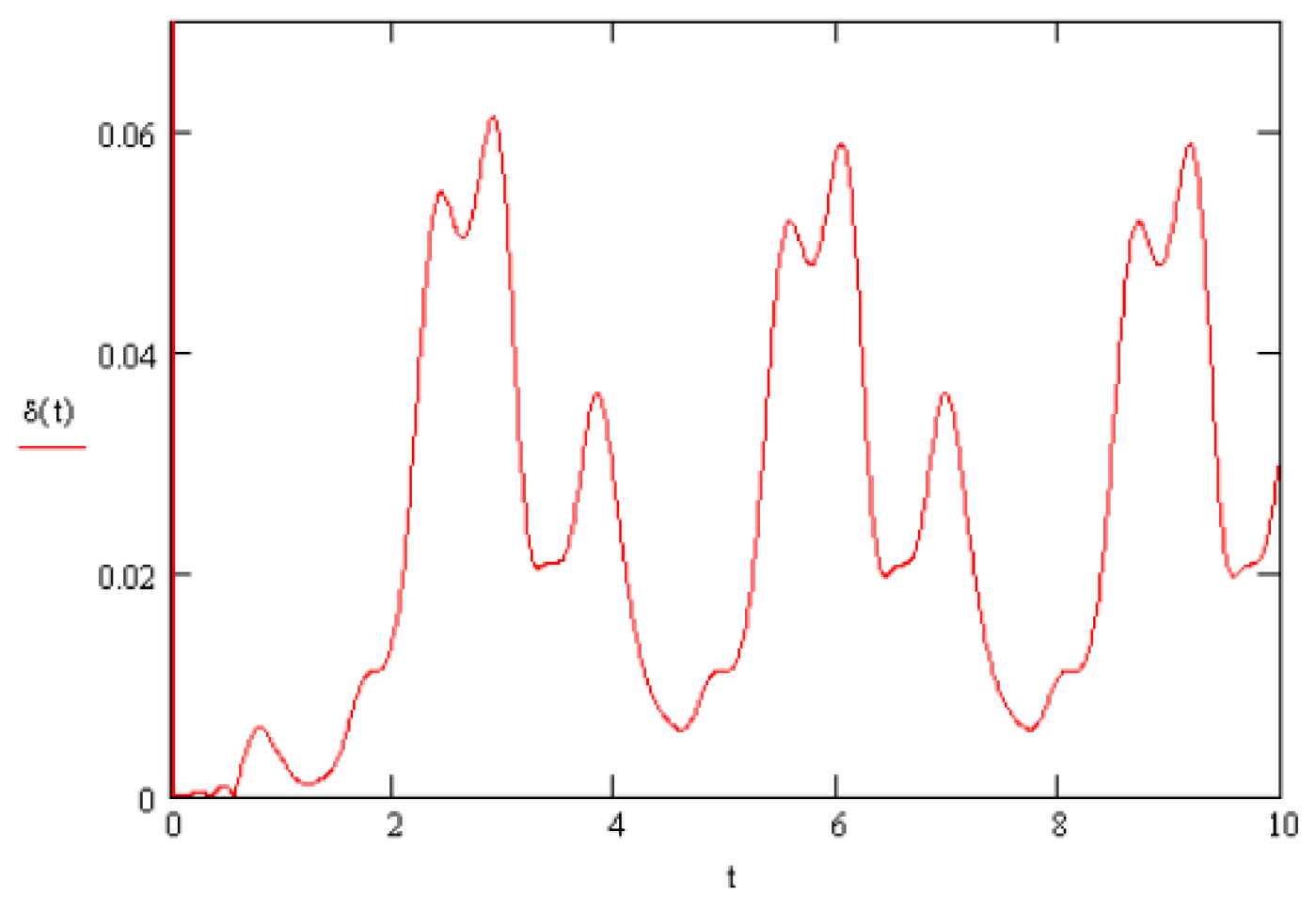

**Fig. 35.** $a = 1, b = 1, c = 0, A = 0.5, B = 0.2, \Omega = 2,$ $\Theta = 10,\ x_0 = 0.$

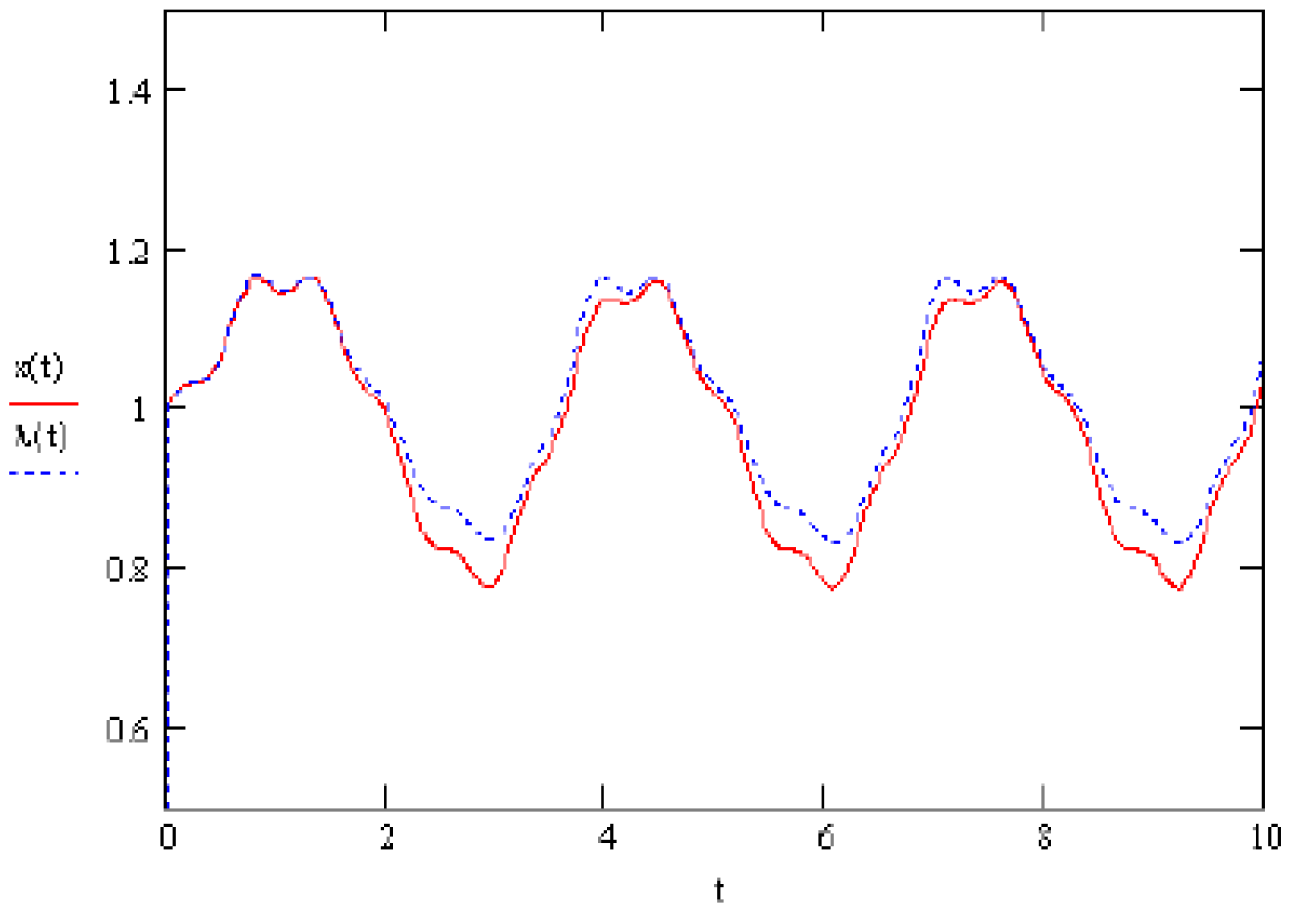

**Fig. 36.** Comparison of classical (red curve) and limiting stochastic (blue curve) dynamics calculated by using **SLDP** : $a = 1, b = 1, c = 0, A = 0.5, B = 0.3, \Omega = 2,$
$\Theta = 20,\ x_0 = 0.$

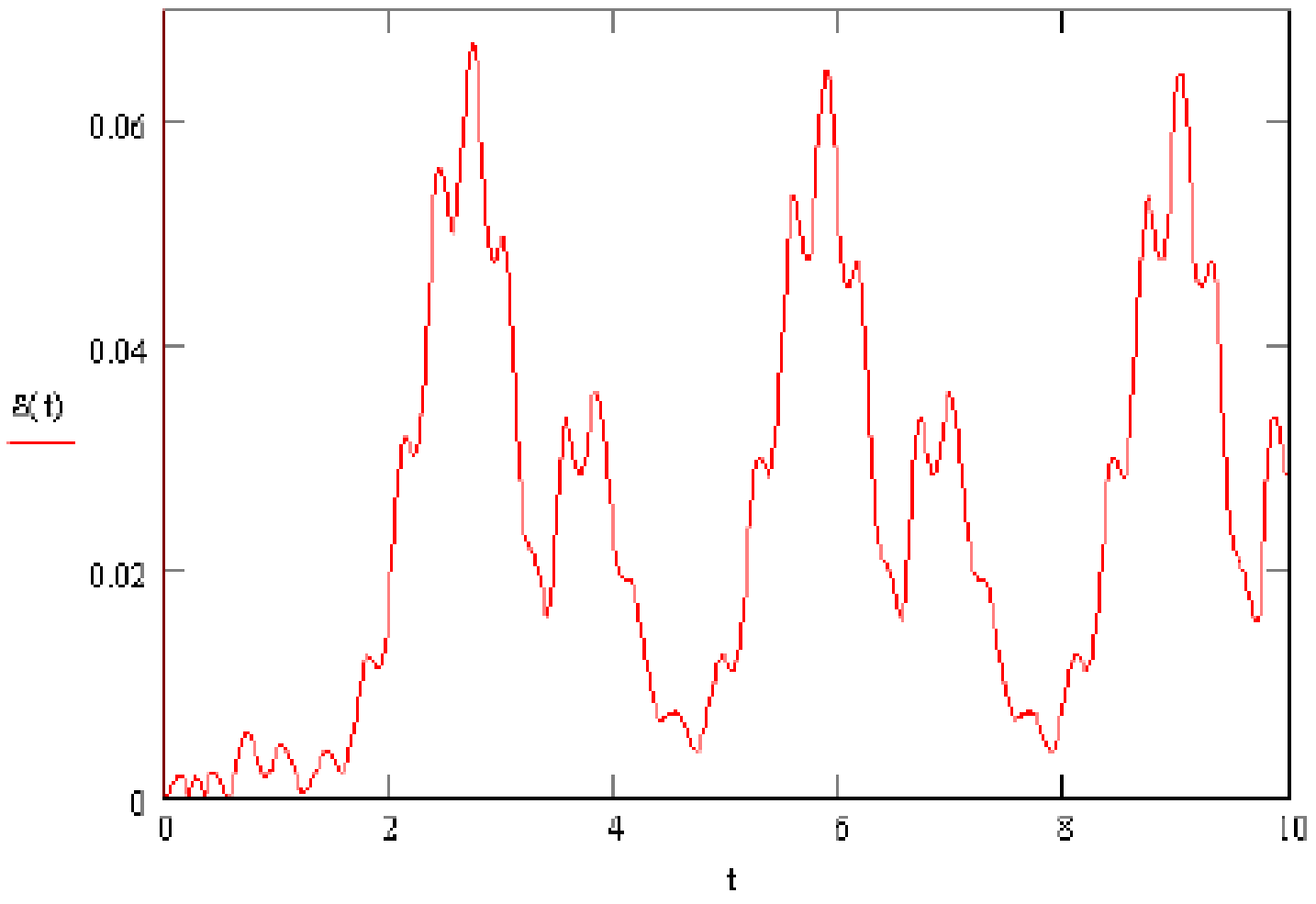

**Fig. 37.** $a = 1, b = 1, c = 0, A = 0.5, B = 0.3, \Omega = 2,$
$\Theta = 20,\ x_0 = 0.$

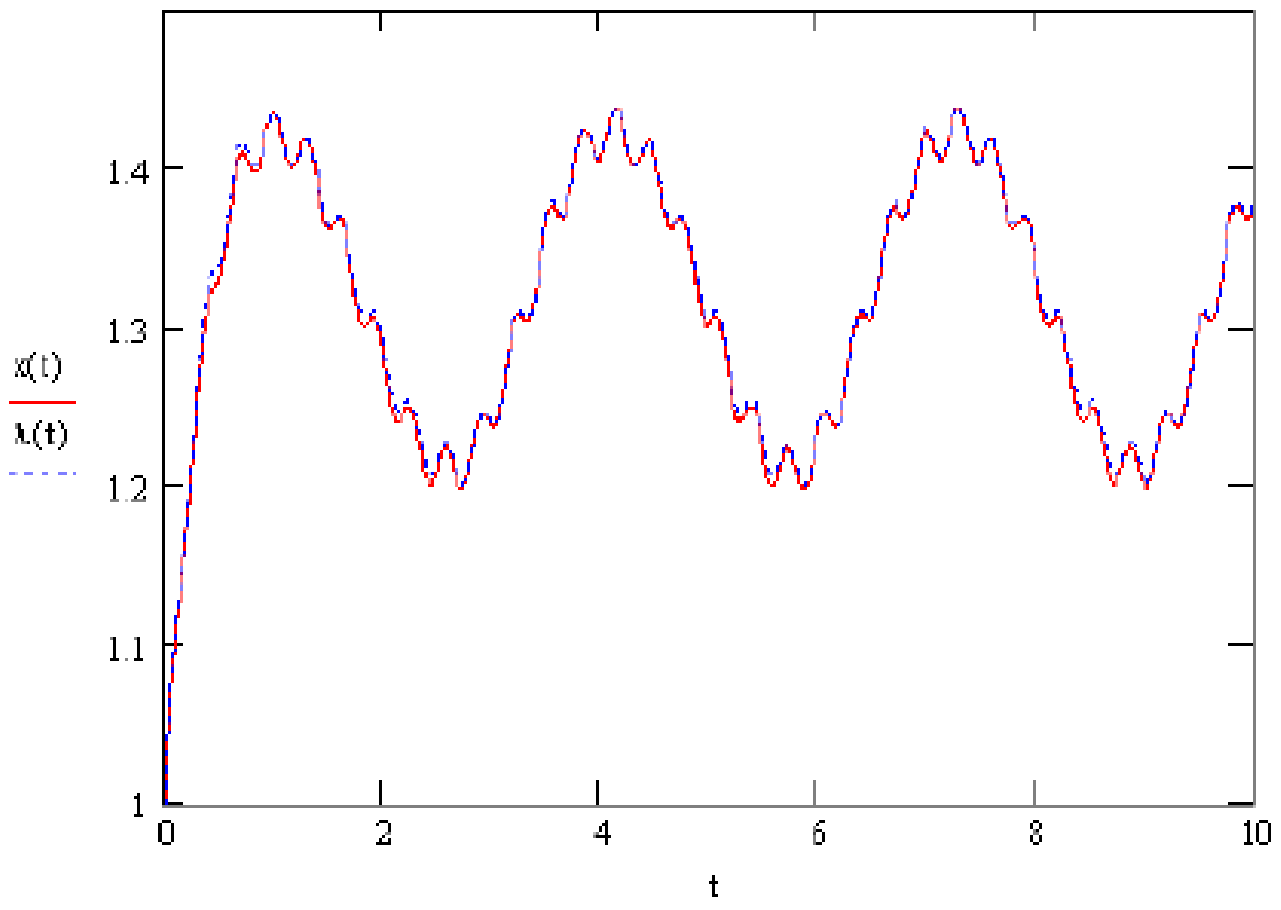

**Fig. 38**. Comparison of classical (red curve) and limiting stochastic (blue curve) dynamics calculated by using **SLDP** : $a = 1, b = 1, c = 1, A = 0.5, B = 0.3, \Omega = 2,$ $\Theta = 20,\ x_0 = 0.$

---

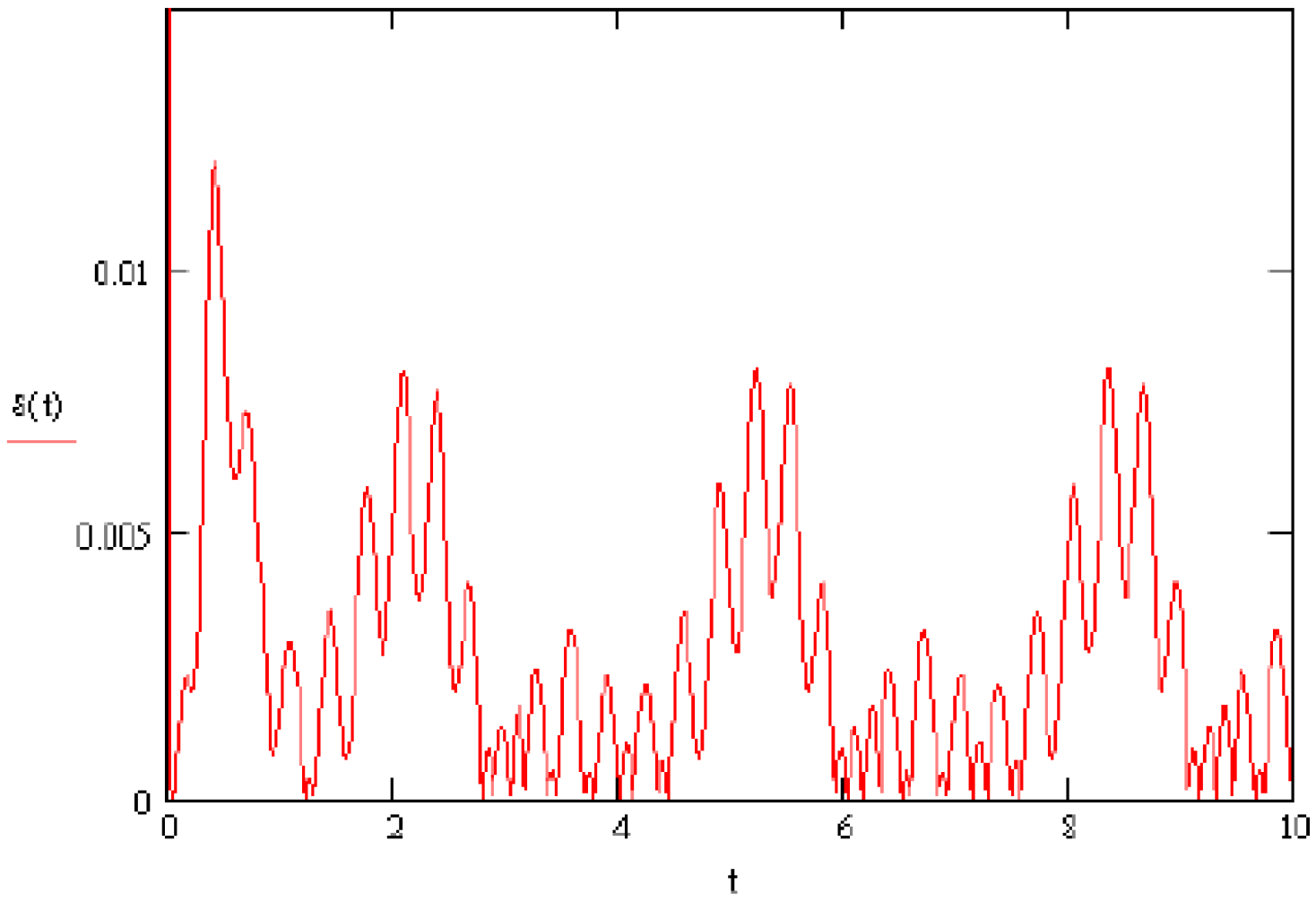

**Fig. 39**. $a = 1, b = 1, c = 1, A = 0.5, B = 0.3, \Omega = 2,$ $\Theta = 20,\ x_0 = 0.$

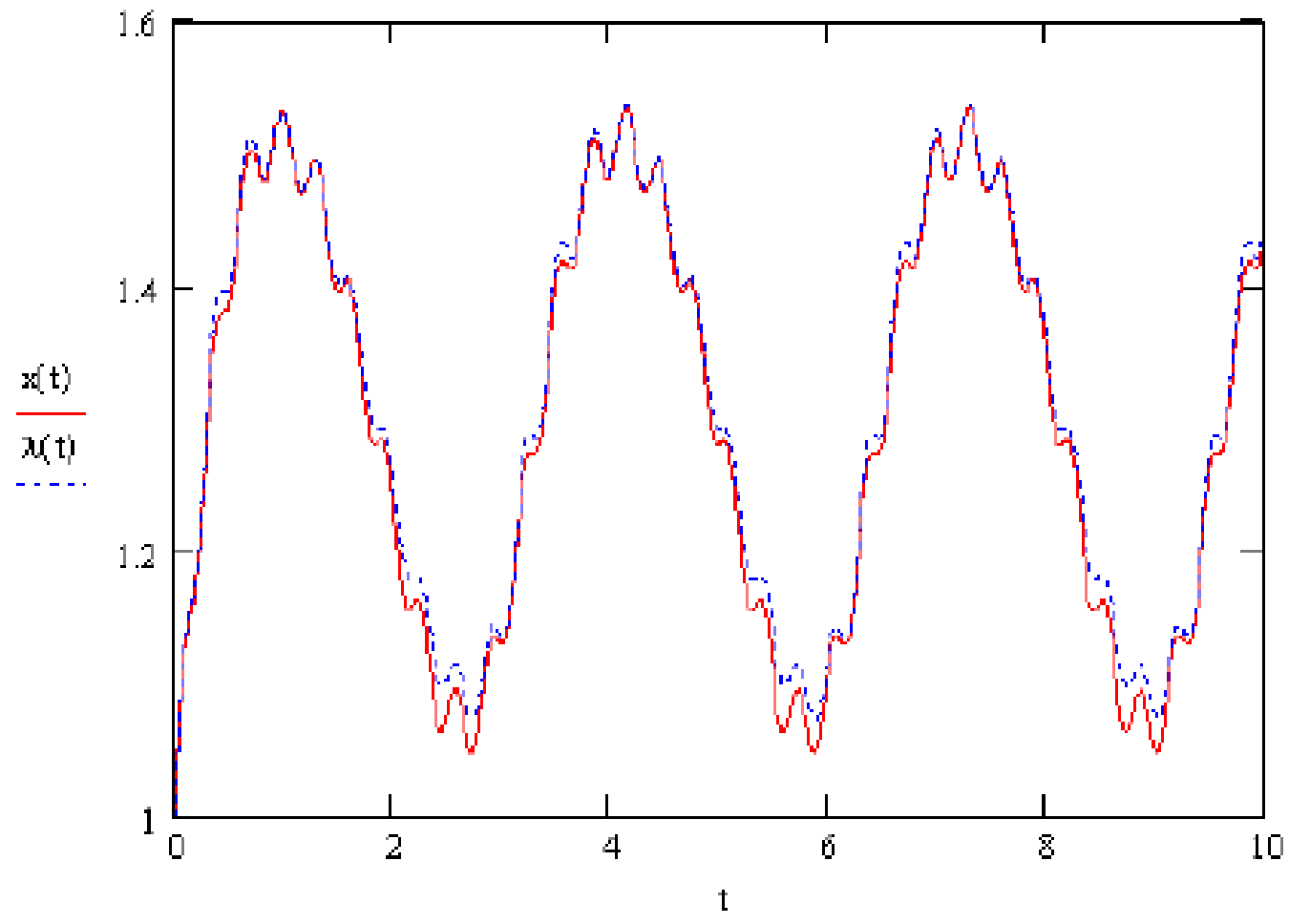

**Fig. 40**. Comparison of classical (red curve) and limiting stochastic (blue curve) dynamics calculated by using **SLDP** : $a = 1, b = 1, c = 1, A = 1, B = 0.5, \Omega = 2,$ $\Theta = 20,\ x_0 = 0.$

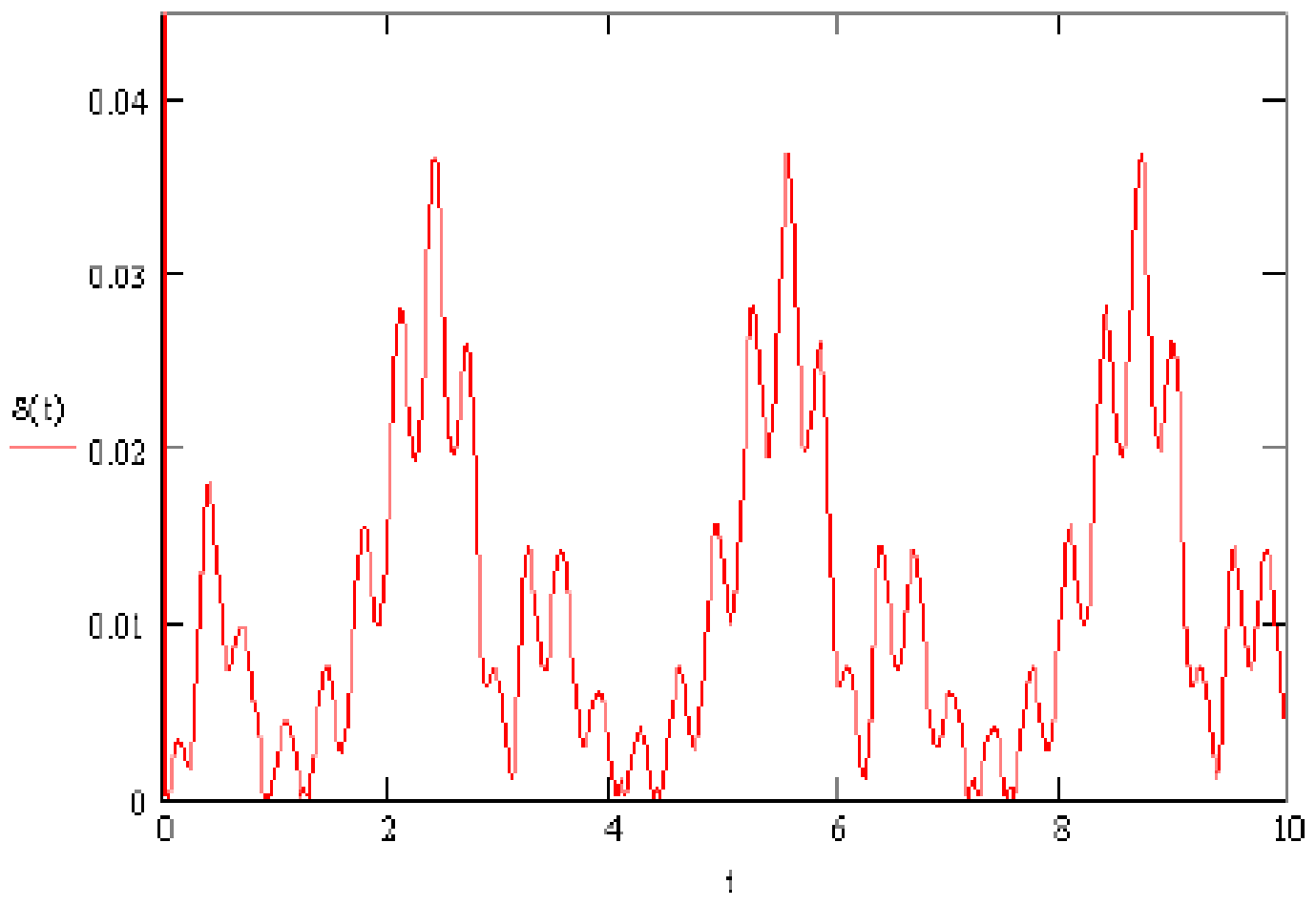

**Fig. 41**. $a = 1, b = 1, c = 1, A = 1, B = 0.5, \Omega = 2,$ $\Theta = 20,\ x_0 = 0.$

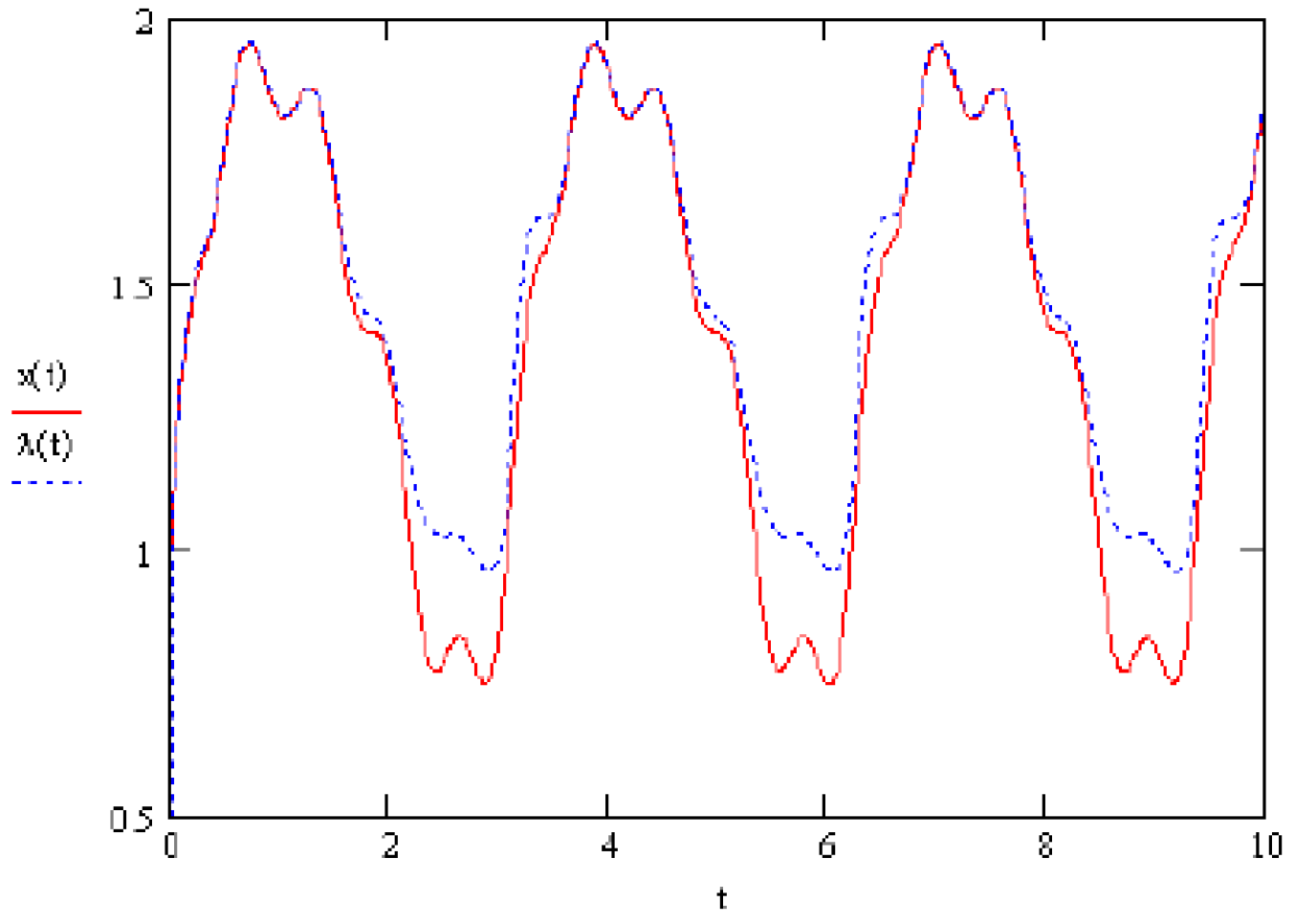

**Fig. 42.** Comparison of classical (redcurve) and limiting stochastic (bluecurve) dynamics calculated by using **SLDP** : $a = 1, b = 1, c = 2, A = 3, B = 1, \Omega = 2,$ $\Theta = 10,\ x_0 = 0.$

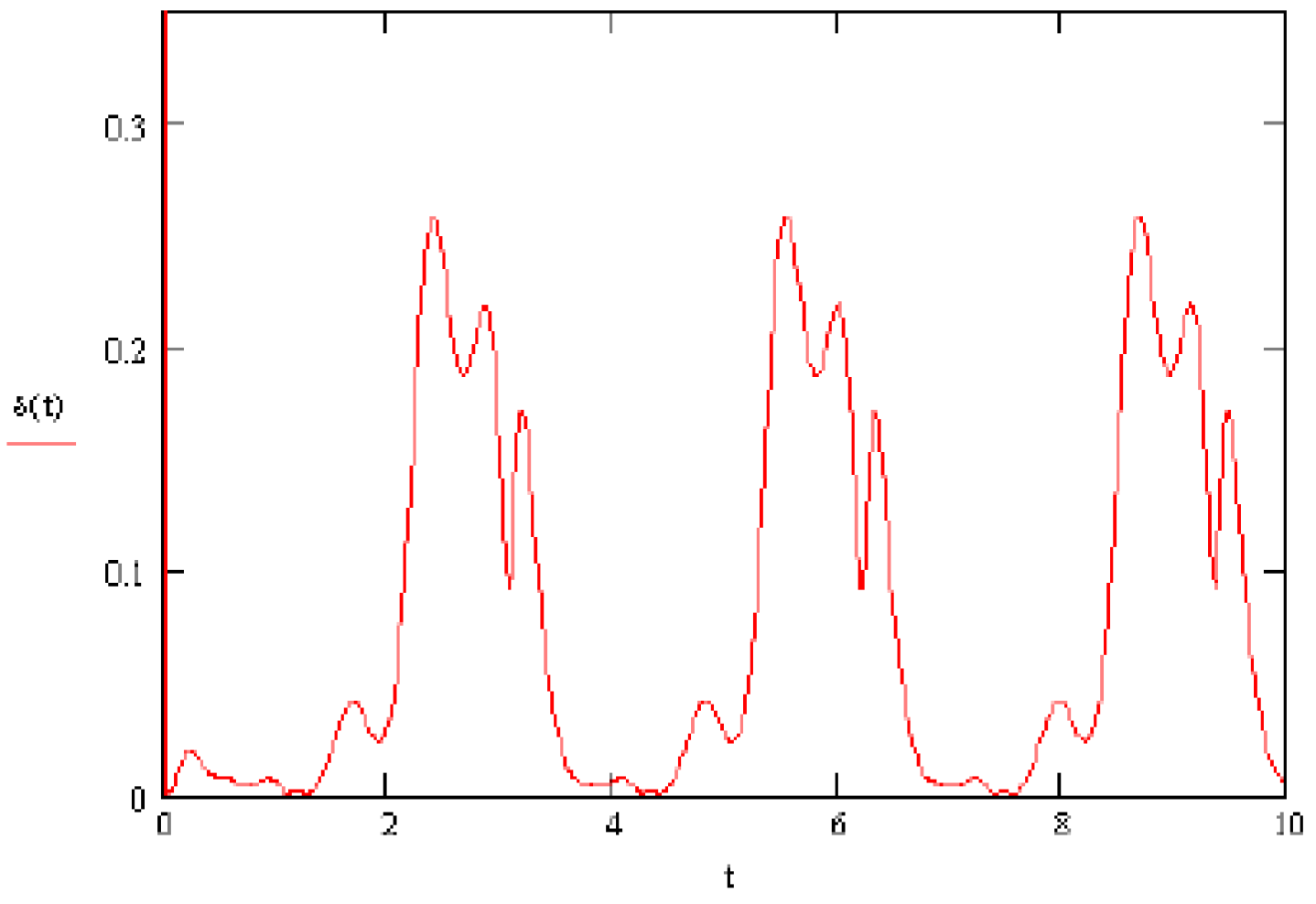

**Fig. 43**. $a = 1, b = 1, c = 1, A = 3, B = 1, \Omega = 2,$

$\Theta = 10,\ x_0 = 0.$

**Examples**.
The stochastic dynamics (2.4.1) we take in the next form

$$\dot{x}(t) = \mathbf{F}(x(t), t) + \sqrt{D}\,\dot{\mathbf{W}}(\omega, t) + \sqrt{\varepsilon}\ w(\varpi, t), \tag{2.4.14}$$

The force field $\mathbf{F}(x,t)$ in (2.4.1) is assumed to derive from a metastable potential which undergoes an arbitrary periodic modulation in time with period $\tau$:

$$\mathbf{F}(x, t + \mathcal{T}) = \mathbf{F}(x, t)\ . \tag{2.4.15}$$

An examples is a static potential $V(x)$, supplemented by an additive sinusoidal and more general driving. The random time-dependent force field $\tilde{\mathbf{F}}(\omega, x, t)$ takes the following form:

$$\tilde{\mathbf{F}}(x,t) = -V'(x) + A\sin(\Omega t) + B\cos(\Theta t) + \sqrt{D}\,\dot{W}(\omega,t) + \sqrt{\varepsilon}\ \dot{w}(\varpi,t),$$

$$\Omega = 2\pi/\mathcal{T}\ . \tag{2.4.16}$$

**2.4.3 Cubic potential**.

As a first example we consider the random force field in (2.4.1) with a cubic metastable potential $V(x)$ as cartooned in Fig.1,

$$V(x) = -\frac{a}{3}x^3 + \frac{b}{2}x^2\ ,\ \ a,b > 0\ . \tag{2.4.17}$$

The time-dependent force field in (2.4.1) takes the following form:

$$\mathbf{F}(x,t) = ax^2 - bx + A\sin(\Omega t) + B\cos(\Theta t)\ . \tag{2.4.18}$$

The stochastic dynamics (2.4.1) takes the following form:

$$\dot{x}(t) = ax^2 - bx + A\sin(\Omega t) + B\cos(\Theta t) + \sqrt{D}\,\dot{W}(\omega,t) + \sqrt{\varepsilon}\ \dot{w}(\varpi t),$$

$$x(0) = x_0\ . \tag{2.4.19}$$

From master equation $(2.3.9)$ we obtain the linear differential master equation

$$\dot{u} = (2a\lambda - b)u + a\lambda^2 - b\lambda + A\sin(\Omega t) + B\cos(\Theta t) + \sqrt{D}\,\dot{W}(\omega,t),$$

$$u(0) = x_0 - \lambda\ . \tag{2.4.20}$$

From Theorem 2.3.4 we have the next transcendental master equation

$$(x_0 - \lambda(t))\exp[(2a\lambda(t) - b)t] +$$

$$(a\lambda^2(t) - b\lambda(t))\int_0^t \exp[(2a\lambda(t) - b)(t-\tau)]d\tau + \qquad (2.4.21)$$

$$\int_0^t \left[A\sin(\Omega\tau) + B\cos(\Theta t) + \sqrt{D}\,\dot{W}(\omega,\tau)\right]\exp[(2a\lambda(t) - b)(t-\tau)]d\tau = 0.$$

$$\int_0^t \exp[(2a\lambda(t) - b)(t-\tau)]dW(\tau) = [W(\tau)\exp[(2a\lambda(t) - b)(t-\tau)]]_0^t -$$

$$\int_0^t W(\tau)d\exp[(2a\lambda(t) - b)(t-\tau)] =$$

$$(2.4.22)$$

$$W(t) - \int_0^t W(\tau)(-(2a\lambda(t) - b))\exp[(2a\lambda(t) - b)(t-\tau)]d\tau =$$

$$W(t) + (2a\lambda(t) - b)\int_0^t W(\tau)\exp[(2a\lambda(t) - b)(t-\tau)]d\tau.$$

From Eq.(2.4.21)-Eq.(2.4.22) we obtain the stochastic transcendental master equation

$$(x_0 - \lambda(t))\exp[(2a\lambda(t) - b)t] +$$

$$(a\lambda^2(t) - b\lambda(t))\int_0^t \exp[(2a\lambda(t) - b)(t-\tau)]d\tau +$$

$$\int_0^t \left[A\sin(\Omega\tau) + B\cos(\Theta t) + \sqrt{D}\,(2a\lambda(t) - b)W(\tau)\right] \times \tag{2.4.23}$$

$$\exp[(2a\lambda(t) - b)(t-\tau)]d\tau +$$

$$\sqrt{D}\,W(t).$$

**Comparison of classical and non-perturbative stochastic dynamics**.

We have compared by $\delta(\omega,t) = |x(\omega,t) - \lambda(\omega,t)|$ the above analytical predictions for the limit (2.4.12) with very accurate numerical results for classical $x(t)$ and limiting stochastic dynamics.

**2.4.4.Duble well potential**.

As a second example we consider a random force field with a duble wel potential

$$V(x) = \frac{a}{4}x^4 - \frac{b}{2}x^2 - cx + \sqrt{D}\,\dot{W}(\omega,t) + \sqrt{\varepsilon}\ w(\varpi t)\ ,\ \ a,b > 0\ . \tag{2.4.24}$$

The time-dependent force field takes the following form:

$$\mathbf{F}(x,t) = -ax^3 + bx + A\sin(\Omega t) + B\cos(\Theta t) + \\ + \sqrt{D}\,\dot{W}(\omega,t) + \sqrt{\varepsilon}\;w(\varpi t) + c\,. \qquad (2.4.25)$$

The stochastic dynamics (2.4.1) takes the following form:

$$\dot{x}(t) = -ax^3 + bx + A\sin(\Omega t) + \sqrt{D}\,\dot{W}(\omega,t) + \sqrt{\varepsilon}\;w(\varpi t)\;, x(0) = x_0. \qquad (2.4.26)$$

From master equation $(2.3.9)$ we obtain the next differential linear master equation

$$\dot{u}(t) = -(3a\lambda^2 - b)u(t) - a\lambda^3 + b\lambda + c + A\sin(\Omega t) + \sqrt{D}\,\dot{W}(\omega,t)\;, \\ u(0) = x_0 - \lambda. \qquad (2.4.27)$$

From Theorem 2.3.4 and master equation (2.4.27) we obtain the next transcendental master equation

$$(x_0 - \lambda(t))\exp[-(3a\lambda^2(t) - b)t] - \\ -(a\lambda^3(t) - b\lambda(t) - c)\int_0^t \exp[-(3a\lambda^2(t) - b)(t-\tau)]d\tau + \\ + A\int_0^t \sin(\Omega\tau)\exp[-(3a\lambda^2(t) - b)(t-\tau)]d\tau + \\ \sqrt{D}\int_0^t \exp[-(3a\lambda^2(t) - b)(t-\tau)]\dot{W}(\omega,\tau)d\tau = 0. \qquad (2.4.28)$$

Note that

$$\int_0^t \exp[-(3a\lambda^2(t) - b)(t-\tau)]d\tau W(\omega,\tau) =$$

$$\exp[-(3a\lambda^2(t) - b)(t-\tau)]W(\omega,\tau)|_0^t - \int_0^t W(\omega,\tau)d\tau \exp[-(3a\lambda^2(t) - b)(t-\tau)] = \qquad (2.4.29)$$

$$W(\omega,t) - W(\omega,0) - (3a\lambda^2(t) - b)\int_0^t W(\omega,\tau)\exp[-(3a\lambda^2(t) - b)(t-\tau)]d\tau.$$

Finally we obtain the next transcendental master equation

$$\begin{aligned}
&(x_0 - \lambda(t))\exp[-(3a\lambda^2(t) - b)t] - \\
&-(a\lambda^3(t) - b\lambda(t) - c)\int_0^t \exp[-(3a\lambda^2(t) - b)(t-\tau)]d\tau + \\
&+ A\int_0^t \sin(\Omega\tau)\exp[-(3a\lambda^2(t) - b)(t-\tau)]d\tau + \qquad (2.4.30)\\
&+ \sqrt{D}\,W(\omega,t) - \sqrt{D}\,(3a\lambda^2(t) - b)\int_0^t W(\omega,\tau)\exp[-(3a\lambda^2(t) - b)(t-\tau)]d\tau = 0.
\end{aligned}$$

### 2.4.5. Comparison of classical and non-perturbative stochastic dynamics.

Now we have compared by $\delta(\omega,t) = |x(\omega,t) - \lambda(\omega,t)|$ the above analytical predictions for the limit (2.4.12) with very accurate numerical results for classical $x(t)$ and limiting stochastic dynamics.

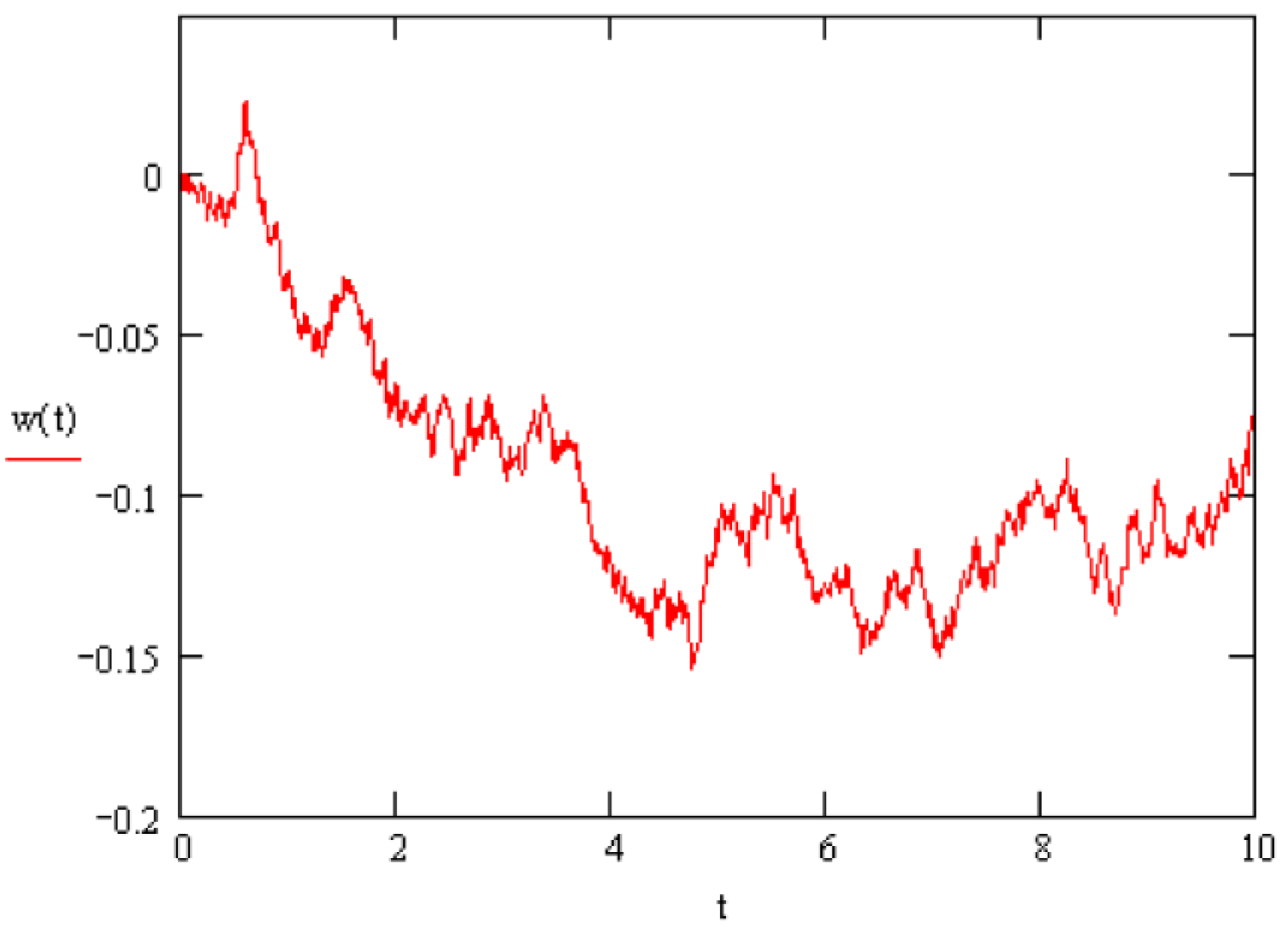

**Fig. 52**. The realization of a Wiener process $\mathbf{w}(t) = \sqrt{D}\,\mathbf{W}(t)$ where $\mathbf{W}(t)$ is a standard Wiener process, $D = 10^{-3}$.

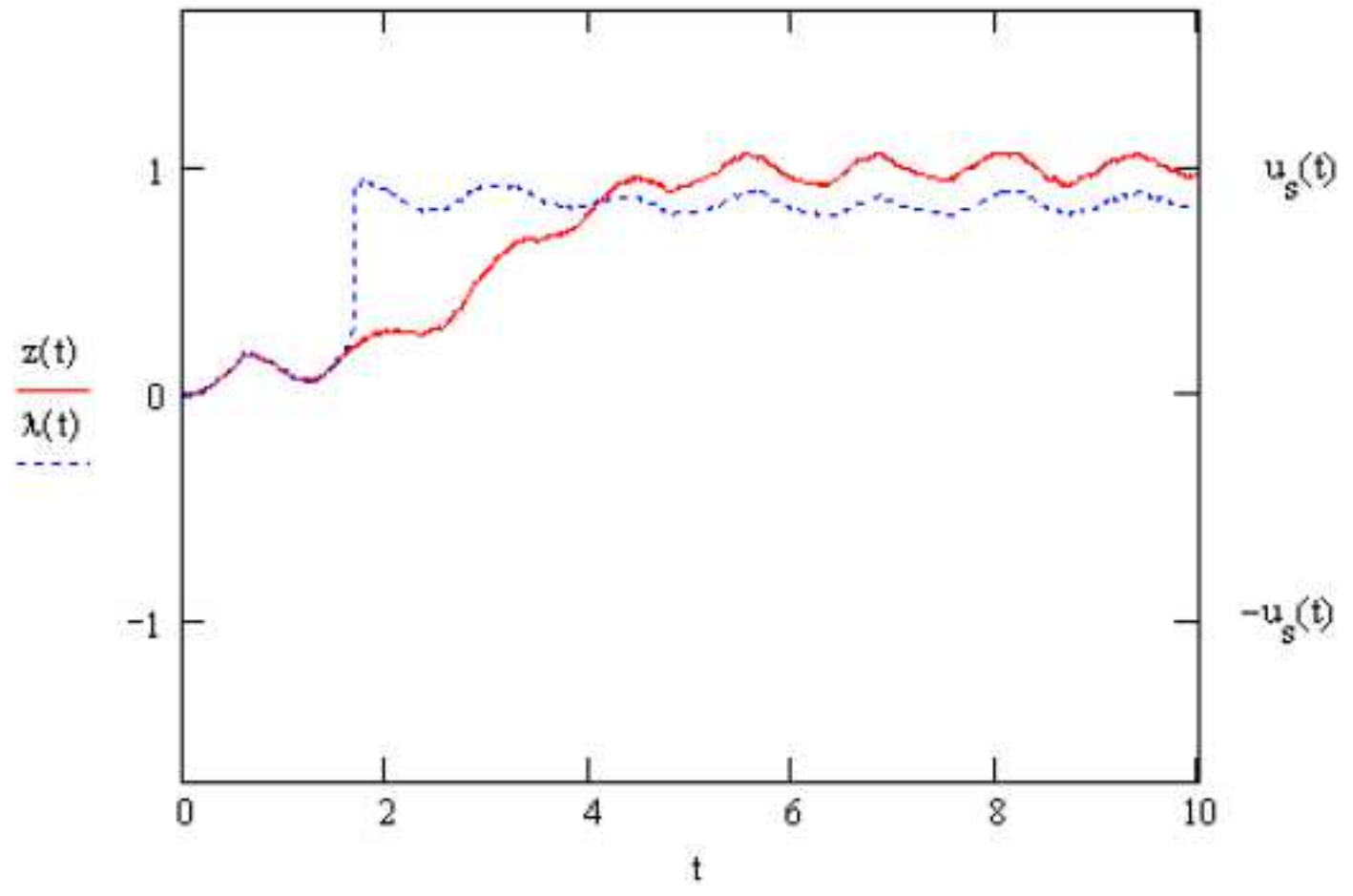

**Fig. 53**. Comparison of classical (red curve) and quasiclassical (blue curve) dynamics calculated from **SLDP**. $a = 1, b = 1, c = 0, A = 0.3, B = 0, \Omega = 5,$

$\Theta = 0, D = 10^{-3}, x_0 = 0.$

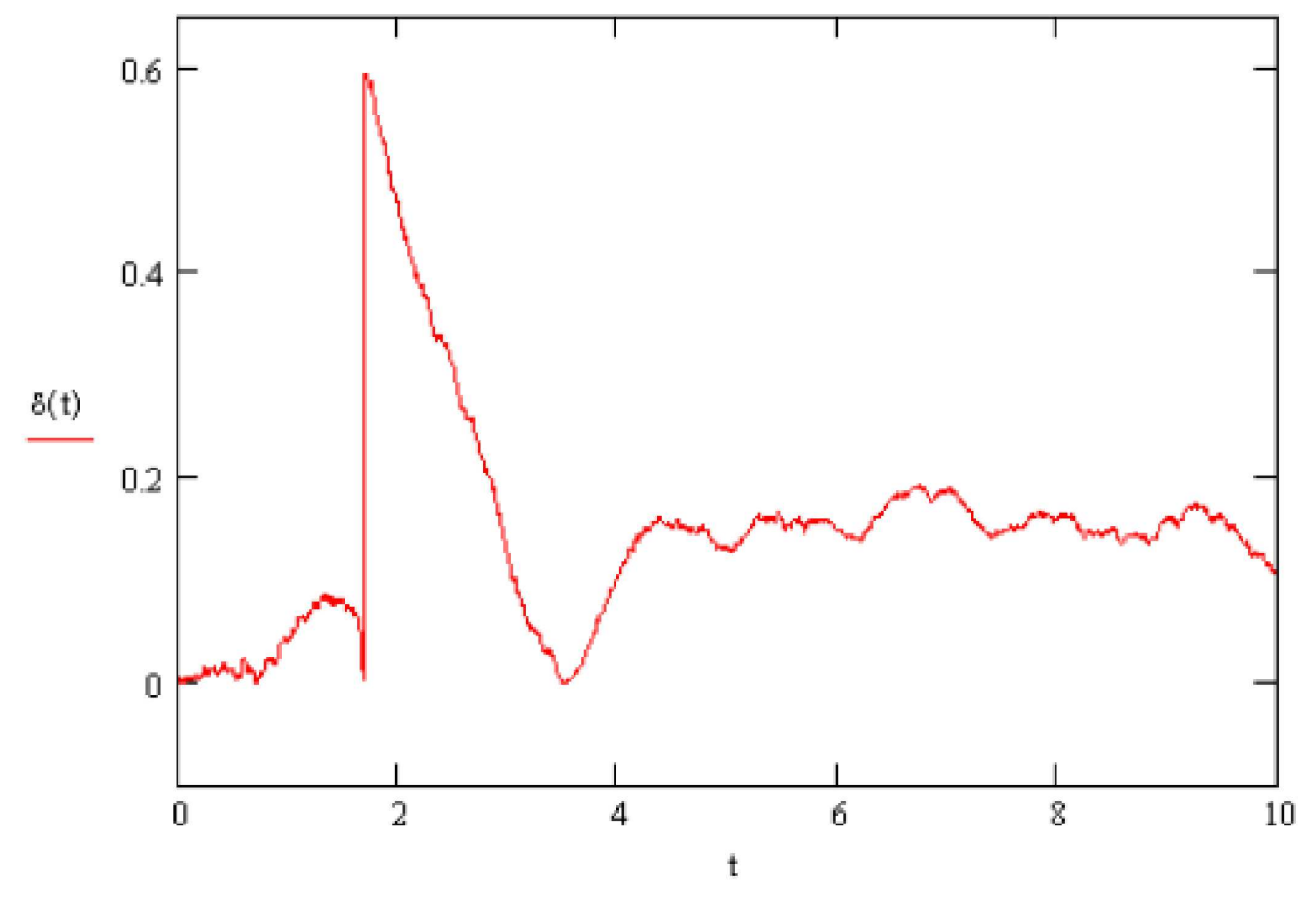

**Fig. 54.** Comparison by function $\delta(\omega, t)$ of a classical (red curve) and quasiclassical (blue curve) dynamics calculated from **SLDP**. $a = 1, b = 1, c = 0, A = 0.3, B = 0,$
$\Omega = 5, \Theta = 0, D = 10^{-3}, x_0 = 0.$

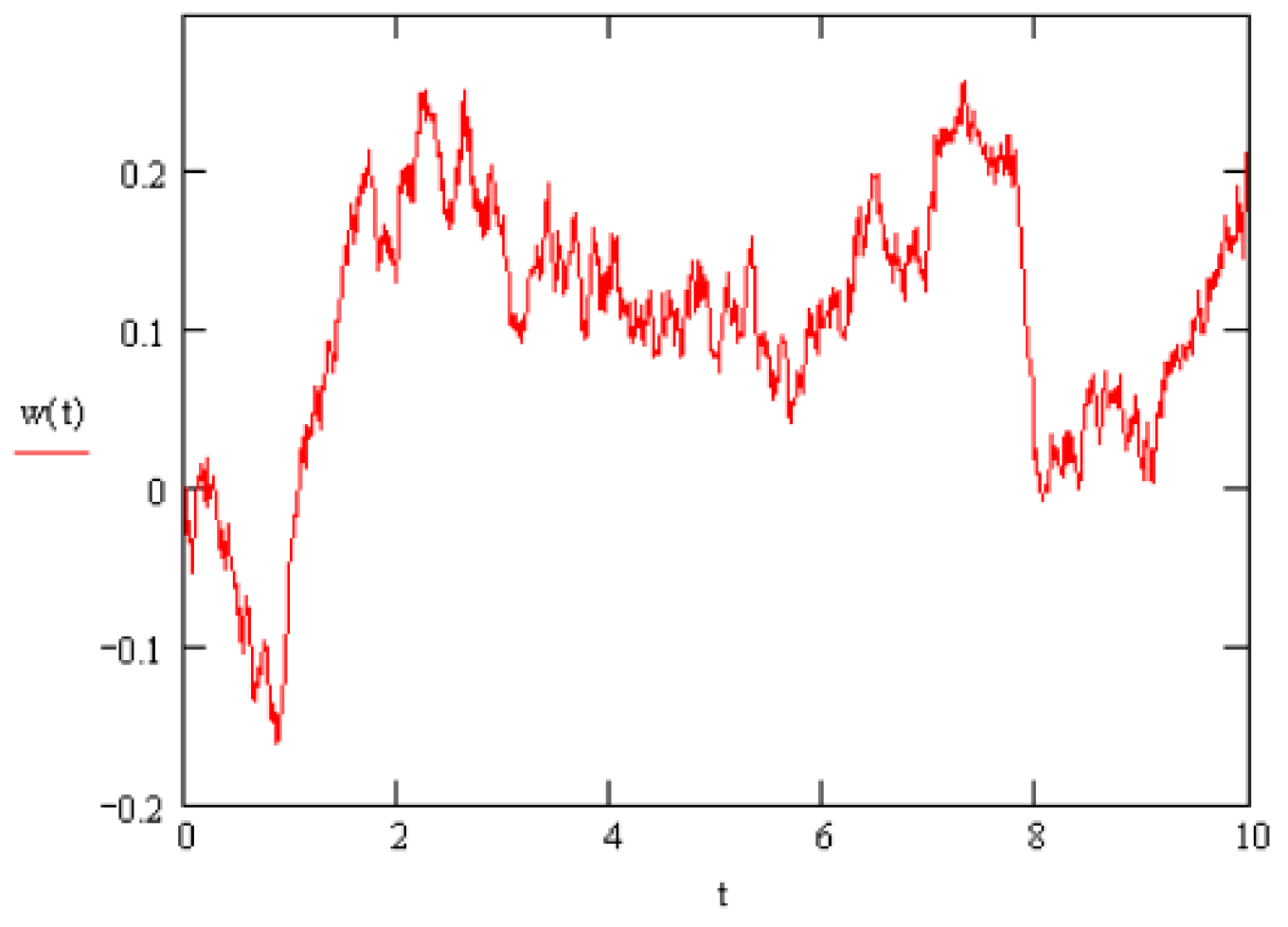

**Fig. 55.** The realization of a Wiener process $\mathbf{w}(t) = \sqrt{D}\,\mathbf{W}(t)$ where $\mathbf{W}(t)$ is a standard Wiener process, $D = 10^{-2}$.

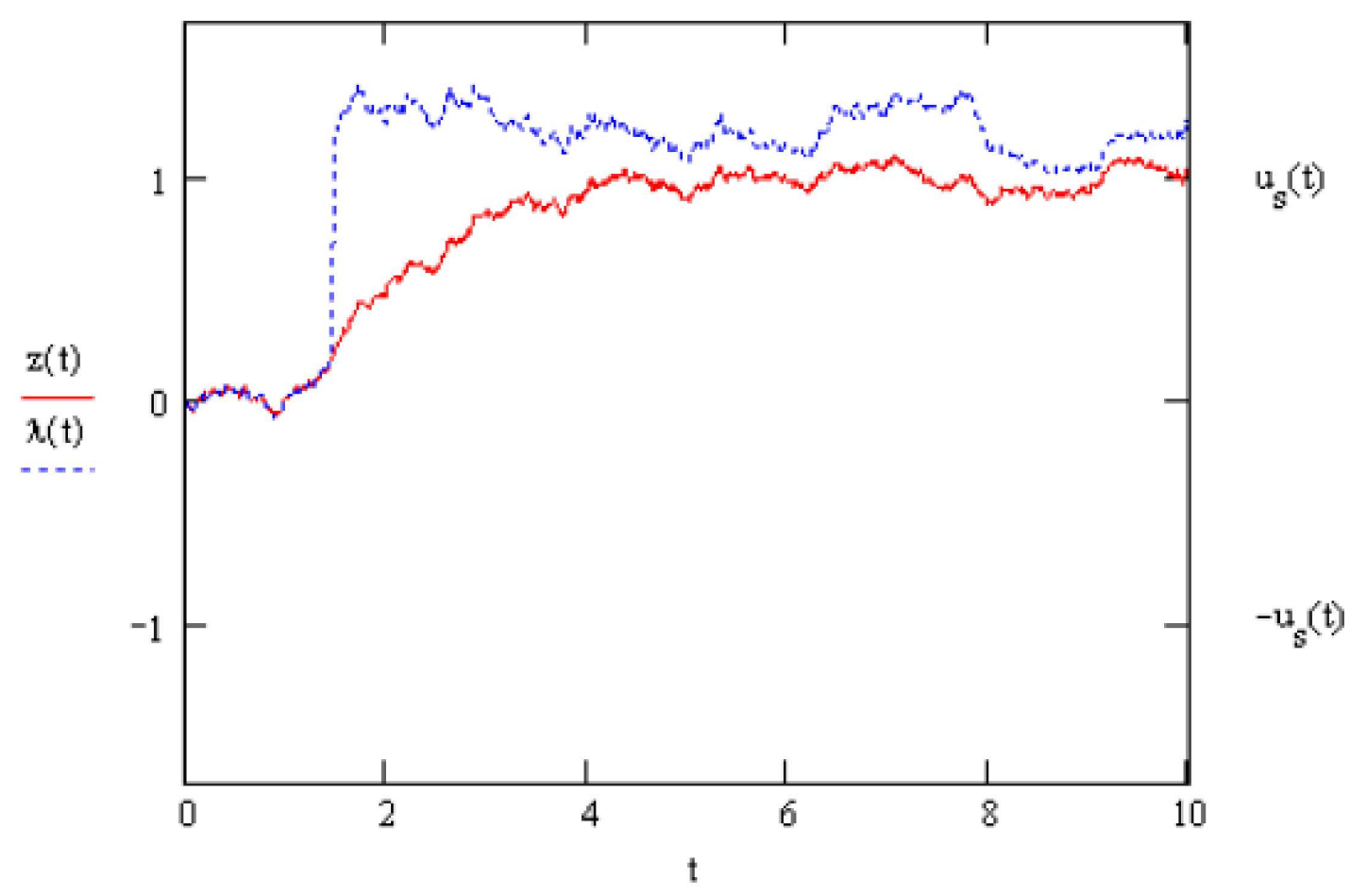

**Fig. 56.** Comparison of classical (red curve) and quasiclassical (blue curve) dynamics calculated from **SLDP**. $a = 1, b = 1, c = 0, A = 0.3, B = 0, \Omega = 5,$ $\Theta = 0, D = 10^{-2}, x_0 = 0.$

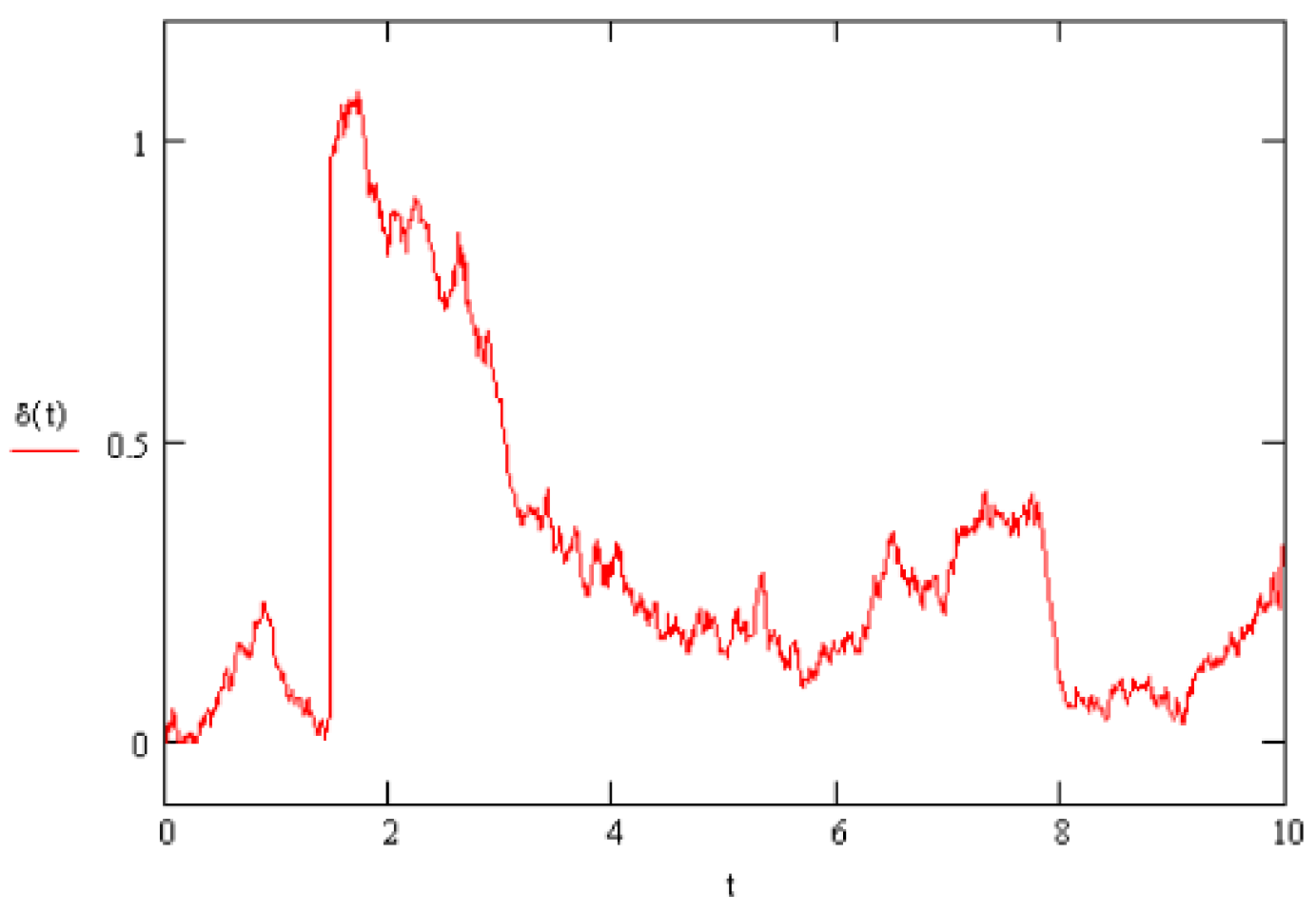

**Fig. 57.** Comparison by function $\delta(\omega, t)$ of a classical (red curve) and quasiclassical (blue curve) dynamics calculated from **SLDP**. $a = 1, b = 1, c = 0, A = 0.3, B = 0,$ $\Omega = 5, \Theta = 0, D = 10^{-2}, x_0 = 0.$

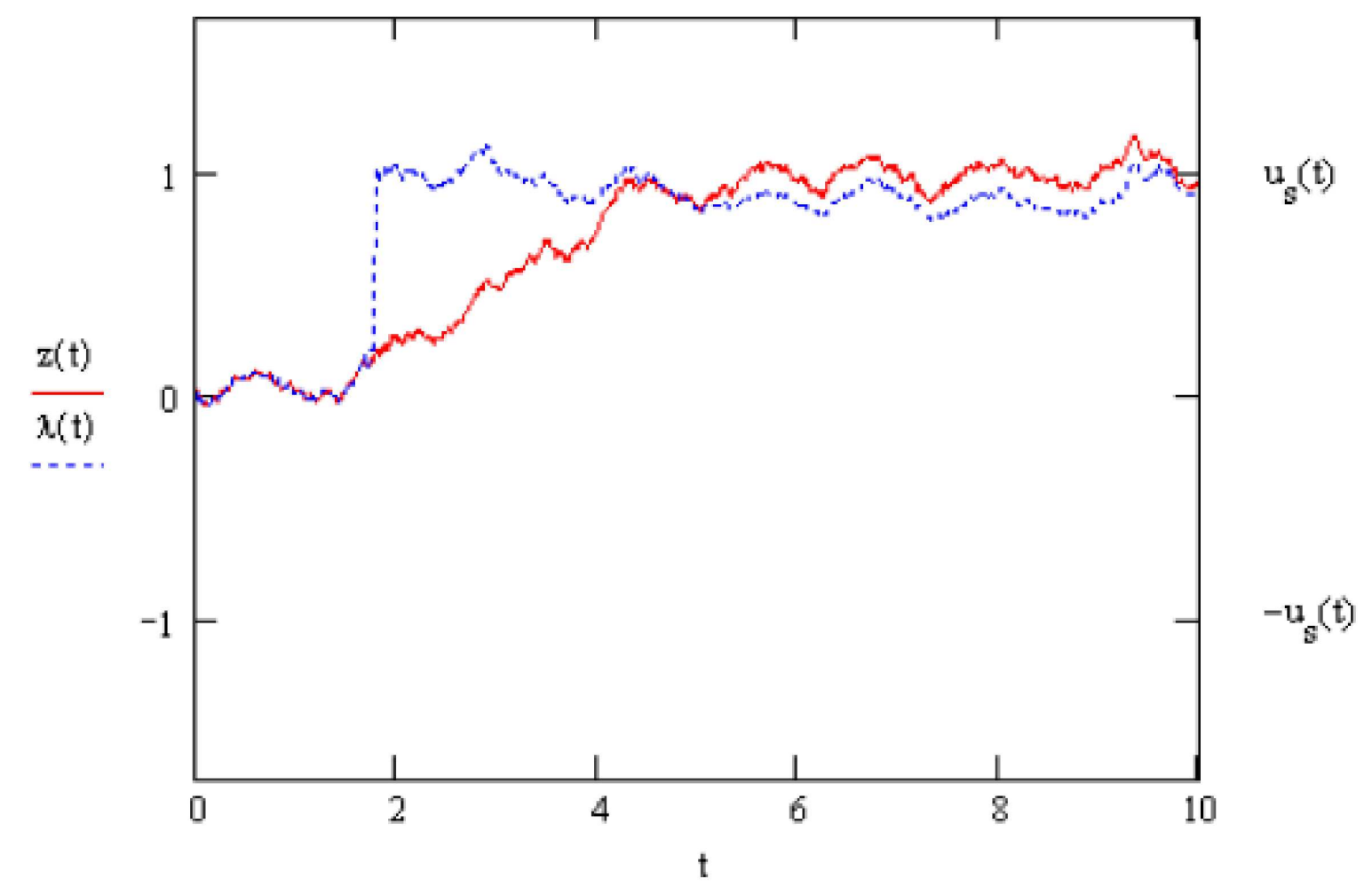

**Fig. 59**. Comparison of classical (red curve) and quasiclassical (blue curve) dynamics calculated from **SLDP**. $a = 1, b = 1, c = 0, A = 0.3, B = 0, \Omega = 5,$ $\Theta = 0, D = 10^{-2}, x_0 = 0.$

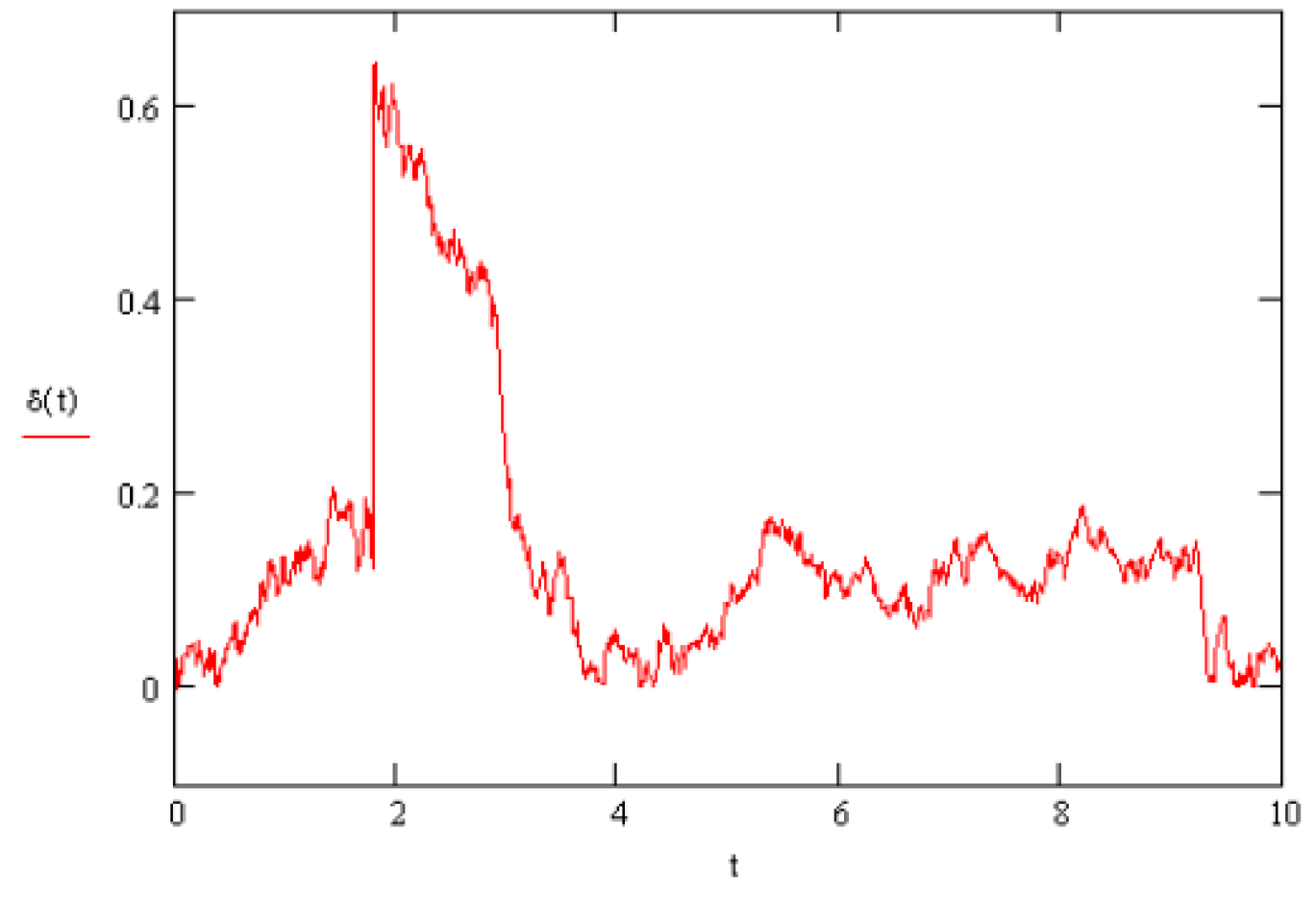

**Fig. 60.** Comparison by function $\delta(\omega, t)$ of classical (red curve) and quasiclassical (blue curve) dynamics calculated from **SLDP.** $a = 1, b = 1, c = 0, A = 0.3, B = 0,$ $\Omega = 5, \Theta = 0, D = 10^{-2}, x_0 = 0.$

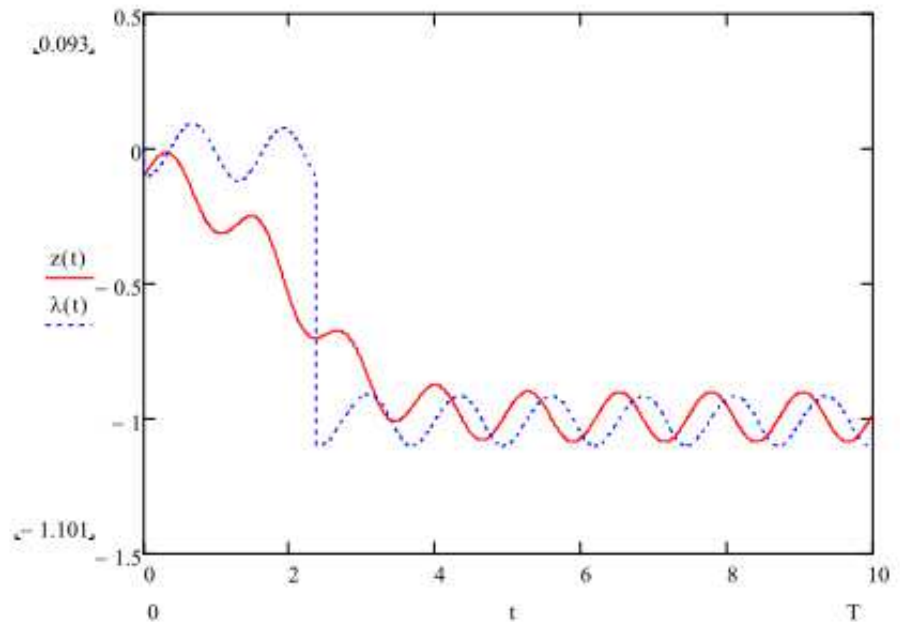

**Fig. 61.** Comparison of classical dynamics (red curve) and quasiclassical (blue curve) dynamics calculated from **SLDP.** $a = 1, b = 1, c = 0, A = 0.5, B = 0,$ $\Omega = 5, \Theta = 0, D = 0, x_0 = -0.1$

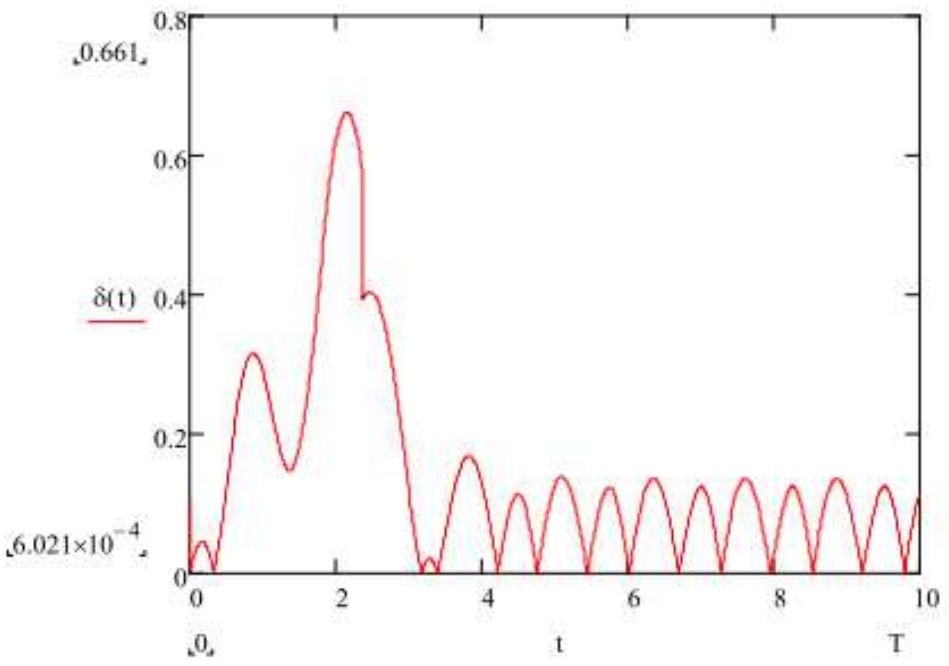

**Fig. 62.** Comparison by function $\delta(\omega, t)$ of classical (red curve) and quasiclassical dynamics calculated from **SLDP**. $a = 1, b = 1, c = 0, A = 0.5, B = 0,$ $\Omega = 5, \Theta = 0, D = 0, x_0 = -0.1$

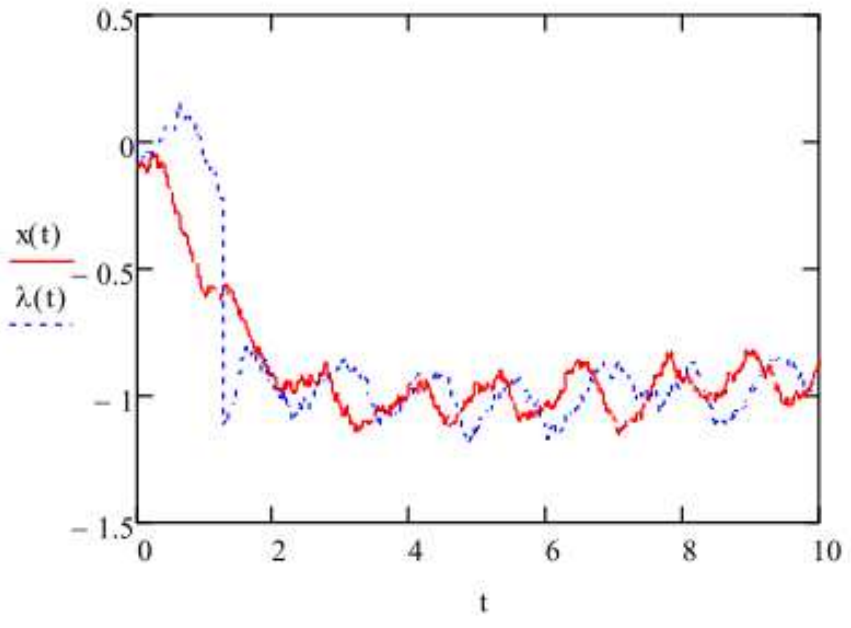

**Fig. 63.** Comparison by function $\delta(\omega, t)$ of classical (red curve) and quasiclassical (blue curve) dynamics calculated from **SLDP**. $a = 1, b = 1, c = 0, A = 0.5, B = 0,$ $\Omega = 5, \Theta = 0, D = 0.01, x_0 = -0.1$

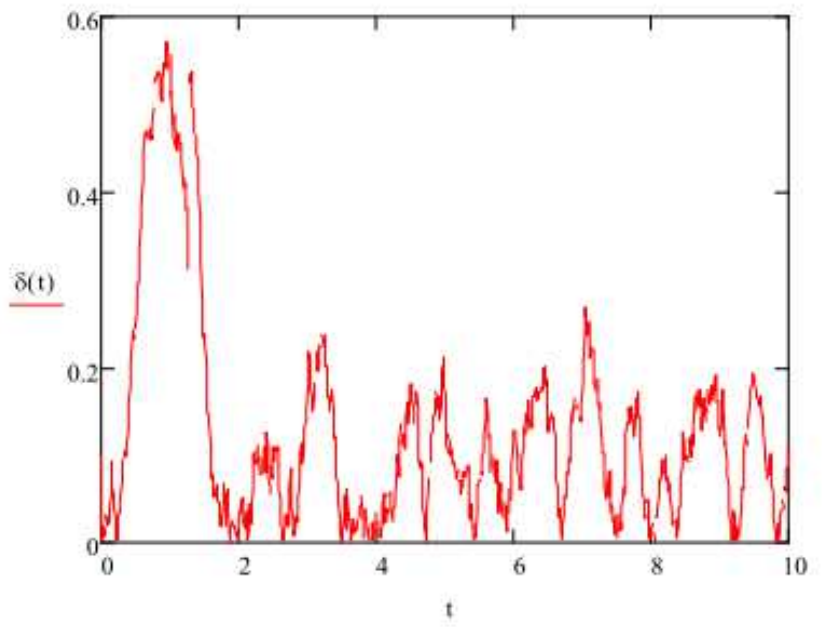

**Fig. 64**. Comparison by function $\delta(\omega, t)$ of classical (red curve) and quasiclassical (blue curve) dynamics calculated from **SLDP**. $a = 1, b = 1, c = 0, A = 0.5, B = 0,$
$\Omega = 5, \Theta = 0, D = 0.01, x_0 = -0.1$

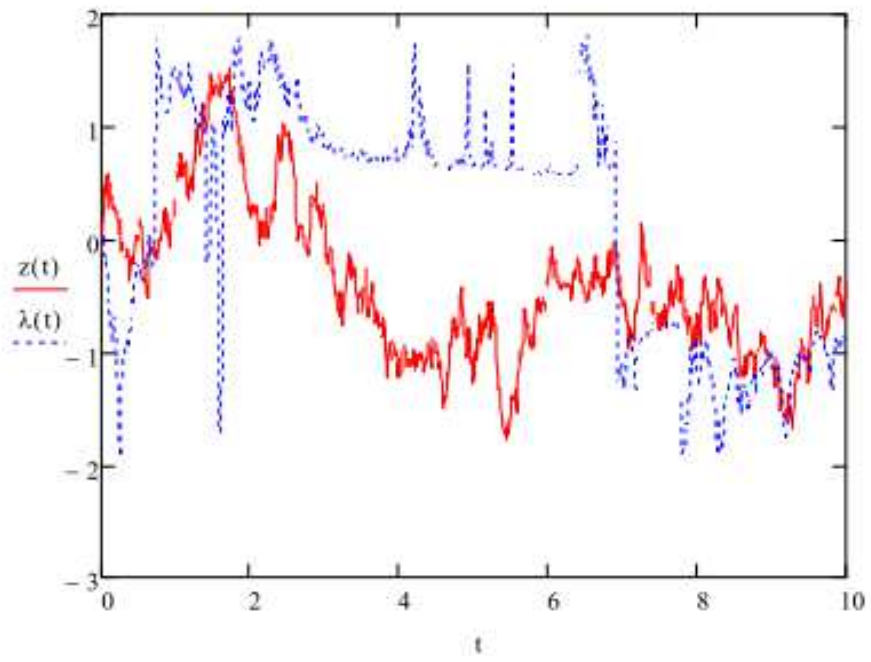

**Fig. 65**. Comparison of classical dynamics (red curve) and quasiclassical (blue curve) dynamics calculated from **SLDP**. $a = 1, b = 1, c = 0, A = 0.5, B = 0,$
$\Omega = 5, \Theta = 0, D = 1, x_0 = -0.1$

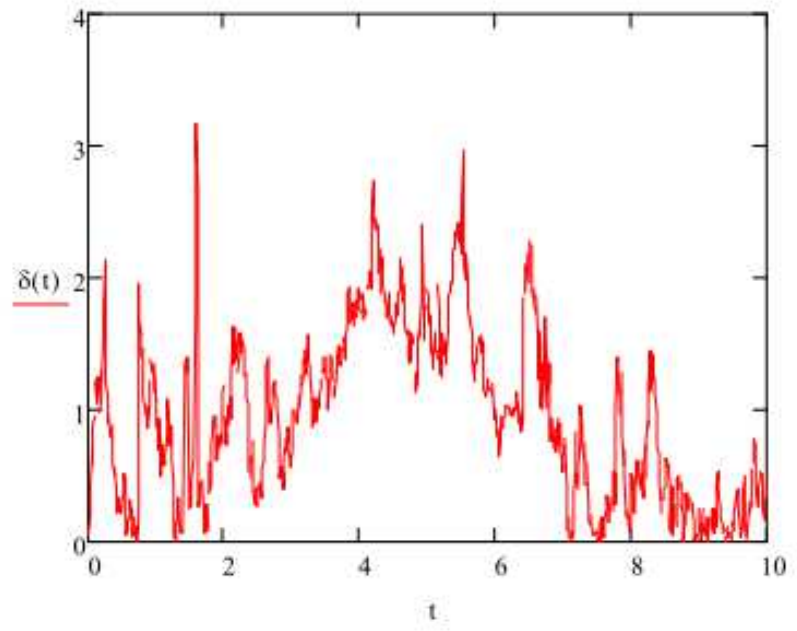

**Fig. 66.** Comparison by function $\delta(\omega, t)$ of classical (red curve) and quasiclassical (blue curve) dynamics calculated from **SLDP**. $a = 1, b = 1, c = 0, A = 0.5, B = 0,$
$\Omega = 5, \Theta = 0, D = 1, x_0 = -0.1$

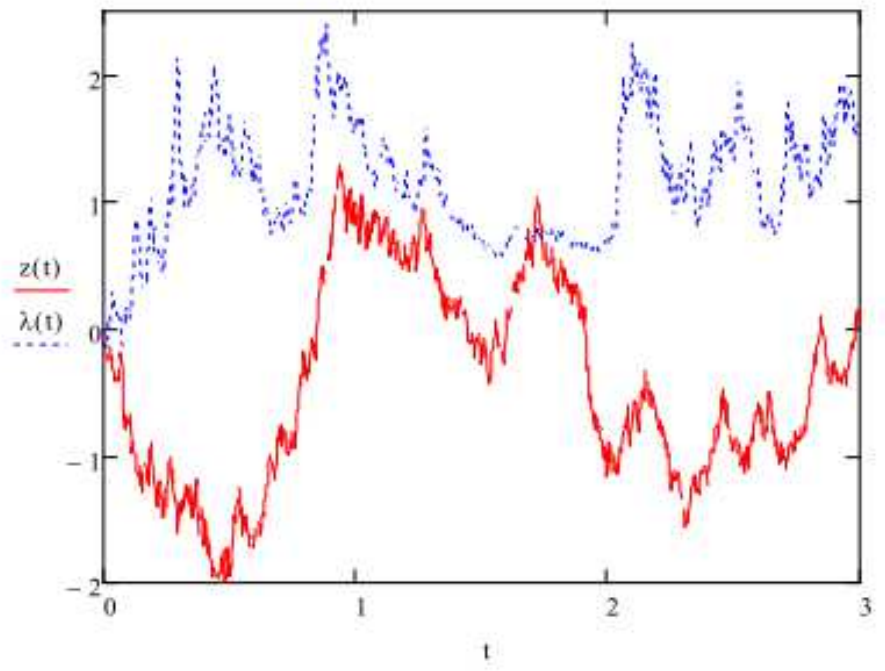

**Fig. 67.** Comparison of classical dynamics (red curve) and quasiclassical (blue curve) dynamics calculated by using **SLDP**. $a = 1, b = 1, c = 0, A = 0.5, B = 0,$
$\Omega = 5, \Theta = 0, D = 2, x_0 = -0.1$

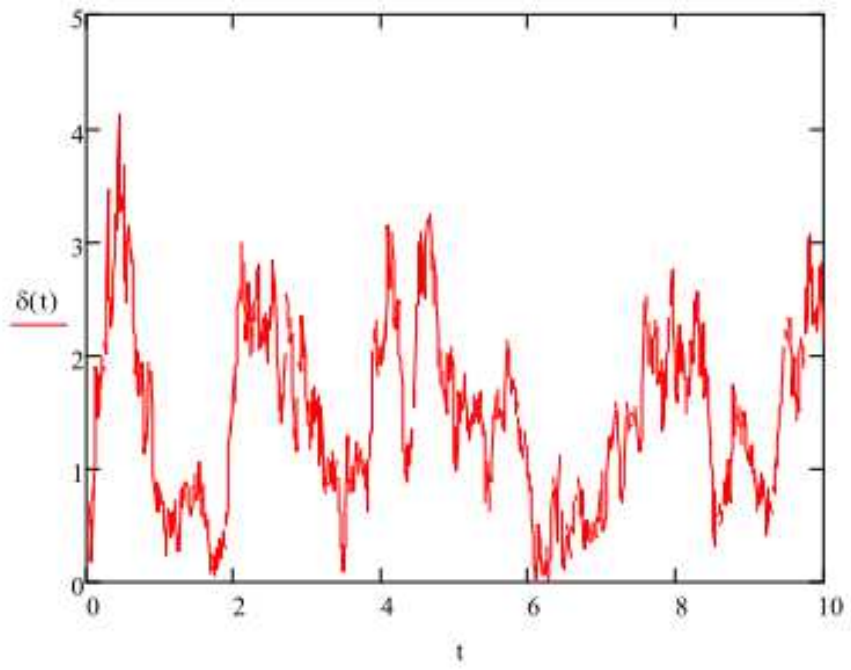

**Fig. 68.** Comparison by function $\delta(\omega, t)$ of classical (red curve) and quasiclassical (blue curve) dynamics calculated from **SLDP**. $a = 1, b = 1, c = 0, A = 0.5, B = 0,$ $\Omega = 5, \Theta = 0, D = 2, x_0 = -0.1$

# **II**. Symmetric external path integral for stochastic equations with additive Gaussian white noise.

## II.4.Master equation for symmetric Colombeau-Feinman path integral.

Let $\zeta_i(t) = \zeta_i(t,\epsilon), i = 1,\ldots,n$ be an Gaussian white noise, which normalized as

$$\langle \xi_i(t)\xi_j(t')\rangle = \epsilon\delta_{ij}\delta(t-t'). \tag{2.1}$$

Then corresponding Ito's SDE of the canonical form is

$$\dot{q}_i(t) = \mathcal{F}_i(\mathbf{q}(t),t) + \xi_i(t), \quad i = 1,2,\ldots d. \tag{2.2}$$

Here $\mathcal{F}_i : \mathbb{R}^d \times [0,\infty) \to \mathbb{R},\ i = 1,2,\ldots d$ is a measurable functions for which there exist constants $C_1$ and $C_2$ such that

$$\sum_{i=1}^{d}|\mathcal{F}_i(\mathbf{q},t)| \leq C_1(1+|\mathbf{q}|),$$

$$\sum_{i=1}^{d}|\mathcal{F}_i(\mathbf{q}_1,t)-\mathcal{F}_i(\mathbf{q}_2,t)| \leq C_2(|\mathbf{q}_1-\mathbf{q}_2|), \tag{2.3}$$

for all $t \in [0,\infty)$ and all $\mathbf{x}$ and $\mathbf{y} \in \mathbb{R}^d$. Ito's SDE (2.1)-(2.2) under assumptions (2.3) is well-known to be equivalent to the Fokker-Planck equation

$$\frac{\partial w(q',t';q'',t'')}{\partial t} = \sum_{i=1}^{d}\frac{\partial^2 w(q',t';q'',t'')}{\partial q_i''\partial q_i''} + \sum_{i=1}^{d}\frac{\partial}{\partial q_i''}[\mathcal{F}_i(\mathbf{q}'',t)w(q',t';q'',t'')],$$

$$w(q',t';q'',t') = w(q',q'') = \delta(q''-q'). \tag{2.4}$$

Here, $w_m(q',t';q'',t'')$ is the probability density that the system will end up at $q''$ at time $t''$ if it started at $q'$ at time $t'$. Fokker-Planck equation (2.4) formally can be solved in terms of Feynman's path integral of the form

$$w_m(q',t';q'',t'') = \int_{\mathbf{q}(t')=q'}^{\mathbf{q}(t'')=q''}[D\mathbf{q}(t)]\exp\left[\int_{t'}^{t''}\mathcal{L}_m(\dot{\mathbf{q}},\mathbf{q},t)dt\right],$$

$$\mathcal{L}_m(\dot{\mathbf{q}},\mathbf{q}) = \frac{1}{2\epsilon}\left\|\dot{\mathbf{q}}(t)-\vec{\mathcal{F}}(\mathbf{q}(t),t,m)\right\|^2 + \mathcal{F}_{i,i}(\mathbf{q}(t),t,m), \tag{2.5}$$

$$\mathcal{F}_{i,j}(\mathbf{q}(t),t,m) = \frac{\partial\mathcal{F}_i(\mathbf{q}(t),t,m)}{\partial q_j}.$$

Let us consider now Colombeau-Ito's SDE

$$(\dot{q}_{i,\delta}(t))_\delta = (\mathcal{F}_{i,\delta}(\mathbf{q}_\delta(t),t))_\delta + \xi_i(t),$$

$$\delta \in (0,1], \tag{2.6}$$

$$i = 1,2,\ldots d\,.$$

**Assumption 1**. We assume now that there exist Colombeau constants $(C_{1,\delta})_\delta$ and $(C_{2,\delta})_\delta$ such that

$$\sum_{i=1}^{d}(|\mathcal{F}_{i,\delta}(\mathbf{q},t)|)_{\delta} \leq ((C_{1,\delta})_{\delta})(1+|\mathbf{q}|),$$

$$\sum_{i=1}^{d}(|\mathcal{F}_{i,\delta}(\mathbf{q}_1,t)-\mathcal{F}_{i,\delta}(\mathbf{q}_2,t)|)_{\delta} \leq ((C_{2,\delta})_{\delta})(|\mathbf{q}_1-\mathbf{q}_2|), \tag{2.7}$$

$$\delta \in (0,1]$$

for all $t \in [0,\infty)$ and all $\mathbf{x}$ and $\mathbf{y} \in \mathbb{R}^d$. Colombeau-Ito's SDE (2.6) under assumptions (2.7) to be equivalent to the Colombeau-Fokker-Planck equation

$$\left(\frac{\partial w_{\delta}(q',t';q'',t'')}{\partial t}\right) = \sum_{i=1}^{d}\left(\frac{\partial^2 w_{\delta}(q',t';q'',t'')}{\partial q_i''\partial q_i''}\right)_{\delta} +$$

$$\sum_{i=1}^{d}\left(\frac{\partial}{\partial q_i''}[\mathcal{F}_{i,\delta}(\mathbf{q}'',t)w_{\delta}(q',t';q'',t'')]\right)_{\delta}, \tag{2.8}$$

$$(w_{\delta}(q',t';q'',t'))_{\delta} = (w_{\delta}(q',q''))_{\delta} = \delta(q''-q').$$

Colombeau-Fokker-Planck equation (2.8) formally can be solved in terms of Feynman's path integral of the form

$$(w_{\delta}(q',t';q'',t''))_{\delta} =$$

$$\int_{\mathbf{q}(t')=q'}^{\mathbf{q}(t'')=q''}[D\mathbf{q}(t)]\big((w_{\delta}(q(t')-q'))_{\delta}\big)\exp\left[\left(\int_{t'}^{t''}\mathcal{L}_{\delta}(\dot{\mathbf{q}},\mathbf{q},t)dt\right)_{\delta}\right],$$

$$(\mathcal{L}_{\delta}(\dot{\mathbf{q}},\mathbf{q}))_{\delta} = \frac{1}{2\epsilon}\left(\left\|\dot{\mathbf{q}}(t)+\overrightarrow{\mathcal{F}}_{\delta}(\mathbf{q}(t),t,m)\right\|^2\right)_{\delta} - \tag{2.9}$$

$$-(\mathcal{F}_{i,i,\delta}(\mathbf{q}(t),t,m))_{\delta},$$

$$(\mathcal{F}_{i,j,\delta}(\mathbf{q}(t),t,m))_{\delta} = \left(\frac{\partial\mathcal{F}_{i,\delta}(\mathbf{q}(t),t,m)}{\partial q_j}\right)_{\delta}.$$

using theory of Colombeau generalized functions [19]. The exact form of $\mathcal{L}_{\delta}$ depends on the convention used in discretizing time when defining the path integral. By using a symmetric discretization one obtain:

$$(w_\delta(q',t';q'',t'',\epsilon))_\delta =$$

$$\left(\lim_{\max \Delta t_i \to 0} \breve{N}_N \int_{\mathbf{q}(t')=q'}^{\mathbf{q}(t'')=q''} \left[\prod_t D\mathbf{q}(t)\right] \times\right.$$

$$\left. w_\delta(q(t') - q') \exp\left[-\sum_t \Delta t \mathcal{L}_\delta\left(\frac{\mathbf{q}(t) - \mathbf{q}(t-\Delta t)}{\Delta t}, \frac{\mathbf{q}(t) + \mathbf{q}(t-\Delta t)}{2}, t\right)\right]\right)_\delta =$$

$$= \left(\lim_{N\to\infty} (4\pi\epsilon)^{-\frac{N}{2}n} \int \frac{dq_0 dq_1 dq_2 \ldots dq_{N-1}}{\sqrt{\Delta t_0 \Delta t_1 \ldots \Delta t_N}} \times\right. \tag{2.10}$$

$$\left. w_\delta(q_0 - q') \exp\left[-\sum_{i=1}^{N-1} \Delta t_i \mathcal{L}_\delta\left(\frac{\mathbf{q}_i - \mathbf{q}_{i-1}}{\Delta t}, \frac{\mathbf{q}_i + \mathbf{q}_{i-1}}{2}, t_i\right)\right]\right)_\delta$$

$$\Delta t_i = t_{i+1} - t_i, \Delta t_i \underset{N\to\infty}{\to} 0,$$

$$q_0 = q', q_{N+1} = q'',$$

$$i = 0,1,\ldots,N+1.$$

**Remark 1**. Note that Colombeau-Fokker-Planck equation (2.8) is just a Euclidean Colombeau-Schrödinger equation, and is well-known that ander long been known assumptions [33], one can transform Colombeau-Schrödinger equation

$$\left(\frac{\partial u_\delta(t,x)}{\partial t}\right)_\delta = \frac{i\epsilon}{2}(\Delta u_\delta(t,x))_\delta - i(V_\delta(t,x)u_\delta(t,x))_\delta, \tag{2.11}$$

$$(u_\delta(0,x))_\delta = (\varphi_\delta(x))_\delta, \delta \in (0,1]$$

into mathematically rigorous path integral by standard method using Trotter's Product Formula [33]. Here $\Delta$ is the Laplace operator $\partial^2/\partial x_1^2 + \cdots + \partial^2/\partial x_d^3, V_\delta$ is a real measurable function on $\mathbb{R}^d, \varphi_\delta, \delta \in (0,1]$ and each $u_\delta, \delta \in (0,1]$, are elements of $\mathcal{L}_2(\mathbb{R}^d)$ and $\epsilon$ is a constant. Let $\mathcal{F}[\circ]$ denote the Fourier transformation,$\mathcal{F}^{-1}[\circ]$ its inverse. We define now as usual [34]

$$(\Delta\phi_\delta)_\delta = \left(\mathcal{F}^{-1}\left[-\|\lambda\|^2\mathcal{F}[\phi_\delta]\right]\right)_\delta, \delta \in (0,1], \tag{2.12}$$

on the domain $D(\Delta)$ of all square-integrable $(\phi_\delta)_\delta$, such that $\left(\mathcal{F}^{-1}\left[-\|\lambda\|^2\mathcal{F}[\phi_\delta]\right]\right)_\delta$ is also square-integrable. (Here $\lambda$ denotes the variable in momentum space and $\|\lambda\|^2 = \lambda_1^2 + \cdots + \lambda_n^2$.) Then $\Delta$ is self-adjoint, and

$$\begin{gathered}(u_\delta(t,x))_\delta = (K_\epsilon^t u_\delta(t,x))_\delta, \\ K_\epsilon^t = \exp\left[\frac{it\epsilon}{2}\Delta\right]\end{gathered} \tag{2.13}$$

is the solution of the Eq.(2.12) for $(V_\delta(t,x))_\delta = 0.$ The operator $V_\delta$ of multiplication by the function $V_\delta(t,x)$, on the domain $D(V_\delta(t,x))$ of all $\phi_\delta$ in $\mathcal{L}_2(\mathbb{R}^d)$ such that $V_\delta(t,x)\phi_\delta$ is also in $\mathcal{L}_2(\mathbb{R}^d)$, is self-adjoint, and

$$\begin{gathered}(u_\delta(t,x))_\delta = (M_{V_\delta}^t\phi_\delta)_\delta, \\ M_{V_\delta}^t = \exp[itV_\delta]\end{gathered} \tag{2.14}$$

is the solution of the Eq.(2.12) with $\epsilon = 0.$ Kato has found conditions under which the operator [33]

$$(\mathfrak{R}_{\epsilon,\delta}u_\delta)_\delta = \frac{i\epsilon}{2}(\Delta u_\delta)_\delta - i(V_\delta u_\delta)_\delta \tag{2.15}$$

and under these conditions if we let

$$U_{\epsilon,V_\delta}^t = \exp[\mathfrak{R}_{\epsilon,\delta}] \tag{2.16}$$

then a theorem of Trotter [33] asserts that for all $\phi_\delta$ in $\mathcal{L}_2(\mathbb{R}^d)$

$$(U_{\epsilon,V_\delta}^t\varphi_\delta)_\delta = \left(\lim_{N\to\infty}\left(\left(K_\epsilon^{\frac{t}{N}}M_{V_\delta}^{\frac{t}{N}}\right)^N\varphi_\delta\right)\right)_\delta. \tag{2.17}$$

This is discussed in detail in [33] (See [33] Appendix B). Using now Eq.(2.12)-Eq.(2.15)

by simple calculation one obtain [33]

$$\left(\left(K_{\epsilon}^{\frac{t}{N}} M_{V_\delta}^{\frac{t}{N}}\right)^{N} \varphi_\delta\right)_\delta = \left(\frac{2\pi i t\epsilon}{N}\right)^{-\frac{1}{2}d\cdot N} \times$$

$$\times\left(\int\ldots\int d^d\mathbf{x}_0\ldots d^d\mathbf{x}_{N-1}\exp[iS_\delta(\mathbf{x}_0,\ldots\mathbf{x}_N;t)]\right)_\delta, \qquad (2.18)$$

$$S_\delta(\mathbf{x}_0,\ldots\mathbf{x}_N;t) = \sum_{i=1}^{N}\left[\frac{\|\mathbf{x}_i-\mathbf{x}_{i-1}\|^2}{(t/N)^2} - V_\delta(\mathbf{x}_i)\right]\frac{t}{N}.$$

where we have set $\mathbf{x}_N = \mathbf{x}$.

Theorem. (**Suzuki-Trotter Formula**) [35],[36]. Let $\{A_j\} = \{A_j\}_{j=1}^{p}$be an family of any bounded operators in an Banach algebra $\Im$ with a norm $\|\circ\|_\Im$. Let $\Phi_n(\{A_j\})$ be a function

$$\Phi_n(\{A_j\}) = \left(\exp\left(\frac{A_1}{n}\right)\ldots\exp\left(\frac{A_p}{n}\right)\right)^n. \qquad (2.19)$$

For any bounded operators $\{A_j\}_{j=1}^{p}$ in a Banach algebra $\Re$ with a norm $\|\circ\|_\Im$ :

$$\lim_{n\to\infty}\left\|\Phi_n(\{A_j\}) - \exp\left(\sum_{j=1}^{p} A_j\right)\right\|_\Im. \qquad (2.20)$$

**Remark 2**. Note that one can transform Colombeau-Fokker-Planck equation (2.8) into mathematically rigorous path integral by standard method using Suzuki -Trotter's Product Formula (2.20). However path integral representation of the solutions of the Colombeau-Fokker-Planck equation (2.8), given by canonical Eq.(2.10), does not valid under canonical assumptions which is discussed above in Remark1.

**Remark3**. Note that formal pseudo-differential operator given by formula [37]-[39]

$$P_\delta^t = \exp\left(\sum_{i=1}^{d}\left[\int_0^t \mathcal{F}_{i,\delta}\left(\overset{2}{\mathbf{q}},\tau\right)d\tau\right]\frac{\overset{1}{\partial}}{\partial q_i}\right), \qquad (2.21)$$

evidently does not define any contraction semi-group on $\mathcal{L}_2(\mathbb{R}^d)$. Nevertheless formal pseudo-differential expression (2.21) define contraction semi-group on an subspace $H_{1,2}^{\infty}(S_R) \subsetneq H^\infty(S_R)$ of a test space $H^\infty(S_R), R = (R_1,\ldots,R_d)$, with a members $\varphi(\mathbf{x}), \mathbf{x} \in \mathbb{R}^d$ such that $\mathcal{F}[\varphi](\xi)$ is supported inside region $S_R = \{\xi \| \xi_i| < R_i; i = 1,\ldots,d\}$ [40]. Pseudo-differential calculus on a test space $H^\infty(S_R)$ is discussed in detail in [40].

**Remark 4**. Note that $\Delta_\delta(x) \in H^\infty(S_R), x \in \mathbb{R}, R = 1/\delta, \delta \in (0,1]$, where

$$\Delta_\delta(x) = \frac{1}{\pi\delta x}\sin\frac{x}{\delta} \tag{2.22}$$

By simple calculation one obtain [40]

$$\mathcal{F}[\Delta_\delta(x)] = \begin{cases} \frac{1}{2\pi}, x \in (-r,r) \\ 0, x \notin (-r,r) \end{cases} \tag{2.23}$$

$$r = 1/\delta.$$

**Assumption 2**. We assume now that

$$(w_\delta(q',t';q'',t'',\epsilon))_{\delta\in(0,1]} = \left(\prod_{i=1}^{d}\Delta_\delta(x_i)\right)_{\delta\in(0,1]} \tag{2.24}$$

We obtain master equation for Langevin equation (2.6) in 1-dimensional case without loss of generality. Langevin equation in 1-dimensional case is

$$\begin{gathered}(\dot{q}_\delta(t))_\delta = -(\mathcal{F}_\delta(q_\delta(t),t))_\delta + \xi(t), \delta \in (0,1], \\ (q_\delta(0))_\delta = q',\end{gathered} \tag{2.25}$$

where we set for short

$$\begin{gathered}\mathcal{F}_\delta(q) = \frac{aq^3}{1+\delta^{2l}q^{2l}} + b\frac{q^2}{1+\delta^{2l}q^{2l}} + c\frac{q}{1+\delta^{2l}q^{2l}} + u(t), \\ l \geq 2, \delta \in (0,1].\end{gathered} \tag{2.26}$$

Now we replace Langevin equation (2.25)-(2.26) by auxiliary Langevin equation such that

$$(\dot{q}_\delta(t))_\delta = -(\mathcal{F}_\delta(q_\delta(t),u_\delta(t),t))_\delta + \xi(t), \delta \in (0,1],$$

$$(\dot{u}_\delta(t))_\delta = \left((q_\delta(t)-\lambda)^2\right)_\delta + \mu\xi(t), \mu \in (0,1], \tag{2.27}$$

$$(q_\delta(0))_\delta = q', (u_\delta(0))_\delta = 0.$$

Here

$$\mathcal{F}_\delta(q,u) = \frac{aq^3}{1+\delta^{2l}q^{2l}+\delta^{2l}u^2} + \frac{bq^2}{1+\delta^{2l}q^{2l}+\delta^{2l}u^2} + \frac{cq}{1+\delta^{2l}q^{2l}+\delta^{2l}u^2} + u(t), \tag{2.28}$$

$$l \geq 2, \delta \in (0,1].$$

**Remark 5**. Below we using the abbreviation $g$ for Colombeau generalized functions $(g_\delta(q))_\delta = \left(\frac{g(q)}{1+1+\delta^{2l}q^{2l}+\delta^{2l}u^2}\right)_\delta$ and let for short

$$\left(\frac{aq^3}{1+\delta^{2l}q^{2l}+\delta^{2l}u^2}\right)_\delta \triangleq aq^3, \left(\frac{bq^2}{1+\delta^{2l}q^{2l}+\delta^{2l}u^2}\right)_\delta \triangleq bq^2, \tag{2.29}$$

$$\left(\frac{cq}{1+\delta^{2l}q^{2l}+\delta^{2l}u^2}\right)_\delta \triangleq cq.$$

Thus we can to rewrite Eq.(2.7) in a short form

$$(\dot{q}_\delta(t))_\delta = -(aq^3 + bq^2 + cq + u(t)) + \xi(t),$$

$$(\dot{u}_\delta(t))_\delta = \left((q_\delta(t)-\lambda)^2\right)_\delta + \mu\xi(t). \tag{2.30}$$

From Eq.(2.30) by using substitution

$$q(t) = v(t) + \lambda \tag{2.31}$$

one obtain

$$
\begin{gathered}
(\dot{v}_\delta(t))_\delta = -\Big(a(v(t)+\lambda)^3 + b(v(t)+\lambda)^2 + c(v(t)+\lambda) + u(t)\Big) + \xi(t), \\
(\dot{u}_\delta(t))_\delta = (v_\delta^2(t))_\delta + \mu\xi(t), \\
(v_\delta(0))_\delta = q' - \lambda, (u_\delta(0))_\delta = 0.
\end{gathered}
\tag{2.32}
$$

From RHS of the Eq.(2.32) one obtain

$$
\begin{gathered}
a(v+\lambda)^3 + b(v+\lambda)^2 + c(v+\lambda) = \\
a(v^3 + 3\lambda v^2 + 3\lambda^2 v + \lambda^3) + b(v^2 + 2\lambda v + \lambda^2) + c(v+\lambda) = \\
av^3 + 3a\lambda v^2 + 3a\lambda^2 v + a\lambda^3 + bv^2 + 2b\lambda v + b\lambda^2 + cv + c\lambda = \\
av^3 + (3a\lambda + b)v^2 + (3a\lambda^2 + 2b\lambda + c)v + a\lambda^3 + b\lambda^2 + c\lambda.
\end{gathered}
\tag{2.33}
$$

Substitution Eq.(2.33) into Eq.(2.32) gives

$$
\begin{gathered}
(\dot{v}_\delta(t))_\delta = -av^3(t) - (3a\lambda + b)v^2(t) - (3a\lambda^2 + 2b\lambda + c)v(t) - \\
-(a\lambda^3 + b\lambda^2 + c\lambda + u(t) + \lambda\delta(t)) + \xi(t), \\
(\dot{u}_\delta(t))_\delta = (v_\delta^2(t))_\delta + \mu\xi(t).
\end{gathered}
\tag{2.34}
$$

Langevin equation (2.34) can be solved in terms of Colombeau-Feynman path integral of the form

$$(w_\delta(t_i, q_i|t_f, q_f))_\delta =$$

$$\int_{q(t_i)=q_i}^{q(t_f)=q_f}[Dq(t)]\int_{u(t_i)=q_i}[Du(t)]\exp\left[-\left(\int_{t_i}^{t_f}\mathcal{L}_\delta(\dot{q}(t), q(t), \dot{u}(t), u(t), t)dt\right)_\delta\right]$$

$$\mathcal{L}_\delta(\dot{q}(t), q(t), \dot{u}(t), u(t), t) =$$

$$\frac{1}{2\epsilon}[\dot{q}(t) + \mathcal{F}_\delta(q(t), u(t))]^2 + \frac{1}{2\mu}[\dot{u}(t) - q^2(t)]^2 - \qquad (2.35)$$

$$-\frac{\partial\mathcal{F}_\delta(q(t), u(t))}{\partial q} - \frac{\partial\mathcal{F}_\delta(q(t), u(t))}{\partial u},$$

$$\mathcal{F}_\delta(q(t), u(t)) = aq^3(t) + (3a\lambda + b)q^2(t) + (3a\lambda^2 + 2b\lambda + c)q(t) +$$

$$+(a\lambda^3 + b\lambda^2 + c\lambda + u(t)).$$

From path integral (2.35) in the limit $\mu \to 0$ one obtain

$$(w_\delta(t_i, q_i|t_f, q_f))_\delta =$$

$$\int_{q(t_i)=q_i}^{q(t_f)=q_f}[Dq(t)]\int_{u(t_i)=q_i}[Du(t)]\exp\left[-\left(\int_{t_i}^{t_f}\mathcal{L}_\delta(\dot{q}(t), q(t), t)dt\right)_\delta\right]$$

$$\mathcal{L}_\delta(\dot{q}(t), q(t)) = \frac{1}{2\epsilon}[\dot{q}(t) + \mathcal{F}_\delta(q(t))]^2 - \partial\mathcal{F}_\delta(q(t))$$

$$\mathcal{F}_\delta(q(t)) = [\mathcal{F}_\delta(q(t), u(t))]_{u(t)=\tilde{u}(t)} = \qquad (2.36)$$

$$[aq^3(t) + (3a\lambda + b)q^2(t) + (3a\lambda^2 + 2b\lambda + c)q(t) + (a\lambda^3 + b\lambda^2 + c\lambda + u(t))]_{u(t)=\tilde{u}(t)},$$

$$\partial\mathcal{F}_\delta(q(t)) = \left[\frac{\partial\mathcal{F}_\delta(q(t), u(t))}{\partial q} + \frac{\partial\mathcal{F}_\delta(q(t), u(t))}{\partial u}\right]_{u(t)=\tilde{u}(t)},$$

$$\tilde{u}(t) = \int_0^t q^2(\tau)d\tau.$$

$$[\dot{q}(t)+\mathcal{F}_{\delta}(q(t))]^2 =$$

$$\{[\dot{q}(t)+(3a\lambda^2+2b\lambda+c)q(t)+(a\lambda^3+b\lambda^2+c\lambda+u(t))]+ \quad (2.37)$$

$$+[aq^3(t)+(3a\lambda+b)q^2(t)]\}^2.$$

By using the replacement

$$g_1(\lambda) = 3a\lambda^2+2b\lambda+c,$$

$$g_2(\lambda,t) = a\lambda^3+b\lambda^2+c\lambda+u(t), \quad (2.38)$$

$$g_3(\lambda) = 3a\lambda+b$$

From Eq.(2.37) one obtain

$$[\dot{q}(t)+\mathcal{F}_{\delta}(q(t))]^2 =$$

$$\{[\dot{q}(t)+g_1(\lambda)q(t)+g_2(\lambda,t)]+[aq^3(t)+g_3(\lambda)q^2(t)]\}^2 =$$

$$[\dot{q}(t)+g_1(\lambda)q+g_2(\lambda,t)]^2+$$

$$2[\dot{q}(t)+g_1(\lambda)q(t)+g_2(\lambda,t)][aq^3(t)+g_3(\lambda)q^2(t)]+$$

$$[aq^3(t)+g_3(\lambda)q^2(t)]^2 = \quad (2.39)$$

$$[\dot{q}(t)+g_1(\lambda)q+g_2(\lambda,t)]^2+$$

$$2a\dot{q}(t)q^3(t)+2g_3(\lambda)\dot{q}(t)q^2(t)+$$

$$2[g_1(\lambda)q(t)+g_2(\lambda,t)][aq^3(t)+g_3(\lambda)q^2(t)]+$$

$$[aq^3(t)+g_3(\lambda)q^2(t)]^2.$$

From Eq.(2.39) one obtain

$$\begin{gathered}2[g_1(\lambda)q(t)+g_2(\lambda,t)][aq^3(t)+g_3(\lambda)q^2(t)]+\\ [aq^3(t)+g_3(\lambda)q^2(t)]^2=\\ 2ag_1(\lambda)q^4(t)+2g_1(\lambda)g_3(\lambda)q^3(t)+2ag_2(\lambda,t)q^3(t)+2g_2(\lambda,t)g_3(\lambda)q^2(t)+\\ a^2q^6(t)+2ag_3(\lambda)q^5(t)+g_3^2(\lambda)q^4(t)=\\ [2g_2(\lambda,t)g_3(\lambda)]q^2(t)+[2g_1(\lambda)g_3(\lambda)+2ag_2(\lambda,t)]q^3(t)+\\ [2ag_1(\lambda)+g_3^2(\lambda)]q^4(t)+[2ag_3(\lambda)]q^5(t)+a^2q^6(t).\end{gathered} \tag{2.40}$$

Using the replacement

$$\begin{gathered}h_2=h_2(\lambda,t)=2g_2(\lambda,t)g_3(\lambda),h_3=h_3(\lambda,t)=2g_1(\lambda)g_3(\lambda)+2ag_2(\lambda,t),\\ h_4=h_4(\lambda)=2ag_1(\lambda)+g_3^2(\lambda),h_5=h_5(\lambda)=2ag_3(\lambda),\end{gathered} \tag{2.41}$$

from Eq.(2.40) one obtain

$$\begin{gathered}2[g_1(\lambda)q(t)+g_2(\lambda,t)][aq^3(t)+g_3(\lambda)q^2(t)]+[aq^3(t)+g_3(\lambda)q^2(t)]^2=\\ h_2(\lambda,t)q^2(t)+h_3(\lambda,t)q^3(t)+h_4(\lambda)q^4(t)+h_5(\lambda)q^5(t)+a^2q^6(t).\end{gathered} \tag{2.42}$$

Substitution Eq.(2.42) into Eq.(2.40) gives

$$
\begin{gathered}
[\dot{q}(t)+\mathcal{F}_{\delta}(q(t))]^2 = \\
[\dot{q}(t)+g_1(\lambda)q(t)+g_2(\lambda,t)]^2 + \\
2g_3(\lambda)\dot{q}(t)q^2(t)+2a\dot{q}(t)q^3(t) + \\
h_2(\lambda,t)q^2(t)+h_3(\lambda,t)q^3(t)+h_4(\lambda)q^4(t)+h_5(\lambda)q^5(t)+a^2q^6(t).
\end{gathered} \tag{2.43}
$$

From Eq.(2.36) we obtain

$$
\begin{gathered}
\left(\mathbf{E}\left[(q_{\delta}(t,\omega,\varepsilon)-\lambda)^2\right]\right)_{\delta} = (\mathbf{E}[v_{\delta}^2(t,\omega,\varepsilon)])_{\delta} = \\
N^{-1}\int_{-\infty}^{+\infty}dq_f\int_{q(0)=q_i}^{q(t_f)=q_f}[Dq(t)]q_f^2\exp\left[\left(\int_0^{t_f}\partial\mathcal{F}_{\delta}(q(t))dt\right)_{\delta}\right]\times \\
\exp\left[-\frac{1}{\varepsilon}\left(\int_0^{t_f}[\dot{q}(t)+\mathcal{F}_{\delta}(q(t))]^2dt\right)_{\delta}\right], \\
\varepsilon = 2\epsilon.
\end{gathered} \tag{2.44}
$$

Substitution Eqs.(2.43) into Eq.(2.44) gives

$$
\begin{gathered}
(\mathbf{E}[v_{\delta}^2(t,\omega,\varepsilon)])_{\delta\in(0,1]} = \\
N^{-1}\int_{-\infty}^{+\infty}dq_f\int_{q(0)=q_i}^{q(t_f)=q_f}[Dq(t)]q_f^2\exp\left[\left(\int_0^{t_f}\frac{\partial\mathcal{F}_{\delta}(q(t))}{\partial q}dt\right)_{\delta}\right]\times \\
\exp\left\{-\frac{1}{\varepsilon}\left(\int_0^{t_f}\left[[\dot{q}(t)+g_1q(t)]^2+2g_2[\dot{q}(t)+g_1q(t)]+g_2^2\right]dt\right)_{\delta}\right\}\times \\
\exp\left\{-\frac{1}{\varepsilon}\left(\int_0^{t_f}[2g_3\dot{q}(t)q^2(t)+2a\dot{q}(t)q^3(t)]dt\right)_{\delta}\right\}\times \\
\exp\left\{-\left(\frac{1}{\varepsilon}\int_0^{t_f}[h_2q^2(t)+h_3q^3(t)+h_4q^4(t)+h_5q^5(t)+a^2q^6(t)]dt\right)_{\delta}\right\}.
\end{gathered} \tag{2.45}
$$

Using replacement $q(t) \to q(t)\sqrt{\varepsilon}$ in the path integral (2.45), we obtain

$$(\mathbf{E}[v_\delta^2(t,\omega,\varepsilon)])_\delta = \widehat{N}^{-1}\int_{-\infty}^{+\infty} dy \int_{q(0)=\frac{q_i}{\sqrt{\varepsilon}}}^{q(t_f)=\frac{y}{\sqrt{\varepsilon}}} [Dq(t)]y^2 \times$$

$$\exp\left[\left(\int_0^{t_f} \partial\mathcal{F}_\delta(\sqrt{\varepsilon}\,q(t))dt\right)_\delta\right]\exp\left\{-\int_0^{t_f}\left[[\dot{q}(t)+g_1q(t)]^2+\right.\right.$$

$$\left.\left.\frac{1}{\sqrt{\varepsilon}}2g_2[\dot{q}(t)+g_1q(t)]+\frac{1}{\varepsilon}g_2^2\right]dt\right)_\delta\Bigg\} \times \tag{2.46}$$

$$\exp\left\{-\left(\int_0^{t_f}[2\sqrt{\varepsilon}\,g_3\dot{q}(t)q^2(t)+2\varepsilon a\dot{q}(t)q^3(t)]dt\right)_\delta\right\} \times$$

$$\exp\left\{-\left(\int_0^{t_f}[h_2q^2(t)+\sqrt{\varepsilon}\,h_3q^3(t)+\right.\right.$$

$$\left.\left.\varepsilon h_4q^4(t)+\varepsilon\sqrt{\varepsilon}\,h_5q^5(t)+a^2\varepsilon^2q^6(t)]dt\right)_\delta\right\}.$$

Let us rewrite now Eq.(2.46) in the form

$$(\mathbf{E}[v_\delta^2(t,\omega)])_\delta = \widehat{N}^{-1}\int_{-\infty}^{+\infty} dy \int_{q(0)=\frac{q_i}{\sqrt{\varepsilon}}}^{q(t_f)=\frac{y}{\sqrt{\varepsilon}}} [Dq(t)]y^2 \times$$

$$\exp\left[\left(\int_0^{t_f}\frac{\partial\mathcal{F}(\sqrt{\varepsilon}\,q(t))}{\partial q}dt\right)_\delta\right]\exp\left\{-\frac{1}{\varepsilon}\left(\int_0^{t_f} g_2^2(\lambda,t)dt\right)_\delta\right\} \times$$

$$\exp\left\{-\left(\int_0^{t_f}[\dot{q}(t)+g_1q(t)]^2dt\right)_\delta\right\} \times \tag{2.47}$$

$$\exp\left\{-\frac{1}{\sqrt{\varepsilon}}\left(\int_0^{t_f}[2g_2[\dot{q}(t)+g_1q(t)]]dt\right)_\delta\right\} \times$$

$$\exp\left\{-\left(\int_0^{t_f} G_\delta[\dot{q}(t),q(t)]dt\right)_\delta\right\},$$

Here

$$(G_\delta[\dot{q}(t),q(t)])_\delta = (2\sqrt{\varepsilon}\,g_3\dot{q}(t)q^2(t) + 2\varepsilon a\dot{q}(t)q^3(t) +$$
$$h_2q^2(t) + \sqrt{\varepsilon}\,h_3q^3(t) + \varepsilon h_4q^4(t) + \varepsilon\sqrt{\varepsilon}\,h_5q^5(t) + a^2\varepsilon^2q^6(t))_\delta. \tag{2.48}$$

**Lemma.2.1.**Let $\frac{1}{p_1} + \frac{1}{p_2} = 1$. Then Hölder's inequality for Colombeau-Feynman path integrals states that:

$$\left(N^{-1}\int_{q(0)=q_i}^{q(t_f)=q_f} |(F_{1,\delta}[\dot{q}(t),q(t)]F_{2,\delta}[\dot{q}(t),q(t)])| \times$$
$$\exp\{(F_{3,\delta}[\dot{q}(t),q(t)])\}[Dq(t)]\right)_\delta \leq$$
$$\left(\left[N^{-1}\int_{q(0)=q_i}^{q(t_f)=q_f} |F_{1,\delta}[\dot{q}(t),q(t)]|^{p_1}\exp\{F_{3,\delta}[\dot{q}(t),q(t)]\}[Dq(t)]\right]^{\frac{1}{p_1}}\right)_\delta \times \tag{2.49}$$
$$\left(\left[N^{-1}\int_{q(0)=0}^{q(t_f)=y} |F_2[\dot{q}(t),q(t)]|^{p_2}\exp\{F_3[\dot{q}(t),q(t)]\}[Dq(t)]\right]^{\frac{1}{p_2}}\right)_{\delta\in(0,1]},$$

where we assume that

$$\left(\left[N^{-1}\int_{q(0)=q_i}^{q(t_f)=q_f} |F_{1,\delta}[\dot{q}(t),q(t)]|^{p_1}\exp\{F_{3,\delta}[\dot{q}(t),q(t)]\}[Dq(t)]\right]^{\frac{1}{p_1}}\right)_\delta < \infty,$$
$$\left(\left[N^{-1}\int_{q(0)=q_i}^{q(t_f)=q_f} |F_{2,\delta}[\dot{q}(t),q(t)]|^{p_1}\exp\{F_{3,\delta}[\dot{q}(t),q(t)]\}[Dq(t)]\right]^{\frac{1}{p_1}}\right)_\delta < \infty \tag{2.50}$$

From (2.46) and (2.48) for $p_2 = \frac{1}{\varepsilon}$, $p_1 = \frac{1}{1-\varepsilon} = 1 + \varepsilon + o(\varepsilon^2) \simeq 1$ we obtain:

$$(\mathbf{E}[v_\delta^2(t,\omega)])_{\delta\in(0,1]} \leq$$

$$\exp\left\{-\frac{1}{\varepsilon}\left(\int_0^{t_f} g_2^2(\lambda,t)dt\right)_\delta\right\} \times$$

$$\left[\widehat{N}^{-1}\int_{-\infty}^{+\infty}dy\int_{q(0)=\frac{q_i}{\sqrt{\varepsilon}}}^{q(t_f)=\frac{y}{\sqrt{\varepsilon}}} y^2\exp\left\{-\left(\int_0^{t_f}[\dot{q}(t)+g_1q(t)]^2dt\right)_\delta\right\}[Dq(t)]\times\right.$$

$$\exp\left[p_1\left(\int_0^{t_f}\partial\mathcal{F}(\sqrt{\varepsilon}\,q(t))dt\right)_\delta\right]\times \tag{2.51}$$

$$\left.\exp\left\{-\frac{p_1}{\sqrt{\varepsilon}}\left(\int_0^{t_f}[2g_2[\dot{q}(t)+g_1q(t)]]dt\right)_\delta\right\}\right]^{\frac{1}{p_1}}\times$$

$$\left[\widehat{N}^{-1}\int_{-\infty}^{+\infty}dy\int_{q(0)=\frac{q_i}{\sqrt{\varepsilon}}}^{q(t_f)=\frac{y}{\sqrt{\varepsilon}}} y^2\exp\left\{-\left(\int_0^{t_f}[\dot{q}(t)+g_1q(t)]^2dt\right)_\delta\right\}[Dq(t)]\times\right.$$

$$\left.\exp\left\{-\frac{1}{\varepsilon}\left(\int_0^{t_f}G_\delta[\dot{q}(t),q(t)]dt\right)_\delta\right\}\right]^{\varepsilon} =$$

$$(\Re_{1,\delta}(\varepsilon,t_f))_\delta\times(\Re_{2,\delta}(\varepsilon,t_f))_\delta.$$

Here

$$\Re_1(\varepsilon,t_f) = \exp\left\{-\frac{1}{\varepsilon}\int_0^{t_f}g_2^2(\lambda,t)dt\right\}\times$$

$$\left[\widehat{N}^{-1}\int_{-\infty}^{+\infty}dy\int_{q(0)=\frac{q_i}{\sqrt{\varepsilon}}}^{q(t_f)=\frac{y}{\sqrt{\varepsilon}}} y^2\exp\left\{-\int_0^{t_f}[\dot{q}(t)+g_1q(t)]^2dt\right\}[Dq(t)]\times\right. \tag{2.52}$$

$$\left.\exp\left[p_1\int_0^{t_f}\frac{\partial\mathcal{F}(\sqrt{\varepsilon}\,q(t))}{\partial q}dt\right]\times\exp\left\{-\frac{p_1}{\sqrt{\varepsilon}}\int_0^{t_f}[2g_2[\dot{q}(t)+g_1q(t)]]dt\right\}\right]^{\frac{1}{p_1}}$$

and

$$(\Re_{2,\delta}(\varepsilon,t_f))_\delta =$$

$$\left[\widehat{N}^{-1}\int_{-\infty}^{+\infty}dy\int_{q(0)=\frac{q_i}{\sqrt{\varepsilon}}}^{q(t_f)=\frac{y}{\sqrt{\varepsilon}}}\exp\left\{-\left(\int_0^{t_f}[\dot{q}(t)+g_1q(t)]^2dt\right)_\delta\right\}[Dq(t)]\times \right. \quad (2.53)$$

$$\left.\exp\left\{-\frac{1}{\varepsilon}\left(\int_0^{t_f}G_\delta[\dot{q}(t),q(t)]dt\right)_\delta\right\}\right]^\varepsilon.$$

Let us now evaluate path integral

$$\left(\widetilde{\Re}_{1,\delta}(\varepsilon,t_f)\right)_\delta =$$

$$\widehat{N}^{-1}\int_{-\infty}^{+\infty}dy\int_{q(0)=\frac{q_i}{\sqrt{\varepsilon}}}^{q(t_f)=\frac{y}{\sqrt{\varepsilon}}}y^2\exp\left\{-\left(\int_0^{t_f}[\dot{q}(t)+g_1q(t)]^2dt\right)_\delta\right\}[Dq(t)]\times$$

$$\exp\left[p_1\left(\int_0^{t_f}\frac{\partial\mathcal{F}_\delta(\sqrt{\varepsilon}\,q(t))}{\partial q}dt\right)_\delta\right]\times \quad (2.54)$$

$$\exp\left\{-\frac{p_1}{\sqrt{\varepsilon}}\left(\int_0^{t_f}[2g_2[\dot{q}(t)+g_1q(t)]]dt\right)_\delta\right\},$$

$$\frac{\partial\mathcal{F}_\delta(\sqrt{\varepsilon}\,q(t))}{\partial q} = 3a\varepsilon q^2(t)+2(3a\lambda+b)\sqrt{\varepsilon}\,q(t)+(3a\lambda^2+2b\lambda+c).$$

Substitution $p_1 = 1+\varepsilon+o(\varepsilon^2)$ into expression (2.54) gives:

$$\left(\widetilde{\Re}_{1,\delta}(\varepsilon,t_f)\right)_{\delta} =$$

$$\widehat{N}^{-1}\int_{-\infty}^{+\infty} dy \int_{q(0)=\frac{q_i}{\sqrt{\varepsilon}}}^{q(t_f)=\frac{y}{\sqrt{\varepsilon}}} y^2 \exp\left\{-\left(\int_0^{t_f}[\dot{q}(t)+g_1 q(t)]^2 dt\right)_{\delta}\right\}[Dq(t)] \times$$

$$\exp[[1+\varepsilon+o(\varepsilon^2)] \tag{2.55}$$

$$\left(\int_0^{t_f}[3a\varepsilon q^2(t)+2(3a\lambda+b)\sqrt{\varepsilon}\,q(t)+(3a\lambda^2+2b\lambda+c)]dt\right)_{\delta}\Bigg] \times$$

$$\exp\left\{-\frac{1+\varepsilon+o(\varepsilon^2)}{\sqrt{\varepsilon}}\left(\int_0^{t_f}[2g_2[\dot{q}(t)+g_1 q(t)]]dt\right)_{\delta}\right\}.$$

From Eq.(2.55) by using replacement $q(t) \to q(t)/\sqrt{\varepsilon}$, we obtain

$$\left(\widetilde{\Re}_{1,\delta}(\varepsilon,t_f)\right)_{\delta\in(0,1]} =$$

$$N^{-1}\int_{-\infty}^{+\infty}dy\int_{q(0)=0}^{q(t_f)=y}\exp\left\{-\frac{1}{\varepsilon}\left(\int_0^{t_f}[\dot{q}(t)+g_1q(t)]^2dt\right)_\delta\right\}\times$$

$$[Dq(t)]y^{2\left[1+\varepsilon+o\left(\varepsilon^2\right)\right]}\times$$

$$\exp[[1+\varepsilon+o(\varepsilon^2)]\times$$

$$\left(\int_0^{t_f}[3aq^2(t)+2(3a\lambda+b)q(t)+(3a\lambda^2+2b\lambda+c)]dt\right)_\delta\Big]\times$$

$$\exp\left\{-\frac{1+\varepsilon+o(\varepsilon^2)}{\varepsilon}\left(\int_0^{t_f}[2g_2[\dot{q}(t)+g_1q(t)]]dt\right)_\delta\right\} = \qquad (2.56)$$

$$N^{-1}\int_{-\infty}^{+\infty}dy\int_{q(0)=0}^{q(t_f)=y}[Dq(t)]y\exp\left\{-\frac{1}{\varepsilon}\left(\int_0^{t_f}[\dot{q}(t)+g_1q(t)]^2dt\right)_\delta\right\}\times$$

$$\exp\left\{-\frac{1}{\varepsilon}\left(\int_0^{t_f}[2g_2[\dot{q}(t)+g_1q(t)]]dt\right)_\delta\right\}\times$$

$$\exp\left\{-\left(\int_0^{t_f}[2g_2[\dot{q}(t)+g_1q(t)]]dt\right)_\delta\right\}\times$$

$$\exp\left\{\left(\int_0^{t_f}[3aq^2(t)+2(3a\lambda+b)q(t)+(3a\lambda^2+2b\lambda+c)]dt\right)_\delta\right\}+O(\varepsilon).$$

From Eq.(2.53) and Eq.(2.56) finally we obtain

$$(\Re_{1,\delta}(\varepsilon,t_f))_\delta =$$

$$\exp\left\{-\frac{1}{\varepsilon}\left(\int_0^{t_f} g_2^2(\lambda,t)dt\right)_\delta\right\} \times \left[\left(\widetilde{\Re}_{1,\delta}(\varepsilon,t_f)\right)_\delta\right]^{1-\varepsilon} =$$

$$N^{-1}\int_{-\infty}^{+\infty} dy \int_{q(0)=0}^{q(t_f)=y} [Dq(t)]y^2 \times$$

$$\exp\left\{-\frac{1}{\varepsilon}\left(\int_0^{t_f}[\dot{q}(t)+g_1(\lambda)q(t)+g_2(\lambda,t)]^2 dt\right)_{\delta\in(0,1]}\right\} \times \tag{2.57}$$

$$\exp\left\{\left(\int_0^{t_f}[-2g_2\dot{q}(t)+3aq^2(t)+2(3a\lambda+b-g_1)q(t)+\right.\right.$$

$$\left.(3a\lambda^2+2b\lambda+c)]dt)_\delta\right\}+O(\varepsilon).$$

In the limit $\varepsilon \to 0$ the path-integral (2.57) is dominated by the minimum of the action

$$(S_\delta[\dot{q}(t),q(t)])_{\delta\in(0,1]} = \left(\int_0^{t_f}[\dot{q}(t)+g_1(\lambda)q(t)+g_2(\lambda,t)]^2 dt\right)_{\delta\in(0,1]}. \tag{2.58}$$

The extremality conditions for the minimizing path $(q_{*,\delta}(t,\lambda))_\delta$ is

$$(\dot{q}_{*,\delta}(t,\lambda))_{\delta\in(0,1]}+(g_1(\lambda)q_{*,\delta}(t,\lambda))_{\delta\in(0,1]}+(g_{2,\delta}(\lambda,t))_{\delta\in(0,1]} = 0. \tag{2.59}$$

In the limit $\varepsilon \to 0,$ the path-integral (2.57) can be evaluated exactly by means of a saddle point approximation about the minimizing path $(q_{*,\delta}(t,\lambda))_\delta$ with the final result is

$$(\Re_{1,\delta}(\varepsilon,t_f))_{\delta\in(0,1]} = (q_{*,\delta}^2(t_f,\lambda)\exp[\psi_\delta(t_f,\lambda)])_\delta; \tag{2.60}$$

Here

$$(\psi_\delta(t_f,\lambda))_\delta =$$

$$\left(\int_0^{t_f}[-2g_2\dot{q}_*(t,\lambda)+3aq_*^2(t,\lambda)+2(3a\lambda+b-g_1)q_*(t,\lambda)+\right. \tag{2.61}$$

$$\left.(3a\lambda^2+2b\lambda+c)dt]\right)_\delta.$$

Let us now evaluate path integral

$$\left(\widetilde{\Re}_{2,\delta}(\varepsilon,t_f)\right)_{\delta\in(0,1]} =$$

$$\widehat{N}^{-1}\int_{-\infty}^{+\infty}dy\int_{q(0)=\frac{q_i}{\sqrt{\varepsilon}}}^{q(t_f)=\frac{y}{\sqrt{\varepsilon}}}\exp\left\{-\left(\int_0^{t_f}[\dot{q}(t)+g_1q(t)]^2dt\right)_\delta\right\}[Dq(t)]\times$$

$$\exp\left\{-\frac{1}{\varepsilon}\left(\int_0^{t_f}G_\delta[\dot{q}(t),q(t)]dt\right)_\delta\right\}, \tag{2.62}$$

$$(G_\delta[\dot{q}(t),q(t)])_{\delta\in(0,1]} = (2\sqrt{\varepsilon}\,g_3\dot{q}(t)q^2(t)+2\varepsilon a\dot{q}(t)q^3(t)+$$

$$h_2q^2(t)+\sqrt{\varepsilon}\,h_3q^3(t)+\varepsilon h_4q^4(t)+\varepsilon\sqrt{\varepsilon}\,h_5q^5(t)+a^2\varepsilon^2q^6(t))_\delta.$$

From (2.40) we obtain

$$\left(\widetilde{\mathfrak{R}}_{2,\delta}(\varepsilon,t_f)\right)_{\delta\in(0,1]} =$$

$$\widehat{N}^{-1}\int_{-\infty}^{+\infty}dy\int_{q(0)=\frac{q_i}{\sqrt{\varepsilon}}}^{q(t_f)=\frac{y}{\sqrt{\varepsilon}}}\exp\left\{-\left(\int_0^{t_f}[\dot{q}(t)+g_1q(t)]^2dt\right)_\delta\right\}[Dq(t)]\times$$

$$\exp\left\{-\frac{1}{\varepsilon}\left(\int_0^{t_f}[2\sqrt{\varepsilon}\,g_3\dot{q}(t)q^2(t)+2\varepsilon a\dot{q}(t)q^3(t)+h_2q^2(t)+\sqrt{\varepsilon}\,h_3q^3(t)+\right.\right.$$

$$\left.\left.\varepsilon h_4q^4(t)+\varepsilon\sqrt{\varepsilon}\,h_5q^5(t)+a^2\varepsilon^2q^6(t)dt]\right\}\right)_{\delta\in(0,1]} = \tag{2.63}$$

$$\widehat{N}^{-1}\int_{-\infty}^{+\infty}dy\int_{q(0)=0}^{q(t_f)=y}\exp\left\{-\left(\int_0^{t_f}[\dot{q}(t)+g_1q(t)]^2dt\right)_{\delta\in(0,1]}\right\}[Dq(t)]\times$$

$$\exp\left\{-\left(\int_0^{t_f}\left[\frac{1}{\sqrt{\varepsilon}}2g_3\dot{q}(t)q^2(t)+2a\dot{q}(t)q^3(t)+\frac{1}{\varepsilon}h_2q^2(t)+\frac{1}{\sqrt{\varepsilon}}h_3q^3(t)+\right.\right.\right.$$

$$\left.\left.\left.h_4q^4(t)+\sqrt{\varepsilon}\,h_5q^5(t)+\varepsilon a^2q^6(t)dt\right]\right)_{\delta\in(0,1]}\right\}.$$

**Lemma.2.1.1**. $\forall t_f\forall\delta\exists\varepsilon$

$$\forall t_f\forall\delta\exists\varepsilon\left[\widetilde{\mathfrak{R}}_{2,\delta}(\varepsilon,t_f)\leq c\right], \tag{2.41}$$

where $c = const$.

**Theorem.2.1.1**.For any $t\in[0,t_f]$ and any parameter $\lambda\in\mathbb{R}$ states that:

$$\liminf_{\varepsilon\to0,\delta\to0}\mathbf{E}[v_\delta^2(t,\omega)]\leq|y(t,\lambda)|\exp[\psi(t)], \tag{2.42}$$

where $y(t,\lambda)$ the solution of the linear differential master equation (2.44).

**Theorem.2.1.2**. For any solution $(x_{\varepsilon,\delta}(t,\omega))_\delta = (x_{\varepsilon,\delta}(t,\omega))_{\delta\in(0,1]}$ of the equation (2.10)□ and any parameter $\lambda\in\mathbb{R}$ states that:

$$\liminf_{\varepsilon\to 0,\delta\to 0} (\mathbf{E}[|x_{\varepsilon,\delta}(t,\omega)-\lambda|]) \leq |y(t,\lambda)|, \tag{2.43}$$

where $y(t,\lambda)$ the solution of the linear differential master equation

$$\dot{y}(t,\lambda) = -(3a\lambda^2 - 2b\lambda^2 + c)y(t,\lambda) - (a\lambda^3 + b\lambda^2 + c\lambda) + u(t) - \lambda\delta(t),$$
$$y(0,\lambda) = 0. \tag{2.44}$$

# **III**. Comparizon with a perturbation theory.

## **III.1**.Path-integrals calculation by using saddle-point approximation.

Let us consider the stochastic dynamics of the form

$$\dot{x}(t,\omega) = F(x(t,\omega),t) + \sqrt{2\epsilon}\ \xi(t,\omega)\ ,x(0,\omega) = x_0,$$
$$\langle\xi(t,\omega)\,\xi(t',\omega)\rangle\ = \delta(t-t'),$$
$$F(x,t) = -V'(x) + A\sin(\Omega t),$$
$$\Omega = \frac{2\pi}{T}. \tag{3.1}$$

Where $V(x)$ is a metastable static potential as cartooned in Fig.3.1.[18]:

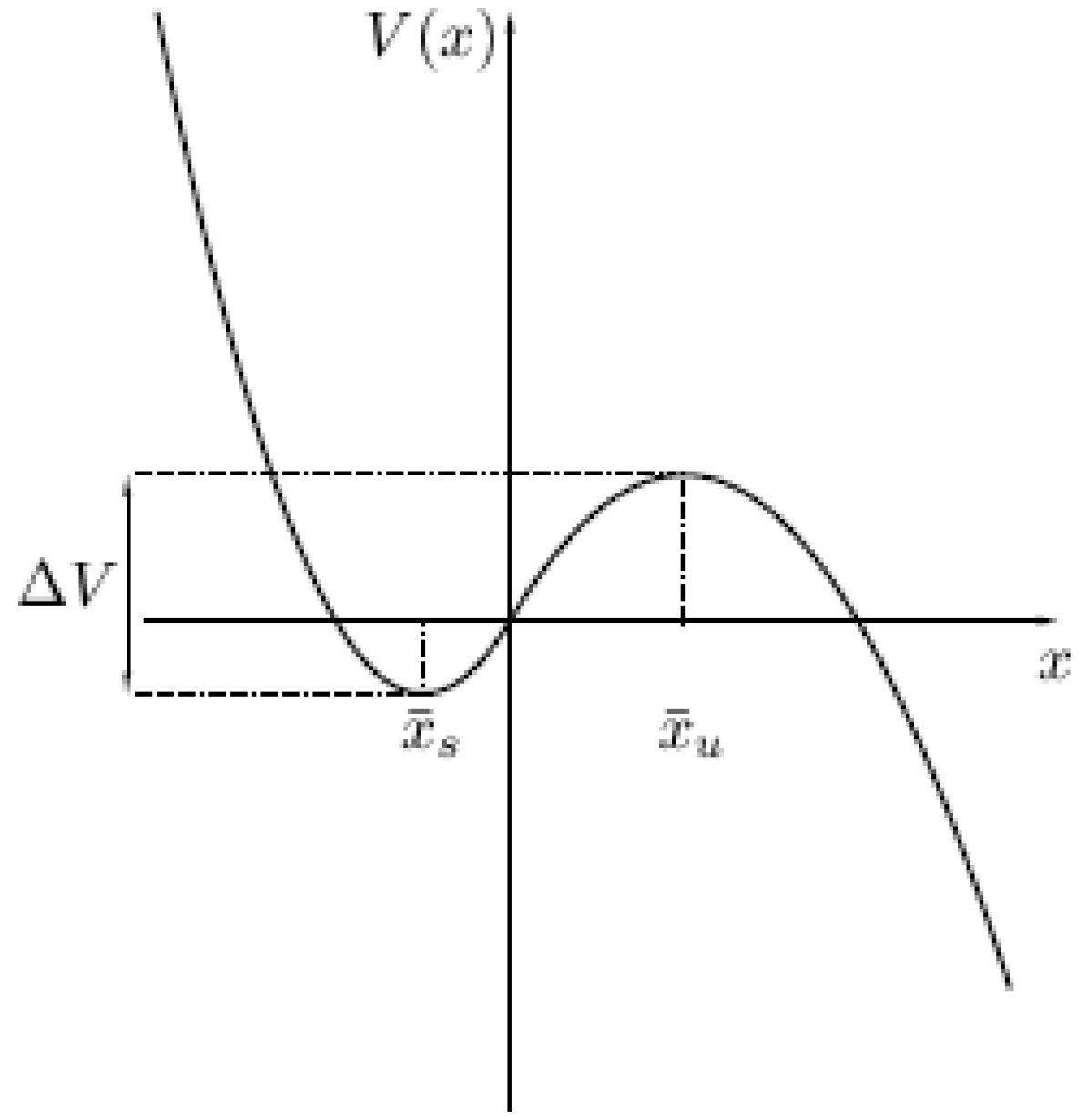

**Fig.** 3.1. Metastable static potential $V(x)$.

Thus corresponding deterministic dynamics with $\epsilon = 0$ has a a stable periodic orbit $x_{\mathbf{s}}(t)$ and an unstable periodic orbit $x_{\mathbf{u}}(t)$, which satisfy following equalities

$$\dot{x}_{\mathbf{s},\mathbf{u}}(t) = F(x_{\mathbf{s},\mathbf{u}}(t), t),$$

$$x_{\mathbf{s},\mathbf{u}}(t + \mathcal{T}) = x_{\mathbf{s},\mathbf{u}}(t)\ , \tag{3.2}$$

where '$\mathbf{s},\mathbf{u}$' means that the index may be either '$\mathbf{s}$' or '$\mathbf{u}$'.Hence,every deterministic trajectory to approach in the long-time limit either the attractor $x_{\mathbf{s}}(t)$ or to diverge towards $x = \infty$, except if it starts exactly at the separatrix $x_{\mathbf{u}}(t)$ between those two basins of attraction and we have, without loss of generality, implicitly restricted ourselves to case such that: $\forall t[x_{\mathbf{u}}(t) > x_{\mathbf{s}}(t)]$. Note that the conditional probability density $p(x_0, t_0|x, t) = p(x, t)$ of stochastic dinamics (3.1) is governed by the canonical Fokker-Planck equation:

$$\frac{\partial}{\partial t}p(x,t) = \frac{\partial}{\partial x}\{-F(x,t)+\epsilon^2\frac{\partial}{\partial x}\}p(x,t),$$
$$p(x,0) = \delta(x-x_0). \tag{3.3}$$

Hence for the conditional expectation $\mathbf{E}[x^2(t,\omega)] = \mathbf{E}[x^2(t,\omega)|x(0,\omega) = x_0]$ we have the canonical path integral representation:

$$\mathbf{E}[x^2(t,\omega)] = \int_{-\infty}^{\infty} x_f^2\, p(x,t)dx_f =$$
$$\int_{-\infty}^{\infty}\int_{x(t_0)=x_0}^{x(t)=x_f} x_f^2 \exp\Big[-\frac{1}{\epsilon}S[\dot{x}(t),x(t)]\Big]dx_f[Dx(t)], \tag{3.4}$$
$$S[\dot{x}(t),x(t)] = \frac{1}{4}\int_0^t[\dot{x}(t')-F(x(t'),t')]dt'.$$

Note that for conditional probability density $p_N(x_0,t_0|x_f,t_f)$ the time discretized path-integral representation is [18]:

$$p_N(x_f,t_f|x_0,t_0) = \underbrace{\int\cdots\int}_{N-1} \exp\left\{-\frac{S_N(x_0,\ldots,x_N)}{\epsilon}\right\}\frac{dx_1\cdots dx_{N-1}}{(4\pi\epsilon\Delta t)^{N/2}}, \tag{3.5}$$

with

$$S_N(x_0,\ldots,x_N) = \sum_{n=0}^{N-1}\frac{\Delta t}{4}\left[\frac{x_{n+1}-x_n}{\Delta t} - F(x_n,t_n)\right]^2, \tag{3.6}$$

where the initial-$x_0$ and end-points $x_N$ are fixed by the prescribed $x_0$ and by the additional constraint $x_N = x_f$. From Eqs.(3.4)-(3.6) we obtain

$$\mathbf{E}[x^2(t,\omega)] = \lim_{\Delta t \to 0} \mathbf{E}[x_N^2(t_N,\omega)] =$$

$$\int_{-\infty}^{\infty} x_N^2 p_N(x_0,t_0|x_N,t)dx_N = \tag{3.7}$$

$$\underbrace{\int \ldots \int}_{N} x_N^2 \exp\left\{-\frac{S_N(x_0,\ldots,x_N)}{\epsilon}\right\} \frac{dx_1 \cdots dx_{N-1}dx_N}{(4\pi\epsilon\Delta t)^{N/2}}.$$

Denoting the global minimum of the discrete-time action $S_N(x_0,\ldots,x_N)$ by $\mathbf{x}^{\#} =$ $= (x_0,x_1^{\#},\ldots,x_N^{\#})$ it follows that $x^{\#}$ satisfies the extremality conditions

$$\frac{\partial S_N(\mathbf{x}^{\#})}{\partial x_n^{\#}} = 0, \tag{3.8}$$

for $n = 1,\ldots,N$, supplemented by the prescribed boundary condition for $n = 0$ :

$$x_0^{\#} = x_0. \tag{3.9}$$

From (3.7) in the limit $\epsilon \to 0$ by using saddle point approximation about the minimizing path $\mathbf{x}^{\#} = (x_0,x_1^{\#},\ldots,x_N^{\#})$ we obtain

$$\mathbf{E}[x_N^2(t_N,\omega)] =$$

$$\int_{-\infty}^{\infty} x_N^2 p_N(x_0,t_0|x_N,t)dx_N =$$

$$\underbrace{\int \ldots \int}_{N} x_N^2 \exp\left\{-\frac{S_N(x_0,\ldots,x_N)}{\epsilon}\right\} \frac{dx_1 \cdots dx_{N-1}dx_N}{(4\pi\epsilon\Delta t)^{N/2}} = \tag{3.10}$$

$$\mathbf{Z}_N(\mathbf{x}^{\#})x_N^{\#2} \exp\left[-\frac{1}{\epsilon}S(\mathbf{x}^{\#})\right] + o(\epsilon),$$

where the prefactor $\mathbf{Z}_N(\mathbf{x}^{\#})$ is given via $N$-dimensional Gaussian integral of the canonical form as

$$\mathbf{Z}_N(\mathbf{x}^{\#}) = \underbrace{\int \ldots \int}_{N} \exp\left\{ -\frac{1}{2\epsilon} \sum_{n,m=1}^{N} y_n \frac{\partial^2 S(\mathbf{x}^{\#})}{\partial x_n^{\#} \partial x_m^{\#}} y_m \right\} \frac{dy_1 \cdots dy_{N-1} dy_N}{(4\pi D \Delta t)^{N/2}}. \qquad (3.11)$$

The Gaussian integral in (3.11) is given via formula

$$Z_N(\mathbf{x}^*) = \left[ 2\Delta t \det\left( 2\Delta t \frac{\partial^2 S(\mathbf{x}^{\#})}{\partial x_n^{\#} \partial x_m^{\#}} \right) \right]^{-\frac{1}{2}}$$

$$n = 1, \ldots, N, m = 1, \ldots, N, \qquad (3.12)$$

Equation (3.12) we rewrite in the following form:

$$Z_N(\mathbf{x}^*) = [2DQ_N^{\#}]^{-1/2}. \qquad (3.13)$$

As demonstrated in [18] quantity $Q_N^{\#}$ in (3.13) can be calculated by using a second order linear recursion procedure given via formulae:

$$\frac{Q_{n+1}^* - 2Q_n^* - Q_{n-1}^*}{\Delta t^2} = 2\frac{Q_n^* F'(x_n^*, t_n) - Q_{n-1}^* F'(x_{n-1}^*, t_{n-1})}{\Delta t} -$$

$$Q_n^* \left[ \frac{x_{n+1}^* - x_n^*}{\Delta t} - F(x_n^*, t_n) \right] F''(x_n^*, t_n) +$$

$$Q_n^* F'(x_n^*, t_n)^2 - Q_{n-1}^* F'(x_{n-1}^*, t_{n-1})^2, \qquad (3.14)$$

$$Q_1^* = \Delta t,$$

$$\frac{Q_2^* - Q_1^*}{\Delta t} = 1 + o(\Delta t)),$$

$$F'(x,t) = \frac{\partial F(x,t)}{\partial x}, F''(x,t) = \frac{\partial^2 F(x,t)}{\partial x^2}.$$

As well known shall see later, we have to leave room for the possibility that even for small noise-strengths $\epsilon \to 0$ more than one (global or local) minimum of the action (3.6) notably contributes to the path-integral expression (3.7). We label those various minima $x_k^{\#}$ by the discrete index $k$. Thus from Eq.(3.10) and Eq.(3.13) we obtain

$$\mathbf{E}[x_N^2(t_N,\omega)] = \int_{-\infty}^{\infty} x_N^2 p_N(x_0,t_0|x_N,t)dx_N =$$

$$= \sum_{k=0}^{M} \frac{(x_{N,k}^{\#})^2 \exp\left[\frac{-S_N(\mathbf{x}_k^{\#})}{\epsilon}\right]}{\sqrt{2Q_{N,k}^{\#}}}[1+o(\epsilon))]. \tag{3.15}$$

The continuous-time limit of the action $S_N$ is

$$S[\dot{x}(t),x(t)] = \frac{1}{4}\int_{t_0}^{t_f}[\dot{x}-F(x,t)]^2dt \tag{3.16}$$

The extremality conditions for the minimizing paths $x_k^{\#}(t)$ in the continuous-time limit are obtained from (3.6) and (3.8) by letting $\Delta t \to 0$ is:

**(I)**

$$\dot{x}^{\#}(t,x_0) = F(x^{\#}(t,x_0),t), x^{\#}(0) = x_0. \tag{3.17}$$

**(II)**

$$\begin{gathered}\ddot{x}_k^{\#}(t,x_k^{\#}(t_f)) = \dot{F}(x_k^{\#}(t,x_k^{\#}(t_f)),t) + F(x_k^{\#}(t,x_k^{\#}(t_f)),t)F'(x_k^{\#}(t,x_k^{\#}(t_f)),t),\\ x_k^{\#}(0) = x_0, x_k^{\#}(t_f) = \widetilde{x}_k,\\ \left.\frac{dS[\dot{x}_k^{\#}(t,x_k^{\#}(t_f)),x_k^{\#}(t,x_k^{\#}(t_f))]}{d(x_k^{\#}(t_f))}\right|_{x_k^{\#}(t_f)=\widetilde{x}_k} = 0.\end{gathered} \tag{3.18}$$

Here we use the canonical definitions $F'(x,t) = \frac{\partial F(x,t)}{\partial x}, \dot{F}(x,t) = \frac{\partial F(x,t)}{\partial t}$.

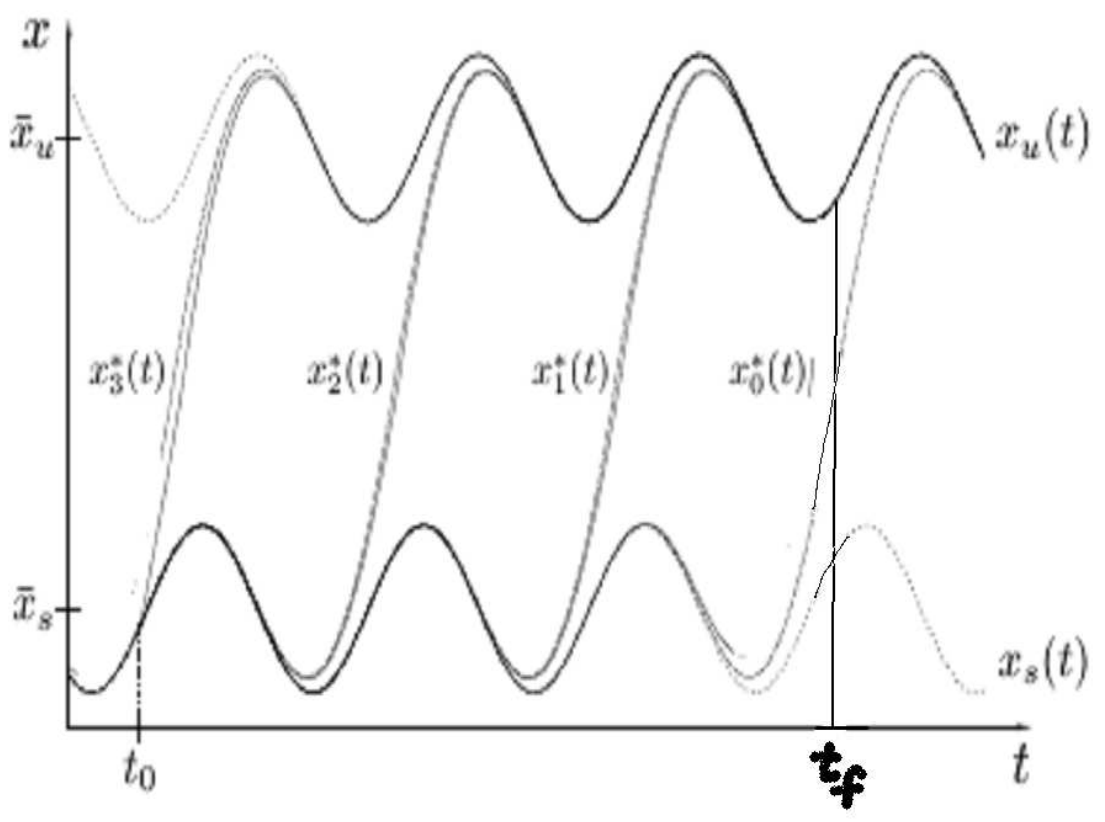

**FIG.** 3.2. The path $x_0^{\#}(t, x_0)$ minimize the action (3.16) and satisfy the extremality conditions (3.17) .
The paths $x_1^{\#}(t, \widetilde{x}_2), x_2^{\#}(t, \widetilde{x}_1), x_3^{\#}(t, \widetilde{x}_0)$ minimize the action (3.16) and satisfy the extremality conditions (3.18) .

Hamiltonian counterpart of the Lagrangian dynamics given by (3.18) is

$$
\begin{aligned}
H(x,p,t) &= p\dot{x} - L = p^2 + pF(x,t), \\
\dot{p}_k^{\#}(t) &= -p_k^{\#}(t)F'(x_k^{\#}(t),t), \\
\dot{x}_k^{\#}(t) &= 2p_k^{\#}(t) + F(x_k^{\#}(t),t).
\end{aligned}
\tag{3.19}
$$

From the last equation of the Eqs.(3.19) one obtain the momentum $p_k^{\#}(t)$ in terms of $x_k^{\#}(t)$ and $\dot{x}_k^{\#}(t)$ :

$$
p_k^{\#}(t) = \frac{1}{2}[\dot{x}_k^{\#}(t) - F(x_k^{\#}(t),t)]. \tag{3.20}
$$

From Eq.(3.16) and Eq.(3.20) we obtain [18]:

$$
\Phi_k(x_f,t_f) = S[x_k^{\#}(t)] = \int_{t_0}^{t_f} p_k^{\#}(t)^2\,dt\,. \tag{3.21}
$$

(**I**) In the limit $\Delta t \to 0$ from the extremality conditions (3.17) for the minimizing paths $x^{\#}(t,x_0)$ and a second order linear recursion procedure given via formulae (3.14) we obtain the second order homogeneous linear differential equation

$$\frac{1}{2}\ddot{Q}_0^{\#}(t,x_0) - \frac{d}{dt}[Q_0^{\#}(t)F'(x^{\#}(t,x_0),t)],$$
$$Q_0^{\#}(t_0,x_0) = 0\ , \quad \dot{Q}_0^{\#}(t_0,x_0) = 1. \tag{3.22}$$

By integration Eq.(3.22) we obtain

$$\frac{1}{2}[\dot{Q}_\varepsilon^{\#}(t) - \dot{Q}_\varepsilon^{\#}(t_0)] = Q_\varepsilon^{\#}(t)F'(x_\varepsilon^{\#}(t,x_0),t) - Q_0^{\#}(t_0)F'(x^{\#}(t,x_0),t),$$

and finally we obtain the first order homogeneous linear differential equation

$$\dot{Q}_0^{\#}(t) = 2Q_0^{\#}(t)F'(x^{\#}(t,x_0),t) + 1.$$

Substitution the extremality conditions (3.17) for the minimizing path $x^{\#}(t,x_0)$ into action $S[\dot{x}(t,x_0),x(t,x_0)]$ expressed by Eq.(3.16) gives

$$S[\dot{x}^{\#}(t,x_0),x^{\#}(t,x_0)] \ \equiv 0. \tag{3.23}$$

Hence in the continuous-time limit $\Delta t \to 0$ the minimizing path $x^{\#}(t,x_0)$ contributes to the path-integral expression (3.7) as

$$\frac{x^{\#}(t_f,x_0)}{\sqrt{2Q_0^{\#}(t_f,x_0)}}[1 + o(\epsilon)]. \tag{3.24}$$

(**II**) In the limit $\Delta t \to 0$ from the extremality conditions (3.18) for the minimizing paths $x_k^{\#}(t,\widetilde{x}_k)$ and a second order linear recursion procedure given via formulae (3.14) we obtain the second order homogeneous linear differential equation:

$$\frac{1}{2}\ddot{Q}_k^{\#}(t) - \frac{d}{dt}[Q_k^{\#}(t)F'(x_k^{\#}(t,\widetilde{x}_k),t)] +$$

$$+Q_k^{\#}(t)\,p_k^{\#}(t)\,F''(x_k^{\#}(t,\widetilde{x}_k),t) = 0\ , \tag{3.25}$$

$$Q_k^{\#}(t_0) = 0\ \ ,\quad \dot{Q}_k^{\#}(t_0) = 1\ .$$

Hence in the continuous-time limit $\Delta t \to 0$ the minimizing paths $x_k^{\#}(t,\widetilde{x}_k)$ contributes to the path-integral expression (3.7) as

$$\sum\nolimits_k \frac{x_k^{\#}(t_f,\widetilde{x}_k)}{\sqrt{2Q_k^{*}(t_f)}}\exp\left[-\frac{\Phi_k(\widetilde{x}_k,t_f)}{\epsilon}\right][1+o(\epsilon)]. \tag{3.26}$$

Summarizing Eq.(3.24) and Eq.(3.26) finally we obtain

$$\mathbf{E}[x_\epsilon(t,\omega)] = \frac{x^{\#}(t_f,x_0)}{\sqrt{2Q_0^{\#}(t_f,x_0)}} +$$

$$\sum\nolimits_k \frac{x_k^{\#}(t_f,\widetilde{x}_k)}{\sqrt{2Q_k^{*}(t_f)]}}\exp\left[-\frac{\Phi_k(\widetilde{x}_k,t_f)}{\epsilon}\right] + [1+\,o(\epsilon)]. \tag{3.27}$$

$$[1+\,o(\epsilon)].$$

$\Phi_k(\widetilde{x}_k,t_f) \neq 0$ consequently in the limit $\epsilon \to 0$ second term $\sum_k(\cdot)$ in Eq.(3.27) is vanishes. Thus one obtain:

$$\mathbf{E}_{\mathbf{p}}(t)= \lim_{\epsilon\to 0}\mathbf{E}[x_\epsilon(t,\omega)] = \frac{x^{\#}(t_f,x_0)}{\sqrt{2Q_0^{\#}(t_f,x_0)}}. \tag{3.28}$$

## III.2.Numerical examples.

An example is a double well potential $V(x)$ as cartooned in Fig.3.1. We consider a force field given by Eq.(1.31) with $B = 0$, i.e.

$$F(x,t) = V'(x) + A\sin(\Omega t) =$$

$$-ax^3 + bx + A\sin(\Omega t), \tag{3.29}$$

$$F'(x,t) = -2ax^2 + b.$$

From Eq.(3.22$'$) and Eq.(3.29) we obtain

$$\dot{Q}_0^{\#}(t,a,b) = 2Q_0^{\#}(t)[-2ax^{\#2}(t) + b] + 2. \tag{3.30}$$

The stochastic dynamics is

$$\dot{x}_\epsilon(t) = -ax_\epsilon^3(t) + bx_\epsilon(t) + A\sin(\Omega t) + \sqrt{2\epsilon}\,\xi(t),$$

$$x(0) = 0, \epsilon \simeq 0. \tag{3.31}$$

Corresponding linear differential master equation is

$$\dot{y} = -(3a\lambda^2 - b)y - (a\lambda^3 - b\lambda) + A\sin(\Omega t),$$

$$y(0) = -\lambda. \tag{3.32}$$

Corresponding transcendental master equation is

$$(x_0 - \lambda(t))\exp[-(3a\lambda^2(t) - b)t] -$$

$$(a\lambda^3(t) - b\lambda(t))\int_0^t \exp[-(3a\lambda^2(t) - b)(t-\tau)]d\tau + \quad (3.33)$$

$$A\int_0^t \sin(\Omega\tau)\exp[-(3a\lambda^2(t) - b)(t-\tau)]d\tau = 0.$$

$$(x_0 - \lambda(t))\exp[-\Theta(\lambda)t] - \Delta(\lambda)\int_0^t \exp[-\Theta(\lambda)(t-\tau)]d\tau +$$

$$A\int_0^t \sin(\Omega\tau)\exp[-\Theta(\lambda)(t-\tau)]d\tau = 0. \quad (3.34)$$

$$\Theta(\lambda) = (3a\lambda^2(t) - b),$$

$$\Delta(\lambda) = a\lambda^3(t) - b\lambda(t)$$

From Eq.(3.34) we obtain

$$(x_0 - \lambda(t))\exp[-\Theta(\lambda)t] - \Delta(\lambda)\int_0^t \exp[-\Theta(\lambda)(t-\tau)]d\tau + \quad (3.35)$$

$$+A\int_0^t \sin(\Omega\tau)\exp[-\Theta(\lambda)(t-\tau)]d\tau = 0.$$

From Eq.(3.35) we obtain

$$(x_0 - \lambda(t)) - \Delta(\lambda)\int\limits_0^t \exp[\Theta(\lambda)\tau]d\tau + A\int\limits_0^t \sin(\Omega\tau)\exp[\Theta(\lambda)\tau]d\tau = 0. \qquad (3.36)$$

From Eq.(3.36) finally we obtain

$$\begin{gathered}(x_0 - \lambda(t)) - \frac{\Delta(\lambda)}{\Theta(\lambda)}\{\exp[\Theta(\lambda)t] - 1\} + \\ A\frac{\exp[\Theta(\lambda)t]}{\Omega^2 + \Theta^2(\lambda)}[\Theta(\lambda)\sin(\Omega t) - \Omega\cos(\Omega t)] + A\frac{\Omega}{\Omega^2 + \Theta^2(\lambda)} = 0.\end{gathered} \qquad (3.37)$$

$$\int e^{ax}\sin nx = \frac{e^{ax}}{a^2 + n^2}(a\sin nx - n\cos nx)$$

$$\mathbf{E_p}(t) = \lim_{\epsilon\to 0}\mathbf{E}[x_\epsilon(t,\omega)] = \frac{x^{\#}(t,x_0)}{\sqrt{2Q_0^{\#}(t,a,b)}}. \qquad (3.34)$$

**Example.3.1.**

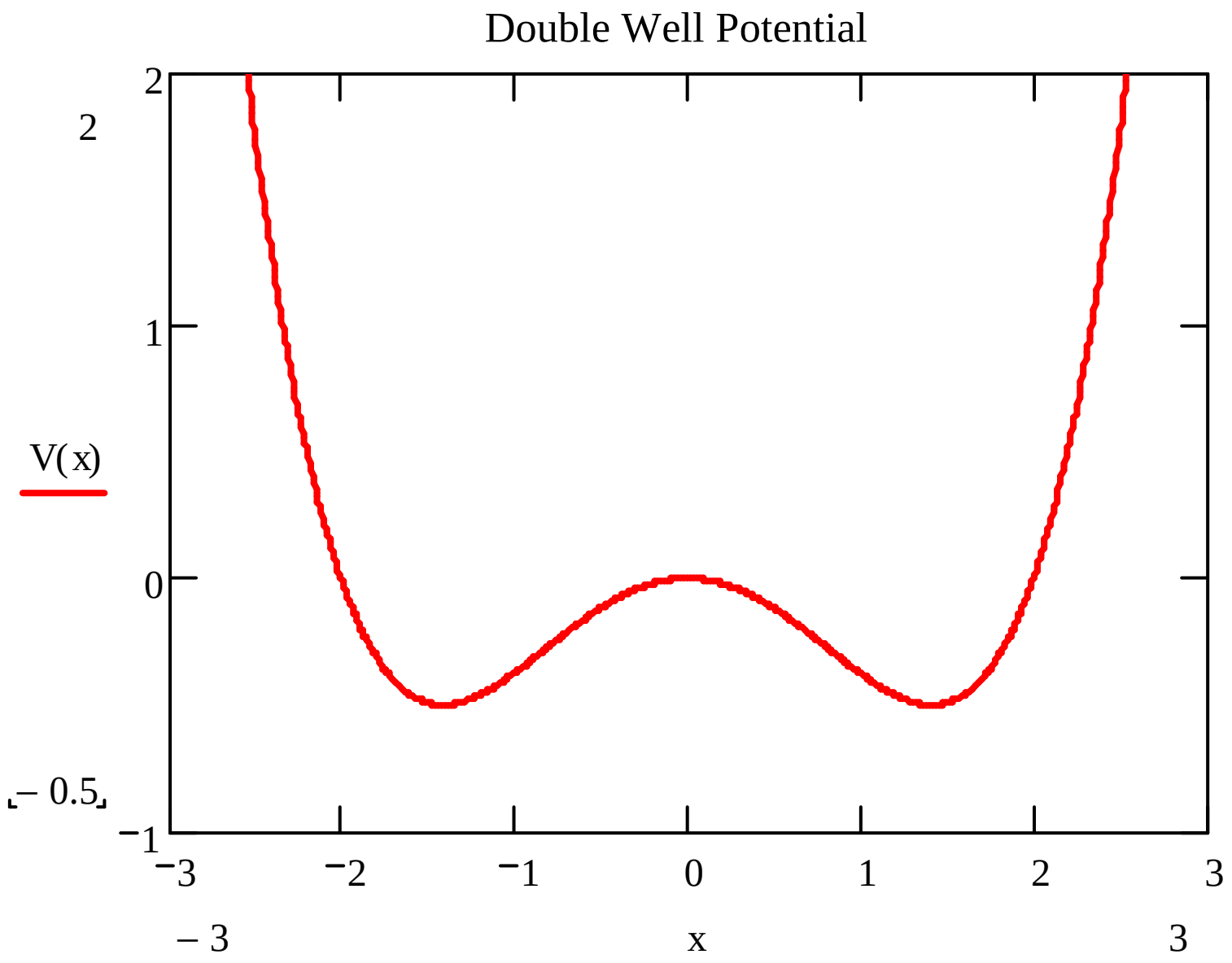

**FIG. 3. 1.** $a = 0.5, b = 1.$

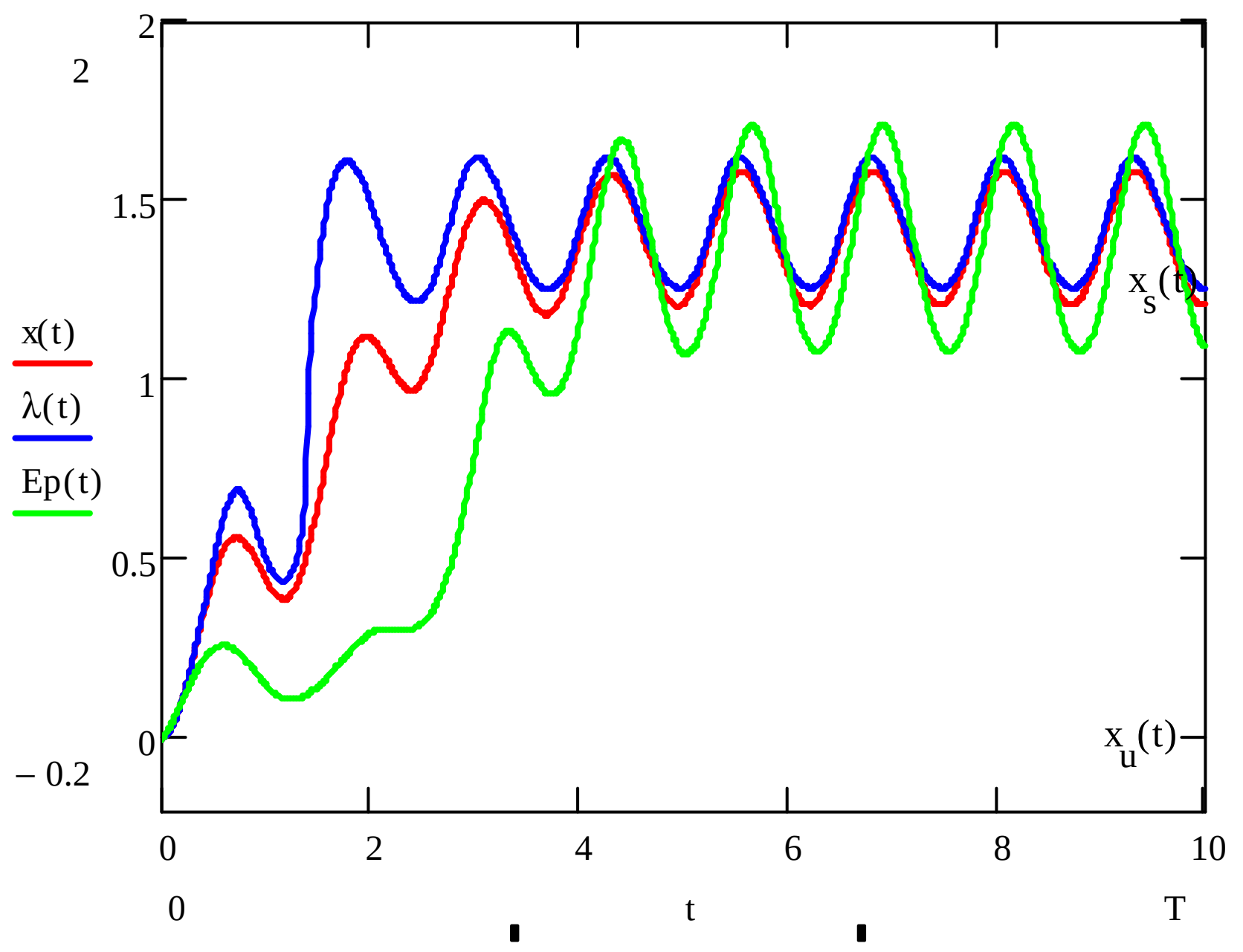

**FIG. 3. 2.** Comparison of classical (red curve $x(t)$) dynamics, stochastic (blue curve $\lambda(t)$) dynamics calculated by using LDP and stochastic (green curve $\mathbf{E_p}(t)$) dynamics calculated by using saddle-point approximation.
$a = 0.5, b = 1, A = 1.$

**Example.3.2.**

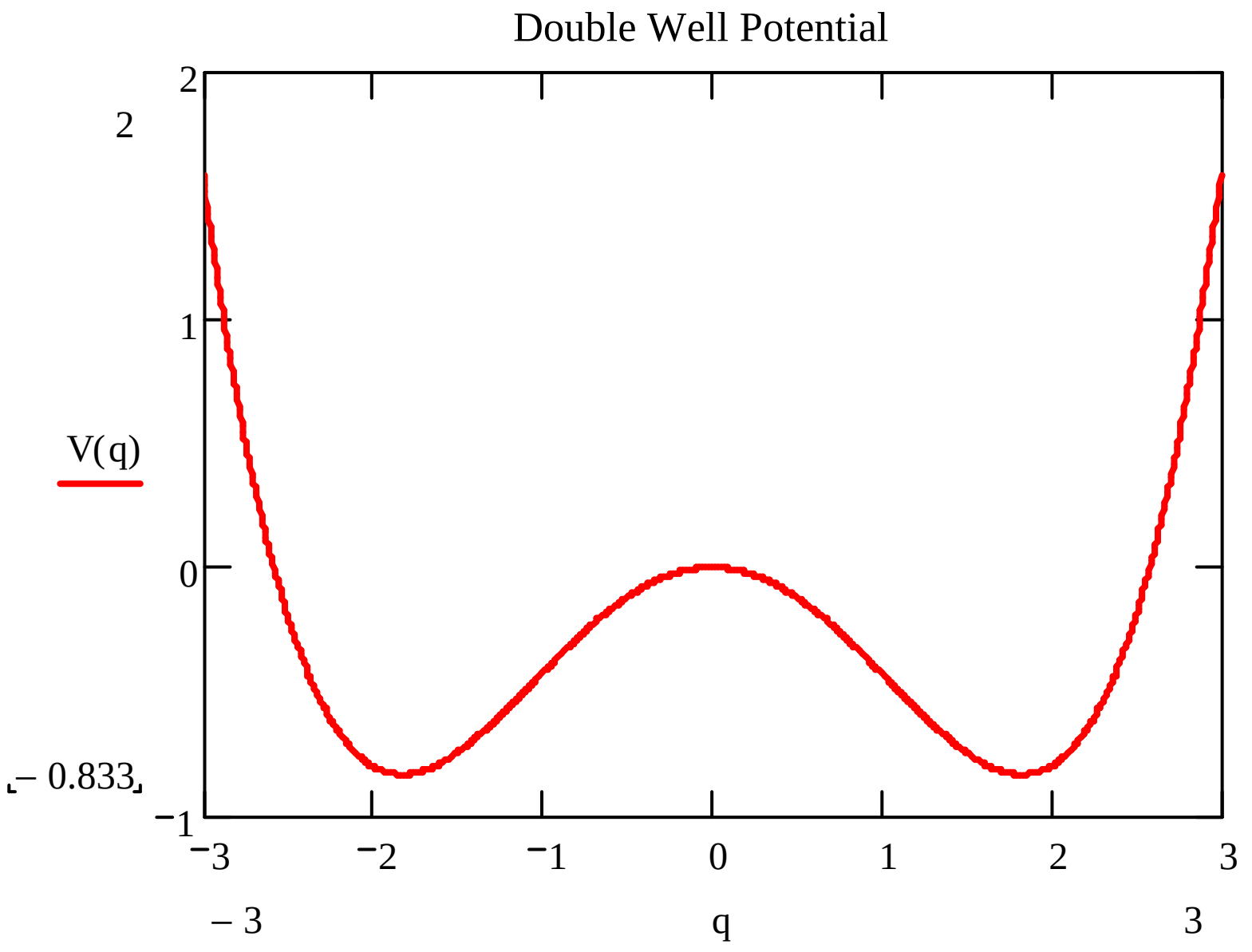

**FIG. 3. 3.** $a = 0.3, b = 1.$

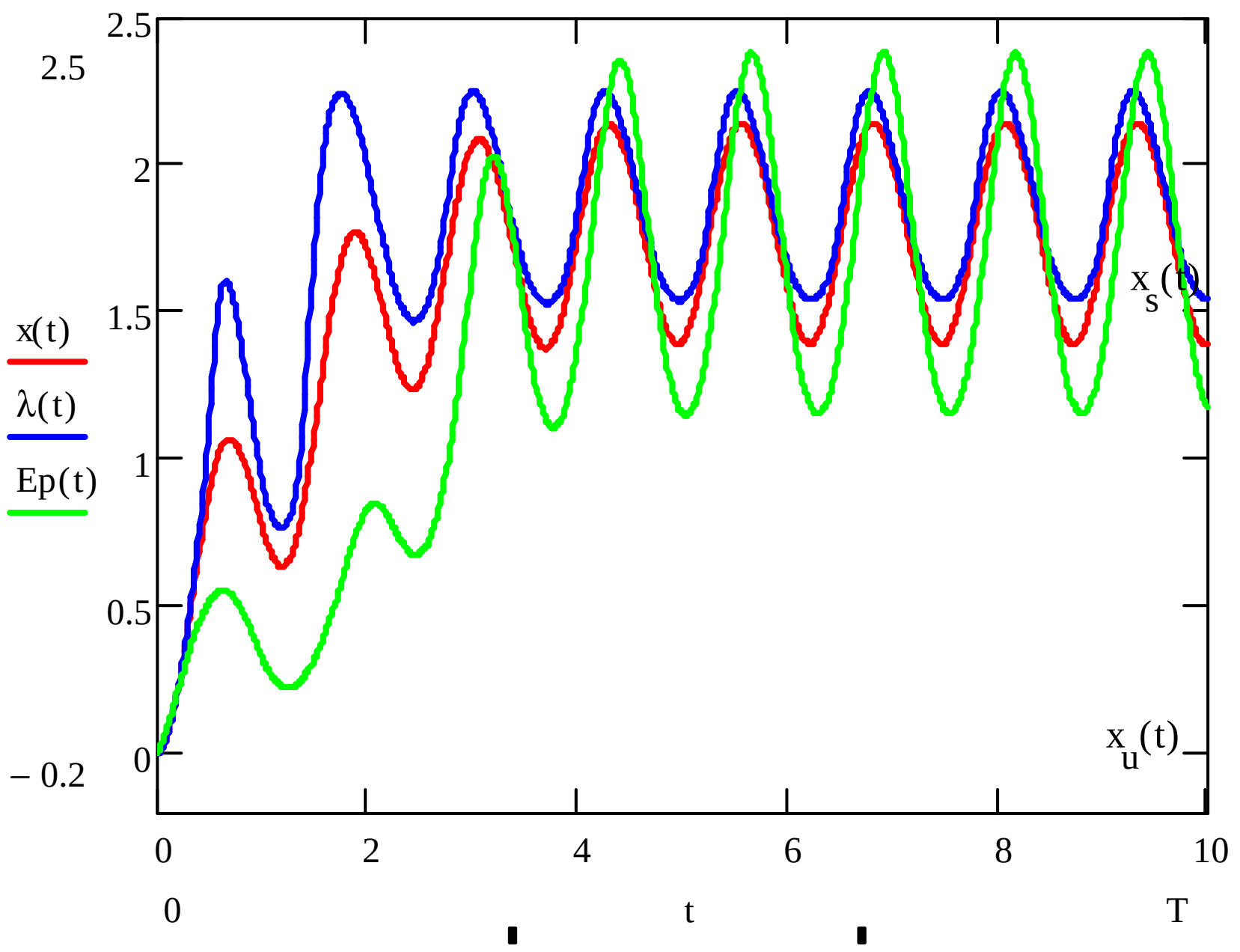

**FIG. 3. 4.** Comparison of classical (red curve $x(t)$) dynamics, stochastic (blue curve $\lambda(t)$) dynamics calculated by using LDP and stochastic (green curve $\mathbf{E_p}(t)$) dynamics calculated by using saddle-point approximation.

$a = 0.3, b = 1, A = 2$

**Example.3.3.**

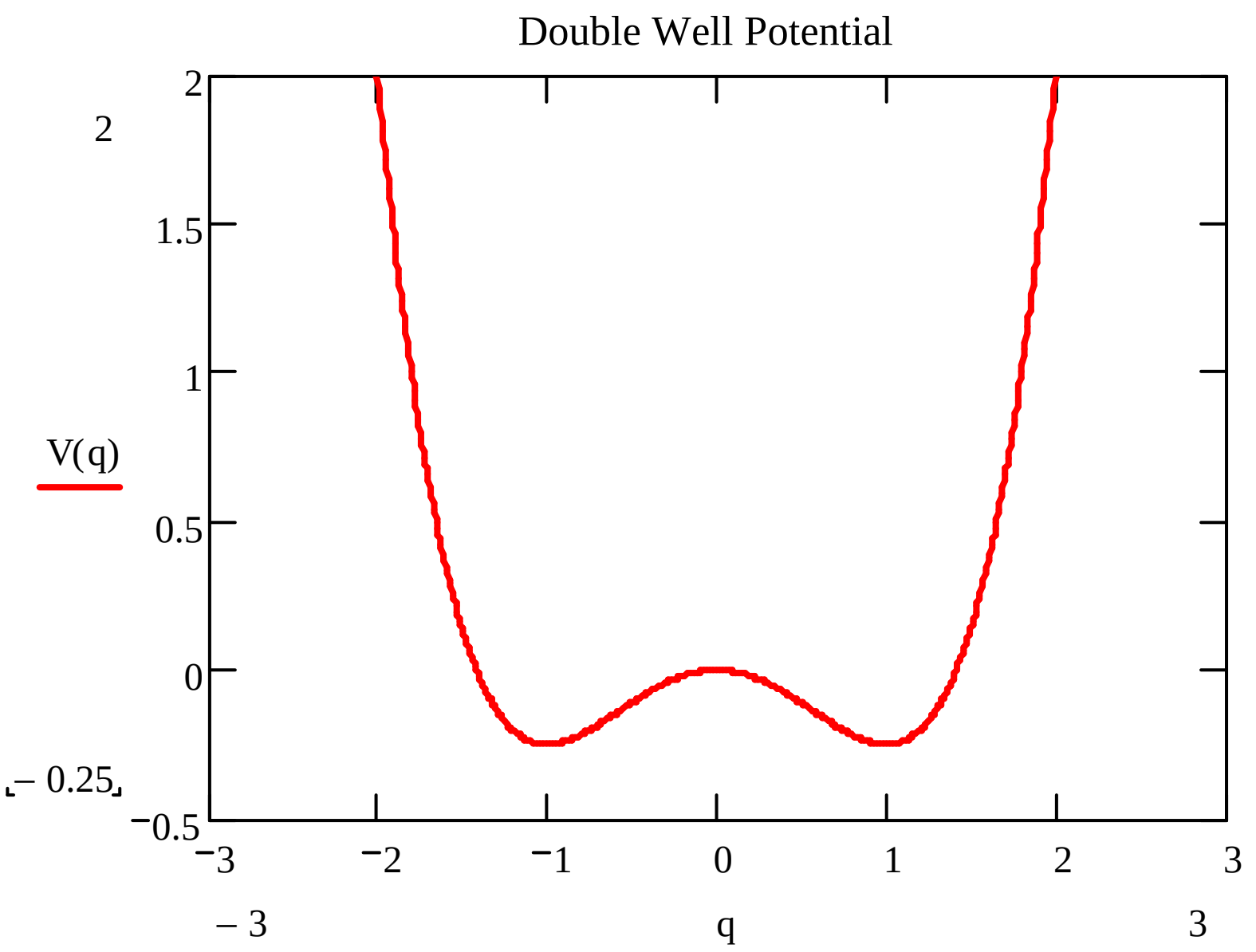

**FIG. 3.5.** $a = b = 1.$

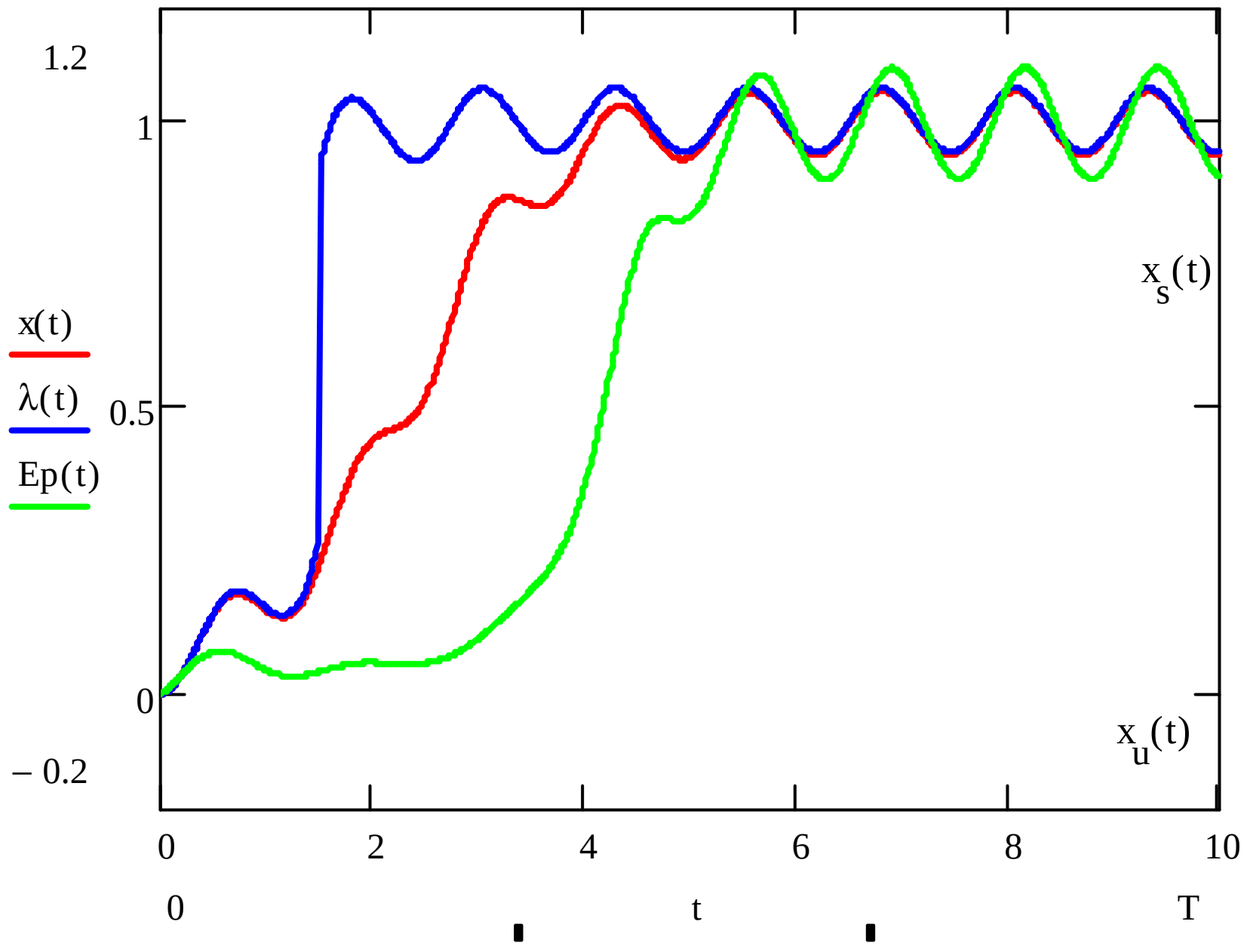

**FIG. 3. 6.** Comparison of classical (red curve $x(t)$) dynamics, stochastic (blue curve $\lambda(t)$) dynamics calculated by using LDP and stochastic (green curve $\mathbf{E_p}(t)$) dynamics calculated by using saddle-point approximation.

$$a = b = 1, A = 0.3.$$

# **IV**.SQM and Nature of quantum jumps in Euclidean quantum mechanics from LDP

Contemporary experiments have found abrupt changes in the intensity of laser light scattered off single ions and atoms. These abrupt changes have been usually interpreted as resulting from Bohr's "quantum jumps" of the ion or atom from one state to another. In Bohr's description, such quantum jumps are not predicted by the WKB approksimation or Maslov asimptotic expansion (i.e.,Maslov canonical operator method) of the solutions of the Schrödinger equation and usually are introduced by additional postulates into quantum theory. We consider here an alternative explanation of quantum jumps using only properties of solutions of the Schrödinger equation beyond WKB (and Maslov canonical operator method) approksimation of the solutions of the Schrödinger equation and avoiding Bohr's added postulate.

As is well known, one can calculate the quantum-mechanical expectation value of an observable, $G(\mathbf{q})$, or other quantities by means of the following path-integral formula,

$$\langle G\rangle = N^{-1}\int G(\mathbf{q}(t))\exp[-\mathbf{S}_E[\mathbf{q}(t)]]\mathbf{D}[\mathbf{q}(t)],$$
$$N = \int \exp[-\mathbf{S}_E[\mathbf{q}(t)]]\mathbf{D}[\mathbf{q}(t)] \tag{4.1}$$

and here we have used Euclidean space-time coordinates obtained by the Wick rotation.

We define Euclidean quantum tradjectory by formulae:

$$1.\langle\mathbf{q}(t)|t_1,\mathbf{q}_1;t_2,\mathbf{q}_2\rangle = N^{-1}\int_{\mathbf{q}(t_1)=\mathbf{q}_1}^{\mathbf{q}(t_2)=\mathbf{q}_2}\mathbf{q}(t)\exp[-\mathbf{S}_E[\mathbf{q}(t)]]\mathbf{D}[\mathbf{q}(t)],$$

$$2.\langle\mathbf{q}(t)|t_1,\mathbf{q}_1\rangle = N^{-1}\int_{\mathbf{q}(t_1)=\mathbf{q}_1}\mathbf{q}(t)\exp[-\mathbf{S}_E[\mathbf{q}(t)]]\mathbf{D}[\mathbf{q}(t)], \tag{4.1$'$}$$

$$3.\langle\mathbf{q}(t)\rangle = N^{-1}\int\mathbf{q}(t)\exp[-\mathbf{S}_E[\mathbf{q}(t)]]\mathbf{D}[\mathbf{q}(t)],$$

$$N = \int\exp[-\mathbf{S}_E[\mathbf{q}(t)]]\mathbf{D}[\mathbf{q}(t)].$$

We use SQM as an abbreviation of the Parisi-Wu stochastic quantization method [51]-[53 ]. As is well known, SQM gives canonical Euclidean quantum mechanics from the thermal equilibrium limit of an hypothetical stochastic process with respect to fictitious time, say $\tau$, other than ordinary time $t$ for particles. For the purpose of making the hypothetical stochastic process, one first introduce an additional dependence of $q$ on $\tau$, and set a Langevin equation [51]-[53 ]

$$\frac{\partial q_i(\tau,t)}{\partial\tau} = -\frac{\delta\mathbf{S}_E[\mathbf{q}(t)]}{\delta q_i(t)}\bigg|_{\mathbf{q}=\mathbf{q}(\tau,t)} + \eta_i(\tau,t,\omega), \tag{4.2}$$

where the $\eta_i, i = 1,\ldots,n$ are Gaussian white noise, i.e.

$$\langle\eta_i(\tau,t,\omega)\rangle = 0,$$
$$\langle\eta_i(\tau,t,\omega)\eta_j(\tau',t',\omega)\rangle = 2\hbar\delta_{ij}\delta(\tau-\tau')\delta(t-t'). \tag{4.3}$$

Solving Eq.(4.2) one obtain $\mathbf{q}(\tau,t)$ and then an arbitrary quantity $G(\mathbf{q}(\tau,t))$ as a function or functional of $\eta_i$'s, whose expectation values, $\langle q_i(\tau,t,\omega)\rangle$, $\langle G(\mathbf{q}(\tau,t,\omega))\rangle$ are given by averaging them over $\eta_i$'s. On the other hand, one can also introduce the probability distribution functional,$\Phi(\mathbf{q},\tau)$ defined by formula

$$\int G(\mathbf{q}(\tau,t))\Phi[\mathbf{q}(\tau,t),\tau]\mathbf{D}[\mathbf{q}(\tau,t)] = \langle G(\mathbf{q}(\tau,t,\omega))\rangle. \tag{4.4}$$

Following the standard procedure [51]-[53 ],one obtain the Fokker-Planck equation

$$\frac{\partial\Phi[\mathbf{q},\tau]}{\partial\tau} = \widehat{F}\Phi[\mathbf{q},\tau]. \tag{4.5}$$

Here $\widehat{F}$ are the Fokker-Planck operator

$$\widehat{F}\Phi[\mathbf{q},\tau] = \hbar\int dt\sum_{i=1}^{n}\frac{\delta}{\delta q_i(t)}\left\{\frac{\delta}{\delta q_i(t)} + \frac{1}{\hbar}\frac{\delta\mathbf{S}_E[\mathbf{q}(t)]}{\delta q_i(t)}\right\}. \tag{4.6}$$

**Remark 4.1**.Note that: (1) If the drift force $\mathbf{K}(\mathbf{q},\tau) = (K_1(\mathbf{q},\tau),\ldots,K_i(\mathbf{q},\tau),\ldots,K_n(\mathbf{q},\tau))$,

$$K_i(\mathbf{q},\tau) = \left(-\frac{\delta\mathbf{S}_E[\mathbf{q}(t)]}{\delta q_i(t)}\right)_{\mathbf{q}=\mathbf{q}(\tau,t)} \tag{4.7}$$

has a damping effect, Eq.(4.5) gives the thermal equilibrium distribution

$$\Phi[\mathbf{q}] = C\exp\left[-\frac{1}{\hbar}\mathbf{S}_E[\mathbf{q}]\right]. \tag{4.8}$$

(2) In this case, in the limit $\tau\to\infty$ the left-hand side of Eq.(4.4) turns into the path integral formula (4.1). Therefore one can to calculate Euclidean quantum tradjectory in

SQM by formulae:

$$1.\lim_{\tau\to\infty}\int_{\mathbf{q}(\tau,t_1)=\mathbf{q}_1}^{\mathbf{q}(\tau,t_2)=\mathbf{q}_2} G(\mathbf{q}(\tau,t))\Phi[\mathbf{q}(\tau,t),\tau]\mathbf{D}[\mathbf{q}(\tau,t)] =$$

$$=\lim_{\tau\to\infty}\mathbf{E}[\widetilde{\mathbf{q}}(\tau,t,\omega)|\widetilde{\mathbf{q}}(\tau,t_1)=\mathbf{q}_1;\widetilde{\mathbf{q}}(\tau,t_2)=\mathbf{q}_2],$$

$$(4.8')$$

$$2.\lim_{\tau\to\infty}\int_{\mathbf{q}(\tau,t_1)=\mathbf{q}_1} G(\mathbf{q}(\tau,t))\Phi[\mathbf{q}(\tau,t),\tau]\mathbf{D}[\mathbf{q}(\tau,t)] =\lim_{\tau\to\infty}\mathbf{E}[\widetilde{\mathbf{q}}(\tau,t,\omega)|\widetilde{\mathbf{q}}(\tau,t_1)=\mathbf{q}_1],$$

$$3.\lim_{\tau\to\infty}\int_{\mathbf{q}(0,t)=\mathbf{q}_0(t)} G(\mathbf{q}(\tau,t))\Phi[\mathbf{q}(\tau,t),\tau]\mathbf{D}[\mathbf{q}(\tau,t)] =\lim_{\tau\to\infty}\mathbf{E}[\widetilde{\mathbf{q}}(\tau,t,\omega)],$$

where $\widetilde{\mathbf{q}}(\tau,t,\omega)$ is the solution of the Langevin equation (4.2).

As a simple examples, we apply SQM to quantization of the scalar field, $q(t)$, described by the Euclidean action:

$$\mathbf{S}_E[q(t)] = \frac{1}{2}\left(\frac{dq}{dt}\right)^2 + V(q). \tag{4.9}$$

The Langevin equation is

$$\frac{\partial q(\tau,t)}{\partial\tau} = \frac{\partial^2 q(\tau,t)}{\partial t^2} - \left(\frac{dV(q)}{dq}\right)_{q=q(\tau,t)} + \eta(\tau,t). \tag{4.10}$$

**Example1**.

As an example we consider a drift force field (4.7) with a duble well potential,i.e.

$$V(q) = \frac{a}{4}q^4 - \frac{b}{2}q^2 - (c + A\sin(\Omega t))q\ ,\ \ a,b > 0\ . \tag{4.11}$$

The Langevin equation is

$$\frac{\partial q(\tau,t)}{\partial\tau} = \frac{\partial^2 q(\tau,t)}{\partial t^2} - aq^3(\tau,t) + bq(\tau,t) + A\sin(\Omega t) + c + \eta(\tau,t). \tag{4.12}$$

From differential master equation () one obtain the linear differential master equation coresponding to Langevin equation (4.12)

$$\frac{\partial u(\tau,t)}{\partial \tau} = \frac{\partial^2 u(\tau,t)}{\partial t^2} - (3a\lambda^2 - b\lambda)u(\tau,t) - a\lambda^3 + b\lambda + A\sin(\Omega t) + c,$$

$$(4.13)$$

$$u(0,t) = x_0 - \lambda.$$

From Eq.(4.13) we obtain

$$\frac{\partial \widetilde{u}(\tau,k)}{\partial \tau} = -[k^2 + (3a\lambda^2 - b\lambda)]\widetilde{u}(t,k)[k] - 2\pi(a\lambda^3 - b\lambda - c)\delta(k) + A\widetilde{\sin(\Omega t)}[k],$$

$$\widetilde{u}(0,t)[k] = 2\pi(x_0 - \lambda)\delta(k),$$

$$(4.14)$$

$$\widetilde{\sin(\Omega t)}[k] = -i\pi[\delta(k+\Omega) - \delta(k-\Omega)].$$

$$2i\sin(\Omega\tau) = (e^{i\Omega\tau} - e^{-i\Omega\tau})$$

$$\frac{1}{2i}(e^{-i\Omega\tau} - e^{i\Omega\tau}) = \sin(-\Omega\tau) = -\sin(\Omega\tau)$$

where $\widetilde{u}(\tau,k)$ is the Fourier transform of $u(\tau,t)$ with respect to $t$. From Eq.(4.14) we obtain

$$\widetilde{u}(\tau,k) = \widetilde{u}(0,t)[k]\exp[-\Theta(k^2,\lambda)\tau] + \Delta(k^2,\lambda)\int_0^{\tau}\exp[-\Theta(k^2,\lambda)(\tau-\bar{\tau})d\bar{\tau}] =$$

$$= \widetilde{u}(0,t)[k]\exp[-\Theta(k^2,\lambda)\tau] + \frac{\Delta(k^2,\lambda)}{\Theta(k^2,\lambda)}\{1 - \exp[-\Theta(k^2,\lambda)\tau]\}, \quad (4.15)$$

$$\Delta(k^2,\lambda) = -2\pi(a\lambda^3 - b\lambda - c)\delta(k) - Ai\pi[\delta(k+\Omega) - \delta(k-\Omega)],$$

$$\Theta(k^2,\lambda) = k^2 + (3a\lambda^2 - b\lambda).$$

Using inverse Fourier transform, from (4.14)-(4.15) one obtain master equation

$$u(\tau,t) =$$

$$= \int e^{itk}\left(\widetilde{u}(0,t)[k]\exp[-\Theta(k^2,\lambda)\tau] + \frac{\Delta(k^2,\lambda)}{\Theta(k^2,\lambda)}\{1-\exp[-\Theta(k^2,\lambda)\tau]\}\right)dk =$$

$$= 2\pi(x_0-\lambda)\int dke^{itk}\delta(k)\exp[-\Theta(k^2,\lambda)\tau] +$$

$$+\int dke^{itk}\frac{\Delta(k^2,\lambda)}{\Theta(k^2,\lambda)}\{1-\exp[-\Theta(k^2,\lambda)\tau]\} =$$

$$2\pi(x_0-\lambda)\exp[-\Theta(0,\lambda)\tau] +$$

$$+\int dke^{itk}\frac{-2\pi(a\lambda^3-b\lambda-c)\delta(k) - i\pi A[\delta(k+\Omega)-\delta(k-\Omega)]}{\Theta(k^2,\lambda)} \times$$

$$\times\{1-\exp[-\Theta(k^2,\lambda)\tau]\} =$$

$$= 2\pi(x_0-\lambda)\exp[-(3a\lambda^2-b\lambda)\tau] - \tag{4.16}$$

$$-2\pi\int dk\delta(k)e^{itk}\frac{a\lambda^3-b\lambda-c}{\Theta(k^2,\lambda)}\{1-\exp[-\Theta(k^2,\lambda)\tau]\} -$$

$$-i\pi A\int dke^{itk}\frac{\delta(k+\Omega)-\delta(k-\Omega)}{\Theta(k^2,\lambda)}\{1-\exp[-\Theta(k^2,\lambda)\tau]\} =$$

$$= 2\pi(x_0-\lambda)\exp[-(3a\lambda^2-b\lambda)\tau] - 2\pi\frac{a\lambda^3-b\lambda-c}{3a\lambda^2-b\lambda}\{1-\exp[-(3a\lambda^2-b\lambda)\tau]\} -$$

$$-i\pi A\frac{e^{it\Omega}-e^{-it\Omega}}{\Omega^2+(3a\lambda^2-b\lambda)} + i\pi A\frac{(e^{it\Omega}-e^{-it\Omega})\exp[-\Theta(\Omega^2,\lambda)\tau]}{\Omega^2+(3a\lambda^2-b\lambda)} =$$

$$2\pi(x_0-\lambda)\exp[-(3a\lambda^2-b\lambda)\tau] - 2\pi\frac{a\lambda^3-b\lambda-c}{3a\lambda^2-b\lambda}\{1-\exp[-(3a\lambda^2-b\lambda)\tau]\} +$$

$$+2\pi A\frac{\sin(\Omega t)}{\Omega^2+(3a\lambda^2-b\lambda)}(1-\exp[-\Theta(\Omega^2,\lambda)\tau]) = 0$$

$$\Theta(\Omega^2,\lambda) = \Omega^2+(3a\lambda^2-b\lambda).$$

Assume now that: (i) $3a\lambda^2-b\lambda > 0,$ then in the limit $\tau\to\infty$ from master equation

(4.16) we obtain

$$-\frac{a\lambda^3 - b\lambda - c}{3a\lambda^2 - b\lambda} + \frac{A\sin(\Omega t)}{\Omega^2 + (3a\lambda^2 - b\lambda)} = 0 \qquad (4.17)$$

(ii) $3a\lambda^2 - b\lambda < 0,$ then in the limit $\tau \to \infty$ from master equation (4.17) we obtain

$$(x_0 - \lambda) + \frac{a\lambda^3 - b\lambda - c}{3a\lambda^2 - b\lambda} - A\frac{\sin(\Omega t)}{\Omega^2 + (3a\lambda^2 - b\lambda)} = 0. \qquad (4.18)$$

**Example2**.The Langevin equation with initial and boundary conditions is

$$\frac{\partial q(\tau,t)}{\partial\tau} = \frac{\partial^2 q(\tau,t)}{\partial t^2} - aq^3(\tau,t) + bq(\tau,t) + A\sin(\Omega t) + c + \eta(\tau,t), q(0,t) = x_0,$$
$$(4.19)$$
$$q(\tau,t_1) = q_1, q(\tau,t_2) = q_2.$$

From differential master equation () one obtain the linear differential master equation coresponding to Langevin equation (4.19)

$$\frac{\partial u(\tau,t)}{\partial\tau} = \frac{\partial^2 u(\tau,t)}{\partial t^2} - (3a\lambda^2 - b\lambda)u(\tau,t) - a\lambda^3 + b\lambda + A\sin(\Omega t) + c,$$

$$u(0,t) = q_1 - \lambda, \qquad (4.20)$$

$$u(\tau,t_1) = q_1 - \lambda, u(\tau,t_2) = q_2 - \lambda.$$

Let $\bar{u}(\tau,t)$ be the solution of the linear boundary problem

$$\frac{\partial^2 \bar{u}(\tau,t)}{\partial t^2} - (3a\lambda^2 - b\lambda)\bar{u}(\tau,t) - a\lambda^3 + b\lambda + A\sin(\Omega t) + c = 0$$
$$(4.21)$$
$$\bar{u}(\tau,t_1) = q_1 - \lambda, \bar{u}(\tau,t_2) = q_2 - \lambda.$$

If we set now $u(\tau,t) = \bar{u}(\tau,t) + u_1(\tau,t),$ then from Eq.(4.20)-Eq.(4.21) we obtain

$$\frac{\partial u_1(\tau,t)}{\partial \tau} = \frac{\partial^2 u_1(\tau,t)}{\partial t^2} - (3a\lambda^2 - b\lambda)u_1(\tau,t),$$

$$u_1(0,t) = 0, u(\tau,t_1) = 0, q(\tau,t_2) = 0. \quad (4.22)$$

Thus $u_1(\tau,t) \equiv 0$ and $u(\tau,t) = \overline{u}(\tau,t)$. We let now $\overline{u}(\tau,t) = \overline{u}(t)$ and rewrite Eq.(4.21) in the form

$$\frac{\partial^2 \overline{u}(t)}{\partial t^2} + (b\lambda - 3a\lambda^2)\overline{u}(t) - a\lambda^3 + b\lambda + A\sin(\Omega t) + c = 0,$$

$$\overline{u}(t_1) = q_1 - \lambda, \overline{u}(t_2) = q_2 - \lambda. \quad (4.23)$$

**Assumption**. We assume now that $b\lambda - 3a\lambda^2 > 0$.
Let us rewrite Eq.(4.23) in a short form

$$\frac{\partial^2 \overline{u}(t)}{\partial t^2} + 2\varpi^2(\lambda)\overline{u}(t) + d(\lambda) - \overline{A}\sin(\Omega t) + c = 0,$$

$$2\varpi^2(\lambda) = b\lambda - 3a\lambda^2 > 0,$$

$$d(\lambda) = -a\lambda^3 + b\lambda + c, \quad (4.24)$$

$$\overline{A} = -A,$$

$$\overline{u}(t_1) = q_1 - \lambda, \overline{u}(t_2) = q_2 - \lambda.$$

Plugging in Eq.(4.24) this trial solution

$$\overline{u}(t) = C + D\sin(\Omega t) \quad (4.25)$$

we obtain

$$-D\Omega^2\sin(\Omega t) + 2\varpi^2(\lambda)(C + D\sin(\Omega t)) + d(\lambda) - \bar{A}\sin(\Omega t) = 0. \tag{4.26}$$

Hence

$$\begin{aligned} 2\varpi^2(\lambda)C + d(\lambda) &= 0, \\ -D\Omega^2 + 2\varpi^2(\lambda)D - \bar{A} &= 0. \end{aligned} \tag{4.27}$$

Solving Eqs.(4.27) we obtain

$$\begin{aligned} C &= -\frac{d(\lambda)}{2\varpi^2(\lambda)}, \\ D &= \frac{A}{(2\varpi^2(\lambda) - \Omega^2)}. \end{aligned} \tag{4.27}$$

Thus general solution of the Eq.(4.24) is

$$u(t,\lambda) = c_1\sin\varpi t + c_2\cos\varpi t + D\sin(\Omega t) + C. \tag{4.28}$$

Substituting Eq.(4.28) into boundary conditions $\bar{u}(0) = q_1 - \lambda, \bar{u}(T) = q_2 - \lambda = \bar{x},$ we obtain

$$\begin{aligned} u(0) &= c_2 + C = q_1 - \lambda, \\ u(T) &= c_1\sin\varpi T + c_2\cos\varpi T + D\sin(\Omega T) + C = \bar{x}. \end{aligned} \tag{4.29}$$

Hence

$$\begin{aligned} c_2 &= -(C + \lambda - q_1), \\ c_1 &= \frac{\bar{x} - C - D\sin(\Omega T) + (C + \lambda - q_1)\cos\varpi T}{\sin\varpi T} \end{aligned} \tag{4.30}$$

and therefore transcendental master equation are

$$u(t,\lambda,\bar{x}) = \frac{\bar{x} - C - D\sin(\Omega T) + (C + \lambda - q_1)\cos\varpi T}{\sin\varpi T}\sin\varpi t -$$
$$-(C + \lambda - q_1)\cos\varpi t +$$

$$+D\sin(\Omega t) + C = 0,$$

$$C = -\frac{d(\lambda)}{2\varpi^2(\lambda)}, D = \frac{\bar{A}}{(2\varpi^2(\lambda) - \Omega^2)}, \bar{x} = q_2 - \lambda, \tag{4.31}$$

$$\varpi(\lambda) = \sqrt{\frac{b\lambda - 3a\lambda^2}{2}}.$$

**Example 2.1**. $a = 2, b = 10, c = 1, A = 3, \Omega = 2, q_1 = 0, q_2 = 1.$

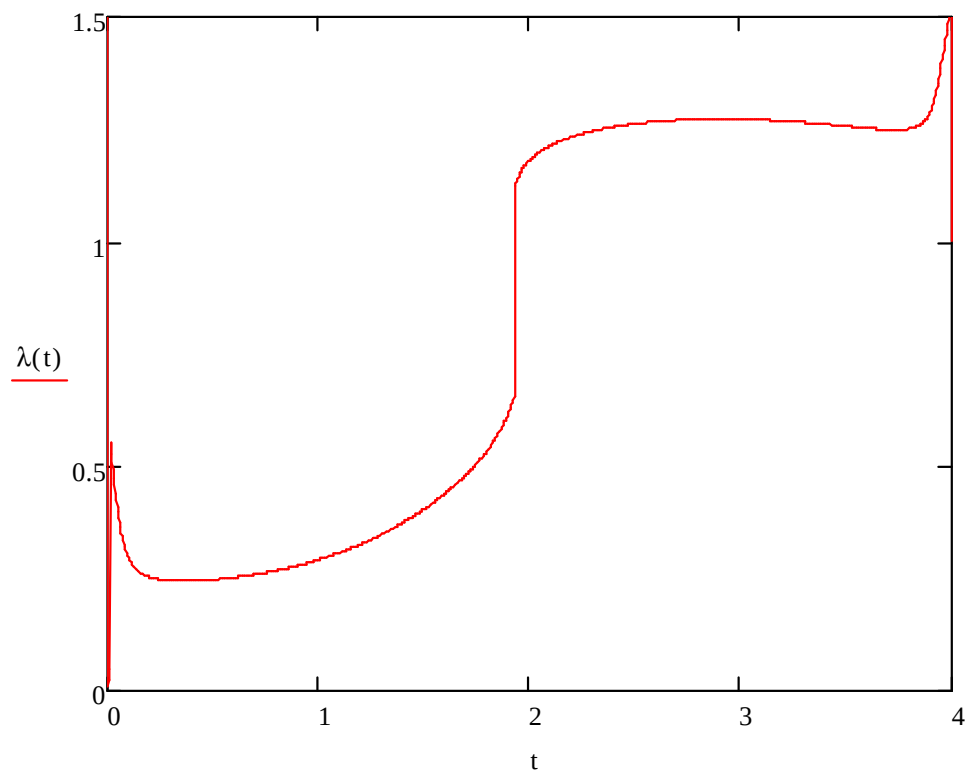

Pic.2.1.1.

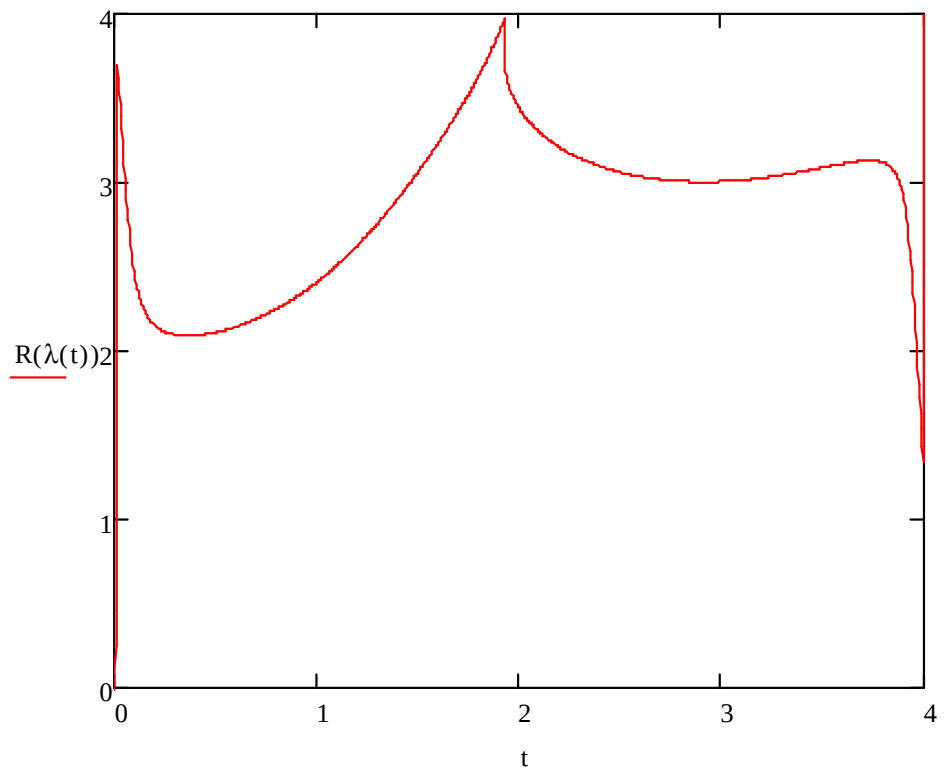

Pic.2.1.2.
$R(\lambda(t)) = b\lambda(t) - 3a\lambda^2(t) > 0.$

**Example 2.2**. $a = 2, b = 10, c = 10, A = 3, \Omega = 2, q_1 = 1, q_2 = 1.$

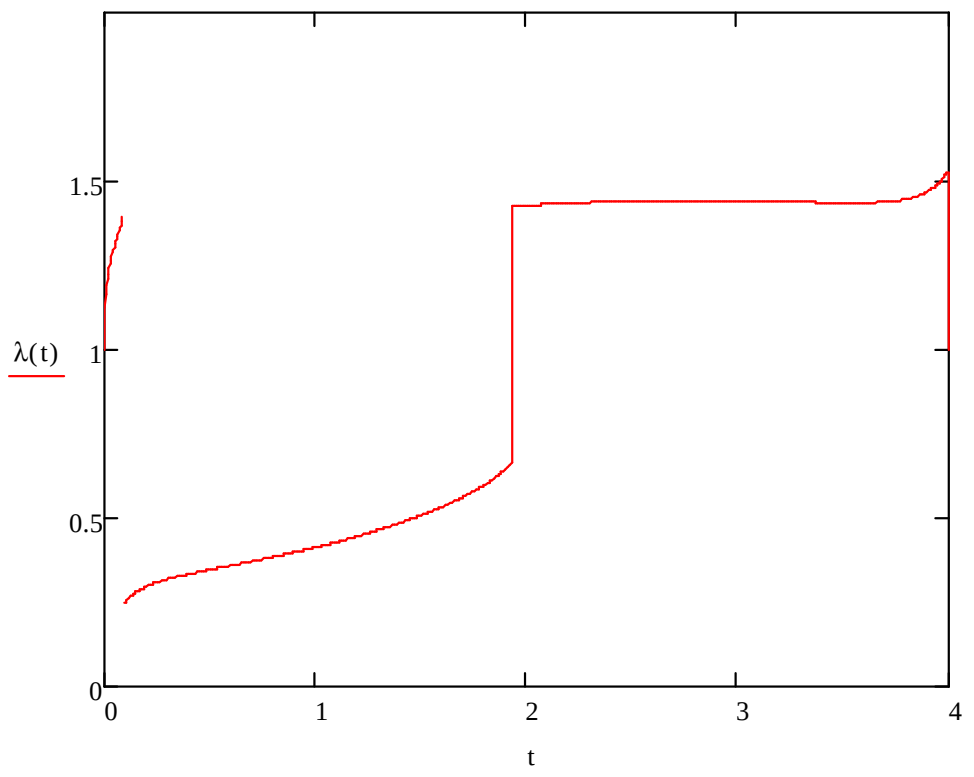

Pic.2.2.1.

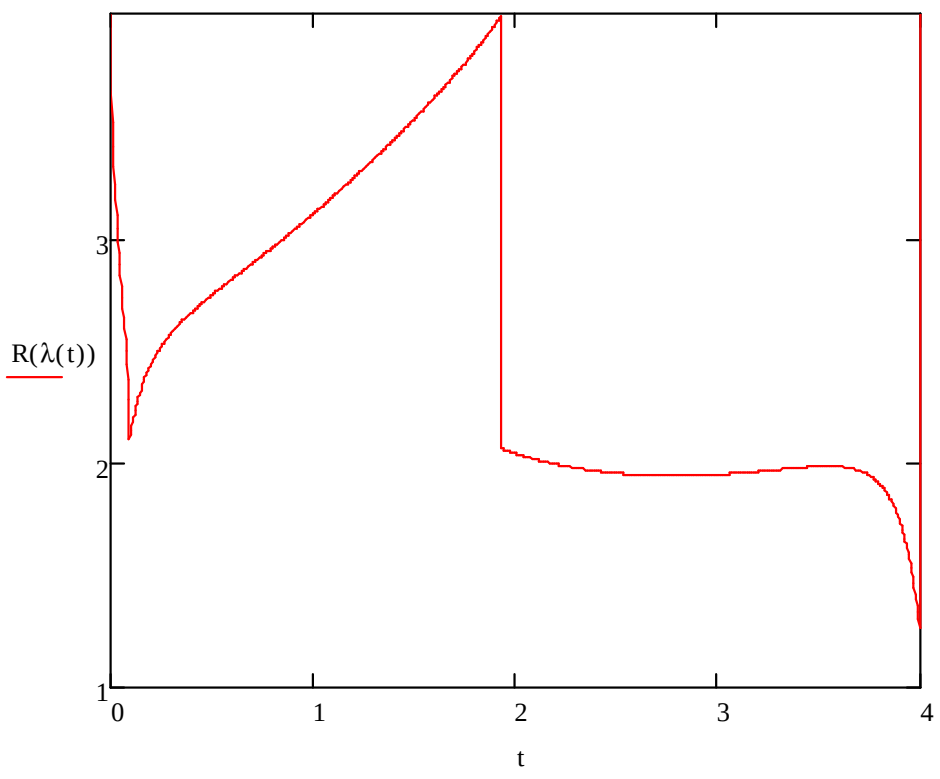

Pic.2.2.2.

$R(\lambda(t)) = b\lambda(t) - 3a\lambda^2(t) > 0.$

**Example 2.3**. $a = 3, b = 10, c = 10, A = 3, \Omega = 2, q_1 = 1, q_2 = 1.$

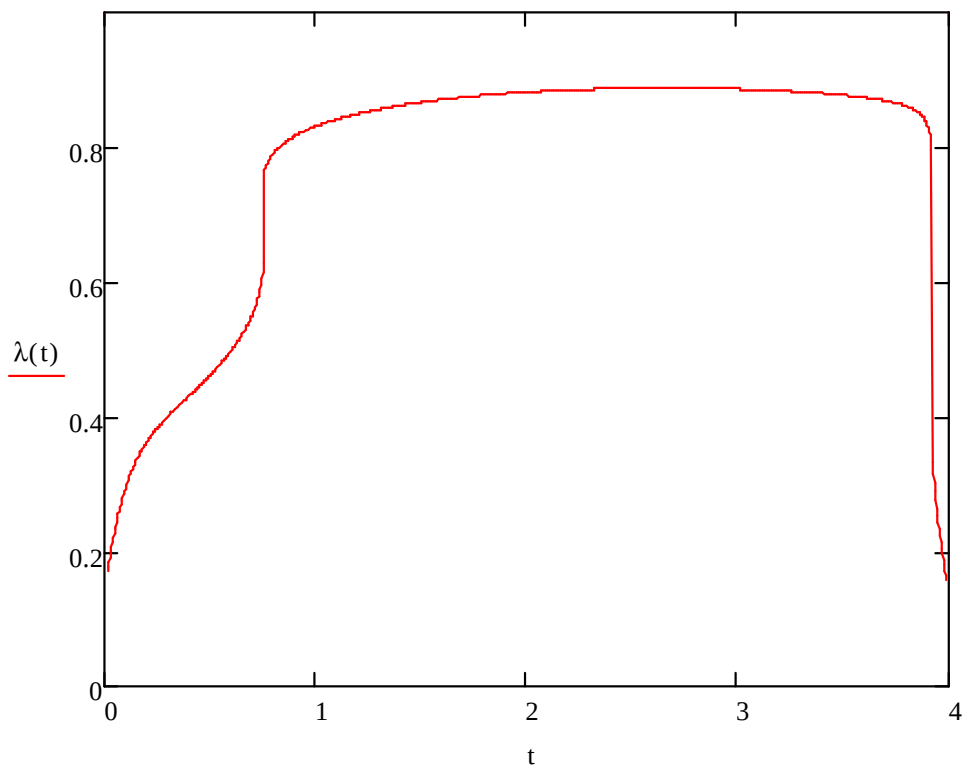

Pic.2.3.1.

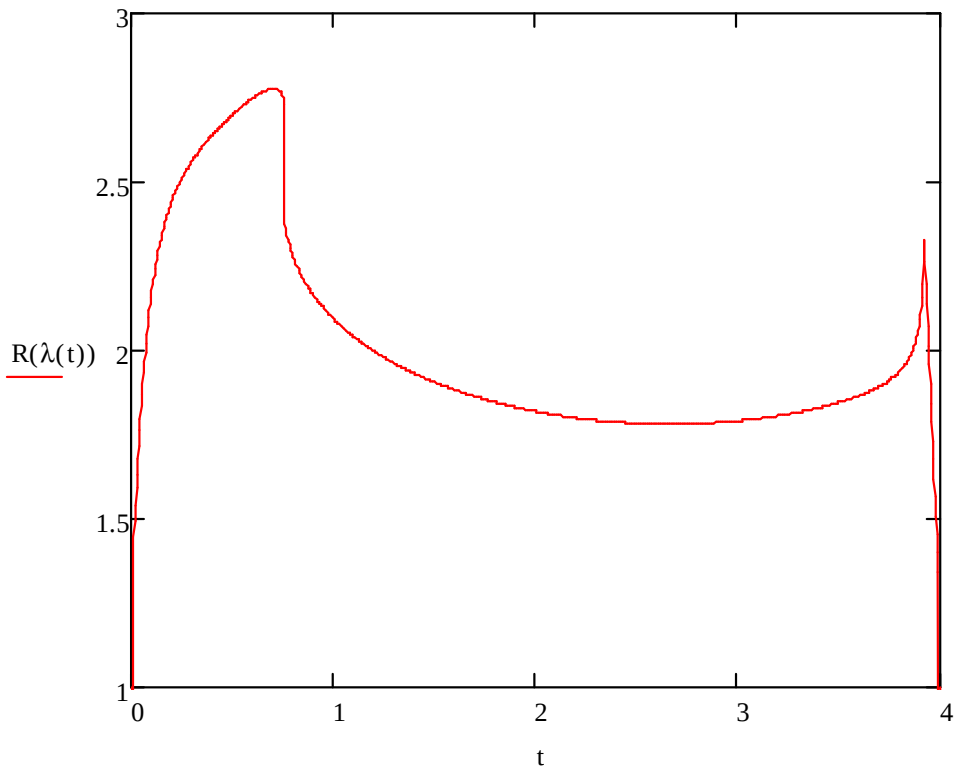

Pic.2.3.2.

$R(\lambda(t)) = b\lambda(t) - 3a\lambda^2(t) > 0.$

(4.0)

(4.0)

## V.Jumps in financial markets from LDP

A classical model of financial market return process, such as the Black-Scholes [54],is the lognormal diffusion process, such that the log-return process has a normal distribution. However, real markets exhibit several deviations from this ideal, although useful, model. The market distribution,sayforstocks, should have several realistic properties not found in the ideal log-normal model: (1)the model must permit large random fluctuations such as cras- hes or sudden upsurges,(2)the log-returndistribution should be skew since large downward outliers are larger than upwardoutliers,and(3)the distribution should beleptokurtic since the mode is usually higher and the tails thickerthan for a normal distribution. For modeling these extra properties phenomenologically, a jump-diffusion process with log-uniform jump- amplitude Poisson process it is usually applied. Let $S(t)$ be the price of a stock or stock fund satisfies a Markov, continuous-time, jump-diffusion stochastic differential equation

$$dS(t) = S(t)[\mu_j dt + \sigma_j dZ(t) + J(Q)dP(t)],$$
$$S(0) = S_0. \tag{5.1}$$

Where $\mu_j$ is the mean return rate, σis the diffusive volatility, $Z(t)$ is a one-dimensional stochastic diffusion process, $J(Q)$ is a log-return mean $\mu_j$ and variance $\sigma_j^2$ random jump-amplitude and $P(t)$ is a simple Poisson jump process with jump rate $\lambda$. The processes $Z(t)$ and $P(t)$ are pairwise independent, while $J(Q)$ is also independent except that it is conditioned on the existence of a jump in $dP(t)$ it is conditioned on the existence of a jump in $dP(t)$.The numerical simulation a jumps in financial markets based on, jump-diffusion stochastic differential equation (5.1), was considered in many papers, see for example [54]-[57].

In contrast with a canonical phenomenological approach we explain jumps phenomena in financial markets from the first principles, without any reference to Poisson jump process. We claim that jumps phenomena in financial markets completely induced by nonlinearity and additive “small“ white noise in corresponding Colombeau-Ito’s stochastic equations.

Let $\Im_i = (\Omega_i, \Sigma_i, \mathbf{P}_i), i = 1,2$ be a probability spaces such that: $\Omega_1 \cap \Omega_2 = \varnothing$. Let $\mathbf{W}(t,\omega)$ be a Wiener process on $\Im_1$ and let $\mathbf{W}(t,\varpi)$ be a Wiener process on $\Im_2$. Let us consider now the family $\left(\mathbf{x}_{t,\varepsilon'}^{x_0;\varepsilon}(\omega,\varpi)\right)_{\varepsilon'}$of the Colombeau generalized stochastic processes which is a solution of the Colombeau-Ito’s stochastic equation with stochastic coefficients:

$$\left(d\mathbf{x}_{t,\varepsilon'}^{x_0,\varepsilon}(\omega,\varpi)\right)_{\varepsilon'} = \left(\mathbf{A}_{\varepsilon'}^{(\mu)}(\mathbf{x}_{t,\varepsilon'}^{x_0,\varepsilon}(\omega,\varpi),t)\right)_{\varepsilon'} + \sqrt{D}\,(\mathbf{w}_{\varepsilon'}(t,\varpi))_{\varepsilon'} + \sqrt{\varepsilon}\,\mathbf{w}(t,\omega),$$

$$\varepsilon,\varepsilon' \in (0,1],\omega \in \Omega_1,\varpi \in \Omega_2. \tag{5.2}$$

Here $\mathbf{w}(t,\varpi)$ and $\mathbf{w}(t,\omega)$ is a white noise on $\mathbb{R}^n$ i.e.,

$$\mathbf{w}(t,\varpi) = \frac{d}{dt}\mathbf{W}(t,\varpi),$$

$$\mathbf{w}(t,\omega) = \frac{d}{dt}\mathbf{W}(t,\omega) \tag{5.3}$$

is almost surely in $D'$,and $\mathbf{w}_{\varepsilon'}(t,\varpi)$ is the smoothed white noise on $\mathbb{R}^n$ i.e.,

$$\mathbf{w}_{\varepsilon'}(t,\varpi) = \langle \mathbf{w}(t,\varpi),\varphi_{\varepsilon'}(s-t)\rangle, \tag{5.4}$$

where $\varphi_{\varepsilon'}(t)$ is a model delta net [23],[24].

**Definition 5.1**. CISDE (5.2) is $\widetilde{\mathbb{R}}^n$-dissipative if there exist random Lyapunov-Colombeau candidate function $(V_{\varepsilon'}(\varpi,\mathbf{x},t))_{\varepsilon'}$

(5.3)

(5.3)

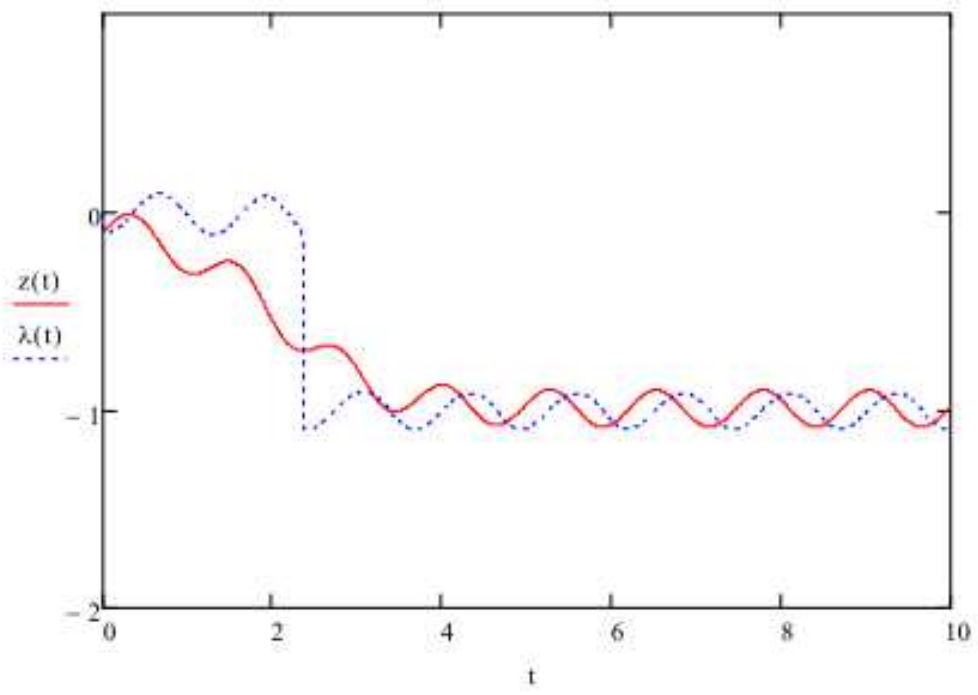

**Figure.5.1.**Comparison of: (1) dynamics with $D = 0$, $\varepsilon = 0$ (red curve) and (2) quasi-classical (blue curve) dynamics in the limit $\varepsilon \to 0$, calculated by using SLDP.

$a = 1, b = 1, c = 0, A = 0.5, B = 0, \Omega = 5, \Theta = 0, D = 0, x_0 = -0.1$

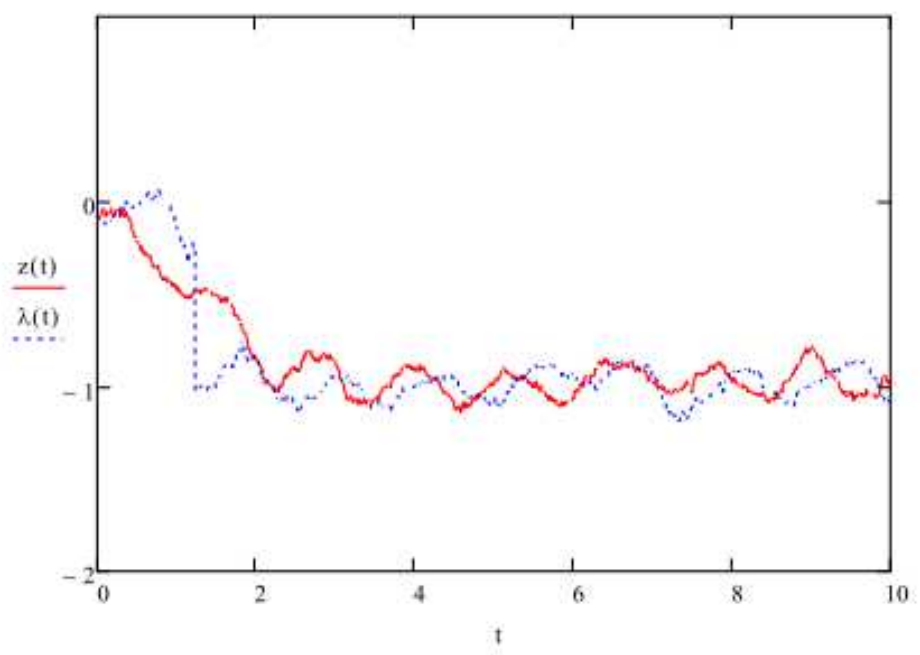

**Figure.5.2.**Comparison of: (1) dynamics with $D = 0$, $\varepsilon = 0$ (red curve) and (2) quasi-classical (blue curve) dynamics in the limit $\varepsilon \to 0$, calculated by using SLDP.

$a = 1, b = 1, c = 0, A = 0.5, B = 0, \Omega = 5, \Theta = 0, D = 0.01, x_0 = -0.1$

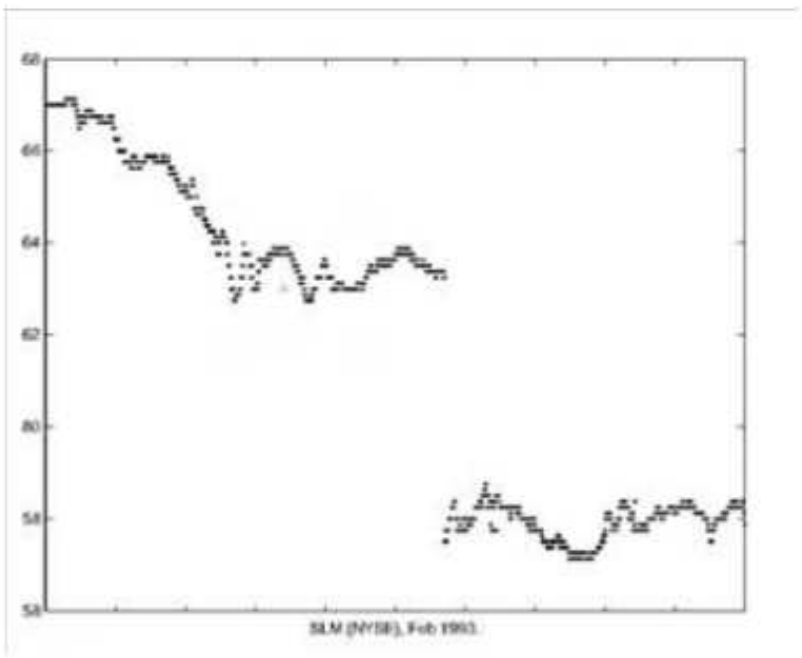

**Figure.5.3**.Evolution of SLM (NYSE) February1993 [54].

Figure 5.3 shows the evolution of SLM over a one month period (February 1993)[54]. The price behavior over this period is clearly dominated by a large downward jump, which accounts for half of the monthly return. If we go down to an intraday scale shown in Figure 5.5, we see that the price moves essentially through jumps.

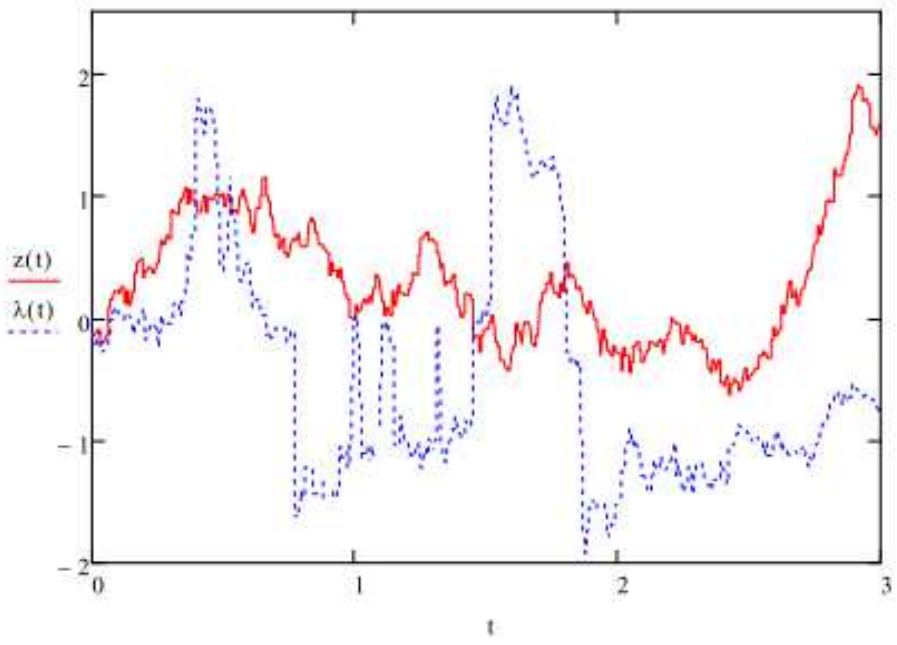

**Figure.5.4**.Comparison of: (1) stochastic dynamics () with $D \neq 0$, $\varepsilon = 0$ (red curve) and (2) quasi-classical stochastic dynamics in the limit $\varepsilon \to 0$, calculated by using SLDP $a = 1, b = 1, c = 0, A = 0.5, B = 0, \Omega = 5, \Theta = 0, D = 1, x_0 = 0.$

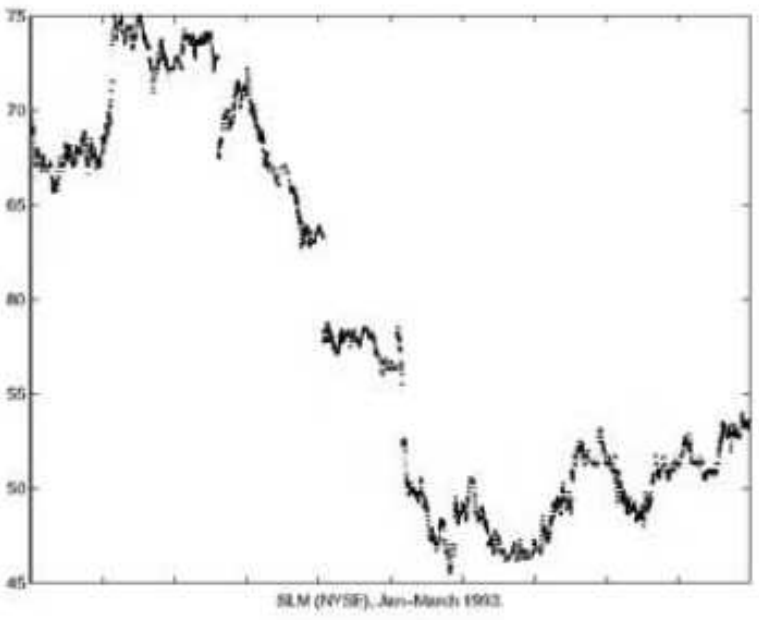

**Figure**.**5**.**5**.Evolution of SLM (NYSE) January-March 1993 [54].

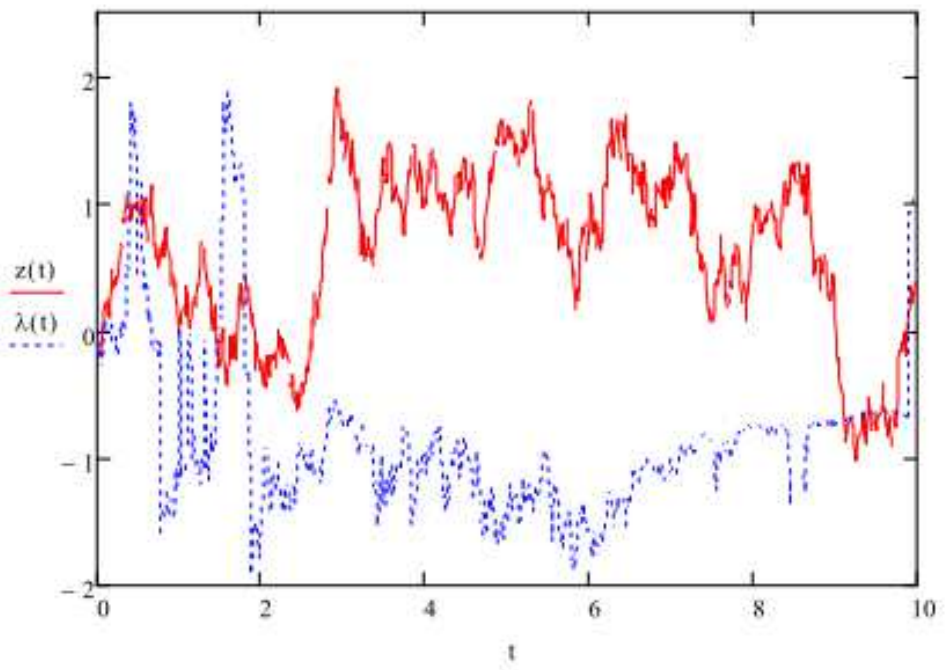

**Figure**.**5**.**6**.Comparison of: (1) stochastic dynamics () with $D \neq 0$, $\varepsilon = 0$ (red curve) and (2) quasi-classical stochastic dynamics in the limit $\varepsilon \to 0$, calculated by using SLDP

$a = 1, b = 1, c = 0, A = 0.5, B = 0, \Omega = 5, \Theta = 0, D = 1, x_0 = 0.$

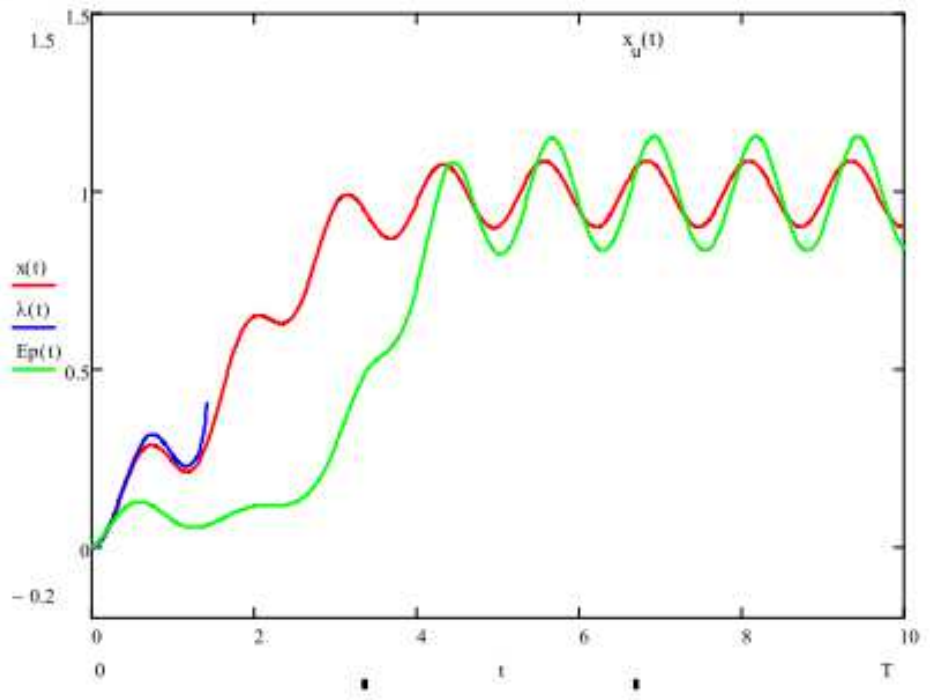

**Figure**.**5**.**7**.Comparison of: (1) dynamics () with $D = 0$,
$\varepsilon = 0$ (red curve), (2) quasi-classical (blue curve) dynamics
in the limit $\varepsilon \to 0$, calculated by using SLDP
and (3) quasi-classical dynamics in the limit $\varepsilon \to 0$,
calculated using saddle-
point approximation (green curve).
$a = 1, b = 1, c = 0, A = 1, B = 0, \Omega = 5, \Theta = 0, D = 0, x_0 = 0.$
$a = 1,$

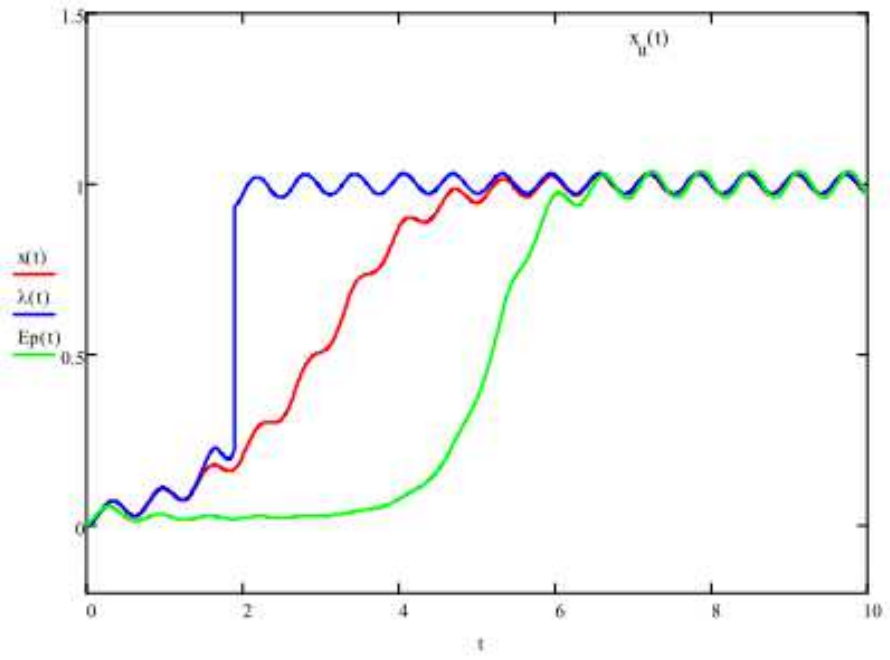

**Figure**.**5**.**8**.Comparison of: (1) dynamics () with $D = 0$,
$\varepsilon = 0$ (red curve), (2) quasi-classical (blue curve) dynamics
in the limit $\varepsilon \to 0$, calculated by using SLDP
and (3) quasi-classical dynamics in the limit $\varepsilon \to 0$,
calculated using saddle-
point approximation (green curve).
$a = 1, b = 1, c = 0, A = 0.3, B = 0, \Omega = 10, \Theta = 0, D = 0, x_0 = 0.$

# Appendix I.

**Proposition 1**.[47] Assume that (1) $\varphi(t),\alpha(t) \in \mathcal{L}_1([0,T]), \sup_{t\in[0,T]}|\varphi(t)| < \infty$, $\sup_{t\in[0,T]}|\alpha(t)| < \infty$ and (2) the inequality

$$\varphi(t) \leq \alpha(t) + L\int_0^t \varphi(s)ds \tag{1}$$

is satisfied. Then the inequality

$$\varphi(t) \leq \alpha(t) + L\int_0^t e^{L(t-s)}\alpha(s)ds. \tag{2}$$

is satisfied.

**Proposition 2**.(**I**) Assume that: (1) let $\eta_n(t), n = 1,2,\ldots$ be the solutions of the Ito's SDE's

$$\boldsymbol{\eta}_n(t) = \boldsymbol{\phi}(t) + \int_0^t \mathbf{A}_n(\boldsymbol{\eta}_n(s),s)ds + \int_0^t \mathbf{B}_n(\boldsymbol{\eta}_n(s),s)d\mathbf{W}(s). \tag{3}$$

and let $\tilde{\eta}_n(t), n = 1,2,\ldots$ be the solutions of the Ito SDE's

$$\tilde{\boldsymbol{\eta}}_n(t) = \boldsymbol{\phi}(t) + \int_0^t \tilde{\mathbf{A}}_n(\tilde{\boldsymbol{\eta}}_n(s),s)ds + \int_0^t \tilde{\mathbf{B}}_n(\tilde{\boldsymbol{\eta}}_n(s),s)d\mathbf{W}(s). \tag{4}$$

Here $\mathbf{A}_n(x,t) = (A_{n,1}(x,t),\ldots,\mathbf{A}_{n,d}(x,t)), x \in \mathbb{R}^d$, $\tilde{\mathbf{A}}_n(x,t) = \left(\tilde{A}_{n,1}(x,t),\ldots,\tilde{\mathbf{A}}_{n,d}(x,t)\right)$, $x \in \mathbb{R}^l$, $\mathbf{B}_n(x,t) = \left\{B_n^{i,j}(x,t)\right\}, i,j = 1,\ldots,l$ is $d \times d$ matrix,$\tilde{\mathbf{B}}_n(x,t) = \left\{\tilde{B}_n^{i,j}(x,t)\right\}, i,j = 1,\ldots,d$ is $l \times l$ matrix,$l \geq d.$

(2) The inequalities

$$(5.1)\ (\|\mathbf{A}_n(x,t)\|^2+\|\mathbf{B}_n(x,t)\|^2)\leq K_n(1+\|x\|^2),$$

$$(\|\widetilde{\mathbf{A}}_n(x,t)\|^2+\|\widetilde{\mathbf{B}}_n(x,t)\|^2)\leq K_n(1+\|x\|^2),$$

$$(5.2)\,\|\mathbf{A}_n(x,t)-\mathbf{A}_n(y,t)\|+\|\mathbf{B}_n(x,t)-\mathbf{B}_n(y,t)\|\leq K_n\|x-y\|,$$

$$\|\widetilde{\mathbf{A}}_n(x,t)-\widetilde{\mathbf{A}}_n(y,t)\|+\|\widetilde{\mathbf{B}}_n(x,t)-\widetilde{\mathbf{B}}_n(y,t)\|\leq K_n\|x-y\|, \tag{5}$$

$$(5.3)\,\|\mathbf{A}_n(x,t)-\widetilde{\mathbf{A}}_n(x,t)\|\leq \delta_{1,n}\|x\|,$$

$$(5.4)\,\|\mathbf{B}_n(x,t)-\widetilde{\mathbf{B}}_n(x,t)\|\leq \delta_{2,n}\|x\|$$

is satisfied. Then the inequality

$$\sup_{0\leq t\leq T}\mathbf{E}\left[\|\boldsymbol{\eta}_n(t)-\widetilde{\boldsymbol{\eta}}_n(t)\|^2\right]\leq e^{L_n}(T\delta_{1,n}^2+\delta_{2,n}^2)\mathbf{E}\left[\int_0^T\left(\|\widetilde{\boldsymbol{\eta}}_n(t)\|^2\right)dt\right], \tag{6}$$

$$L_n=3(1+T)K_n^2$$

is satisfied.

**(II)** Let $\boldsymbol{\tau}_{U_n}=\tau_n(\omega)$ be the random variable equal to the time at which the sample function of the process $\widetilde{\boldsymbol{\eta}}_n(t)$ first leaves the bounded neighborhood $U_n\ni 0$,and let $\tau_n(t)=\min(\tau_n(\omega),t)$. Assume that: (1) $\forall n\in\mathbb{N}: U_n\subset U_{n+1}$ and $\bigcup_{n\in\mathbb{N}}U_n=\mathbb{R}^d$,

(2) $\sup_{n\in\mathbb{N}}\left(\mathbf{E}\left[\int_0^T\left(\|\widetilde{\boldsymbol{\eta}}_n(t)\|^2\right)dt\right]\right)<\infty.$

Then the inequality

$$\sup_{0\leq t\leq T}\mathbf{E}\left[\|\boldsymbol{\eta}_n(\tau_n(t))-\widetilde{\boldsymbol{\eta}}_n(\tau_n(t))\|^2\right]\leq$$

$$\leq e^{L_n}(T\delta_{1,n}^2+\delta_{2,n}^2)\left\{\sup_{n\in\mathbb{N}}\left(\mathbf{E}\left[\int_0^T\left(\|\widetilde{\boldsymbol{\eta}}_n(t)\|^2\right)dt\right]\right)\right\}, \tag{6'}$$

$$L_n=3(1+T)K_n^2.$$

is satisfied.

**Proof.**(**I**) From Eq.(3) and Eq.(4) one obtain

$$\eta_n(t) - \tilde{\eta}_n(t) =$$

$$\int_0^t \mathbf{A}_n(\eta_n(s), s)ds + \int_0^t \mathbf{B}_n(\eta_n(s), s)d\mathbf{W}(s) - \int_0^t \tilde{\mathbf{A}}_n(\tilde{\eta}_n(s), s)ds - \int_0^t \tilde{\mathbf{B}}_n(\tilde{\eta}_n(s), s)d\mathbf{W}(s) =$$

$$= \zeta_n(t) + \int_0^t [\mathbf{A}_n(\eta_n(s), s) - \mathbf{A}_n(\tilde{\eta}_n(s), s)]ds + \tag{7}$$

$$+ \int_0^t [\mathbf{B}_n(\eta_n(s), s) - \mathbf{B}_n(\tilde{\eta}_n(s), s)]d\mathbf{W}(s).$$

Here

$$\zeta_n(t) = \int_0^t \left[\mathbf{A}_n(\tilde{\eta}_n(s), s) - \tilde{\mathbf{A}}_n(\tilde{\eta}_n(s), s)\right]ds + \int_0^t \left[\mathbf{B}_n(\tilde{\eta}_n(s), s) - \tilde{\mathbf{B}}_n(\tilde{\eta}_n(s), s)\right]d\mathbf{W}(s). \tag{8}$$

From Eq.(8) and inequalities (5.1)-(5.2) one obtain the inequality

$$\mathbf{E}\left[\|\eta_n(t) - \tilde{\eta}_n(t)\|^2\right] \leq 3\mathbf{E}\left[\|\zeta_n(t)\|^2\right] + L_n \int_0^t \mathbf{E}\left[\|\eta_n(s) - \tilde{\eta}_n(s)\|^2\right]ds, \tag{9}$$

$$L_n = 3(1+T)K_n^2.$$

Using Proposition 1, from inequality (9) one obtain the inequality

$$\mathbf{E}\left[\|\eta_n(t) - \tilde{\eta}_n(t)\|^2\right] \leq 3\mathbf{E}\left[\|\zeta_n(t)\|^2\right] + L_n \int_0^t e^{L_n(t-s)}\mathbf{E}\left[\|\zeta_n(s)\|^2\right]ds. \tag{10}$$

From inequality (5.3) one obtain the inequality

$$\sup_{0\leq t\leq T}\left\|\mathbf{E}\left[\int_0^t\left[\mathbf{A}_n(\tilde{\boldsymbol{\eta}}_n(s),s)-\tilde{\mathbf{A}}_n(\tilde{\boldsymbol{\eta}}_n(s),s)\right]ds\right]\right\|^2\leq$$

$$\leq T\int_0^T\mathbf{E}\left[\left\|\mathbf{A}_n(\tilde{\boldsymbol{\eta}}_n(s),s)-\tilde{\mathbf{A}}_n(\tilde{\boldsymbol{\eta}}_n(s),s)\right\|^2\right]ds\leq T\delta_{1,n}^2\left[\int_0^T\mathbf{E}\left[\|\tilde{\boldsymbol{\eta}}_n(s)\|^2\right]ds\right]. \tag{11}$$

From inequality (5.4) one obtain the inequality

$$\mathbf{E}\left[\sup_{0\leq t\leq T}\left\|\int_0^t\left[\mathbf{B}_n(\tilde{\boldsymbol{\eta}}_n(s),s)-\tilde{\mathbf{B}}_n(\tilde{\boldsymbol{\eta}}_n(s),s)\right]d\mathbf{W}(s)\right\|^2\right]\leq$$

$$\leq 4\mathbf{E}\left[\int_0^T\left[\left\|\mathbf{B}_n(\tilde{\boldsymbol{\eta}}_n(s),s)-\tilde{\mathbf{B}}_n(\tilde{\boldsymbol{\eta}}_n(s),s)\right\|^2\right]ds\right]\leq\delta_{2,n}^2\left[\int_0^T\mathbf{E}\left[\|\tilde{\boldsymbol{\eta}}_n(s)\|^2\right]ds\right]. \tag{12}$$

From Eq.(8) and inequalities (10)-(12) one obtain the inequality

$$\sup_{0\leq t\leq T}\mathbf{E}\left[\|\boldsymbol{\zeta}_n(t)\|^2\right]\leq T\delta_{1,n}^2\left[\int_0^T\mathbf{E}\left[\|\tilde{\boldsymbol{\eta}}_n(s)\|^2\right]ds\right]+\delta_{2,n}^2\left[\mathbf{E}\int_0^T\left[\|\tilde{\boldsymbol{\eta}}_n(s)\|^2\right]ds\right]. \tag{13}$$

Substitution the inequality (13) into inequality (10) gives

$$\sup_{0\leq t\leq T}\mathbf{E}\left[\|\boldsymbol{\eta}_n(t)-\tilde{\boldsymbol{\eta}}_n(t)\|^2\right]\leq e^{L_n}(T\delta_{1,n}^2+\delta_{2,n}^2)\mathbf{E}\left[\int_0^T\left[\|\tilde{\boldsymbol{\eta}}_n(s)\|^2\right]ds\right]. \tag{14}$$

The inequality (14) completed the proof.

**Proof**.(**II**) Similarity to proof of the statement (**I**).

Assume that: (1) let $\mathbf{x}_n(t), n=1,2,\ldots$ be the solutions of the Ito's SDE's

$$\mathbf{x}_n(t)=\boldsymbol{\phi}(t)+\int_0^t\mathbf{A}_n(\mathbf{x}_n(s),s)ds+\int_0^t\mathbf{B}_n(\mathbf{x}_n(s),s)d\mathbf{W}(s). \tag{15}$$

(2) The inequalities

$$1.\ (\|\mathbf{A}_n(x,t)\|^2 + \|\mathbf{B}_n(x,t)\|^2) \leq K_n(1+\|x\|^2),$$
$$2.\ \|\mathbf{A}_n(x,t) - \mathbf{A}_n(y,t)\| + \|\mathbf{B}_n(x,t) - \mathbf{B}_n(y,t)\| \leq K_n\|x-y\| \tag{16}$$

are valid only in every cylinder $U_{R_n} \times I_\infty$, with $K_n = K_n(R_n)$.

(3) $\forall n,m : \mathbf{A}_n(\mathbf{x},t) = \mathbf{A}_m(\mathbf{x},t)$ if $n \leq m, \|\mathbf{x}\| \leq n$,

(4) $\forall n,m : \mathbf{B}_n(\mathbf{x},t) = \mathbf{B}_m(\mathbf{x},t)$ if $n \leq m, \|\mathbf{x}\| \leq n$.

Therefore there exists a sequence of Markov processes $\mathbf{x}_n(t,\omega)$ corresponding to the functions $\mathbf{A}_n(\mathbf{x},t)$ and $\mathbf{B}_m(\mathbf{x},t)$.

**Assumption1**.Suppose now that the distribution of $\mathbf{x}_n(0,\omega), m = 1,2,\ldots$ has compact support in $\mathbb{R}^n$.

Then as, well known, that the first exit random times $\boldsymbol{\tau}_m(\omega)$ of the processes $\mathbf{x}_m(t,\omega)$, $m = 1,2,\ldots$ from the set $\|\mathbf{x}\| \leq n$ are identical for $m \geq n$ [46]-[47].Let this common value be $\boldsymbol{\tau}_n(\omega)$. Itis also clear that the processes themselves coincide up to time $\boldsymbol{\tau}_n(\omega)$, i.e.

$$\forall n \geq m : \mathbf{P}\left[\sup_{0\leq t\leq \boldsymbol{\tau}_n(\omega)} \|\mathbf{x}_m(t,\omega) - \mathbf{x}_n(t,\omega)\| > 0\right] = 0. \tag{17}$$

**Definition1**. (i) Let $\boldsymbol{\tau}_\infty(\omega)$ denote the (finite or infinite) limit of the monotone increasing sequence $\boldsymbol{\tau}_\infty(\omega)$ as $n \to \infty$.We call the random variable $\boldsymbol{\tau}_\infty(\omega)$ the first exit time from every bounded domain, or briefly the explosion time.

(ii) We now define a new stochastic process $\mathbf{x}(t,\omega)$ by setting [46]-[47]:

$$\mathbf{x}(t,\omega) = \mathbf{x}_n(t,\omega) \text{ if } t < \boldsymbol{\tau}_n(\omega). \tag{18}$$

It well known, that this is always a Markov process for $t < \boldsymbol{\tau}_n(\omega)$ [46].We also can to define a new stochastic process $\mathbf{x}(t,\omega)$ by setting

$$\mathbf{x}(t,\omega) = \mathbf{P}\text{-}\lim_{n\to\infty} \mathbf{x}_n(t,\omega) \tag{19}$$

if finite or infinite limit in RHS of Eq.(19) exist.

(iii) In general case we set

$$(\mathbf{x}_{\varepsilon'}(t,\omega))_{\varepsilon'} = (\mathbf{x}_n(t,\omega))_n, n = 1/\varepsilon'. \tag{20}$$

We note that the Colombeau-Ito's equation

$$(\mathbf{x}_{\varepsilon'}(t,\omega))_{\varepsilon'} = (\mathbf{x}_{\varepsilon'}(0,\omega))_{\varepsilon'} + \left(\int_0^t \mathbf{A}_n(\mathbf{x}_{\varepsilon'}(s,\omega),s)ds\right)_{\varepsilon'} + \left(\int_0^t \mathbf{B}_n(\mathbf{x}_{\varepsilon'}(s,\omega),s)dW(s)\right)_{\varepsilon'} \tag{21}$$

is satisfied for all $t \in [0,\boldsymbol{\tau}_\infty(\omega))$.

(iv) Markov process $\mathbf{x}(t,\omega)$ is regular if for all $s < \infty, \mathbf{x} \in \mathbb{R}^n$

$$\mathbf{P}^{s,\mathbf{x}}\{\boldsymbol{\tau}_\infty(\omega) = \infty\} = 1. \tag{22}$$

Let $G^n(\mathbb{R}^n)$ be $\underbrace{G(\mathbb{R}^n) \times \ldots \times G(\mathbb{R}^n)}_{n}$ and $G^n_{P,r}(E)$ be $\underbrace{G_{P,r}(E) \times \ldots \times G_{P,r}(E)}_{n}$. Here $G(\mathbb{R}^n)$ is algebra of Colombeau generalized function, $G_{P,r}(E) = \mathcal{F}_{P,r}(E)/K_{P,r}(E)$ is the Colombeau type algebra [48], [49],[50] and $E$ is an appropriate algebra of functions, which is a locally convex vector space over field $\mathbb{C}$.

**Assumption2**. We assume now that :(1) $\forall\epsilon = (\epsilon_1,\ldots,\epsilon_k) \in (0,1]^n$ and $\forall t \in [0,\infty)$

$$\begin{gathered}(\mathbf{A}_{\varepsilon'}(\mathbf{x},t,\epsilon))_{\varepsilon'} = (A_{1,\varepsilon'}(\mathbf{x},t,\epsilon),\ldots,A_{n,\varepsilon'}(\mathbf{x},t,\epsilon))_{\varepsilon'} \in G^n(\mathbb{R}^n),\\ \text{or}\\ (\mathbf{A}_{\varepsilon'}(\mathbf{x},t,\epsilon))_{\varepsilon'} = (A_{1,\varepsilon'}(\mathbf{x},t,\epsilon),\ldots,A_{n,\varepsilon'}(\mathbf{x},t,\epsilon))_{\varepsilon'} \in G_{P,r}(E).\end{gathered} \tag{23}$$

(2) $\forall\epsilon = (\epsilon_1,\ldots,\epsilon_k) \in (0,1]^n$ there exist infinite Colombeau constants $(C^\epsilon_{\varepsilon'})_{\varepsilon'}$ and $(D^\epsilon_{\varepsilon'})_{\varepsilon'}$ such that

$$\begin{gathered}1.\ (\|\mathbf{A}_{\varepsilon'}(x,t,\epsilon)\|^2)_{\varepsilon'} \leq (C^\epsilon_{\varepsilon'})_{\varepsilon'}(1+\|x\|^2),\\ 2.\ (\|\mathbf{A}_{\varepsilon'}(x,t,\epsilon) - \mathbf{A}_{\varepsilon'}(y,t,\epsilon)\|)_{\varepsilon'} \leq (D^\epsilon_{\varepsilon'})_{\varepsilon'}\|x-y\|,\end{gathered} \tag{24}$$

for all $t \in [0,\infty)$ and for all $x \in \mathbb{R}^n$ and for all $y \in \mathbb{R}^n$.

**Definition2**.[21-24] 1.Let $\mathfrak{I} = (\Omega,\Sigma.P)$ be a probability space. Let $\mathcal{E}R$ be the space of nets $(X_\varepsilon(\omega))_\varepsilon$ of measurable functions on $\Omega$. Let $\mathcal{E}R_M$ be the space of nets $(X_\varepsilon)_\varepsilon \in \mathcal{E}R, \varepsilon \in (0,1]$, with the propertythat for almost all $\omega \in \Omega$ there exist constants $r, C > 0$ and $\varepsilon_0 \in (0,1]$ such that $|(X_\varepsilon)_\varepsilon| \leq C\varepsilon^{-r}, \varepsilon \leq \varepsilon_0$.

2.Let $NR$ is the space of nets $(X_\varepsilon)_\varepsilon \in \mathcal{E}R, \varepsilon \in (0,1]$, with the property that for almost all $\omega \in \Omega$ and all $b \in \mathbb{R}_+$ there exist constants $C > 0$ and $\varepsilon_0 \in (0,1]$ such that $|(X_\varepsilon)_\varepsilon)_\varepsilon| \leq C\varepsilon^b$,
$\varepsilon \leq \varepsilon_0$. The differential algebra $GR$ of Colombeau generalized random variables is the factor algebra $GR = \mathcal{E}R/NR$.

Let us consider now a family $\left(\mathbf{x}^{x_0,\varepsilon}_{t,\epsilon,\varepsilon'}(\omega)\right)_{\varepsilon'}$ of the solutions Colombeau-Ito's SDE:

$$\left(d\mathbf{x}^{x_0,\varepsilon}_{t,\epsilon,\varepsilon'}(\omega)\right)_{\varepsilon'} = \left(\mathbf{A}_{\varepsilon'}(\mathbf{x}^{x_0,\varepsilon}_{t,\epsilon,\varepsilon'}(\omega), t, \epsilon)\right)_{\varepsilon'} + \sqrt{\varepsilon}\, d\mathbf{W}(t),$$

$$\left(\mathbf{x}^{x_0,\varepsilon}_{0,\epsilon,\varepsilon'}(\omega)\right)_{\varepsilon'} = \left(\mathbf{x}^{x_0}_{0,\varepsilon'}(\omega)\right)_{\varepsilon'} \in GR \tag{25}$$

$$\left(\mathbf{E}\left[\mathbf{x}^{x_0}_{0,\varepsilon'}(\omega)\right]\right)_{\varepsilon'} = x_0.$$

Here (i) $\mathbf{W}(t)$ is $n$-dimensional Brownian motion,(ii) $\forall t \in [0,T] : \mathbf{A}_{\varepsilon'}(x,t,\epsilon) \in G^n(\mathbb{R}^n)$ $\left(\text{or } \mathbf{A}_{\varepsilon'}(\mathbf{x},t,\epsilon) \in G_{P,r}(E)\right), \mathbf{A}_0(\mathbf{x},t,\epsilon) \equiv \mathbf{A}_{\varepsilon'=0}(\mathbf{x},t,\epsilon) : \mathbb{R}^n \to \mathbb{R}^n$ is a polynomial on variable $\mathbf{x} = (x_1,\ldots,x_n)$,i.e.,

$$A_{i,0}(\mathbf{x},t,\epsilon) = \sum_{\alpha,|\alpha|\leq r} a^{\alpha}_{i,0}(t,\epsilon)x^{\alpha}, \tag{26}$$

$\alpha = (i_1,\ldots,i_n), |\alpha| = \sum_{j=1}^{n} i_j, 0 \leq i_j \leq p,$ or

(iii) $\mathbf{A}_0(\mathbf{x},t,\epsilon) \equiv \mathbf{A}_{\varepsilon'=0}(\mathbf{x},t,\epsilon) : \mathbb{R}^n \to \mathbb{R}^n$ is $\mathbb{R}$-analytic function on variable $\mathbf{x} = (x_1,\ldots,x_n)$,i.e.,

$$A_{i,0}(\mathbf{x},t,\epsilon) = \sum_{r=1}^{\infty} \sum_{\alpha,|\alpha|\leq r} a^{\alpha}_{i,0}(t,\epsilon)x^{\alpha}, \tag{27}$$

and

(iv)

$$\lim_{\|\mathbf{x}\|\to\infty} \frac{\mathbf{A}_0(\mathbf{x},t,\epsilon)}{\|\mathbf{x}\|} = \infty \tag{28}$$

(v)

$$A_{i,\varepsilon'}(\mathbf{x}(t),t,\epsilon) = A_{i,0}(\mathbf{x}_{\varepsilon'}(t),t,\epsilon)$$

$$\mathbf{x}_{\varepsilon'}(t) = (x_{1,\varepsilon'}(t),\ldots,x_{n,\varepsilon'}(t)) \tag{29}$$

Here

$$x_{i,\varepsilon'}(t) = \begin{cases} \dfrac{x_i(t)}{1+(\varepsilon')^{2l}x_i^{2l}(t)}, \\ \text{or} \\ x_i(t)\theta_{\epsilon_i}[x_i(t)] \end{cases} \tag{30}$$

Here $\theta_{\epsilon_i}[z] \in C^\infty(\mathbb{R}), Supp(\theta_{\epsilon_i}[z]) \subseteq [-v_1(\epsilon_i), v_1(\epsilon_i)]$

$$\begin{cases} \theta_{\epsilon_i}[z] = 1 \Leftrightarrow z \in [-v(\epsilon_i), v(\epsilon_i)] \subsetneq [-v_1(\epsilon_i), v_1(\epsilon_i)], \\ \theta_{\epsilon_i}[z] = 0 \Leftrightarrow z \in \mathbb{R}\backslash[-v_1(\epsilon_i), v_1(\epsilon_i)], \\ 0 \leq \theta_{\epsilon_i}[z] \leq 1 \Leftrightarrow z \in [-v_1(\epsilon_i), v_1(\epsilon_i)]\backslash[-v(\epsilon_i), v(\epsilon_i)]. \end{cases} \tag{31}$$

**Remark 5**.By classical theorem of existence [46-47] and enequality (24) for every Colombeau generalized random variable $\left(\mathbf{x}_{0,\varepsilon'}^{x_0}(\omega)\right)_{\varepsilon'} \in GR$ independent of the process $\mathbf{W}(t)$ there exist Colombeau generalized stochastic proces $\left(\mathbf{x}_{t,\epsilon,\varepsilon'}^{x_0,\varepsilon}(\omega)\right)_{\varepsilon'}$ such that $\left(\mathbf{x}_{0,\epsilon,\varepsilon'}^{x_0,\varepsilon}(\omega)\right)_{\varepsilon'} = \left(\mathbf{x}_{0,\varepsilon'}^{x_0}(\omega)\right)_{\varepsilon'}$ and $\left(\mathbf{x}_{t,\epsilon,\varepsilon'}^{x_0,\varepsilon}(\omega)\right)_{\varepsilon'}$ is the solution of the Colombeau-Ito's SDE (25),which is an almost surely continuous Colombeau generalized stochastic process and is unique up to equivalence

$$\left(\mathbf{P}\left[\,\left\|\mathbf{x}_{1,t,\epsilon,\varepsilon'}^{x_0,\varepsilon}(\omega) - \mathbf{x}_{2,t,\epsilon,\varepsilon'}^{x_0,\varepsilon}(\omega)\right\| > 0\right]\right)_{\varepsilon'} = 0, \tag{32}$$

for all t$\in[0,\infty)$.

**Remark 6**.One can to construct a sequence of Colombeau generalized functions $(\mathbf{A}_{\varepsilon',n}(\mathbf{x}(t),t,\epsilon))_{\varepsilon'}$ such that for $\|x\| < n \in \mathbb{R}_+$ :

$$(\mathbf{A}_{\varepsilon',n}(\mathbf{x}(t),t,\epsilon))_{\varepsilon'} = (\mathbf{A}_{\varepsilon'}(\mathbf{x}(t),t,\epsilon))_{\varepsilon'},$$

$$\varepsilon' \in (0,1], \epsilon \in (0,1]^n, \tag{33}$$

and therefore for each $(\mathbf{A}_{\varepsilon',n}(\mathbf{x}(t),t,\epsilon))_{\varepsilon'}$ satisfy conditions (16) everywhere in $\mathbb{R}^n$. By Theorem of existence [46-47], there exists a sequence of Colombeau generalized stochastic processes $\left(\mathbf{x}^{x_0,\varepsilon}_{t,\epsilon,\varepsilon',n}(\omega)\right)_{\varepsilon'}, \varepsilon' \in (0,1]$, corresponding to sequence of Colombeau generalized functions $(\mathbf{A}_{\varepsilon',n}(\mathbf{x}(t),t,\epsilon))_{\varepsilon'}$. Suppose now that for each $\varepsilon' \in (0,1], \epsilon \in (0,1]^n$ the distribution of $\mathbf{x}^{x_0}_{0,\varepsilon'}(\omega)$ has compact support in $\mathbb{R}^n$. Then there exit times of the processes $\mathbf{x}^{x_0,\varepsilon}_{t,\epsilon,\varepsilon',m}(\omega)$, $m \in \mathbb{R}^n$ fromthe set $\|x\| < n$ are identical for $m \geq n$. Let this common value be $\tau_{\varepsilon'}(\omega,\varepsilon,\epsilon)$. It is also clear that the processes $\left(\mathbf{x}^{x_0,\varepsilon}_{t,\epsilon,\varepsilon',n}(\omega)\right)_{\varepsilon'}$ and $\left(\mathbf{x}^{x_0,\varepsilon}_{t,\epsilon,\varepsilon',m}(\omega)\right)_{\varepsilon'}$ themselves coincide up to generalized time $(\tau_{\varepsilon'}(\omega,\varepsilon,\epsilon))_{\varepsilon'}$ i.e.,

$$\left(\mathbf{P}\left[\sup_{0 \leq t \leq \tau_{\varepsilon'}(\omega,\varepsilon,\epsilon)} \left\| \mathbf{x}^{x_0,\varepsilon}_{t,\epsilon,\varepsilon',n}(\omega) - \mathbf{x}^{x_0,\varepsilon}_{t,\epsilon,\varepsilon',m}(\omega) \right\| > 0\right]\right)_{\varepsilon'} = 0, \tag{34}$$

for all $m \geq n$.

**Definition3**. (i) Let $\tau_{\varepsilon'}(\omega,\varepsilon,\epsilon)$ denote the (finite or infinite) limit of the monotone increasing sequence $\tau_{\varepsilon',n}(\omega,\varepsilon,\epsilon)$ as $n \to \infty$. We call the generalized random variable $(\tau_{\varepsilon'}(\omega,\varepsilon,\epsilon))_{\varepsilon'}$ the first exit time of the sample function from every bounded domain, or briefly the generalized explosion time.

(ii) We now define Colombeau generalized stochastic process $\left(\mathbf{x}^{x_0,\varepsilon}_{t,\epsilon,\varepsilon'}(\omega)\right)_{\varepsilon'}$ by setting

$$\mathbf{x}^{x_0,\varepsilon}_{t,\epsilon,\varepsilon'}(\omega) = \mathbf{x}^{x_0,\varepsilon}_{t,\epsilon,\varepsilon',n}(\omega) \text{ for } t = t(\omega) < \tau_{\varepsilon',n}(\omega,\varepsilon,\epsilon). \tag{35}$$

That this is always a Markov process for $t = t(\omega) < (\tau_{\varepsilon',n}(\omega,\varepsilon,\epsilon))_{\varepsilon'}$.

(iii) Colombeau generalized stochastic process $\left(\mathbf{x}^{x_0,\varepsilon}_{t,\epsilon,\varepsilon'}(\omega)\right)_{\varepsilon'}, \varepsilon' \in (0,1], \epsilon \in (0,1]^n$, defined by setting (35) on the random generalizeed interval $[0,(\tau_{\varepsilon'}(\omega,\varepsilon,\epsilon))_{\varepsilon'}]$ is *regular*, if for any $s < \infty, \mathbf{x} \in \mathbb{R}^n$

$$(\mathbf{P}^{s,\mathbf{x}}(\tau_{\varepsilon',n}(\omega,\varepsilon,\epsilon) = \infty))_{\varepsilon'} = 1 \tag{36}$$

(iv) Colombeau generalized stochastic process $\left(\mathbf{x}^{x_0,\varepsilon}_{t,\epsilon,\varepsilon'}(\omega)\right)_{\varepsilon'}, \varepsilon' \in (0,1], \epsilon \in (0,1]^n$, defined by setting (35) is a *strongly regular* if for any $s < \infty, \mathbf{x} \in \mathbb{R}^d, \varepsilon' \in [0,1], \boldsymbol{\epsilon} \in (0,1]^d$

$$(\mathbf{P}^{s,\mathbf{x}}(\tau_{\varepsilon',n}(\omega,\varepsilon,\epsilon) = \infty))_{\varepsilon'} = 1 \tag{37}$$

**Remark7**.Wenote that: (iii) does not imply (iv).

**Definition3$'$**.Let $(\tau_{\varepsilon',n}(\omega,\varepsilon,\epsilon,t))_{\varepsilon'}$ be $(\tau_{\epsilon,\varepsilon',n}(\omega,\varepsilon,\epsilon,t))_{\varepsilon'} = (\min\{\tau_{\varepsilon',n}(\omega,\varepsilon,\epsilon),t\})_{\varepsilon'}$.

**Proposition 3**.We set now $\theta_{\epsilon_i}[z] \equiv 1, i = 1,\ldots,d.$Assume that: (1) Colombeau generalized stochastic process $\left(\mathbf{x}^{x_0,\varepsilon}_{t,\epsilon,\varepsilon'}(\omega)\right)_{\varepsilon'}$, $t \in [0,T], \varepsilon' \in (0,1], \epsilon \in (0,1]^n$, defined by setting (35) is a strongly regular, (2) $\sup_{n\in\mathbb{N}}\left(\int_0^T \mathbf{E}\left[\left\|\mathbf{x}^{\mathbf{x}_0,\varepsilon}_{s,\epsilon,\varepsilon'=0,n}(\omega)\right\|^2\right]ds\right) < \infty$. Let $\Delta$ be a set of the all $\mathbb{R}_+$- valued sequence $\{\varepsilon'_n\} = \{\varepsilon'_n\}_{n=1}^{\infty}$ such that $\lim_{n\to\infty}\varepsilon'_n = 0$. Let $\left\{\mathbf{x}^{x_0,\varepsilon}_{t,\epsilon,\varepsilon'_n}(\omega)\right\}$ be a sequence Colombeau generalized stochastic proceses defined by setting

$$\mathbf{x}^{x_0,\varepsilon}_{t,\epsilon,\varepsilon'_n}(\omega) = \mathbf{x}^{x_0,\varepsilon}_{t,\epsilon,\varepsilon'_n,n}(\omega) \text{ for } t < \tau_{\varepsilon'_n,n}(\omega,\varepsilon,\epsilon).$$

Then $\forall t \in [0,T], \forall\delta > 0$

$$(38.1)$$

$$\inf_{\{\varepsilon'_n\}\in\Delta}\left[\lim_{n\to\infty}\left(\sup_{t\in\left[0,\tau_{\varepsilon',n}(\omega,\varepsilon,\epsilon,T)\right]}\mathbf{E}\left[\left\|\mathbf{x}^{\mathbf{x}_0,\varepsilon}_{\tau_{\epsilon,\varepsilon',n}(\omega,\varepsilon,\epsilon,T),\epsilon,\varepsilon'_n}(\omega) - \mathbf{x}^{\mathbf{x}_0,\varepsilon}_{\tau_{\epsilon,\varepsilon',n}(\omega,\varepsilon,\epsilon,T),\epsilon,\varepsilon'=0,n}(\omega)\right\|^2\right]\right)\right] = 0,$$

$$(38.2)$$

$$\inf_{\{\varepsilon'_n\}\in\Delta}\left(\lim_{n\to\infty}\mathbf{P}\left\{\sup_{t\in\left[0,\tau_{\varepsilon',n}(\omega,\varepsilon,\epsilon,T)\right]}\left\|\mathbf{x}^{\mathbf{x}_0,\varepsilon}_{\tau_{\epsilon,\varepsilon',n}(\omega,\varepsilon,\epsilon,T),\epsilon,\varepsilon'_n}(\omega) - \mathbf{x}^{\mathbf{x}_0,\varepsilon}_{\tau_{\epsilon,\varepsilon',n}(\omega,\varepsilon,\epsilon,T),\epsilon,\varepsilon'=0,n}(\omega)\right\| > \delta\right\}\right) = 0, \quad (38)$$

$$(38.3)\ \liminf_{\varepsilon'\to 0}\left(\sup_{t\in[0,T]}\mathbf{E}\left[\left\|\mathbf{x}^{\mathbf{x}_0,\varepsilon}_{t,\epsilon,\varepsilon'}(\omega) - \mathbf{x}^{\mathbf{x}_0,\varepsilon}_{t,\epsilon,\varepsilon'=0}(\omega)\right\|^2\right]\right) = 0,$$

$$(38.4)\ \liminf_{\varepsilon'\to 0}\mathbf{P}\left\{\sup_{t\in[0,T]}\left\|\mathbf{x}^{\mathbf{x}_0,\varepsilon}_{t,\epsilon,\varepsilon'}(\omega) - \mathbf{x}^{\mathbf{x}_0,\varepsilon}_{t,\epsilon,\varepsilon'=0}(\omega)\right\| > \delta\right\} = 0.$$

**Proof**. We willing to chose an sequences $\{\varepsilon'_n\}_{n=1}^{\infty}$ and $\{\epsilon_n\}_{n=1}^{\infty}$ such that: (1) $\lim_{n\to\infty}\varepsilon'_n = 0, \lim_{n\to\infty}\epsilon_n = 0$ and (2) the all conditions of the Proposition 2 is satisfied and $\lim_{n\to\infty}[\delta^2_{1,n}(\varepsilon'_n,\epsilon_n) + \delta^2_{2,n}(\varepsilon'_n)] = 0$. Therefore from Proposition 2(II) we obtain

$$\sup_{0\leq t\leq\tau_{\varepsilon_n',n}(\omega,\varepsilon,\epsilon,T)}\mathbf{E}\left[\left\|\mathbf{x}^{\mathbf{x}_0,\varepsilon}_{\tau_{\epsilon,\varepsilon',n}(\omega,\varepsilon,\epsilon,T),\epsilon,\varepsilon_n'}(\omega)-\mathbf{x}^{\mathbf{x}_0,\varepsilon}_{\tau_{\epsilon,\varepsilon',n}(\omega,\varepsilon,\epsilon,T),\epsilon,\varepsilon'=0,n}(\omega)\right\|^2\right]\leq$$

$$\leq e^{L_n}[\delta_{1,n}^2(\varepsilon_n')+\delta_{2,n}^2(\varepsilon_n')]\mathbf{E}\left[\int_0^{\tau_{\varepsilon_n',n}(\omega,\varepsilon,\epsilon,T)}\left[\left\|\mathbf{x}^{\mathbf{x}_0,\varepsilon}_{s,\epsilon,\varepsilon'=0,n}(\omega)\right\|^2\right]ds\right]\leq$$

$$\leq e^{L_n}[\delta_{1,n}^2(\varepsilon_n')+\delta_{2,n}^2(\varepsilon_n')]\mathbf{E}\left[\int_0^T\left[\left\|\mathbf{x}^{\mathbf{x}_0,\varepsilon}_{s,\epsilon,\varepsilon'=0,n}(\omega)\right\|^2\right]ds\right]\leq$$

$$e^{L_n}[\delta_{1,n}^2(\varepsilon_n')+\delta_{2,n}^2(\varepsilon_n')]T\left[\int_0^T\mathbf{E}\left[\left\|\mathbf{x}^{\mathbf{x}_0,\varepsilon}_{s,\epsilon,\varepsilon'=0,n}(\omega)\right\|^2\right]ds\right].$$

Where $L_n = 3(1+T)K_n^2$. Note that we willing to chose the sequence $\{\varepsilon_n'\}_{n=1}^{\infty}$ such that $\lim_{n\to\infty}e^{L_n}[\delta_{1,n}^2(\varepsilon_n')+\delta_{2,n}^2(\varepsilon_n')] = 0$. Therefore we obtain

$$\lim_{n\to\infty}\left(\sup_{0\leq t\leq\tau_{\varepsilon_n',n}(\omega,\varepsilon,\epsilon,T)}\mathbf{E}\left[\left\|\mathbf{x}^{\mathbf{x}_0,\varepsilon}_{t,\epsilon,\varepsilon_n'}(\omega)-\mathbf{x}^{\mathbf{x}_0,\varepsilon}_{t,\epsilon,\varepsilon'=0,n}(\omega)\right\|^2\right]\right) = 0.$$

Using now condition (1) we obtain (38.1). By definitions (38.1) imply (38.2).

Let us consider now a family $\mathbf{x}^{\mathbf{x}_0,\varepsilon}_{t,\epsilon,\varepsilon'}(\omega)$ of the solutions of the Colombeau SDE:

$$\left(d\mathbf{x}^{x_0,\varepsilon}_{t,\epsilon,\varepsilon'}(\omega)\right)_{\varepsilon'} = \left(\mathbf{A}_{\varepsilon'}(\mathbf{x}^{x_0,\varepsilon}_{t,\epsilon,\varepsilon'}(\omega),t,\epsilon,\omega)\right)_{\varepsilon'}+\sqrt{\varepsilon}\,d\mathbf{W}(t),$$
$$\left(\mathbf{x}^{x_0,\varepsilon}_{0,\epsilon,\varepsilon'}(\omega)\right)_{\varepsilon'} = \left(\mathbf{x}^{x_0}_{0,\varepsilon'}(\omega)\right)_{\varepsilon'}\in GR,\left(\mathbf{E}\left[\mathbf{x}^{x_0}_{0,\varepsilon'}(\omega)\right]\right)_{\varepsilon'} = x_0. \tag{39}$$

Here $\mathbf{W}(t)$ is n-dimensional Brownian motion, and (1) $\forall\epsilon\in(0,1]^n,\ \forall t\in[0,T]$ and for almost al $\omega\in\Omega:(\mathbf{A}_{\varepsilon'}(\mathbf{x},t,\epsilon,\omega))_{\varepsilon'}\in G^n(\mathbb{R}^n)$, (2) $\forall t\in[0,T]$ :
$\mathbf{A}_0(\mathbf{x},t,0) = \mathbf{A}_{\varepsilon'=0}(\mathbf{x},t,\epsilon=0,\omega)$ is a polynomial vector-function on a variable $\mathbf{x} = (x_1,\ldots,x_n)$ i.e.,

$$A_{i,0}(\mathbf{x},t,0) = \textstyle\sum_{\alpha,|\alpha|\leq r}A^{\alpha}_{i,0}(t)x^{\alpha},\alpha = (i_1,\ldots,i_n),|\alpha| = \sum_{j=1}^n i_j,0\leq i_j\leq p. \tag{40}$$

And

$$A_{i,\varepsilon'}(\mathbf{x}(t),t,\epsilon,\omega) = A_{i,0}(\mathbf{x}_{\varepsilon',\epsilon}(t,\omega),t,\epsilon), \tag{41}$$

where $\mathbf{x}_{\varepsilon',\epsilon}(t,\omega) = (x_{1,\varepsilon',\epsilon}(t,\omega),\ldots,x_{n,\varepsilon',\epsilon}(t,\omega))$ and

$$x_{i,\varepsilon',\epsilon}(t,\omega) = \frac{x_i(t)}{1+\varepsilon' x_i^{2l}(t) + \varepsilon'\left[\epsilon_i \int\limits_0^t \theta_{\epsilon_i}[x_i(\tau)]x_i^{2l}(\tau)d\tau + \sqrt{\delta}\,W_i(t)\right]^2}. \tag{42}$$

$i = 1,..,n.$ Now we let

$$u_i(t) = \epsilon_i \int\limits_0^t \theta_{\epsilon_i}[x_i(\tau)]x_i^{2l}(\tau)d\tau + \sqrt{\delta}\,W_i(t) \tag{43}$$

and rewrite Eq.(39) of the canonical Colombeau-Ito form:

$$\begin{gathered}
\left(d\mathbf{x}_{t,\epsilon,\varepsilon'}^{x_0,\varepsilon}(\omega)\right)_{\varepsilon'} = \left(\mathbf{A}_{\varepsilon'}(\mathbf{x}_{t,\epsilon,\varepsilon'}^{x_0,\varepsilon}(\omega),\mathbf{u}_{t,\epsilon,\varepsilon'}^{\delta}(\omega),t)\right)_{\varepsilon'} + \sqrt{\varepsilon}\,d\mathbf{W}(t),\\
\mathbf{u}_{t,\epsilon,\varepsilon'}^{\delta}(\omega) = \left(u_{1,t,\epsilon_i,\varepsilon'}^{\delta}(\omega),\ldots,u_{n,t,\epsilon_i,\varepsilon'}^{\delta}(\omega)\right),\\
du_{i,t,\epsilon,\varepsilon'}^{\delta}(\omega) = \left(\theta_{\epsilon_i}\left[u_{i,t,\epsilon,\varepsilon'}^{\delta}(\omega)\right]\left(u_{i,t,\epsilon,\varepsilon'}^{\delta}(\omega)\right)^{2l}\right)_{\varepsilon'} + \sqrt{\delta}\,W_i(t),\\
\left(\mathbf{x}_{0,\epsilon,\varepsilon'}^{x_0,\varepsilon}(\omega)\right)_{\varepsilon'} = \left(\mathbf{x}_{0,\varepsilon'}^{x_0}(\omega)\right)_{\varepsilon'} \in GR, \left(\mathbf{E}\left[\mathbf{x}_{0,\varepsilon'}^{x_0}(\omega)\right]\right)_{\varepsilon'} = \mathbf{x}_0.
\end{gathered} \tag{44}$$

**Proposition 4**. Let us consider a pair of the Colombeau-Ito's SDE:

$$\left(d\mathbf{x}_{t,\varepsilon'}^{x_0,\varepsilon}(\omega)\right)_{\varepsilon'} = \left(\mathbf{A}_{\varepsilon'}^{(\mu)}(\mathbf{x}_{t,\varepsilon'}^{x_0,\varepsilon}(\omega),t)\right)_{\varepsilon'} + \sqrt{\varepsilon}\,d\mathbf{W}(t),$$

$$\mu = 1,2, \tag{45}$$

$$\left(\mathbf{x}_{0,\varepsilon'}^{x_0,\varepsilon}(\omega)\right)_{\varepsilon'} = \left(\mathbf{x}_{0,\varepsilon'}^{x_0}(\omega)\right)_{\varepsilon'} \in GR, \left(\mathbf{E}\left[\mathbf{x}_{0,\varepsilon'}^{x_0}(\omega)\right]\right)_{\varepsilon'} = x_0.$$

Assume now that: (1) Conditions (24) and is satisfied. (2) For a given $N > 0, \forall \mathbf{x} \in \mathbb{R}^n$ such that $\|\mathbf{x}\| \leq N : \mathbf{A}_{\varepsilon'}^{(1)}(\mathbf{x},t) = \mathbf{A}_{\varepsilon'}^{(2)}(\mathbf{x},t)$.

Let $\mathbf{x}_{\mu,t,\epsilon,\varepsilon'}^{x_0,\varepsilon}(\omega), \mu = 1,2$ be a pair of the solutions of the Colombeau-Ito's SDE (45) and let $\mathcal{F}_{\varepsilon',\mu}^{N}(\omega)$, $\mu = 1,2$ be a set

$$\mathcal{F}_{\varepsilon',\mu}^{N}(\omega) \triangleq \left\{t | \sup_{0\leq s\leq t} \left\|\mathbf{x}_{\mu,t,\epsilon,\varepsilon'}^{x_0,\varepsilon}(\omega)\right\| \leq N\right\}. \tag{46}$$

We let now

$$\left(\tau_{\varepsilon',\mu}^{N}(\omega)\right)_{\varepsilon'} = \left(\sup\left\{t | t \in \mathcal{F}_{\varepsilon',\mu}^{N}(\omega)\right\}\right)_{\varepsilon'}. \tag{47}$$

Then $\forall \varepsilon' \in (0,1]$ :

$$\mathbf{P}\left\{\tau_{\varepsilon',1}^{N}(\omega) = \tau_{\varepsilon',2}^{N}(\omega)\right\} = 1 \tag{48}$$

and

$$\mathbf{P}\left\{\sup_{0\leq t\leq \tau_{\varepsilon',1}^{N}} \left\|\mathbf{x}_{1,t,\epsilon,\varepsilon'}^{x_0,\varepsilon}(\omega) - \mathbf{x}_{2,t,\epsilon,\varepsilon'}^{x_0,\varepsilon}(\omega)\right\| = 0\right\} = 1. \tag{49}$$

**Proof**. A proof of this statement, complete similarly, to a classical case. For example see [47] ,chapt.2, subsect.6, theorem 2.

Let us rewrite now Eq.(44) in the next form (with $\theta_{\epsilon_i}\left[u_{i,\tau,\epsilon,\varepsilon'}^{\delta}(\omega)\right] \equiv 1, i = 1,2,\ldots,n$ )

$$\left(\mathbf{x}^{x_0,\varepsilon}_{t,\epsilon,\varepsilon'}(\omega)\right)_{\varepsilon'} = (\mathbf{x}^{x_0}_{\varepsilon'}(\omega))_{\varepsilon'} + \left(\int_0^t \mathbf{A}_{\varepsilon'}(\mathbf{x}^{x_0,\varepsilon}_{\tau,\epsilon,\varepsilon'}(\omega),\mathbf{u}^{\delta}_{\tau,\epsilon,\varepsilon'}(\omega),\tau)d\tau\right)_{\varepsilon'} + \sqrt{\varepsilon}\,\mathbf{W}(t),$$

$$\mathbf{u}^{\delta}_{t,\epsilon,\varepsilon'}(\omega) = \left(u^{\delta}_{1,t,\epsilon_i,\varepsilon'}(\omega),\ldots,u^{\delta}_{n,t,\epsilon_i,\varepsilon'}(\omega)\right), \tag{50}$$

$$u^{\delta}_{i,t,\epsilon,\varepsilon'}(\omega) = \left(\int_0^t \left(u^{\delta}_{i,\tau,\epsilon,\varepsilon'}(\omega)\right)^{2l} d\tau\right)_{\varepsilon'} + \sqrt{\delta}\,W_i(t),$$

$$\left(\mathbf{x}^{x_0,\varepsilon}_{0,\epsilon,\varepsilon'}(\omega)\right)_{\varepsilon'} = (\mathbf{x}^{x_0}_{\varepsilon'}(\omega))_{\varepsilon'} \in GR, \left(\mathbf{E}\left[\mathbf{x}^{x_0}_{0,\varepsilon'}(\omega)\right]\right)_{\varepsilon'} = \mathbf{x}_0.$$

Let $G_N(\mathbf{y})$ be a function: (i) $G_N(\mathbf{y}) = \mathbf{y}$ if $\|\mathbf{y}\| \leq N,$ (ii) $G_N(\mathbf{y}) = \mathbf{0}$ if $\|\mathbf{y}\| > N.$ We set now

$$\mathbf{A}^N_{\varepsilon'}(\mathbf{x},\mathbf{u},t,\epsilon) = \mathbf{A}_{\varepsilon'}(G_N(\mathbf{x}),G_N(\mathbf{u}),t,\epsilon). \tag{51}$$

Let

$$\left(\mathbf{y}^{x_0,\varepsilon}_{t,\epsilon,\varepsilon'}(\omega,\delta,N)\right)_{\varepsilon'} = \left\{\left(\mathbf{x}^{x_0,\varepsilon}_{t,\epsilon,\varepsilon'}(\omega,\delta,N)\right)_{\varepsilon'}, \left(\mathbf{u}^{x_0,\varepsilon}_{t,\epsilon,\varepsilon'}(\omega,\delta,N)\right)_{\varepsilon'}\right\} \tag{52}$$

be a family of the solution of the Colombeau-Ito’s SDE:

$$\left(\mathbf{x}^{x_0,\varepsilon}_{t,\epsilon,\varepsilon'}(\omega,\delta,N)\right)_{\varepsilon'} = (\mathbf{x}^{x_0}_{\varepsilon'}(\omega))_{\varepsilon'} + \left(\int_0^t \mathbf{A}_{\varepsilon'}(\mathbf{x}^{x_0,\varepsilon}_{\tau,\epsilon,\varepsilon'}(\omega,\delta,N),\mathbf{u}^{\delta}_{\tau,\epsilon,\varepsilon'}(\omega),\tau)d\tau\right)_{\varepsilon'} +$$

$$+\sqrt{\varepsilon}\,\mathbf{W}(t),$$

$$\mathbf{u}^{\delta}_{t,\epsilon,\varepsilon'}(\omega) = \left(u^{\delta}_{1,t,\epsilon_i,\varepsilon'}(\omega),\ldots,u^{\delta}_{n,t,\epsilon_i,\varepsilon'}(\omega)\right), \tag{53}$$

$$u^{\delta}_{i,t,\epsilon,\varepsilon'}(\omega) = \left(\int_0^t \left(u^{\delta}_{i,\tau,\epsilon,\varepsilon'}(\omega)\right)^{2l} d\tau\right)_{\varepsilon'} + \sqrt{\delta}\,W_i(t),$$

$$(\mathbf{x}^{x_0}_{\varepsilon'}(\omega))_{\varepsilon'} \in GR, \left(\mathbf{E}\left[\mathbf{x}^{x_0}_{0,\varepsilon'}(\omega)\right]\right)_{\varepsilon'} = \mathbf{x}_0.$$

**Definition 4**. (1) Let $\left(\mathbf{y}^{x_0,\varepsilon}_{t,\epsilon,\varepsilon'}(\omega,\delta,N)\right)_{\varepsilon'}$ be a sequence of the solution of the Colombeau-Ito's SDE (53). Let $\mathcal{F}^N_{\varepsilon',\epsilon}(\omega,\delta)$, $\mu = 1,2$ be a set

$$\mathcal{F}^N_{\varepsilon',\mu}(\omega,\delta) \triangleq \left\{t|\sup_{0\leq s\leq t} \left\|\mathbf{x}^{x_0,\varepsilon}_{\mu,t,\epsilon,\varepsilon'}(\omega,\delta)\right\| \leq N\right\}. \quad (54)$$

We let now

$$\left(\tau^N_{\varepsilon',\epsilon}(\omega,\delta)\right)_{\varepsilon'} = \left(\sup\left\{t|t \in \mathcal{F}^N_{\varepsilon',\epsilon}(\omega,\delta)\right\}\right)_{\varepsilon'}. \quad (55)$$

and

$$\left(\tau^\infty_{\varepsilon',\epsilon}(\omega,\delta)\right)_{\varepsilon'} = \left(\lim_{N\to\infty} \tau^N_{\varepsilon',\epsilon}(\omega,\delta)\right)_{\varepsilon'}. \quad (56)$$

(3) Let $\widetilde{\mathbf{y}}^{x_0,\varepsilon}_{t,\epsilon,\varepsilon'}(\omega,\delta,N)$ be a net of the stochastic processes defined by setting

$$\widetilde{\mathbf{y}}^{x_0,\varepsilon}_{t,\epsilon,\varepsilon'}(\omega,\delta) = \widetilde{\mathbf{y}}^{x_0,\varepsilon}_{t,\epsilon,\varepsilon'}(\omega,\delta,N) \text{ iff } t < \tau^N_{\varepsilon',\epsilon}(\omega,\delta). \quad (57)$$

(4) Let $\left(\widetilde{\mathbf{y}}^{x_0,\varepsilon}_{t,\epsilon,\varepsilon'}(\omega,\delta)\right)_{\varepsilon'}$ be Colombeau generalized stochastic process defined by setting

$$\left(\mathbf{y}^{x_0,\varepsilon}_{t,\epsilon,\varepsilon'}(\omega,\delta)\right)_{\varepsilon'} = \left(\widetilde{\mathbf{y}}^{x_0,\varepsilon}_{t,\epsilon,\varepsilon'}(\omega,\delta)\right)_{\varepsilon'}. \quad (58)$$

**Definition5**. Let $\left(\mathbf{y}^{x_0,\varepsilon}_{t,\epsilon,\varepsilon'}(\omega,\delta)\right)_{\varepsilon'}$ be a family of the solutions Colombeau-Ito's SDE (53).

(1) A family $\left(\mathbf{y}^{x_0,\varepsilon}_{t,\epsilon,\varepsilon'}(\omega,\delta)\right)_{\varepsilon'}$ $\varepsilon,\varepsilon' \in (0,1], \epsilon \in (0,1]^n$ is regular if

$$\left(\lim_{c\to\infty} \mathbf{P}\left\{\left\|\mathbf{y}^{x_0,\varepsilon}_{t,\epsilon,\varepsilon'}(\omega,\delta)\right\| > c\right\}\right)_{\varepsilon'} = 0. \quad (59)$$

Or in the next equivalent form

$$\left(\mathbf{P}\left\{\tau^{\infty}_{\varepsilon',\epsilon}(\omega,\delta) = \infty\right\}\right)_{\varepsilon'} = 1,$$
$$\varepsilon,\varepsilon' \in (0,1], \epsilon \in (0,1]^n. \tag{60}$$

(2) A family $\left(\mathbf{y}^{x_0,\varepsilon}_{t,\epsilon,\varepsilon'}(\omega,\delta)\right)_{\varepsilon'}$ $\varepsilon,\varepsilon' \in (0,1], \epsilon \in (0,1]^n$ is a strongly regular if $\forall\varepsilon\forall\varepsilon'[\varepsilon,\varepsilon' \in (0,1]], \forall\epsilon[\epsilon \in (0,1]^n]$ :

$$\left(\lim_{c\to\infty} \mathbf{P}\left\{\left\|\mathbf{y}^{x_0,\varepsilon}_{t,\epsilon,\varepsilon'}(\omega,\delta)\right\| > c\right\}\right)_{\varepsilon'} = 0. \tag{61}$$

Or in the next equivalent form

$$\left(\mathbf{P}\left\{\tau^{\infty}_{\varepsilon',\epsilon}(\omega,\delta) = \infty\right\}\right)_{\varepsilon'} = 1,$$
$$\varepsilon' \in (0,1], \epsilon \in (0,1]^n. \tag{62}$$

**Definition 6**. Let $\left(\mathbf{y}^{x_0,\varepsilon}_{t,\epsilon,\varepsilon'}(\omega,\delta)\right)_{\varepsilon'}$ be a family of the solutions Colombeau-Ito's SDE (53).

(1) A family $\left(\mathbf{y}^{x_0,\varepsilon}_{t,\epsilon,\varepsilon'}(\omega,\delta)\right)_{\varepsilon'}$ $\varepsilon,\varepsilon' \in (0,1], \epsilon \in (0,1]^n$ is a non-regular if

$$\exists t'\forall t > t' : \left(\lim_{c\to\infty} \mathbf{P}\left\{\left\|\mathbf{y}^{x_0,\varepsilon}_{t,\epsilon,\varepsilon'}(\omega,\delta)\right\| > c\right\}\right)_{\varepsilon'} \neq 0. \tag{63}$$

Or in the next equivalent form

$$\left(\mathbf{P}\left\{\tau^{\infty}_{\varepsilon',\epsilon}(\omega,\delta) < \infty\right\}\right)_{\varepsilon'} = 1,$$
$$\varepsilon,\varepsilon' \in (0,1], \epsilon \in (0,1]^n. \tag{64}$$

**Proposition 5**. Assume that Colombeau generalized stochastic process $\left(\mathbf{y}^{x_0,\varepsilon}_{t,\epsilon,\varepsilon'}(\omega,\delta)\right)_{\varepsilon'}$ defined by setting (58) is a strongly regular.Then (1)

$$\liminf_{\varepsilon'\to 0,\epsilon\to 0}\liminf_{\delta\to 0}\mathbf{E}\left[\left\|\mathbf{y}^{x_0,\varepsilon}_{t,\epsilon,\varepsilon'}(\omega,\delta)-\mathbf{y}^{x_0,\varepsilon}_t(\omega)\right\|^2\right]=0, \quad (65)$$

(2) $\forall\sigma>0$ :

$$\liminf_{\varepsilon'\to 0,\epsilon\to 0}\liminf_{\delta\to 0}\mathbf{P}\left\{\left\|\mathbf{y}^{x_0,\varepsilon}_{t,\epsilon,\varepsilon'}(\omega,\delta)-\mathbf{y}^{x_0,\varepsilon}_t(\omega)\right\|>\sigma\right\}=0, \quad (66)$$

where $\mathbf{y}^{x_0,\varepsilon}_t(\omega)=\mathbf{y}^{x_0,\varepsilon}_{t,\epsilon=0,\varepsilon'=0}(\omega,\delta=0)$.

Proof. Immediately follows from Proposition 4 and Proposition 3.

**Proposition 6**. Let $\left(\mathbf{y}^{x_0,\varepsilon}_{t,\epsilon,\varepsilon'}(\omega,\delta)\right)_{\varepsilon'}$ be a family of the solutions Colombeau-Ito’s SDE (53) with $\theta_\epsilon[z]\equiv 1$. A family $\left(\mathbf{y}^{x_0,\varepsilon}_{t,\epsilon,\varepsilon'}(\omega,\delta)\right)_{\varepsilon'},\varepsilon,\varepsilon'\in(0,1],\epsilon\in(0,1]^n$ is regular.

**Proof**. Assume that: process $\left(\mathbf{y}^{x_0,\varepsilon}_{t,\epsilon,\varepsilon'}(\omega,\delta)\right)_{\varepsilon'}$ is a non-regular. Therefore

$$\begin{gathered}\left(\mathbf{P}\left\{\tau^{\infty}_{\varepsilon',\epsilon}(\omega,\delta)<\infty\right\}\right)_{\varepsilon'}=1,\\ \varepsilon,\varepsilon'\in(0,1],\epsilon\in(0,1]^n.\end{gathered} \quad (71)$$

and consequently

$$\left(\mathbf{P}\left\{\left\|\mathbf{y}^{x_0,\varepsilon}_{\tau^{\infty}_{\varepsilon',\epsilon}(\omega,\delta),\epsilon,\varepsilon'}(\omega,\delta)\right\|=\infty\right\}\right)_{\varepsilon'}>0. \quad (72)$$

But other hand from Eq.(53) we obtain

$$\left(\mathbf{x}^{x_0,\varepsilon}_{\tau^{\infty}_{\varepsilon',\epsilon}(\omega,\delta),\epsilon,\varepsilon'}(\omega,\delta,N)\right)_{\varepsilon'} = (\mathbf{x}^{x_0}_{\varepsilon'}(\omega))_{\varepsilon'} +$$

$$\left(\int\limits_0^{\tau^{\infty}_{\varepsilon',\epsilon}(\omega,\delta)} \mathbf{A}_{\varepsilon'}(\mathbf{x}^{x_0,\varepsilon}_{\widetilde{\tau},\epsilon,\varepsilon'}(\omega,\delta,N),\mathbf{u}^{\delta}_{\widetilde{\tau},\epsilon,\varepsilon'}(\omega),\tau)d\widetilde{\tau}\right)_{\varepsilon'} +$$

$$+\sqrt{\varepsilon}\,\mathbf{W}\left(\tau^{\infty}_{\varepsilon',\epsilon}(\omega,\delta)\right), \tag{73}$$

$$\mathbf{u}^{\delta}_{\widetilde{\tau},\epsilon,\varepsilon'}(\omega) = \left(u^{\delta}_{1,\widetilde{\tau},\epsilon_i,\varepsilon'}(\omega),\ldots,u^{\delta}_{n,\widetilde{\tau},\epsilon_i,\varepsilon'}(\omega)\right),$$

$$u^{\delta}_{i,t,\epsilon,\varepsilon'}(\omega) = \left(\int\limits_0^{\tau^{\infty}_{\varepsilon',\epsilon}(\omega,\delta)} \left(u^{\delta}_{i,\tau,\epsilon,\varepsilon'}(\omega)\right)^{2l} d\tau\right)_{\varepsilon'} + \sqrt{\delta}\,W_i\left(\tau^{\infty}_{\varepsilon',\epsilon}(\omega,\delta)\right),$$

$$(\mathbf{x}^{x_0}_{\varepsilon'}(\omega))_{\varepsilon'} \in GR, \left(\mathbf{E}\left[\mathbf{x}^{x_0}_{0,\varepsilon'}(\omega)\right]\right)_{\varepsilon'} = \mathbf{x}_0.$$

and therefore (72)-(73) gives

$$\left(\mathbf{P}^{0,x_0}\left\{\int\limits_0^{\tau^{\infty}_{\varepsilon',\epsilon}(\omega,\delta)} \mathbf{A}_{\varepsilon'}(\mathbf{x}^{x_0,\varepsilon}_{\widetilde{\tau},\epsilon,\varepsilon'}(\omega,\delta,N),\mathbf{u}^{\delta}_{\widetilde{\tau},\epsilon,\varepsilon'}(\omega),\tau)d\widetilde{\tau}\right\} = \infty\right)_{\varepsilon'} = 0. \tag{74}$$

And therefore

$$\left(\mathbf{P}^{0,x_0}\left\{\mathbf{x}^{x_0,\varepsilon}_{\tau^{\infty}_{\varepsilon',\epsilon}(\omega,\delta),\epsilon,\varepsilon'}(\omega,\delta,N)=\infty\right\}\right)_{\varepsilon'}=$$

$$\left(\mathbf{P}^{0,x_0}\left\{\mathbf{x}^{x_0}_{\varepsilon'}(\omega)+\int_0^{\tau^{\infty}_{\varepsilon',\epsilon}(\omega,\delta)}\mathbf{A}_{\varepsilon'}(\mathbf{x}^{x_0,\varepsilon}_{\tilde{\tau},\epsilon,\varepsilon'}(\omega,\delta,N),\mathbf{u}^{\delta}_{\tilde{\tau},\epsilon,\varepsilon'}(\omega),\tau)d\tilde{\tau}\right.\right.$$

$$\left.\left.+\sqrt{\varepsilon}\,\mathbf{W}\left(\tau^{\infty}_{\varepsilon',\epsilon}(\omega,\delta)\right)=\infty\right\}\right)_{\varepsilon'}=0,$$

$$\mathbf{u}^{\delta}_{\tilde{\tau},\epsilon,\varepsilon'}(\omega)=\left(u^{\delta}_{1,\tilde{\tau},\epsilon_i,\varepsilon'}(\omega),\ldots,u^{\delta}_{n,\tilde{\tau},\epsilon_i,\varepsilon'}(\omega)\right),\tag{75}$$

$$\left(\mathbf{P}^{0,0}\left\{u^{\delta}_{i,t,\epsilon,\varepsilon'}(\omega)=\infty\right\}\right)_{\varepsilon'}=$$

$$\left(\mathbf{P}^{0,0}\left\{\int_0^{\tau^{\infty}_{\varepsilon',\epsilon}(\omega,\delta)}\left(u^{\delta}_{i,\tau,\epsilon,\varepsilon'}(\omega)\right)^{2l}d\tau+\sqrt{\delta}\,W_i\left(\tau^{\infty}_{\varepsilon',\epsilon}(\omega,\delta)\right)=\infty\right\}\right)_{\varepsilon'}=0.$$

Thus

$$\left(\mathbf{P}^{0,x_0}\left\{\mathbf{y}^{x_0,\varepsilon}_{\tau^{\infty}_{\varepsilon',\epsilon}(\omega,\delta),\epsilon,\varepsilon'}(\omega,\delta)\right\}=\infty\right)_{\varepsilon'}=0.\tag{76}$$

But this is the contradiction. This contradiction completed the proof.

**Definition 7**.CISDE (25) is $\widetilde{\mathbb{R}}$-dissipative if there exist Lyapunov candidate function $(V_{\varepsilon'}(\mathbf{x},t))_{\varepsilon'}:\widetilde{\mathbb{R}}^n\times[0,T]\to\widetilde{\mathbb{R}}$ and positive infinite Colombeau constants $\widetilde{C}=(C_{\varepsilon'})_{\varepsilon'}\in\widetilde{\mathbb{R}}_+$ and $\tilde{r}=(r_{\varepsilon'})_{\varepsilon'}\in\widetilde{\mathbb{R}}_+$ such that:

(1) $\forall\varepsilon'\in(0,1]:V^*_{\varepsilon'}=\lim_{R\to\infty}(\inf_{\|\mathbf{x}\|>R}V_{\varepsilon'}(\mathbf{x},t))=\infty$, and

(2) $\forall(\mathbf{x}_{\varepsilon'})_{\varepsilon'}[(\|\mathbf{x}_{\varepsilon'}\|)_{\varepsilon'}>\tilde{r}]$ the inequality

$$\left[(\dot{V}_{\varepsilon'}(\mathbf{x}_{\varepsilon'},t;\mathbf{A}_{\varepsilon'}))_{\varepsilon'}\right]\leq\widetilde{C}[(V_{\varepsilon'}(\mathbf{x}_{\varepsilon'},t))_{\varepsilon'}]\tag{77}$$

is satisfied. Here

$$\left[(\dot{V}_{\varepsilon'}(\mathbf{x}_{\varepsilon'},t;\mathbf{A}_{\varepsilon'}))_{\varepsilon'}\right] = \left[\left(\frac{\partial V_{\varepsilon'}(\mathbf{x}_{\varepsilon'},t)}{\partial t}\right)_{\varepsilon'}\right] + \sum_{i=1}^{n}\left[\left(\frac{\partial V_{\varepsilon'}(\mathbf{x}_{\varepsilon'},t)}{\partial x_{i,\varepsilon'}}\right)_{\varepsilon'}\right]. \quad (78)$$

Or in the next equivalent form:

CISDE (25) is $\widetilde{\mathbb{R}}$-dissipative if there exist Lyapunov candidate function $(V_{\varepsilon'}(\mathbf{x},t))_{\varepsilon'} : \widetilde{\mathbb{R}}^n \times [0,T] \to \widetilde{\mathbb{R}}$ and positive infinite Colombeau constants $\widetilde{C} = (C_{\varepsilon'})_{\varepsilon'} \in \widetilde{\mathbb{R}}_+$ and $\widetilde{r} = (r_{\varepsilon'})_{\varepsilon'} \in \widetilde{\mathbb{R}}_+$ such that:

(1) $\forall\varepsilon' \in (0,1] : V^*_{\varepsilon'} = \lim_{R\to\infty}(\inf_{\|\mathbf{x}\|>R} V_{\varepsilon'}(\mathbf{x},t)) = \infty$, and

(2′) $\forall\varepsilon' \in (0,1]\forall\mathbf{x}_{\varepsilon'}[(\mathbf{x}_{\varepsilon'} \in \mathbb{R}^n) \wedge (\|\mathbf{x}_{\varepsilon'}\| > r_{\varepsilon'})]$ the inequality

$$(\dot{V}_{\varepsilon'}(\mathbf{x}_{\varepsilon'},t;\mathbf{A}_{\varepsilon'}))_{\varepsilon'} \leq (C_{\varepsilon'})_{\varepsilon'}(V_{\varepsilon'}(\mathbf{x}_{\varepsilon'},t))_{\varepsilon'} \quad (79)$$

is satisfied. Here

$$(\dot{V}_{\varepsilon'}(\mathbf{x}_{\varepsilon'},t;\mathbf{A}_{\varepsilon'}))_{\varepsilon'} = \left(\frac{\partial V_{\varepsilon'}(\mathbf{x}_{\varepsilon'},t)}{\partial t}\right)_{\varepsilon'} + \sum_{i=1}^{n}\left(\frac{\partial V_{\varepsilon'}(\mathbf{x}_{\varepsilon'},t)}{\partial x_{i,\varepsilon'}}\right)_{\varepsilon'}. \quad (80)$$

**Definition 8**.CISDE (25) is a strongly $\widetilde{\mathbb{R}}$-dissipative if there exist Lyapunov candidate function $(V_{\varepsilon'}(\mathbf{x},t))_{\varepsilon'} : \widetilde{\mathbb{R}}^n \times [0,T] \to \widetilde{\mathbb{R}}$ and positive finite Colombeau constants $\widetilde{C} = (C_{\varepsilon'})_{\varepsilon'} \in \widetilde{\mathbb{R}}_+$ and $\widetilde{r} = (r_{\varepsilon'})_{\varepsilon'} \in \widetilde{\mathbb{R}}_+$ such that:

(1) $\forall\varepsilon' \in (0,1] : V^*_{\varepsilon'} = \lim_{R\to\infty}(\inf_{\|\mathbf{x}\|>R} V_{\varepsilon'}(\mathbf{x},t)) = \infty$, and

(2′) $\forall\varepsilon' \in (0,1]\forall\mathbf{x}_{\varepsilon'}[(\mathbf{x}_{\varepsilon'} \in \mathbb{R}^n) \wedge (\|\mathbf{x}_{\varepsilon'}\| > r_{\varepsilon'})]$ the inequality

$$(\dot{V}_{\varepsilon'}(\mathbf{x}_{\varepsilon'},t;\mathbf{A}_{\varepsilon'}))_{\varepsilon'} \leq (C_{\varepsilon'})_{\varepsilon'}(V_{\varepsilon'}(\mathbf{x}_{\varepsilon'},t))_{\varepsilon'} \quad (81)$$

is satisfied. Here

$$(\dot{V}_{\varepsilon'}(\mathbf{x}_{\varepsilon'},t;\mathbf{A}_{\varepsilon'}))_{\varepsilon'} = \left(\frac{\partial V_{\varepsilon'}(\mathbf{x}_{\varepsilon'},t)}{\partial t}\right)_{\varepsilon'} + \sum_{i=1}^{n}\left(\frac{\partial V_{\varepsilon'}(\mathbf{x}_{\varepsilon'},t)}{\partial x_{i,\varepsilon'}}\right)_{\varepsilon'}. \quad (82)$$

**Proposition 6**. Let $\left(\mathbf{x}^{x_0,\varepsilon}_{t,\epsilon,\varepsilon'}(\omega)\right)_{\varepsilon'}$ be generalized stochastic process satisfying Colombeau-Ito's SDE (25) on the time interval $[s,T]$ and let $(\tau^U_{\varepsilon'}(\omega))_{\varepsilon'}$ be a generalized random variable equal to the time at which the sample function of the generalized stochastic process $\left(\mathbf{x}_{t,\epsilon,\varepsilon'}^{x_0,\varepsilon}(\omega)\right)_{\varepsilon'}$ first leaves the bounded neighborhood $U$, and let

$$\left(\tau_{\epsilon,\varepsilon'}^{U}(\omega,t)\right)_{\varepsilon'} = \left(\min\left\{\tau_{\epsilon,\varepsilon'}^{U}(\omega),t\right\}\right)_{\varepsilon'}. \tag{83}$$

Suppose moreover that

$$\forall\varepsilon' \in (0,1], \forall\epsilon \in (0,1]^n : \mathbf{P}\left\{\mathbf{x}_{t,\epsilon,\varepsilon'}^{x_0,\varepsilon}(\omega) \in D\right\} = 1. \tag{84}$$

Then

$$\left(\mathbf{E}\left[V_{\varepsilon'}\left(\mathbf{x}_{\tau_{\epsilon,\varepsilon'}^{U}(\omega,t),\epsilon,\varepsilon'}^{x_0,\varepsilon}(\omega),\tau_{\epsilon,\varepsilon'}^{U}(\omega,t)\right) - V_{\varepsilon'}\left(\mathbf{x}_{s,\epsilon,\varepsilon'}^{x_0,\varepsilon}(\omega),s\right)\right]\right)_{\varepsilon'} =$$

$$\left(\mathbf{E}\left[\int_{s}^{\tau_{\epsilon,\varepsilon'}^{U}(\omega,t)} \dot{V}_{\varepsilon'}\left(\mathbf{x}_{u,\epsilon,\varepsilon'}^{x_0,\varepsilon}(\omega),u\right)du\right]\right)_{\varepsilon'}. \tag{85}$$

**Proof**. Similarly to the proof of the corresponding classical result, see [48] Lemma 3.2.

**Proposition 7**. Assume that: (i) for CISDE (25) the inequalities (24) is satisfied and (ii) CISDE (35)-(36) is $\widetilde{\mathbb{R}}$-dissipative.

Then: (1) Colombeau generalized stochastic process $\left(\mathbf{x}_{t,\epsilon,\varepsilon'}^{x_0,\varepsilon}(\omega)\right)_{\varepsilon'}$, $\varepsilon' \in (0,1]$, $\epsilon \in (0,1]^n$ defined by setting (35) is regular, and (2) the inequality

$$\left(\mathbf{E}\left[V_{\varepsilon'}\left(\mathbf{x}_{t,\epsilon,\varepsilon'}^{x_0,\varepsilon}(\omega),t\right)\right]\right)_{\varepsilon'} \leq \left(\mathbf{E}\left[V_{\varepsilon'}\left(\mathbf{x}_{t_0,\epsilon,\varepsilon'}^{x_0,\varepsilon}(\omega),t_0\right)\right]\right)_{\varepsilon'}\{\exp[(C_{\varepsilon'})_{\varepsilon'}(t-t_0)]\},$$
$$\varepsilon' \in (0,1], \epsilon \in (0,1]^n \tag{86}$$

is satisfied.

**Proof**. From (81)-(82) it follows that the Colombeau generalized function

$$(W_{\varepsilon'}(\mathbf{x}_{\varepsilon'},t))_{\varepsilon'} = (V_{\varepsilon'}(\mathbf{x}_{\varepsilon'},t))_{\varepsilon'}\exp[-(C_{\varepsilon'})_{\varepsilon'}(t-t_0)] \tag{87}$$

is satisfies the inequality

$$\left(\dot{W}_{\varepsilon'}(\mathbf{x}_{\varepsilon'},t)\right)_{\varepsilon'} \leq 0. \tag{88}$$

Hence, by Proposition 6, for $(\tau_{\epsilon,\varepsilon',n}(\omega,t))_{\varepsilon'} = (\min\{\tau_{\epsilon,\varepsilon',n}(\omega),t\})_{\varepsilon'}$ we have

$$\begin{gathered}\left(\mathbf{E}\left[V_{\varepsilon'}\left(\mathbf{x}^{x_0,\varepsilon}_{t,\epsilon,\varepsilon'}(\omega),\tau_{\epsilon,\varepsilon',n}(\omega,t)\right)\right]\right)_{\varepsilon'}\{\exp[-(C_{\varepsilon'})_{\varepsilon'}(\tau_{\epsilon,\varepsilon',n}(\omega,t)-t_0)]\} - \\ -\left(\mathbf{E}\left[V_{\varepsilon'}\left(\mathbf{x}^{x_0,\varepsilon}_{t_0,\epsilon,\varepsilon'}(\omega),t_0\right)\right]\right)_{\varepsilon'} \leq \\ \leq \left(\mathbf{E}\left[\int\limits_{t_0}^{\tau_{\epsilon,\varepsilon',n}(\omega,t)}\dot{W}_{\varepsilon'}\left(\mathbf{x}^{x_0,\varepsilon}_{u,\epsilon,\varepsilon'}(\omega),u\right)du\right]\right)_{\varepsilon'} \leq 0.\end{gathered} \tag{89}$$

This, together with the inequalities

$$\tau_{\epsilon,\varepsilon',n}(\omega,t) \leq t, 0 \leq (V_{\varepsilon'}(\mathbf{x}_{\varepsilon'},t))_{\varepsilon'}, \tag{90}$$

implies

$$\begin{gathered}\left(\mathbf{E}\left[V_{\varepsilon'}\left(\mathbf{x}^{x_0,\varepsilon}_{t,\epsilon,\varepsilon'}(\omega),\tau_{\epsilon,\varepsilon',n}(\omega,t)\right)\right]\right)_{\varepsilon'} \leq \\ \leq \left(\mathbf{E}\left[V_{\varepsilon'}\left(\widetilde{\mathbf{x}}^{x_0,\varepsilon}_{t_0,\epsilon,\varepsilon'}(\omega),t_0\right)\right]\right)_{\varepsilon'}\{\exp[(C_{\varepsilon'})_{\varepsilon'}(\tau_{\epsilon,\varepsilon',n}(\omega,t)-t_0)]\}.\end{gathered} \tag{91}$$

From (91) one derive the estimate

$$(\mathbf{P}\{\tau_{\epsilon,\varepsilon',n}(\omega) < t\})_{\varepsilon'} \leq \exp[(C_{\varepsilon'})_{\varepsilon'}(t-t_0)]\frac{\left(\mathbf{E}\left[V_{\varepsilon'}\left(\mathbf{x}^{x_0,\varepsilon}_{t,\epsilon,\varepsilon'}(\omega),\tau_{\epsilon,\varepsilon',n}(\omega,t)\right)\right]\right)_{\varepsilon'}}{\left(\inf_{\|\mathbf{x}\|\geq n,u\geq t_0}V_{\varepsilon'}(\mathbf{x},u)\right)_{\varepsilon'}} \tag{92}$$

Letting $n \to \infty$ and making use of the Definition 7 we now get (62).

**Proposition 8**. Assume that: (i) for CISDE (25) the inequalities (24) is satisfied and (ii) CISDE (35)-(36) is a strongly $\widetilde{\mathbb{R}}$-dissipative.

Then: (1) Colombeau generalized stochastic process $\left(\mathbf{x}^{x_0,\varepsilon}_{t,\epsilon,\varepsilon'}(\omega)\right)_{\varepsilon'}$, $\varepsilon' \in [0,1]$, $\epsilon \in [0,1]^n$ defined by setting (35) is regular, and (2) the inequality

$$\left(\mathbf{E}\left[V_{\varepsilon'}\left(\mathbf{x}^{x_0,\varepsilon}_{t,\epsilon,\varepsilon'}(\omega),t\right)\right]\right)_{\varepsilon'} \leq \left(\mathbf{E}\left[V_{\varepsilon'}\left(\mathbf{x}^{x_0,\varepsilon}_{t_0,\epsilon,\varepsilon'}(\omega),t_0\right)\right]\right)_{\varepsilon'}\{\exp[(C_{\varepsilon'})_{\varepsilon'}(t-t_0)]\},$$

$$\varepsilon' \in [0,1], \epsilon \in [0,1]^n \tag{93}$$

is satisfied.

**Proof**. Similarity to the proof of the Proposition 7.

**Theorem 1**.We set now $\theta_{\epsilon_i}[z] \equiv 1, i = 1,2,\ldots,n$ For any solution $\mathbf{x}^{x_0,\varepsilon}_{t,\epsilon,\varepsilon'}(\omega,\delta)$ of a strongly $\widetilde{\mathbb{R}}$-dissipative CISDE (44) and any $\mathbb{R}$-valued parameters $\lambda_1,\ldots,\lambda_n$, there exist finite Colombeau constant $\tilde{C}' = [(C_{\varepsilon'})_{\varepsilon'}] > 0$, such that $\forall \boldsymbol{\lambda}[\lambda = (\lambda_1,\ldots,\lambda_n)]$, the inequality

$$\liminf_{\varepsilon\to 0,\epsilon\to 0,\varepsilon'\to 0,\frac{\varepsilon'}{\varepsilon}\to 0}\left(\liminf_{\delta\to 0} \mathbf{E}\left\|\mathbf{x}^{x_0,\varepsilon}_{t,\epsilon,\varepsilon'}(\omega,\delta)-\boldsymbol{\lambda}\right\|^2\right) \leq \tilde{C}'\|\mathbf{U}(t,\boldsymbol{\lambda})\|^2. \tag{94}$$

Or in the next equivalent form: for a sufficiently small $\epsilon \approx 0$ and for a sufficiently small $\varepsilon \approx 0$ and $\varepsilon' \approx 0$ such that $\varepsilon'/\varepsilon \approx 0$, the inequality

$$\left(\liminf_{\delta\to 0} \mathbf{E}\left\|\mathbf{x}^{x_0,\varepsilon}_{t,\epsilon,\varepsilon'}(\omega,\delta)-\boldsymbol{\lambda}\right\|^2\right)_{\varepsilon'} \leq ((C_{\varepsilon'})_{\varepsilon'})\|\mathbf{U}(t,\boldsymbol{\lambda})\|^2 \tag{95}$$

is satisfied. Here the vector-function $\mathbf{U}(t,\boldsymbol{\lambda}) = \{U_1(t,\boldsymbol{\lambda}),\ldots,U_2(t,\boldsymbol{\lambda})\}$ is the solutionof the differential master equation:

$$\dot{\mathbf{U}}(t,\boldsymbol{\lambda}) = \mathbf{J}[\mathbf{A}_0(\boldsymbol{\lambda},t)]\mathbf{U}(t,\boldsymbol{\lambda}) + \mathbf{A}_0(\boldsymbol{\lambda},t), \mathbf{U}(t,\boldsymbol{\lambda}) = x_0 - \boldsymbol{\lambda}, \tag{96}$$

where $\mathbf{J}[\mathbf{A}_0(\boldsymbol{\lambda},t)]$ is a Jacobian i.e., $\mathbf{J}$ is $n \times n$-matrix

$$\mathbf{J}[\mathbf{A}_0(\boldsymbol{\lambda},t)] = \left.\frac{\partial \mathbf{A}_0, i(\mathbf{x},t)}{\partial x_j}\right|_{\mathbf{x}=\boldsymbol{\lambda}}. \tag{97}$$

**Proof**.We let now

$$\mathbf{x}^{x_0,\varepsilon}_{t,\epsilon,\varepsilon'}(\omega,\delta) - \boldsymbol{\lambda} = \mathbf{y}^{x_0,\varepsilon}_{t,\epsilon,\varepsilon'}(\omega,\delta). \tag{98}$$

Replacement

$$\mathbf{x}^{x_0,\varepsilon}_{t,\epsilon,\varepsilon'}(\omega,\delta) = \mathbf{y}^{x_0,\varepsilon}_{t,\epsilon,\varepsilon'}(\omega,\delta) + \boldsymbol{\lambda} \tag{99}$$

into Eq.(44) gives

$$\begin{gathered}
\left(d\mathbf{y}^{x_0,\varepsilon}_{t,\epsilon,\varepsilon'}(\omega)\right)_{\varepsilon'} = \left(\mathbf{A}_{\varepsilon'}(\mathbf{y}^{x_0,\varepsilon}_{t,\epsilon,\varepsilon'}(\omega) + \boldsymbol{\lambda}, \mathbf{u}^{\delta}_{t,\epsilon,\varepsilon'}(\omega), t)\right)_{\varepsilon'} + \sqrt{\varepsilon}\, d\mathbf{W}(t), \\
\mathbf{u}^{\delta}_{t,\epsilon,\varepsilon'}(\omega) = \left(u^{\delta}_{1,t,\epsilon_i,\varepsilon'}(\omega), \ldots, u^{\delta}_{n,t,\epsilon_i,\varepsilon'}(\omega)\right), \\
du^{\delta}_{i,t,\epsilon,\varepsilon'}(\omega) = \left(\left(u^{\delta}_{i,t,\epsilon,\varepsilon'}(\omega)\right)^{2l}\right)_{\varepsilon'} + \sqrt{\delta}\, W_i(t), \\
\left(\mathbf{y}^{x_0,\varepsilon}_{0,\epsilon,\varepsilon'}(\omega)\right)_{\varepsilon'} = \left(\mathbf{y}^{x_0}_{0,\varepsilon'}(\omega) + \boldsymbol{\lambda}\right)_{\varepsilon'} \in GR, \left(\mathbf{E}\left[\mathbf{y}^{x_0}_{0,\varepsilon'}(\omega)\right]\right)_{\varepsilon'} = \mathbf{y}_0 + \boldsymbol{\lambda}.
\end{gathered} \tag{100}$$

Thus we need to estimate the quantity

$$\liminf_{\varepsilon\to0,\epsilon\to0,\varepsilon'\to0,\frac{\varepsilon'}{\varepsilon}\to0} \left(\liminf_{\delta\to0} \mathbf{E}\left[\left\|\mathbf{y}^{x_0,\varepsilon}_{t,\epsilon,\varepsilon'}(\omega,\delta)\right\|^2\right]\right). \tag{101}$$

Application of the Theorem II.4 (see Appendix II) to Eq.(44) gives the inequality (95) directly.

**Theorem 2**.Main result.(**Strong large deviations principle**) Assume that CISDE (25)-(31) is a strongly $\widetilde{\mathbb{R}}$-dissipative. Then:

(1) for any solution $\left(\mathbf{x}^{x_0,\varepsilon}_{t,\epsilon,\varepsilon'}(\omega)\right)_{\varepsilon'}, \varepsilon' \in (0,1]$, $\epsilon \in (0,1]^n$ of a strongly $\widetilde{\mathbb{R}}$-dissipative CISDE(25)-(31) and any $\mathbb{R}$-valued parameters $\lambda_1,\ldots,\lambda_n$, there exist finite Colombeau constant $\tilde{C}' = [(C_{\varepsilon'})_{\varepsilon'}] > 0$, such that $\forall\boldsymbol{\lambda}[\lambda = (\lambda_1,\ldots,\lambda_n)]$, the inequality

$$\liminf_{\varepsilon\to0,\epsilon\to0,\varepsilon'\to0,\frac{\varepsilon'}{\varepsilon}\to0} \mathbf{E}\left\|\mathbf{x}^{x_0,\varepsilon}_{t,\epsilon,\varepsilon'}(\omega) - \boldsymbol{\lambda}\right\|^2 \leq \tilde{C}' \|\mathbf{U}(t,\boldsymbol{\lambda})\|^2. \tag{102}$$

is satisfied. Or in the next equivalent form: for a sufficiently small $\epsilon \approx 0$ and for a sufficiently small $\varepsilon \approx 0$ and $\varepsilon' \approx 0$ such that $\varepsilon'/\varepsilon \approx 0$, the inequality

$$\left(\mathbf{E}\left\|\mathbf{x}^{x_0,\varepsilon}_{t,\epsilon,\varepsilon'}(\omega) - \boldsymbol{\lambda}\right\|^2\right)_{\varepsilon'} \leq ((C_{\varepsilon'})_{\varepsilon'}) \|\mathbf{U}(t,\boldsymbol{\lambda})\|^2 \tag{103}$$

is satisfied.

(2) for any solution $\left(\mathbf{x}^{x_0,\varepsilon}_{t,\epsilon,\varepsilon'}(\omega)\right)_{\varepsilon'}, \varepsilon' \in (0,1],\ \epsilon \in (0,1]^n$ of a strongly $\widetilde{\mathbb{R}}$-dissipative CISDE(25)-(31) and any $\mathbb{R}$-valued parameters $\lambda_1,\ldots,\lambda_n,$ there exist finite Colombeau constant $\tilde{C}' = [(C_{\varepsilon'})_{\varepsilon'}] > 0,$ such that $\forall\boldsymbol{\lambda}[\lambda = (\lambda_1,\ldots,\lambda_n)],$ the inequality

$$\liminf_{\varepsilon\to 0} \mathbf{E}\left\|\mathbf{x}^{x_0,\varepsilon}_{t,\epsilon=0,\varepsilon'=0}(\omega) - \boldsymbol{\lambda}\right\|^2 \leq \tilde{C}'\|\mathbf{U}(t,\boldsymbol{\lambda})\|^2. \tag{104}$$

is satisfied. Here the vector-function $\mathbf{U}(t,\boldsymbol{\lambda}) = \{U_1(t,\boldsymbol{\lambda}),\ldots,U_2(t,\boldsymbol{\lambda})\}$ is the solutionof the differential master equation:

$$\dot{\mathbf{U}}(t,\boldsymbol{\lambda}) = \mathbf{J}[\mathbf{A}_0(\boldsymbol{\lambda},t)]\mathbf{U}(t,\boldsymbol{\lambda}) + \mathbf{A}_0(\boldsymbol{\lambda},t), \mathbf{U}(t,\boldsymbol{\lambda}) = x_0 - \boldsymbol{\lambda}, \tag{105}$$

where $\mathbf{J}[\mathbf{A}_0(\boldsymbol{\lambda},t)]$ is a Jacobian i.e., $\mathbf{J}$ is $n \times n$-matrix

$$\mathbf{J}[\mathbf{A}_0(\boldsymbol{\lambda},t)] = \left.\frac{\partial \mathbf{A}_0, i(\mathbf{x},t)}{\partial x_j}\right|_{\mathbf{x}=\boldsymbol{\lambda}}. \tag{106}$$

**Proof 1**.From the equality

$$\begin{gathered}\left(\mathbf{E}\left\|\mathbf{x}^{x_0,\varepsilon}_{t,\epsilon,\varepsilon'}(\omega) - \boldsymbol{\lambda}\right\|^2\right)_{\varepsilon'} = \left(\mathbf{E}\left\|\mathbf{x}^{x_0,\varepsilon}_{t,\epsilon,\varepsilon'}(\omega) - \mathbf{x}^{x_0,\varepsilon}_{t,\epsilon,\varepsilon'}(\omega,\delta)\right\|^2\right)_{\varepsilon'} + \\ \left(\mathbf{E}\left\|\mathbf{x}^{x_0,\varepsilon}_{t,\epsilon,\varepsilon'}(\omega,\delta) - \boldsymbol{\lambda}\right\|^2\right)_{\varepsilon'},\end{gathered} \tag{107}$$

by using triangle inequality, one obtain

$$\begin{gathered}\left(\sqrt{\mathbf{E}\left\|\mathbf{x}^{x_0,\varepsilon}_{t,\epsilon,\varepsilon'}(\omega) - \boldsymbol{\lambda}\right\|^2}\right)_{\varepsilon'} \leq \left(\sqrt{\mathbf{E}\left\|\mathbf{x}^{x_0,\varepsilon}_{t,\epsilon,\varepsilon'}(\omega) - \mathbf{x}^{x_0,\varepsilon}_{t,\epsilon,\varepsilon'}(\omega,\delta)\right\|^2}\right)_{\varepsilon'} + \\ \left(\sqrt{\mathbf{E}\left\|\mathbf{x}^{x_0,\varepsilon}_{t,\epsilon,\varepsilon'}(\omega,\delta) - \boldsymbol{\lambda}\right\|^2}\right)_{\varepsilon'}.\end{gathered} \tag{108}$$

Therefore statement (1) immediately follows from Proposition 5, Proposition 2 and Theorem1.

**2**.From the equality

$$\left(\mathbf{E}\left\|\mathbf{x}_{t,\epsilon=0,\varepsilon'=0}^{x_0,\varepsilon}(\omega)-\boldsymbol{\lambda}\right\|^2\right)_{\varepsilon'}=\left(\mathbf{E}\left\|\mathbf{x}_{t,\epsilon=0,\varepsilon'=0}^{x_0,\varepsilon}(\omega)-\mathbf{x}_{t,\epsilon,\varepsilon'}^{x_0,\varepsilon}(\omega)\right\|^2\right)_{\varepsilon'}+ \left(\mathbf{E}\left\|\mathbf{x}_{t,\epsilon,\varepsilon'}^{x_0,\varepsilon}(\omega,\delta)-\boldsymbol{\lambda}\right\|^2\right)_{\varepsilon'}, \tag{109}$$

using triangle inequality, one obtain

$$\left(\sqrt{\mathbf{E}\left\|\mathbf{x}_{t,\epsilon=0,\varepsilon'=0}^{x_0,\varepsilon}(\omega)-\boldsymbol{\lambda}\right\|^2}\right)_{\varepsilon'}=\left(\sqrt{\mathbf{E}\left\|\mathbf{x}_{t,\epsilon=0,\varepsilon'=0}^{x_0,\varepsilon}(\omega)-\mathbf{x}_{t,\epsilon,\varepsilon'}^{x_0,\varepsilon}(\omega)\right\|^2}\right)_{\varepsilon'}+ \left(\sqrt{\mathbf{E}\left\|\mathbf{x}_{t,\epsilon,\varepsilon'}^{x_0,\varepsilon}(\omega)-\boldsymbol{\lambda}\right\|^2}\right)_{\varepsilon'}. \tag{110}$$

Therefore statement (2) immediately follows from Proposition 5, Proposition 2 and statement (1).

# Appendix II.

**Definition 1**. [25] Let $\Im=(\Omega,\Sigma,\mathbf{P})$ be a probability space. Let $\mathcal{E}R$ be the space of nets $(X_{\varepsilon'}(\omega))_{\varepsilon'}$ of measurable functions on Ω. Let $\mathcal{E}R_M$ be the space of nets $(X_{\varepsilon'}(\omega))_{\varepsilon'}\in\mathcal{E}R$, $\varepsilon\in(0,1]$, with the property that for almost all $\omega\in\Omega$ there exist constants $r,C>0$ and $\varepsilon_0\in(0,1]$ such that $X_{\varepsilon'}(\omega)\leq C(\varepsilon')^{-r},\varepsilon'\leq\varepsilon_0$.

**Definition 2**.Let $\xi$ be a distribution $\xi\in D'$. Distribution $\xi$ is the generalized probability density of net $(X_{\varepsilon'}(\omega))_{\varepsilon'}\in\mathcal{E}R_M,\varepsilon'\in(0,1]$ iff $\forall f\in D:(E(X_{\varepsilon'}(\omega)))_{\varepsilon'}=\xi(f)$.

Let us consider now Colombeau-Ito’s SDE:

$$\begin{gathered}\left(d\mathbf{x}_{t'',\varepsilon'}^{\mathbf{x}_0,\varepsilon}(\omega)\right)_{\varepsilon'}=\left(\mathbf{b}_{\varepsilon'}\left(\mathbf{x}_{t'',\varepsilon'}^{\mathbf{x}_0,\varepsilon}(\omega),t''\right)\right)_{\varepsilon'}+\sqrt{\varepsilon}\,\mathbf{W}(t'',\omega),t''\in[t',T],\varepsilon,\varepsilon'\in(0,1]\\ \left(\mathbf{E}\left[f\left(\mathbf{x}_{t',\varepsilon'}^{\mathbf{x}_0,0}(\omega)\right)\right]\right)_{\varepsilon'}=f(\mathbf{x}_0),\mathbf{x}_0=\mathbf{q}'\in\mathbb{R}^n,f\in C_0^\infty(\mathbb{R}^n),\mathbf{x}_0\in supp(f).\end{gathered} \tag{1}$$

Here $\mathbf{W}(t)$ is $n$-dimensional Brownian motion, and $\forall t\in[0,T]:(b_{\varepsilon'}(x,t))_{\varepsilon'}\in G^n(\mathbb{R}^n)$, $b_0(\bullet,t)\equiv b_{\varepsilon'=0}(\bullet,t):\mathbb{R}^n\to\mathbb{R}^n$ is a polynomial on variable $x=(x_1,\ldots,x_n)$ i.e., $b_{0,i}(x,t)=\sum_{|\alpha|\leq r}b_{0,i}^{\alpha}(t)x^{\alpha},\alpha=(i_1,\ldots,i_n),|\alpha|=\sum_{j=1}^{m}i_j,0\leq i_j\leq p$.

**Assumption1**.We assume now that there exist Colombeau constants $(C_{\varepsilon'})_{\varepsilon'}$ and $(D_{\varepsilon'})_{\varepsilon'}$ such that

$$\begin{gathered}1.\ (\|b_{\varepsilon'}(x,t)\|)_{\varepsilon'}\leq((C_{\varepsilon'})_{\varepsilon'})(1+\|x\|),\\ 2.\ (\|b_{\varepsilon'}(x,t)-b_{\varepsilon'}(y,t)\|)_{\varepsilon'}\leq((D_{\varepsilon'})_{\varepsilon'})\|x-y\|\end{gathered} \tag{2}$$

for all $t \in [0,\infty)$ and all $x$ and $y \in \mathbb{R}^n$.

It well-known that ander Assumption1, Colombeau-Ito's SDE(1) is equivalent to the Colombeau-Fokker-Planck equation

$$\left(\frac{\partial p^{\varepsilon}_{\varepsilon'}(\mathbf{q}',t'|\mathbf{q}'',t'')}{\partial t''}\right)_{\varepsilon'} = \varepsilon\sum_{i=1}^{n}\left(\frac{\partial^2 p^{\varepsilon}_{\varepsilon'}(\mathbf{q}',t'|\mathbf{q}'',t'')}{\partial q''_i \partial q''_i}\right)_{\varepsilon'} - \\ -\sum_{i=1}^{n}\left(\frac{\partial}{\partial q''_i}(b_{i,}(\mathbf{q}'',t'')p^{\varepsilon}_{\varepsilon'}(\mathbf{q}',t'|\mathbf{q}'',t''))\right)_{\varepsilon'}. \tag{3}$$

Withinitial condition of the form

$$(p^{\varepsilon}_{\varepsilon'}(\mathbf{q}',t'|\mathbf{q}'',t''))_{\varepsilon'} = (p_{\varepsilon'}(\mathbf{q}''-\mathbf{q}'))_{\varepsilon'} = \delta(\mathbf{q}''-\mathbf{q}') \tag{4}$$

Colombeau PDE (3)-(4) can be solved formally in terms of Feynman path integral of the form [22-23]:

$$p^{\varepsilon}_{\varepsilon'}(q',t'|q'',t'') = \left(\lim_{\Delta t\to 0} I_{N,\varepsilon'}(q',t'|q'',t'')\right)_{\varepsilon'}. \tag{5}$$

Here

$$(I_{N,\varepsilon'}(q',t'|q'',t''))_{\varepsilon'} = \widehat{\mathbf{N}}_N \left( \int_{-\infty}^{\infty} d\mathbf{q}_0 \int_{-\infty}^{\infty} d\mathbf{q}_1 \ldots \int_{-\infty}^{\infty} d\mathbf{q}_m \ldots \int_{-\infty}^{\infty} d\mathbf{q}_{N-1} p_{\varepsilon'}(q_0 - q') \times \right.$$

$$\left. \left[ \times \exp\left[ \tfrac{1}{2\varepsilon} S_{\varepsilon'}(d\mathbf{q}_0, d\mathbf{q}_1, \ldots, d\mathbf{q}_m, \ldots, \mathbf{q}_{N-1}, \mathbf{q}_N, \varepsilon) \right] \right] \right)_{\varepsilon'},$$

$$\mathbf{q}_N = q'', d\mathbf{q}_m = \prod dq_{j,m}, m = 0,1,\ldots,N, \Delta t = (t'' - t')/N, t_m = m\Delta t,$$

$$S_{\varepsilon'}(d\mathbf{q}_0, d\mathbf{q}_1, \ldots, d\mathbf{q}_m, \ldots, \mathbf{q}_{N-1}, \mathbf{q}_N, \varepsilon) = \tag{6}$$

$$\Delta t \sum_{m=1}^{N} \mathcal{L}_{\varepsilon'}\left( \frac{\mathbf{q}_m - \mathbf{q}_{m-1}}{\Delta t}, \frac{\mathbf{q}_m - \mathbf{q}_{m-1}}{2}, t_m \right),$$

$$\mathcal{L}_{\varepsilon'}\left( \frac{\mathbf{q}_m - \mathbf{q}_{m-1}}{\Delta t}, \frac{\mathbf{q}_m - \mathbf{q}_{m-1}}{2}, t_m \right) =$$

$$= \left\| \frac{\mathbf{q}_m - \mathbf{q}_{m-1}}{\Delta t} + \mathbf{b}_{\varepsilon'}\left( \frac{\mathbf{q}_m - \mathbf{q}_{m-1}}{2}, t_m \right) \right\|^2 - \varepsilon \sum_{i=1}^{n} b_{i,i,\varepsilon'}\left( \frac{\mathbf{q}_m - \mathbf{q}_{m-1}}{2}, t_m \right),$$

$$b_{i,i,\varepsilon'}(\mathbf{q},t) = \frac{\partial b_{i,i\varepsilon'}(\mathbf{q},t)}{\partial q_i}, \varepsilon, \varepsilon' \in (0,1].$$

Here $p_{\varepsilon'}^{\varepsilon}(q',t'|q'',t'')$ is the generalized probability density that the system (1) will end up at $q''$ at time $t''$ if it started at $q'$ at time $t'$ and $\widehat{\mathbf{N}}_N$ is the usual overall normalization of the path integral

$$\widehat{\mathbf{N}}_N = (2\pi\varepsilon\Delta t)^{-\frac{nN}{2}}. \tag{7}$$

**Remark1**.Note that Colombeau-Fokker-Planck equation (4) is just a Euclidean Colombeau-Schrodinger equation, and is well-known that one can transform Colombeau-Schrodinger equation

$$\left(\frac{\partial u_{\varepsilon'}(x,t)}{\partial t}\right)_{\varepsilon'} = \frac{i\varepsilon}{2}(\Delta u_{\varepsilon'}(x,t))_{\varepsilon'} - i(V_{\varepsilon'}(x)u_{\varepsilon'}(x,t))_{\varepsilon'},$$
$$(u_{\varepsilon'}(x,0))_{\varepsilon'} = (\varphi_{\varepsilon'}(x))_{\varepsilon'}, \varepsilon' \in (0,1] \tag{8}$$

into mathematically rigorous path integral by standard method using Trotter's Product Formula [33].Here $\Delta$ is the Laplace operator $\partial^2/(\partial x_1^2) + \cdots + \partial^2/(\partial x_n^2), V_{\varepsilon'}(x)$ is a real measurable function on $\mathbb{R}^n$, $u_{\varepsilon'}(x,t)$ and each $u_{\varepsilon'}(x,t)$ are elements of $\mathcal{L}_2(\mathbb{R}^n)$ and $\varepsilon$ is a constant. Let $\mathcal{F}$ denote the Fourier transformation, $\mathcal{F}^{-1}$ its inverse.

We define now as usual [33]

$$(\Delta\varphi_{\varepsilon'}(x))_{\varepsilon'} = \left(\mathcal{F}^{-1}\left[-\|\lambda\|^2\mathcal{F}[\varphi_{\varepsilon'}]\right]\right)_{\varepsilon'}, \varepsilon' \in (0,1] \tag{9}$$

on the domain $D(\Delta)$ of all square-integrable $(\varphi_{\varepsilon'}(x))_{\varepsilon'}$ such that $\left(\mathcal{F}^{-1}\left[-\|\lambda\|^2\mathcal{F}[\varphi_{\varepsilon'}]\right]\right)_{\varepsilon'}$ is also square-integrable. (Here $\lambda$ denotes the variable in momentum space and $\|\lambda\|^2 = \lambda_1^2 + \cdots + \lambda_n^2$. Then $\Delta$ is self-adjoint, and

$$(u_{\varepsilon'}(x,t))_{\varepsilon'} = (K_\varepsilon^t \varphi_{\varepsilon'}(x))_{\varepsilon'},$$
$$K_\varepsilon^t = \exp\left[\frac{it\varepsilon}{2}\Delta\right] \tag{10}$$

is the solution of the Eq.(8) for $(V_{\varepsilon'}(x))_{\varepsilon'} = 0$. The operator $V_{\varepsilon'}$ of multiplication by the function $V_{\varepsilon'}(x,t)$, on the domain $D(V_{\varepsilon'}(x))$ of all $\varphi_{\varepsilon'}(x)$ in $\mathcal{L}_2(\mathbb{R}^n)$ such that $V_{\varepsilon'}(x)\varphi_{\varepsilon'}(x)$ is also in $\mathcal{L}_2(\mathbb{R}^n)$, is self-adjoint, and

$$(u_{\varepsilon'}(x,t))_{\varepsilon'} = \left(M_{V_{\varepsilon'}}^t \varphi_{\varepsilon'}(x)\right)_{\varepsilon'},$$
$$M_{V_{\varepsilon'}}^t = \exp[-itV_{\varepsilon'}(x)] \tag{11}$$

is the solution of the Eq.(8) with $\varepsilon = 0$. Kato has found conditions under which the operator $\Re_{\varepsilon,\varepsilon'}$ is self-adjoint [33]

$$(\Re_{\varepsilon,\varepsilon'}u_{\varepsilon'})_{\varepsilon'} = \frac{i\varepsilon}{2}(\Delta u_{\varepsilon'})_{\varepsilon'} - i(V_{\varepsilon'}u_{\varepsilon'})_{\varepsilon'}, \tag{12}$$

And under these conditions if we let

$$U^t_{V_{\varepsilon'}} = \exp[t\Re_{\varepsilon,\varepsilon'}] \tag{13}$$

Then a theorem of Trotter [33] asserts that for all $\varphi_{\varepsilon'}(x)$ in $\mathcal{L}_2(\mathbb{R}^n)$

$$\left(U^t_{V_{\varepsilon'}}\varphi_{\varepsilon'}(x)\right)_{\varepsilon'} = \left(\lim_{N\to\infty}\left[K_{\varepsilon}^{\frac{t}{N}} M_{V_{\varepsilon'}}^{\frac{t}{N}}\varphi_{\varepsilon'}(x)\right]\right)_{\varepsilon'} \tag{14}$$

This is discussed in detail in [33] (See [33] Appendix B). Using now Eq.(9)-Eq.(14) by simple calculation one obtain [33]

$$\left(K_{\varepsilon}^{\frac{t}{N}} M_{V_{\varepsilon'}}^{\frac{t}{N}}\varphi_{\varepsilon'}(x)\right)_{\varepsilon'} =$$

$$= \left(\frac{2\pi it\varepsilon}{N}\right)^{-\frac{t}{2}nN}\left(\int\ldots\int d^n x_0\ldots d^n x_{N-1}\exp[iS_{\varepsilon'}(x_0\ldots x_{N-1},x_N,t)]\right)_{\varepsilon'}, \tag{15}$$

$$S_{\varepsilon'}(x_0\ldots x_{N-1},x_N,t) = \sum_{i=1}^{N}\left[\frac{1}{2\varepsilon}\frac{\|x_i - x_{i-1}\|^2}{(t/N)^2} - V_{\varepsilon'}(x_i)\right]\frac{t}{N},$$

where we have set $x_N = x$.

**Theorem 1**.(**Suzuki**-**Trotter Formula**) [35]-[36]. Let $\{A_j\}_{j=1}^{j=p}$ be an family of any bounded operator in an Banach algebra **ß** with a norm $\|\circ\|_{\text{ß}}$. Let $\Phi_n(\{A_j\}_{j=1}^{j=p})$ be a function

$$\Phi_n(\{A_j\}_{j=1}^{j=p}) = (\exp(A_{1/n})\ldots\exp(A_{p/n}))^n \tag{16}$$

For any bounded operators $\{A_j\}_{j=1}^{j=p}$ in a Banach algebra **ß**:

$$lim_{n\to\infty}\|\Phi_n(\{A_j\}_{j=1}^{j=p}) - \exp(\textstyle\sum_{j=1}^{p}A_j)\|_{\text{ß}} = 0. \tag{17}$$

Let $(P^t_{\varepsilon'})_{\varepsilon'}$ be a formal Colombeau pseudo-differential operator (see[42]), given by formula

$$(P_{\varepsilon'}^t)_{\varepsilon'} = \left(\exp\left(\sum_{i=1}^{n} b_{i,\varepsilon'}(x,t)\frac{\partial}{\partial x_i}\right)\right)_{\varepsilon'} \quad (18)$$

**Remark 2**. Note that formal Colombeau pseudo-differential operator [29], given by formula (18), evidently does not define any contraction Colombeau semi-group on$\mathcal{L}_2(\mathbb{R}^n)$.Even in the case when a functions $(b_{i,\varepsilon'}(x,t))_{\varepsilon'}$ , $i = 1,\ldots,n$ is the Colombeau constants $\left[(b_{i,\varepsilon'})_{\varepsilon'}\right] \in \widetilde{\mathbb{R}},\ i = 1,\ldots,n$ formal Colombeau pseudo-differential operator, given by formula

$$(P_{\varepsilon'}^t)_{\varepsilon'} = \left(\exp\left(\sum_{i=1}^{n} b_{i,\varepsilon'}\frac{\partial}{\partial x_i}\right)\right)_{\varepsilon'}; \quad (19)$$

Let $H^{\infty}(\mathbf{S_R})$ be a test space $H^{\infty}(S_{\mathbf{R}}), \mathbf{R} = (R_1,\ldots,R_n)$, with a members $\varphi(x), x \in \mathbb{R}^n$ such that Fourier transform $\mathcal{F}[\varphi](\xi)$ is supported inside region $\mathbf{S_R} = \{\xi \| \xi_i| < R_i : i = 1,\ldots,n\}$ [40] and let $H^{-\infty}(\mathbf{S_R})$ be a corresponding dual space$H^{-\infty}(S_{\mathbf{R}})$. Pseudo-differential calculus on a test space $H^{\infty}(\mathbf{S_R})$ and on dual space $H^{-\infty}(S_{\mathbf{R}})$ is discussed in detail in [40].

**Theorem 2**. [40]. Let

$$D = (D_1,\ldots,D_n), D_i = \partial/(\partial x_i) \triangleq \partial x_i \quad (20)$$

and

$$A(D) = \textstyle\sum_{|\alpha|=0}^{M} a_\alpha D^\alpha, a_\alpha \in \mathbb{C}, M \leq \infty. \quad (21)$$

Then (1) $A(D)\varphi(x) \in H^{\infty}(S_{\mathbf{R}})$ if $\varphi(x) \in H^{\infty}(S_{\mathbf{R}})$,

(2) $A(D)\Upsilon(x) \in H^{-\infty}(S_{\mathbf{R}})$ if $\Upsilon(x) \in H^{-\infty}(S_{\mathbf{R}})$,

(3) Operator $A(D)$ is bounded on $H^{\infty}(S_{\mathbf{R}})$,

(4) Operator $A(D)$ is bounded on $H^{-\infty}(S_{\mathbf{R}})$.

**Remark 3**. (1) From Theorem 1and Theorem 2 one obtain, that:

(1) Operator $\sum_{i=1}^{n} b_{i,\varepsilon'}\partial x_i, \varepsilon \in (0,1]$ is bounded on $H^{\infty}(S_{\mathbf{R}})$.

(2) The Colombeau generalized function $(u_{\varepsilon'}(q',t'|q'',t''))_{\varepsilon'}$ given by formula

$$(u_{\varepsilon'}(\mathbf{q}',t'|\mathbf{q}'',t''))_{\varepsilon'} = (p^{\varepsilon}_{\varepsilon'}(\mathbf{q}',t'|\mathbf{q}'',t''))_{\varepsilon'} = \left(P^{t''-t'}_{\varepsilon'}\varphi_{\varepsilon'}(\mathbf{q}''-\mathbf{q}')\right)_{\varepsilon'} =$$

$$(\exp(-[t''-t']\textstyle\sum_{i=1}^{n} b_{i,\varepsilon'}\partial q''_i)\varphi_{\varepsilon'}(\mathbf{q}''-\mathbf{q}'))_{\varepsilon'} =$$

$$\left(\mathcal{F}^{-1}\left[\exp\left(-[t''-t']\textstyle\sum_{i=1}^{n} b_{i,\varepsilon'}(i\xi_i)\right)\mathcal{F}[\varphi_{\varepsilon'}(\mathbf{q}''-\mathbf{q}')](\boldsymbol{\xi})\right]\right)_{\varepsilon'} = \qquad (21)$$

$$\frac{1}{(2\pi)^n}\left(\int_{-\infty}^{\infty} d^n\boldsymbol{\xi}\exp\left(i(\mathbf{q}'',\boldsymbol{\xi})-(t''-t')\textstyle\sum_{i=1}^{n} b_{i,\varepsilon'}(i\xi_i)\right)\mathcal{F}[\varphi_{\varepsilon'}(\mathbf{q}''-\mathbf{q}')](\boldsymbol{\xi})\right)_{\varepsilon'}$$

is the solution of the Colombeau-Fokker-Planck equation (3) with initial condition $\varphi_{\varepsilon'}(\mathbf{q}''-\mathbf{q}') \in H^{\infty}(S_{\mathbf{R}})$ for the case: $\varepsilon = 0$ and $(b_{i,\varepsilon'}(\mathbf{q},t))_{\varepsilon'} = (b_{i,\varepsilon'})_{\varepsilon'} \in \widetilde{\mathbb{R}}, i = 1,\dots,n$ is the Colombeau constants, and

(3) $\forall\mathbf{q}'\forall t'\forall t''\left[u_{\varepsilon'}(\mathbf{q}',t'|\mathbf{q}'',t'') \in H^{\infty}(S_{\mathbf{R}})\right]$.

(**II**) From Theorem 1and Theorem 2 one obtain, that:

(1) Operator $\Delta$ is bounded on $H^{\infty}(S_{\mathbf{R}})$.

(2) The Colombeau generalized function $(u_{\varepsilon'}(\mathbf{q}',t'|\mathbf{q}'',t''))_{\varepsilon'}$ given by formula

$$(u_{\varepsilon'}(\mathbf{q}',t'|\mathbf{q}'',t''))_{\varepsilon'} = K_{\varepsilon}^{(t''-t')}\varphi_{\varepsilon'}(\mathbf{q}''-\mathbf{q}'),$$

$$K_{\varepsilon}^{(t''-t')} = \exp\left[\frac{\varepsilon}{2}\left(t''-t'\right)\Delta\right], \qquad (22)$$

$$(u_{\varepsilon'}(\mathbf{q}',t'|\mathbf{q}'',t''))_{\varepsilon'} = \left(\mathcal{F}^{-1}\left[\exp\left(-\frac{\varepsilon}{2}(t''-t')\|\boldsymbol{\xi}\|^2\right)\mathcal{F}[\varphi_{\varepsilon'}(\mathbf{q}''-\mathbf{q}')](\boldsymbol{\xi})\right]\right)_{\varepsilon'}$$

is the solution of the Colombeau-Fokker-Planck equation (3) with initial condition $\varphi_{\varepsilon'}(\mathbf{q}''-\mathbf{q}') \in H^{\infty}(S_{\mathbf{R}})$ for the case: $(b_{i,\varepsilon'})_{\varepsilon'} = 0$

(3) $\forall\mathbf{q}'\forall t'\forall t''\left[u_{\varepsilon'}(\mathbf{q}',t'|\mathbf{q}'',t'') \in H^{\infty}(S_{\mathbf{R}})\right]$.

(**III**) From Theorem 1-2 one obtain, that: operator

$$\Re_{\varepsilon,\varepsilon'} = \textstyle\sum_{i=1}^{n} b_{i,\varepsilon'}\frac{\partial}{\partial q''_i} + \varepsilon\Delta \qquad (23)$$

is bounded on $H^{\infty}(S_{\mathbf{R}})$. If we let now

$$U_{\varepsilon,\varepsilon'}^{(t''-t')} = \exp\left[\left(t'' - t'\right)\Re_{\varepsilon,\varepsilon'}\right] \tag{24}$$

then Theorem 1 and Theorem 2 asserts that for all $\varphi_{\varepsilon'}(\mathbf{q}'' - \mathbf{q}') \in H^{\infty}(S_{\mathbf{R}})$

$$\left(U_{\varepsilon,\varepsilon'}^{(t''-t')}\varphi_{\varepsilon'}(\mathbf{q}'' - \mathbf{q}')\right)_{\varepsilon'} = \left(\lim_{N\to\infty}\left[K_{\varepsilon}^{\frac{t''-t'}{N}} P_{\varepsilon}^{\frac{t''-t'}{N}}\right]^N \varphi_{\varepsilon'}(\mathbf{q}'' - \mathbf{q}')\right)_{\varepsilon'}, \tag{25}$$

where the limit is calculated by norm in $H^{\infty}(S_{\mathbf{R}})$. From Eq.(24)-Eq.(25) by simple calculation one obtain

$$\left[K_{\varepsilon}^{\frac{t''-t'}{N}} P_{\varepsilon}^{\frac{t''-t'}{N}}\right]^N \varphi_{\varepsilon'}(\mathbf{q}'' - \mathbf{q}') = \frac{1}{(2\pi)^{\frac{nN}{2}}} \int_{-\infty}^{\infty} d\mathbf{q} \int_{-\infty}^{\infty} d\boldsymbol{\xi} \times$$

$$\exp\left\{i\sum_{m=0}^{N}(\mathbf{q}_{m+1} - \mathbf{q}_m)\boldsymbol{\xi}_m + \frac{t''-t'}{N}\sum_{i=1}^{n} b_{i,\varepsilon'}\xi_{i,m} - \frac{t''-t'}{N}\frac{\varepsilon}{2}\sum_{m=0}^{N}\|\boldsymbol{\xi}_m\|^2\right\}\varphi_{\varepsilon'}(\mathbf{q}_0 - \mathbf{q}'). \tag{26}$$

Here $d\mathbf{q} = d\mathbf{q}_0 \ldots d\mathbf{q}_m \ldots d\mathbf{q}_{N-1}, d\boldsymbol{\xi} = d\boldsymbol{\xi}_0 \ldots d\boldsymbol{\xi}_m \ldots d\boldsymbol{\xi}_N, \mathbf{q}_N = \mathbf{q}'', d\mathbf{q}_m = \prod_{j=1}^{n} dq_{j,m}$, $d\boldsymbol{\xi}_m = \prod_{j=1}^{n} d\xi_{j,m}, m = 0,\ldots,N.$ Integratingon on variable $\boldsymbol{\xi}$ gives

$$\left[K_{\varepsilon}^{\frac{t''-t'}{N}} P_{\varepsilon}^{\frac{t''-t'}{N}}\right]^N \varphi_{\varepsilon'}(\mathbf{q}'' - \mathbf{q}') = I_{N,\varepsilon'}(\mathbf{q}', t'|\mathbf{q}'', t'') =$$

$$= \frac{1}{(2\pi)^{\frac{nN}{2}}} \int_{-\infty}^{\infty} d\mathbf{q} \exp\left\{\frac{t''-t'}{N}\frac{\varepsilon}{2}\sum_{i=1}^{n}\sum_{m=0}^{N}\left[\frac{(q_{i,m+1} - q_{i,m})}{\frac{t''-t'}{N}} - b_{i,\varepsilon'}\right]^2\right\}\varphi_{\varepsilon'}(\mathbf{q}_0 - \mathbf{q}'). \tag{27}$$

Finally we obtain

$$\left(U_{\varepsilon,\varepsilon'}^{(t''-t')}\varphi_{\varepsilon'}(\mathbf{q}'' - \mathbf{q}')\right)_{\varepsilon'} = \left(\lim_{N\to\infty} I_{N,\varepsilon'}(\mathbf{q}', t'|\mathbf{q}'', t'')\right)_{\varepsilon'}. \tag{28}$$

**Remark 4**. Note that $\delta_{\varepsilon'}(x) \in H^{\infty}(S_{\mathbf{R}}), R = 1/\varepsilon', x \in \mathbb{R}, \varepsilon' \in (0,1]$, where

$$\delta_{\varepsilon'}(x) = \frac{1}{\pi\varepsilon' x}\sin\left(\frac{x}{\varepsilon'}\right). \tag{29}$$

By simple calculation one obtain [40]

$$\mathcal{F}[\delta_{\varepsilon'}(x)] = \begin{cases} \frac{1}{2\pi}, x \in (-r,r) \\ 0, x \notin (-r,r) \end{cases} \tag{30}$$

$$r = 1/\varepsilon'.$$

**Assumption 2**.We assume now that

$$(p_{\varepsilon'}(\mathbf{q}'' - \mathbf{q}'))_{\varepsilon'} = \prod_{i=1}^{n}(\delta_{\varepsilon'}(q_i'' - q_i'))_{\varepsilon'}. \tag{31}$$

**Remark 4**. Note that $(p_{\varepsilon'}(\mathbf{q}'' - \mathbf{q}'))_{\varepsilon'} = \delta(\mathbf{q}'' - \mathbf{q}')$.

**Definition 3**.We let now $n = 1$. A tagged partition of the real line $\mathbb{R} = (-\infty, +\infty)$ is a finite sequence $-\infty = x_0 < x_1 < x_2 < \cdots < x_{p-1} < x_p = +\infty$. This partitions the open interval $(-\infty, +\infty)$ into $n$ sub-intervals $J_r = [x_{r-1}, x_r], r = 2, \ldots, p, J_1 = (-\infty, x_1], J_p = [x_p, +\infty)$ indexed by $r = 1, \ldots, p$. Let $b_{\varepsilon'}\ (q,t,r)$ be a quantity

$$b_{\varepsilon'}(t,r) = \sup_{q \in J_r} b_{\varepsilon'}(q,t) \tag{32}$$

and let $\partial b_{\varepsilon'}\ (t,r)$ be a quantity

$$\partial b_{\varepsilon'}(t,r) = \sup_{q \in J_r} \frac{\partial b_{\varepsilon'}(q,t)}{\partial q} \tag{33}$$

Let $\breve{b}_{\varepsilon'}(q\ ,t)$ be a function

$$\breve{b}_{\varepsilon'}(q,t) = \sum_{r=1}^{p} \mathbf{1}_{J_r}\big(q\big) b_{\varepsilon'}(t,r) \tag{34}$$

Let $\partial\breve{b}_{\varepsilon'}(q,t)$ be a function

$$\partial\breve{b}_{\varepsilon'}(q,t) = \sum_{r=1}^{p} \mathbf{1}_{J_r}(q)\partial b_{\varepsilon'}(q,t,r) \tag{35}$$

Here $\mathbf{1}_{J_r}\big(q\big)$ is indicator function of a subset $J_r = [x_{r-1}, x_r]$.

**Definition 4**. Let $H^{\infty}(S_{\mathbf{R}}, \breve{S}_{\mathbf{U}}), \mathbf{R} = (R_1,\ldots,R_n), \mathbf{U} = (U_1,\ldots,U_n)$ be a test space with a members $\varphi(x,p), x \in \mathbb{R}^n_x, p \in \mathbb{R}^n_p$ such that $\forall \breve{x}, \breve{x} \in \breve{S}_{\mathbf{U}} \subseteq S_{\mathbf{U}} = \{x | |x_i| < U_i, i = 1,\ldots,n\}$ function $\varphi(\breve{x},p)$ is supported inside region $S_{\mathbf{R}} = \{p | |p_i| < R_i, i = 1,\ldots,n\}$,i.e $\forall \breve{x}, \breve{x} \in \breve{S}_{\mathbf{U}}$ :

$$\mathcal{F}^{-1}[\varphi(\breve{x},p)](\breve{x},x) \in H^{\infty}(S_{\mathbf{R}}) \tag{36}$$

**Definition 5**.(1) We let

$$\mathcal{F}^{\#}[\psi(x)](p) = \varphi(x,p) \tag{37}$$

iff there exist an function $\varphi(x,p) \in H^{\infty}(S_{\mathbf{R}}, \breve{S}_{\mathbf{U}})$ such that

$$\mathcal{F}^{-1}[\varphi(x,p)](x,x) = \psi(x). \tag{38}$$

(2) We let now $H^{\infty}(S_{\mathbf{R}}, \breve{S}_{\mathbf{U},p})$ if $\breve{S}_{\mathbf{U}} = S_{\mathbf{U}} \backslash \{x_0,\ldots,x_p\}$.

**Remark 5**.(**I**) From Theorem 1-2 one obtain, that:

(1) Operator $\breve{b}_{\varepsilon'}(q,t)\partial q = \sum_{r=1}^{p} \mathbf{1}_{J_r}\big(q\big) b_{\varepsilon'}(t,r)\partial q, \varepsilon' \in (0,1]$ is bounded on $H^{\infty}(S_{\mathbf{R}}, \breve{S}_{\mathbf{U},p})$.

(2) The Colombeau generalized function $(u_{\varepsilon',p}(q',t'|q'',t''))_{\varepsilon'}$ given by formula

$$(u_{\varepsilon',p}(q',t'|q'',t''))_{\varepsilon'} = \left(P^{t''-t'}_{\varepsilon',p}\varphi_{\varepsilon'}(q''-q')\right)_{\varepsilon'} =$$

$$(\exp(-[t''-t']\textstyle\sum_{r=1}^{p}\mathbf{1}_{J_r}(q'')b_{\varepsilon'}(t,r)\partial q'')\varphi_{\varepsilon'}(q''-q'))_{\varepsilon'} =$$

$$\left(\prod_{r=1}^{p}\exp(-[t''-t']\mathbf{1}_{J_r}(q'')b_{\varepsilon'}(t,r)\partial q'')\varphi_{\varepsilon'}(q''-q')\right)_{\varepsilon'} =$$

$$\mathcal{F}^{-1}[\exp(-[t''-t']\mathbf{1}_{J_p}(q'')b_{\varepsilon'}(t,p)(i\xi)) \times$$

$$\times\mathcal{F}^{\#}[\mathcal{F}^{-1}\exp(-[t''-t']\mathbf{1}_{J_{p-1}}(q'')b_{\varepsilon'}(t,p-1)(i\xi))\ldots \tag{39}$$

$$\ldots\mathcal{F}^{-1}[\exp(-[t''-t']b_{\varepsilon'}(t,0)(i\xi))\mathcal{F}^{\#}[\varphi_{\varepsilon'}(q''-q')](\boldsymbol{\xi})] =$$

$$\frac{1}{2\pi}\left(\int_{-\infty}^{\infty}d\xi\exp\left(i\xi\left(\textstyle\sum_{r=1}^{p}\mathbf{1}_{J_r}(q'')q''\right)-(t''-t')\textstyle\sum_{r=1}^{p}\mathbf{1}_{J_r}(q'')b_{\varepsilon'}(t,r)(i\xi)\right)\times\right.$$

$$\left.\times\mathcal{F}^{\#}[\varphi_{\varepsilon'}(q''-q')](\xi)\right)_{\varepsilon'}$$

is the solution (except points $\{x_0,\ldots,x_p\}$) of the Colombeau-Fokker-Planck equation (3) with initial condition $\varphi_{\varepsilon'}(\mathbf{q}''-\mathbf{q}') \in H^{\infty}(S_{\mathbf{R}},\check{S}_{\mathbf{U},p})$ for the case:

$$\left(b_{\varepsilon'}(q,t)\right)_{\varepsilon'} = \left(\check{b}_{\varepsilon'}(q,t)\right)_{\varepsilon'} = \sum_{r=1}^{p}\mathbf{1}_{J_r}\left(q\right)(b_{\varepsilon'}(t,r))_{\varepsilon'} \tag{40}$$

(3) $\forall q'\forall t'\forall t''\left[u_{\varepsilon',p}(q',t'|q'',t'') \in H^{\infty}(S_{\mathbf{R}},\check{S}_{\mathbf{U},p})\right]$.

(**II**) From Theorem 1 and Theorem 2 one obtain, that:

(1) Operator $\Delta$ is bounded on $H^{\infty}(S_{\mathbf{R}},\check{S}_{\mathbf{U},p})$.

(2) The Colombeau generalized function $(u_{\varepsilon',p}(q',t'|q'',t''))_{\varepsilon'}$ given by formula

$$(u_{\varepsilon',p}(q',t'|q'',t''))_{\varepsilon'} = K_{\varepsilon,p}^{(t''-t')}\varphi_{\varepsilon'}(q''-q'),$$

$$K_{\varepsilon,p}^{(t''-t')} = \exp\left[\frac{\varepsilon}{2}\left(t''-t'\right)\Delta\right], \quad (41)$$

$$(u_{\varepsilon',p}(q',t'|\mathbf{q}'',t''))_{\varepsilon'} = \left(\mathcal{F}^{-1}\left[\exp\left(-\frac{\varepsilon}{2}(t''-t')\|\boldsymbol{\xi}\|^2\right)\mathcal{F}^{\#}[\varphi_{\varepsilon'}(q''-q')](\boldsymbol{\xi})\right]\right)_{\varepsilon'}$$

is the solution of the Colombeau-Fokker-Planck equation (3) with initial condition $\varphi_{\varepsilon'}(q''-q') \in H^{\infty}(S_{\mathbf{R}}, \check{S}_{\mathbf{U},p})$ for the case: $(b_{\varepsilon'}(q,t))_{\varepsilon'} = 0$

(3) $\forall q' \forall t' \forall t'' \left[u_{\varepsilon'}(q',t'|q'',t'') \in H^{\infty}(S_{\mathbf{R}}, \check{S}_{\mathbf{U},p})\right]$.

(**III**) From Theorem 1-2 one obtain, that: operator

$$\Re_{\varepsilon,\varepsilon',p} = \sum_{r=1}^{p} \mathbf{1}_{J_r}(q'')b_{\varepsilon'}(t,r)\frac{\partial}{\partial q''} + \varepsilon\Delta = P_{\varepsilon',p}^{(t''-t')} + \varepsilon\Delta \quad (42)$$

is bounded on $H^{\infty}(S_{\mathbf{R}}, \check{S}_{\mathbf{U},p})$. If we let now

$$U_{\varepsilon,\varepsilon',p}^{(t''-t')} = \exp\left[\left(t''-t'\right)\Re_{\varepsilon,\varepsilon',p}\right], \quad (43)$$

then Theorem 1 and Theorem 2 asserts that for all $\varphi_{\varepsilon'}(\mathbf{q}''-\mathbf{q}') \in H^{\infty}(S_{\mathbf{R}}, \check{S}_{\mathbf{U},p})$

$$\left(U_{\varepsilon,\varepsilon',p}^{(t''-t')}\varphi_{\varepsilon'}(q''-q')\right)_{\varepsilon'} = \left(\lim_{N\to\infty}\left[K_{\varepsilon,p}^{\frac{t''-t'}{N}} P_{\varepsilon',p}^{\frac{t''-t'}{N}}\right]^{N}\varphi_{\varepsilon'}(q''-q')\right)_{\varepsilon'}, \quad (44)$$

where the limit is calculated by norm in $H^{\infty}(S_{\mathbf{R}}, \check{S}_{\mathbf{U},p})$. From Eq.(39)-Eq.(44) by simple calculation one obtain

$$\left[K_{\varepsilon,p}^{\frac{t''-t'}{N}}P_{\varepsilon',p}^{\frac{t''-t'}{N}}\right]^N\varphi_{\varepsilon'}(q''-q') = \frac{1}{(2\pi)^{\frac{nN}{2}}}\int\limits_{-\infty}^{\infty}d\mathbf{q}\int\limits_{-\infty}^{\infty}d\boldsymbol{\xi}\times$$

$$\exp\left\{i\sum_{m=0}^{N}(q_{m+1}-q_m)\xi_m+\frac{t''-t'}{N}\sum_{r=1}^{p}\mathbf{1}_{J_r}(q_m)b_{\varepsilon'}(t,r)\xi_m-\frac{t''-t'}{N}\frac{\varepsilon}{2}\sum_{m=0}^{N}\xi_m^2\right\}\times \tag{45}$$

$$\times\varphi_{\varepsilon'}(q_0-q').$$

Here $d\mathbf{q} = dq_0\ldots dq_m\ldots dq_{N-1}, d\xi = d\xi_0\ldots d\xi_m\ldots d\xi_N, q_N = q'', m = 0,\ldots,N.$ Integrating on variable $\boldsymbol{\xi}$ gives

$$\left(U_{\varepsilon,\varepsilon',p}^{(t''-t')}\varphi_{\varepsilon'}(q''-q')\right)_{\varepsilon'} =$$

$$\left[K_{\varepsilon,p}^{\frac{t''-t'}{N}}P_{\varepsilon',p}^{\frac{t''-t'}{N}}\right]^N\varphi_{\varepsilon'}(q''-q') = I_{N,\varepsilon',p}(q',t'|q'',t'') =$$

$$= \frac{1}{(2\pi)^{\frac{nN}{2}}}\int\limits_{-\infty}^{\infty}d\mathbf{q}\exp\left\{\frac{t''-t'}{N}\frac{\varepsilon}{2}\sum_{m=0}^{N}\left[\frac{(q_{m+1}-q_m)}{\frac{t''-t'}{N}}-\sum_{r=1}^{p}\mathbf{1}_{J_r}(q_m)b_{\varepsilon'}(t,r)\right]^2\right\}\times \tag{46}$$

$$\times\varphi_{\varepsilon'}(q_0-q').$$

From Eq.(46) we obtain

$$(u_{\varepsilon',p}(q',t'|q'',t''))_{\varepsilon'} = \left(U_{\varepsilon,\varepsilon',p}^{(t''-t')}\varphi_{\varepsilon'}(q''-q')\right)_{\varepsilon'},$$

$$\left(U_{\varepsilon,\varepsilon',p}^{(t''-t')}\varphi_{\varepsilon'}(q''-q')\right)_{\varepsilon'} = \left(\lim_{N\to\infty}I_{N,\varepsilon',p}(q',t'|q'',t'')\right)_{\varepsilon'}, \tag{47}$$

where the limit is calculated by norm in $H^{\infty}(S_{\mathbf{R}},\check{S}_{\mathbf{U},p})$. Let $\delta_p = \max_{1\leq r\leq p}\{\delta_r|r = 1,\ldots,p\}$, $\delta_r = |x_{r-1}-x_r|$. We assume that: (1) $\delta_p \to 0$ if $p\to\infty$, (2) $x_0\to-\infty$ if $p\to\infty$, (3) $x_p\to+\infty$ if $p\to\infty$.

Finally we obtain

$$(u_{\varepsilon'}(q',t'|q'',t''))_{\varepsilon'} = \left(\lim_{p\to\infty} U_{\varepsilon,\varepsilon',p}^{(t''-t')}\varphi_{\varepsilon'}(q''-q')\right)_{\varepsilon'}, \tag{48}$$

where the limit is calculated by norm $\|u\|_{W_\omega^{k,p}(\mathbb{R})}$

$$\|u\|_{W_\omega^{k,p}} = \left(\sum_{|\alpha|\leq k}\int_{\mathbb{R}}|D^\alpha u(x)|\omega(x)dx\right)^{\frac{1}{p}} < \infty$$

of the weighted Sobolev space $W_\omega^{2,2}(\mathbb{R}) = W^{2,2}(\mathbb{R},\omega)$ [43].

**Theorem 3**. Assume that $\varphi_{\varepsilon'}(q''-q') \in H^\infty(S_{\mathbf{R}})\cap W_\omega^{2,2}(\mathbb{R}), \varepsilon' \in (0,1], \omega = \omega(q'')$.

Then: (1) $\forall\varepsilon' \in (0,1]$ there exist $p_0$ such that $\forall p_1\forall p_2[(p_1 \geq p_0)\wedge(p_2 \geq p_0)]$ the inequality

$$\|u_{\varepsilon',p_1}(q',t'|q'',t'') - u_{\varepsilon',p_1p_2}(q',t'|q'',t'')\|_{W_\omega^{2,2}(\mathbb{R})} \leq \frac{t''-t'}{p_1}C_1\exp[C_2(t''-t')]\|\varphi_{\varepsilon'}\|_{W_\omega^{2,2}}$$

holds for each $t'' \in [t',\infty)$.

(2) The Colombeau generalized function $(u_{\varepsilon',p}(q',t'|q'',t''))_{\varepsilon'}$ given by formula (48) is the solution of the Colombeau-Fokker-Planck equation (3) (except a set Lebesgue measure zero) with initial condition $(\varphi_{\varepsilon'}(q''-q'))_{\varepsilon'}$ such that $\forall\varepsilon' \in (0,1]$ : $\varphi_{\varepsilon'}(q''-q') \in H^\infty(S_{\mathbf{R}})\cap W_\omega^{2,2}(\mathbb{R})$.

Let us consider now $n$-dimensional case. We shall at first be working with rectangular parallelograms in $\mathbb{R}^n$, those parallelograms whose edges are mutually orthogonal. Actually, we shall be even more restrictive, and consider only those whose edges are in the directions of the coordinate axes.

**Definition 6**.We call them special rectangles. Each of these may be expressed as a Cartesian product of intervals in $\mathbb{R}$.

$\mathbf{I} = [a_1,b_1]\times\ldots\times[a_n,b_n] = \{q|q \in \mathbb{R}^n, a_i \leq q_i \leq b_i, i = 1,\ldots,n\}$

And we let $\mathbf{I}_\infty = \mathbb{R}^n\backslash\mathbf{I}$.

**Definition 7**. We define a partition of $\mathbf{I}$ to be a collection of non-overlapping special rectangles $\mathbf{I}_1,\mathbf{I}_2,\ldots,\mathbf{I}_r,..,\mathbf{I}_p$ whose union is $\mathbf{I}$. “Non-overlapping” requires that the interiors of these rectangles are mutually disjoint.

**Definition 8**.Let $b_{i,\varepsilon'}(t,r), i = 1,\ldots,n, \mathbf{q} =(q_1,\ldots,q_n)$ be a quantity

$$b_{i,\varepsilon'}(t,r) = \sup_{\mathbf{q}\in\mathbf{I}_r} b_{i,\varepsilon'}(\mathbf{q},t) \tag{49}$$

and let $\partial q_j b_{i,\varepsilon'}(t,r), i = 1,\dots,n, j = 1,\dots,n$ be a quantity

$$\partial q_j b_{i,\varepsilon'}(t,r) = \sup_{\mathbf{q}\in\mathbf{I}_r} \frac{\partial b_{i,\varepsilon'}(\mathbf{q},t)}{\partial q_j}. \tag{50}$$

Let $\breve{b}_{i,\varepsilon'}(\mathbf{q},t)$ be a function

$$\breve{b}_{i,\varepsilon'}(\mathbf{q},t) = \sum_{r=1}^{p} \mathbf{1}_{\mathbf{I}_r}(\mathbf{q}) b_{i,\varepsilon'}(t,r) \tag{51}$$

Here $\mathbf{1}_{\mathbf{I}_r}(\mathbf{q})$ is indicator function of a subset $\mathbf{I}_r \subset \mathbf{I}$.

**Definition 9**.We let now $H_n^\infty(S_{\mathbf{R}}, \breve{S}_{\mathbf{U},p})$ if $\breve{S}_{\mathbf{U}} = \backslash\{\partial\mathbf{I}_0,\dots,\partial\mathbf{I}_p\}$

**Remark 6**.(**I**) From Theorem 1-2 one obtain, that:

(1) Operator $\breve{b}_{\varepsilon'}(\mathbf{q},t)\partial\mathbf{q} = \sum_{i=1}^{n} \breve{b}_{i,\varepsilon'}(\mathbf{q},t)\partial q_i, \varepsilon' \in (0,1]$ is bounded on $H_n^\infty(S_{\mathbf{R}}, \breve{S}_{\mathbf{U},p})$.

(2)The Colombeau generalized function $(u_{\varepsilon',p}(\mathbf{q}',t'|\mathbf{q}'',t''))_{\varepsilon'}$ given by formula

$$\begin{gathered}
(u_{\varepsilon',p}(\mathbf{q}',t'|\mathbf{q}'',t''))_{\varepsilon'} = \left(P_{\varepsilon',p}^{t''-t'} \varphi_{\varepsilon'}(\mathbf{q}''-\mathbf{q}')\right)_{\varepsilon'} = \\
\left(\exp\left(-[t''-t']\breve{b}_{\varepsilon'}(\mathbf{q},t)\partial\mathbf{q}\right)\varphi_{\varepsilon'}(\mathbf{q}''-\mathbf{q}')\right)_{\varepsilon'} = \\
\left(\exp\left(-[t''-t']\sum_{i=1}^{n}\sum_{r=1}^{p}\mathbf{1}_{\mathbf{I}_r}(\mathbf{q}'')b_{i,\varepsilon'}(t,r)\partial q_i''\right)\varphi_{\varepsilon'}(\mathbf{q}''-\mathbf{q}')\right)_{\varepsilon'} = \\
\left(\prod_{i=1}^{n}\prod_{r=1}^{p}\exp(-[t''-t']\mathbf{1}_{\mathbf{I}_r}(\mathbf{q}'')b_{\varepsilon'}(t,r)\partial q'')\varphi_{\varepsilon'}(\mathbf{q}''-\mathbf{q}')\right)_{\varepsilon'}
\end{gathered} \tag{52}$$

is the solution (except points $\mathbf{q}'' \in \bigcup_{r=1}^{p} I_r$) of the Colombeau-Fokker-Planck equation (3) with initial condition $\varphi_{\varepsilon'}(\mathbf{q}''-\mathbf{q}') \in H_n^\infty(S_{\mathbf{R}}, \breve{S}_{\mathbf{U},p})$ for the case $\varepsilon = 0$ and

$$b_{i,\varepsilon'}(\mathbf{q}'',t'') = \sum_{r=1}^{p}\mathbf{1}_{\mathbf{I}_r}(\mathbf{q}'')b_{i,\varepsilon'}(t,r), i = 1,\dots,n. \tag{53}$$

(3) We note that:

$$(\exp(-[t''-t']\sum_{r=1}^{p}\mathbf{1}_{J_r}(q'')b_{i,\varepsilon'}(t,r)\partial q_i'')\varphi_{\varepsilon'}(\mathbf{q}''-\mathbf{q}'))_{\varepsilon'} =$$

$$(\mathcal{F}^{-1}[\exp(-[t''-t']\sum_{r=1}^{p}\mathbf{1}_{J_r}(q'')b_{i,\varepsilon'}(t,r)(i\xi_i))\mathcal{F}^{\#}[\varphi_{\varepsilon'}(\mathbf{q}''-\mathbf{q}')]])_{\varepsilon'} = \tag{54}$$

(3) $\forall\mathbf{q}'\forall t'\forall t''\left[u_{\varepsilon',p}(\mathbf{q}',t'|\mathbf{q}'',t'') \in H_n^{\infty}(S_{\mathbf{R}},\check{S}_{\mathbf{U},p})\right]$.

(**II**) From Theorem 1 and Theorem 2 one obtain, that:
(1) Operator $\Delta$ is bounded on $H_n^{\infty}(S_{\mathbf{R}},\check{S}_{\mathbf{U},p})$.
(2) The Colombeau generalized function $(u_{\varepsilon',p}(\mathbf{q}',t'|\mathbf{q}'',t''))_{\varepsilon'}$ given by formula

$$(u_{\varepsilon',p}(\mathbf{q}',t'|\mathbf{q}'',t''))_{\varepsilon'} = K_{\varepsilon,p}^{(t''-t')}\varphi_{\varepsilon'}(\mathbf{q}''-\mathbf{q}'),$$

$$K_{\varepsilon,p}^{(t''-t')} = \exp\left[\frac{\varepsilon}{2}\left(t''-t'\right)\Delta\right], \tag{55}$$

$$(u_{\varepsilon',p}(q',t'|\mathbf{q}'',t''))_{\varepsilon'} = \left(\mathcal{F}^{-1}\left[\exp\left(-\frac{\varepsilon}{2}(t''-t')\|\boldsymbol{\xi}\|^2\right)\mathcal{F}^{\#}[\varphi_{\varepsilon'}(\mathbf{q}''-\mathbf{q}')](\boldsymbol{\xi})\right]\right)_{\varepsilon'}$$

is the solution of the Colombeau-Fokker-Planck equation (3) with initial condition $(\varphi_{\varepsilon'}(\mathbf{q}''-\mathbf{q}'))_{\varepsilon'}$ such that $\forall\varepsilon' \in [0,1) : \varphi_{\varepsilon'}(\mathbf{q}''-\mathbf{q}') \in H^{\infty}(S_{\mathbf{R}},\check{S}_{\mathbf{U},p})$ for the case: $(b_{\varepsilon'}(\mathbf{q},t))_{\varepsilon'} = 0$

(3) $\forall\mathbf{q}'\forall t'\forall t''\left[u_{\varepsilon'}(\mathbf{q}',t'|\mathbf{q}'',t'') \in H^{\infty}(S_{\mathbf{R}},\check{S}_{\mathbf{U},p})\right]$.

(**III**) From Theorem 1-2 one obtain, that: operator

$$\Re_{\varepsilon,\varepsilon',p} = \sum_{i=1}^{n}\sum_{r=1}^{p}\mathbf{1}_{\mathbf{I}_r}(\mathbf{q}'')b_{\varepsilon'}(t,r)\frac{\partial}{\partial q_i''} + \varepsilon\Delta = P_{\varepsilon',p}^{(t''-t')} + \varepsilon\Delta \tag{56}$$

is bounded on $H_n^{\infty}(S_{\mathbf{R}},\check{S}_{\mathbf{U},p})$. If we let now

$$U_{\varepsilon,\varepsilon',p}^{(t''-t')} = \exp\left[\left(t''-t'\right)\Re_{\varepsilon,\varepsilon',p}\right] \tag{57}$$

then Theorem 1 and Theorem 2 asserts that for all $\varphi_{\varepsilon'}(\mathbf{q}''-\mathbf{q}') \in H^{\infty}(S_{\mathbf{R}},\check{S}_{\mathbf{U},p})$

$$\left(U^{(t''-t')}_{\varepsilon,\varepsilon',p}\varphi_{\varepsilon'}(\mathbf{q}''-\mathbf{q}')\right)_{\varepsilon'} = \left(\lim_{N\to\infty}\left[K^{\frac{t''-t'}{N}}_{\varepsilon,p}P^{\frac{t''-t'}{N}}_{\varepsilon',p}\right]^{N}\varphi_{\varepsilon'}(\mathbf{q}''-\mathbf{q}')\right)_{\varepsilon'}, \tag{58}$$

where the limit is calculated by norm in $H^{\infty}(S_{\mathbf{R}},\check{S}_{\mathbf{U},p})$. From Eq.(54)-Eq.(59) by simple calculation one obtain

$$\left[K^{\frac{t''-t'}{N}}_{\varepsilon,p}P^{\frac{t''-t'}{N}}_{\varepsilon',p}\right]^{N}\varphi_{\varepsilon'}(\mathbf{q}''-\mathbf{q}') = \frac{1}{(2\pi)^{\frac{nN}{2}}}\int_{-\infty}^{\infty}d\mathbf{q}\int_{-\infty}^{\infty}d\boldsymbol{\xi}\times$$

$$\exp\left\{i\sum_{m=0}^{N}(\mathbf{q}_{m+1}-\mathbf{q}_m)\boldsymbol{\xi}_m+\frac{t''-t'}{N}\sum_{i=1}^{n}\sum_{r=1}^{p}\mathbf{1}_{I_r}(\mathbf{q}_m)b_{i,\varepsilon'}(t,r)\xi_{i,m}-\right. \tag{59}$$

$$\left.-\frac{t''-t'}{N}\frac{\varepsilon}{2}\sum_{m=0}^{N}\boldsymbol{\xi}_m^2\right\}\varphi_{\varepsilon'}(\mathbf{q}_0-\mathbf{q}').$$

Here $d\mathbf{q}=d\mathbf{q}_0\ldots d\mathbf{q}_m\ldots d\mathbf{q}_{N-1}, d\boldsymbol{\xi}=d\boldsymbol{\xi}_0\ldots d\boldsymbol{\xi}_m\ldots d\boldsymbol{\xi}_N, \mathbf{q}_N=\mathbf{q}'', d\mathbf{q}_m=\prod_{i=1}^{n}q_{i,m}$, $d\boldsymbol{\xi}_m=\prod_{i=1}^{n}\xi_{i,m}, m=0,\ldots,N.$ Integrating on variable $\boldsymbol{\xi}$ gives

$$\left(U^{(t''-t')}_{\varepsilon,\varepsilon',p}\varphi_{\varepsilon'}(\mathbf{q}''-\mathbf{q}')\right)_{\varepsilon'} =$$

$$\left[K^{\frac{t''-t'}{N}}_{\varepsilon,p}P^{\frac{t''-t'}{N}}_{\varepsilon',p}\right]^{N}\varphi_{\varepsilon'}(\mathbf{q}''-\mathbf{q}') = I_{N,\varepsilon',p}(\mathbf{q}',t'|\mathbf{q}'',t'') =$$

$$= \frac{1}{(2\pi)^{\frac{nN}{2}}}\int_{-\infty}^{\infty}d\mathbf{q}\times \tag{60}$$

$$\exp\left\{\frac{t''-t'}{N}\frac{\varepsilon}{2}\sum_{m=0}^{N}\sum_{i=1}^{n}\left[\frac{(q_{i,m+1}-q_{i,m})}{\frac{t''-t'}{N}}-\sum_{r=1}^{p}\mathbf{1}_{I_r}(\mathbf{q}_m)b_{i,\varepsilon'}(t,r)\right]^{2}\right\}\times$$

$$\times\varphi_{\varepsilon'}(\mathbf{q}_0-\mathbf{q}').$$

where the limit is calculated by norm in $H^{\infty}(S_{\mathbf{R}},\check{S}_{\mathbf{U},p})$. Let $\delta_p=\max_{1\le r\le p}\{\delta_r|r=1,\ldots,p\}$, $\delta_r = diam(\mathbf{I}_r)$. We assume that: (1) $\delta_p \to 0$ if $p \to \infty$, (2) $a_i \to -\infty, i = 1,\dots,n$ if $p \to \infty$, (3) $b_i \to +\infty, i = 1,\dots,n$ if $p \to \infty$.

Finally we obtain

$$(u_{\varepsilon'}(\mathbf{q}',t'|\mathbf{q}'',t''))_{\varepsilon'} = \left(\lim_{p\to\infty} U_{\varepsilon,\varepsilon',p}^{(t''-t')} \varphi_{\varepsilon'}(\mathbf{q}''-\mathbf{q}')\right)_{\varepsilon'}, \tag{61}$$

where the limit is calculated by norm $\|u\|_{W_\omega^{k,p}(\mathbb{R}^n)}$

$$\|u\|_{W_\omega^{k,p}} = \left(\sum_{|\alpha|\leq k} \int_{\mathbb{R}^n} |D^\alpha u(x)|\omega(x) d^n x\right)^{\frac{1}{p}} < \infty$$

of the weighted Sobolev space $W_\omega^{2,2}(\mathbb{R}^n) = W^{2,2}(\mathbb{R}^n, \omega)$ [43].

**Theorem 4**. Assume that $\varphi_{\varepsilon'}(\mathbf{q}'' - \mathbf{q}') \in H^\infty(S_{\mathbf{R}}) \cap W_\omega^{2,2}(\mathbb{R}^n), \varepsilon' \in (0,1], \omega = \omega(q'')$.

Then: (1) $\forall \varepsilon' \in (0,1]$ there exist $p_0$ such that $\forall p_1 \forall p_2 [(p_1 \geq p_0) \wedge (p_2 \geq p_0)]$ the inequality

$$\|u_{\varepsilon',p_1}(\mathbf{q}',t'|\mathbf{q}'',t'') - u_{\varepsilon',p_1 p_2}(\mathbf{q}',t'|\mathbf{q}'',t'')\|_{W_\omega^{2,2}(\mathbb{R}^n)} \leq \frac{t''-t'}{p_1} C_1 \exp[C_2(t''-t')] \|\varphi_{\varepsilon'}\|_{W_\omega^{2,2}(\mathbb{R}^n)}$$

holds for each $t'' \in [t', \infty)$.

(2) The Colombeau generalized function $(u_{\varepsilon',p}(\mathbf{q}',t'|\mathbf{q}'',t''))_{\varepsilon'}$ given by formula (48) is the solution of the Colombeau-Fokker-Planck equation (3) (except a set Lebesgue measure zero) with initial condition $(\varphi_{\varepsilon'}(\mathbf{q}''-\mathbf{q}'))_{\varepsilon'}$ such that $\forall \varepsilon' \in (0,1]$ : $\varphi_{\varepsilon'}(\mathbf{q}''-\mathbf{q}') \in H^\infty(S_{\mathbf{R}}) \cap W_\omega^{2,2}(\mathbb{R}^n)$.

**Remark 8**.The continuous-time conditional probability when $p \to \infty$ in (266) is symbolically indicated by the path-integral expression [22-23]:

$$(p_{\varepsilon'}^{\varepsilon}(\mathbf{q}',t'|\mathbf{q}'',t''))_{\varepsilon'} = \int_{\mathbf{q}(t')=\mathbf{q}'}^{\mathbf{q}(t'')=\mathbf{q}''} \mathbf{D}[\mathbf{q}(t)] \exp\left[-\frac{1}{2\varepsilon}(\mathbf{S}_{\varepsilon'}(\dot{\mathbf{q}}(t),\mathbf{q}(t),t;\varepsilon))_{\varepsilon'}\right]. \tag{62}$$

Here

$$(\mathbf{S}_{\varepsilon'}(\dot{\mathbf{q}}(t),\mathbf{q}(t),t;\varepsilon))_{\varepsilon'} = \left(\int_{t'}^{t''}\mathcal{L}_{\varepsilon'}(\dot{\mathbf{q}}(t),\mathbf{q}(t),t;\varepsilon)dt\right)_{\varepsilon'} \quad (63)$$

is the continuous-time limit of the discrete action with

$$\mathcal{L}_{\varepsilon'}(\dot{\mathbf{q}}(t),\mathbf{q}(t),t;\varepsilon) = \|\dot{\mathbf{q}}(t)-\mathbf{b}_{\varepsilon'}(\mathbf{q}(t),t;\varepsilon)\|^2 - \varepsilon\sum_{i=1}^{n} b_{i,i,\varepsilon'}(\mathbf{q}(t),t;\varepsilon),$$
$$b_{i,i,\varepsilon'}(\mathbf{q}(t),t;\varepsilon) = \frac{\partial b_{i,\varepsilon'}(\mathbf{q}(t),t;\varepsilon)}{\partial q_i},\varepsilon,\varepsilon' \in (0,1] \quad (64)$$

as the Lagrangian. From Eq. (62) one obtain

$$\left(\mathbf{E}\left[\mathbf{x}_{t'',\varepsilon'}^{\mathbf{q}',\varepsilon}(\omega)\right]^2\right)_{\varepsilon'} = \int_{-\infty}^{\infty} d\mathbf{q}''\|\mathbf{q}''\|^2(p_{\varepsilon'}^{\varepsilon}(\mathbf{q}',t'|\mathbf{q}'',t''))_{\varepsilon'} = \quad 65$$
$$\left(\lim_{p\to\infty}\lim_{\Delta t\to 0}\breve{\mathbf{I}}_{N,\varepsilon,\varepsilon',p}(\mathbf{q}',t',t'')\right)_{\varepsilon'}.$$

Here

$$\breve{\mathbf{I}}_{N,\varepsilon,\varepsilon',p}(\mathbf{q}',t',t'') = \int_{-\infty}^{\infty} d\mathbf{q}''\|\mathbf{q}''\|^2(\mathbf{I}_{N,\varepsilon,\varepsilon',p}(\mathbf{q}',t'|\mathbf{q}'',t''))_{\varepsilon'} \quad (66)$$

**Remark 9**.Note that for any fixed $N$ in the limit $\varepsilon \to 0$ only one unique minimizing path $\{\breve{\mathbf{q}}_0,\breve{\mathbf{q}}_1,\ldots,\breve{\mathbf{q}}_{N-1},\breve{\mathbf{q}}_N\}$ significantly contribute to the multiple integral $\breve{\mathbf{I}}_{N,\varepsilon',p}(\mathbf{q}',t',t'')$ given by expression (66).The extremality conditions for this minimizing path is

$$\overline{\nabla}\mathbf{q}(t_m) = \mathbf{b}_{\varepsilon'}(\mathbf{q}(t_{m-1}),t_m), m = 1,\ldots,N, \quad (67)$$

with a boundary condition

$$\mathbf{q}(t') = \breve{\mathbf{q}}_0 = \mathbf{q}'. \quad (68)$$

Here $\bar{\nabla}$ is a conjugate of the difference operator $\nabla$ defined by [44]:

$$\nabla\mathbf{q}(t_m) = \frac{\mathbf{q}(t_{m+1}) - \mathbf{q}(t_m)}{\Delta t}, N \geq m \geq 0,$$
$$\bar{\nabla}\mathbf{q}(t_m) = \frac{\mathbf{q}(t_m) - \mathbf{q}(t_{m-1})}{\Delta t}, N+1 \geq m \geq 1. \tag{69}$$

From Eq.(62)-Eq.(63) one obtain

$$\left(\mathbf{E}\left[\mathbf{x}^{\mathbf{q}',\varepsilon}_{t'',\varepsilon'}(\omega)\right]^2\right)_{\varepsilon'} = \int_{-\infty}^{\infty} d\mathbf{q}'' \|\mathbf{q}''\|^2 (p^{\varepsilon}_{\varepsilon'}(\mathbf{q}',t'|\mathbf{q}'',t''))_{\varepsilon'} =$$
$$\int_{-\infty}^{\infty} d\mathbf{q}'' \|\mathbf{q}''\|^2 \int_{\mathbf{q}(t')=\mathbf{q}'}^{\mathbf{q}(t'')=\mathbf{q}''} [\mathbf{D}\mathbf{q}(t)] \exp\left[-\frac{1}{2\varepsilon}\left(\int_{t'}^{t''} \mathcal{L}_{\varepsilon'}(\dot{\mathbf{q}}(t),\mathbf{q}(t),t;\varepsilon)dt\right)_{\varepsilon'}\right] \tag{70}$$

Let us consider now the quantity

$$(p^{\varepsilon}_{\varepsilon'}(\mathbf{q}',t'|\mathbf{q}'',t'';L,m'))_{\varepsilon'} = \left(\lim_{\Delta t\to 0} \check{\mathbf{I}}^{L}_{m',N,\varepsilon,\varepsilon'}(\mathbf{q}',t'|\mathbf{q}'',t'')\right)_{\varepsilon'}. \tag{71}$$

Here

$$\lim_{\Delta t\to 0} \check{\mathbf{I}}^{L}_{m',N,\varepsilon,\varepsilon'}(\mathbf{q}',t'|\mathbf{q}'',t'') = \check{\mathbf{N}}_N \int_{-L}^{L} d\mathbf{q}_1 \ldots \int_{-L}^{L} d\mathbf{q}_{m'} \int_{-\infty}^{\infty} d\mathbf{q}_{m'+1} \ldots \int_{-\infty}^{\infty} d\mathbf{q}_{N-1} \times$$
$$\times \exp\left[-\frac{1}{2\varepsilon}(\mathbf{S}_{\varepsilon'}(\mathbf{q}_0,\mathbf{q}_1,\ldots,\mathbf{q}_{m'},\mathbf{q}_{m'+1},\ldots,\mathbf{q}_{N-1},\mathbf{q}_N;\varepsilon))\right] \tag{72}$$

and $m \ll N, L \gg 1$.

The quantity defined byEq.(71)-Eq. (72) is symbolically indicated by the path- integral expression

$$(p^{\varepsilon}_{\varepsilon'}(\mathbf{q}',t'|\mathbf{q}'',t'';L,m'))_{\varepsilon'} = \int_{\mathbf{q}(t')=\mathbf{q}'}^{\mathbf{q}(t'')=\mathbf{q}''} \mathbf{D}[\mathbf{q}(t);L,m'] \exp\left[-\frac{1}{2\varepsilon}(\mathbf{S}_{\varepsilon'}(\dot{\mathbf{q}}(t),\mathbf{q}(t),t;\varepsilon))_{\varepsilon'}\right]. \tag{73}$$

Using Eq.(71)-Eq.(73) we define the quantity

$$\left(\mathbf{E}_{L,m'}\left[\mathbf{x}_{t'',\varepsilon'}^{\mathbf{q}',\varepsilon}(\omega)\right]^2\right)_{\varepsilon'} = \int_{-\infty}^{\infty} d\mathbf{q}'' \|\mathbf{q}''\|^2 (p_{\varepsilon'}^{\varepsilon}(\mathbf{q}',t'|\mathbf{q}'',t'';L,m'))_{\varepsilon'} = \tag{74}$$

$$= \lim_{\Delta t\to 0} \breve{\mathbf{I}}_{N,\varepsilon,\varepsilon'}(\mathbf{q}',t'|\mathbf{q}'',t'';L,m')$$

Here

$$\lim_{\Delta t\to 0} \breve{\mathbf{I}}_{m',N,\varepsilon,\varepsilon'}^{L}(\mathbf{q}',t'|\mathbf{q}'',t'') = \int_{-\infty}^{\infty} d\mathbf{q}'' \|\mathbf{q}''\|^2 (\mathbf{I}_{N,\varepsilon,\varepsilon',p}(\mathbf{q}',t'|\mathbf{q}'',t'';L,m'))_{\varepsilon'} \tag{75}$$

The quantity (74) is symbolically indicated by the path-integral expression

$$\left(\mathbf{E}_{L,m'}\left[\mathbf{x}_{t'',\varepsilon'}^{\mathbf{q}',\varepsilon}(\omega)\right]^2\right)_{\varepsilon'} = \int_{-\infty}^{\infty} d\mathbf{q}'' \|\mathbf{q}''\|^2 (p_{\varepsilon'}^{\varepsilon}(\mathbf{q}',t'|\mathbf{q}'',t'';L,m'))_{\varepsilon'} =$$

$$\int_{-\infty}^{\infty} d\mathbf{q}'' \|\mathbf{q}''\|^2 \int_{\mathbf{q}(t')=\mathbf{q}'}^{\mathbf{q}(t'')=\mathbf{q}''} [\mathbf{D}\mathbf{q}(t);L,m'] \exp\left[-\frac{1}{2\varepsilon}\left(\int_{t'}^{t''} \mathcal{L}_{\varepsilon'}(\dot{\mathbf{q}}(t),\mathbf{q}(t),t;\varepsilon)dt\right)_{\varepsilon'}\right] = \tag{76}$$

$$\int_{\mathbf{q}(t')=\mathbf{q}'} [\mathbf{D}\mathbf{q}(t);L,m'] \|\mathbf{q}(t'')\|^2 \exp\left[-\frac{1}{2\varepsilon}\left(\int_{t'}^{t''} \mathcal{L}_{\varepsilon'}(\dot{\mathbf{q}}(t),\mathbf{q}(t),t;\varepsilon)dt\right)_{\varepsilon'}\right].$$

From Eq.(205) and Eq.(279) we obtain

$$\mathbf{I}_{m',N,\varepsilon,\varepsilon'}(\mathbf{q}',t'|\mathbf{q}'',t'') = \mathbf{I}_{m',N,\varepsilon,\varepsilon'}^{L}(\mathbf{q}',t'|\mathbf{q}'',t'') + \mathbf{\Theta}_{m',N,\varepsilon,\varepsilon'}^{L}(\mathbf{q}',t'|\mathbf{q}'',t''). \tag{77}$$

Here

$$\Theta^{L}_{m',N,\varepsilon,\varepsilon'}(\mathbf{q}',t'|\mathbf{q}'',t'') = \breve{\mathbf{N}}_N \int\limits_{\mathbb{R}^n\backslash\varpi_L} d\mathbf{q}_1 \ldots \int\limits_{\mathbb{R}^n\backslash\varpi_L} d\mathbf{q}_{m'} \int\limits_{-\infty}^{\infty} d\mathbf{q}_{m'+1} \ldots \int\limits_{-\infty}^{\infty} d\mathbf{q}_{N-1} \times$$

$$\exp\left[-\frac{1}{2\varepsilon}(\mathbf{S}_{\varepsilon'}(\mathbf{q}_0,\mathbf{q}_1,\ldots,\mathbf{q}_{m'},\mathbf{q}_{m'+1},\ldots,\mathbf{q}_{N-1},\mathbf{q}_N;\varepsilon))\right], \quad (78)$$

$$\varpi_L = [-L,L]^n.$$

From Eq.(77) and Eq.(78) we obtain

$$\breve{\mathbf{I}}_{m',N,\varepsilon,\varepsilon'}(\mathbf{q}',t',t'') = \breve{\mathbf{I}}^{L}_{m',N,\varepsilon,\varepsilon'}(\mathbf{q}',t',t'') + \breve{\Theta}^{L}_{m',N,\varepsilon,\varepsilon'}(\mathbf{q}',t',t''). \quad (79)$$

Here

$$\breve{\Theta}^{L}_{m',N,\varepsilon,\varepsilon'}(\mathbf{q}',t',t'') = \breve{\mathbf{N}}_N \int\limits_{\mathbb{R}^n\backslash\varpi_L} d\mathbf{q}_1 \ldots \int\limits_{\mathbb{R}^n\backslash\varpi_L} d\mathbf{q}_{m'} \int\limits_{-\infty}^{\infty} d\mathbf{q}_{m'+1} \ldots \int\limits_{-\infty}^{\infty} d\mathbf{q}_{N-1} \int\limits_{-\infty}^{\infty} d\mathbf{q}_N \times$$

$$\|\mathbf{q}_N\|^2 \exp\left[-\frac{1}{2\varepsilon}(\mathbf{S}_{\varepsilon'}(\mathbf{q}_0,\mathbf{q}_1,\ldots,\mathbf{q}_{m'},\mathbf{q}_{m'+1},\ldots,\mathbf{q}_{N-1},\mathbf{q}_N;\varepsilon))\right]. \quad (80)$$

**Remark 10**. We note that: $\forall\varepsilon',\varepsilon' \neq 0$ there exist parameter $L = L(\varepsilon')$ such that $\forall\mathbf{q}_m(\|\mathbf{q}_m\| \geq L)$ the inequality

$$\|\mathbf{b}_{\varepsilon'}(\mathbf{q}_m,t_m)\|^2 \leq \|\mathbf{q}_m\|^{-q}, q \geq 2 \quad (81)$$

satisfied.

**Lemma1**.

$$(82)$$

**Proof**. Using inequality (81), we willing to choose parameter $L$ such that the equality

$$\mathbf{S}_{\varepsilon'}(\mathbf{q}_0,\mathbf{q}_1,\ldots,\mathbf{q}_{m'},\mathbf{q}_{m'+1},\ldots,\mathbf{q}_{N-1},\mathbf{q}_N;\varepsilon) = \Delta t\sum_{m=1}^{N}\left\|\frac{\mathbf{q}_m-\mathbf{q}_{m-1}}{\Delta t}\right\|^2 + +O((t''-t')L^{-q}) \simeq \frac{1}{\Delta t}\sum_{m=1}^{N}\|\mathbf{q}_m-\mathbf{q}_{m-1}\|^2 \quad (83)$$

is satisfied. From Eq.(80) and Eq.(83) we obtain

$$\breve{\boldsymbol{\Theta}}^{L}_{m',N,\varepsilon,\varepsilon'}(\mathbf{q}',t',t'') = \breve{\mathbf{N}}_N\int_{\mathbb{R}^n\setminus\varpi_L}d\mathbf{q}_1\ldots\int_{\mathbb{R}^n\setminus\varpi_L}d\mathbf{q}_{m'}\int_{-\infty}^{\infty}d\mathbf{q}_{m'+1}\ldots\int_{-\infty}^{\infty}d\mathbf{q}_{N-1}\int_{-\infty}^{\infty}d\mathbf{q}_N\times \|\mathbf{q}_N\|^2\exp\left[-\frac{1}{2\varepsilon}(\mathbf{S}_{\varepsilon'}(\mathbf{q}_0,\mathbf{q}_1,\ldots,\mathbf{q}_{m'},\mathbf{q}_{m'+1},\ldots,\mathbf{q}_{N-1},\mathbf{q}_N;\varepsilon))\right] \simeq \simeq \breve{\mathbf{N}}_N\int_{\mathbb{R}^n\setminus\varpi_L}d\mathbf{q}_1\ldots\int_{\mathbb{R}^n\setminus\varpi_L}d\mathbf{q}_{m'}\int_{-\infty}^{\infty}d\mathbf{q}_{m'+1}\ldots\int_{-\infty}^{\infty}d\mathbf{q}_{N-1}\int_{-\infty}^{\infty}d\mathbf{q}_N\times \|\mathbf{q}_N\|^2\exp\left[-\frac{1}{2\varepsilon\Delta t}\sum_{m=1}^{N}\|\mathbf{q}_m-\mathbf{q}_{m-1}\|^2\right]. \quad (84)$$

From Eq.(84) we obtain the inequality

$$\breve{\boldsymbol{\Theta}}^{L}_{m',N,\varepsilon,\varepsilon'}(\mathbf{q}',t',t'') \leq \breve{\mathbf{N}}_N\int_{-\infty}^{\infty}d\mathbf{q}_1\ldots\int_{\mathbb{R}^n\setminus\varpi_L}d\mathbf{q}_{m'}\int_{-\infty}^{\infty}d\mathbf{q}_{m'+1}\ldots\int_{-\infty}^{\infty}d\mathbf{q}_{N-1}\int_{-\infty}^{\infty}d\mathbf{q}_N\times \|\mathbf{q}_N\|^2\exp\left[-\frac{1}{2\varepsilon\Delta t}\sum_{m=1}^{N}\|\mathbf{q}_m-\mathbf{q}_{m-1}\|^2\right]. \quad (85)$$

Let $\Im^{L}_{m',N,\varepsilon,\varepsilon'}(\mathbf{q}',t',t'')$ be the multiple integral:

$$\Im^{L}_{m',N,\varepsilon,\varepsilon'}(\mathbf{q}',t',t'') = \breve{\mathbf{N}}_N \int_{-\infty}^{\infty} d\mathbf{q}_1 \ldots \int_{-\infty}^{\infty} d\mathbf{q}_{m'-1} \int_{\mathbb{R}^n \backslash \varpi_L} d\mathbf{q}_{m'} \int_{-\infty}^{\infty} d\mathbf{q}_{m'+1} \ldots \int_{-\infty}^{\infty} d\mathbf{q}_{N-1} \int_{-\infty}^{\infty} d\mathbf{q}_N \times$$

$$\|\mathbf{q}_N\|^2 \exp\left[ -\frac{1}{2\varepsilon\Delta t} \sum_{m=1}^{N} \|\mathbf{q}_m - \mathbf{q}_{m-1}\|^2 \right] =$$

$$\breve{\mathbf{N}}_N \int_{-\infty}^{\infty} d\mathbf{q}_1 \ldots \int_{-\infty}^{\infty} d\mathbf{q}_{m'-1} \exp\left[ -\frac{1}{2\varepsilon\Delta t} \sum_{m=1}^{m'} \|\mathbf{q}_m - \mathbf{q}_{m-1}\|^2 \right] \times \tag{86}$$

$$\times \int_{\mathbb{R}^n \backslash \varpi_L} d\mathbf{q}_{m'} \int_{-\infty}^{\infty} d\mathbf{q}_{m'+1} \ldots \int_{-\infty}^{\infty} d\mathbf{q}_{N-1} \int_{-\infty}^{\infty} d\mathbf{q}_N \|\mathbf{q}_N\|^2 \exp\left[ -\frac{1}{2\varepsilon\Delta t} \sum_{m=m'+1}^{N} \|\mathbf{q}_m - \mathbf{q}_{m-1}\|^2 \right]$$

By simple canonical observation (see [45] chapt.3) we obtain

$$\int_{-\infty}^{\infty} d\mathbf{q}_1 \ldots \int_{-\infty}^{\infty} d\mathbf{q}_{m'-1} \exp\left[ -\frac{1}{2\varepsilon\Delta t} \sum_{m=1}^{m'} \|\mathbf{q}_m - \mathbf{q}_{m-1}\|^2 \right] = \tag{87}$$

$$(2\pi\varepsilon\Delta t)^{\frac{n(m'-1)}{2}} \exp\left[ -\frac{1}{2\varepsilon\Delta t(m'-1)} \|\mathbf{q}_{m'} - \mathbf{q}_0\|^2 \right].$$

Substitution Eq.(87) into Eq.(86) gives

$$\Im^{L}_{m',N,\varepsilon,\varepsilon'}(\mathbf{q}',t',t'') = \breve{\mathbf{N}}_{N,m'} \int_{\mathbb{R}^n \backslash \varpi_L} d\mathbf{q}_{m'} \exp\left[ -\frac{1}{2\varepsilon\Delta t(m'-1)} \|\mathbf{q}_{m'} - \mathbf{q}_0\|^2 \right] \times \tag{88}$$

$$\int_{-\infty}^{\infty} d\mathbf{q}_{m'+1} \ldots \int_{-\infty}^{\infty} d\mathbf{q}_{N-1} \int_{-\infty}^{\infty} d\mathbf{q}_N \|\mathbf{q}_N\|^2 \exp\left[ -\frac{1}{2\varepsilon\Delta t} \sum_{m=m'+1}^{N} \|\mathbf{q}_m - \mathbf{q}_{m-1}\|^2 \right].$$

Here

$$\breve{\mathbf{N}}_{N,m'} = (2\pi\varepsilon\Delta t)^{\frac{n(m'-1)}{2}} \breve{\mathbf{N}}_N. \tag{89}$$

From Eq.(88) we obtain

$$\Im^{L}_{m',N,\varepsilon,\varepsilon'}(\mathbf{q}',t',t'') = \check{\mathbf{N}}_{N,m'} \int_{\mathbb{R}^n\backslash\varpi_L} d\mathbf{q}_{m'} \exp\left[-\frac{1}{2\varepsilon\Delta t(m'-1)}\|\mathbf{q}_{m'}-\mathbf{q}_0\|^2\right]\times$$

$$\int_{-\infty}^{\infty} d\mathbf{q}_N \|\mathbf{q}_N\|^2 \exp\left[-\frac{1}{2\varepsilon\Delta t(N-m'+1)}\|\mathbf{q}_N-\mathbf{q}_{m'}\|^2\right] = \tag{90}$$

$$\check{\mathbf{N}}_{N,m'}(2\pi\varepsilon\Delta t)^{\frac{n(N-m'+1)}{2}} \int_{\mathbb{R}^n\backslash\varpi_L} d\mathbf{q}_{m'} \|\mathbf{q}_{m'}\|^2 \exp\left[-\frac{1}{2\varepsilon\Delta t(m'-1)}\|\mathbf{q}_{m'}-\mathbf{q}_0\|^2\right].$$

**Assumption**. We assume now that $\mathbf{q}_0 \notin \mathbb{R}^n\backslash\varpi_L$.
From Eq.(90) using Laplace approximation [25] we obtain

$$\Im^{L}_{m',N,\varepsilon,\varepsilon'}(\mathbf{q}',t',t'') = \tag{91}$$

**Theorem 5**.(**Hölder's inequality**) Let $r_1 = p, r_2 = q \in [1,\infty]$, with $1/p + 1/q = 1$ and let $\Upsilon, \Im_i : (C^1([t',t'']))^n \to \mathbb{R}$ be an functional such that $\Upsilon = \Upsilon(\dot{\mathbf{q}}(t),\mathbf{q}(t),t',t'')$, $\Im_i = \Im_i(\dot{\mathbf{q}}(t),\mathbf{q}(t),t',t''), i = 1,2,3, \Im_3 = \Im_1\Im_2, \Upsilon(\dot{\mathbf{q}}(t),\mathbf{q}(t),t',t'') > 0.$ Let $\|\Im_i(\mathbf{q}',t',t'';L,m')\|^{\Upsilon}_{r_i}$, $i = 1,2,3$ be the path integral

$$\|\Im_i(\mathbf{q}',t',t'';L,m')\|^{\Upsilon}_{r_i} =$$

$$\left(\int_{-\infty}^{\infty} d\mathbf{q}'' \int_{\mathbf{q}(t')=\mathbf{q}'}^{\mathbf{q}(t'')=\mathbf{q}''} [\mathbf{D}\mathbf{q}(t);L,m']\Upsilon(\dot{\mathbf{q}}(t),\mathbf{q}(t),t',t'')\Im_i^{r_i}(\dot{\mathbf{q}}(t),\mathbf{q}(t),t',t'')\right)^{1/r_i} < \infty, \tag{92}$$

$$\Upsilon(\dot{\mathbf{q}}(t),\mathbf{q}(t),t',t'') > 0.$$

Then

$$\|\Im_3(\mathbf{q}',t',t'';L,m')\|^{\Upsilon}_1 \leq \|\Im_3(\mathbf{q}',t',t'';L,m')\|^{\Upsilon}_p \|\Im_3(\mathbf{q}',t',t'';L,m')\|^{\Upsilon}_q. \tag{93}$$

From Theorem 5 we obtain.

**Corolarly1**.Assume that: (1) $1/p + 1/q = 1$, (2) $I_p = I_p(\mathbf{q}', t', t''; L, m')$ where

$$I_p(\mathbf{q}', t', t''; L, m') = \int_{-\infty}^{\infty} d\mathbf{q}'' \int_{\mathbf{q}(t')=\mathbf{q}'}^{\mathbf{q}(t'')=\mathbf{q}''} [\mathbf{D}\mathbf{q}(t); L, m'] \exp\left( -\int_{t'}^{t''} [\dot{\mathbf{q}}(t)]^2 dt \right) \times$$
$$\exp\left( p \int_{t'}^{t''} G_1(\dot{\mathbf{q}}(t), \mathbf{q}(t), t) dt \right) < \infty, \tag{94}$$

(3) $I_q = I_q(\mathbf{q}', t', t''; L, m')$ where

$$I_q(\mathbf{q}', t', t''; L, m') = \int_{-\infty}^{\infty} d\mathbf{q}'' \int_{\mathbf{q}(t')=\mathbf{q}'}^{\mathbf{q}(t'')=\mathbf{q}''} [\mathbf{D}\mathbf{q}(t); L, m'] \exp\left( -\int_{t'}^{t''} [\dot{\mathbf{q}}(t)]^2 dt \right) \times$$
$$\exp\left( q \int_{t'}^{t''} G_2(\dot{\mathbf{q}}(t), \mathbf{q}(t), t) dt \right) < \infty. \tag{95}$$

Then inequality

$$\int_{-\infty}^{\infty} d\mathbf{q}'' \int_{\mathbf{q}(t')=\mathbf{q}'}^{\mathbf{q}(t'')=\mathbf{q}''} [\mathbf{D}\mathbf{q}(t); L, m'] \exp\left( -\int_{t'}^{t''} [\dot{\mathbf{q}}(t)]^2 dt \right) \exp\left( p \int_{t'}^{t''} G_1(\dot{\mathbf{q}}(t), \mathbf{q}(t), t) dt \right) \times$$
$$\times \exp\left( q \int_{t'}^{t''} G_2(\dot{\mathbf{q}}(t), \mathbf{q}(t), t) dt \right) \leq [I_p]^{1/p} + [I_q]^{1/q} \tag{96}$$

is satisfied.

**Theorem 6**. (1) Let $\mathbf{b}(\mathbf{x}, t)$ be an vector function, where $b_i(\mathbf{x}, t), i = 1, \ldots, n$ is a polinomial on variable $\mathbf{x}$. Let $\widehat{\mathbf{b}}(\mathbf{x}, t)$ be the linear part of the $\mathbf{b}(\mathbf{x}, t)$ i.e.,

$$\widehat{b}_i(\mathbf{x}, t) = \sum_{\alpha, |\alpha| \leq 1} b_i^{\alpha}(t) \mathbf{x}^{\alpha}, i = 1, \ldots, n. \tag{97}$$

Let $\widehat{\mathcal{L}}(\dot{\mathbf{q}}, \mathbf{q}, t)$ be the Lagrangian

$$\widehat{\mathcal{L}}(\dot{\mathbf{q}},\mathbf{q},t) = \left\| \dot{\mathbf{q}} - \widehat{\mathbf{b}}(\mathbf{q},t) \right\|^2 - \varepsilon \sum_{i=1}^{n} \widehat{b}_{i,i}(\mathbf{q},t). \quad (98)$$

Here

$$\widehat{b}_{i,i}(\mathbf{q},t) = \frac{\partial \widehat{b}_i(\mathbf{q},t)}{\partial q_i}, i = 1,\ldots,n. \quad (99)$$

And let $\mathcal{L}_{\varepsilon'}(\dot{\mathbf{q}},\mathbf{q},t)$ be

$$\mathcal{L}_{\varepsilon'}(\dot{\mathbf{q}},\mathbf{q},t) = \|\dot{\mathbf{q}}(t) - \mathbf{b}(\mathbf{q}_{\varepsilon'}(t),t)\|^2 - \varepsilon \sum_{i=1}^{n} b_{i,i}(\mathbf{q}_{\varepsilon'}(t),t). \quad (100)$$

Here $\mathbf{q}_{\varepsilon'}(t) = (\mathbf{q}_{1,\varepsilon'}(t),\ldots,\mathbf{q}_{i,\varepsilon'}(t),\ldots,\mathbf{q}_{n,\varepsilon'}(t))$ and

$$\mathbf{q}_{i,\varepsilon'}(t) = \frac{\mathbf{q}_i(t)}{1 + \left[ \varepsilon' \mathbf{q}_i^2(t) + \int_{t'}^{t''} \mathbf{q}_i^2(t)dt \right]^l} \quad (101)$$

$$l \geq 3.$$

(2) Let $\mathbf{x}_{t'',\varepsilon'}^{\mathbf{x}_0,\varepsilon}(\omega)$ be the solution of the Colombeau-Ito's SDE given by Eq.(1). Then there exist finite Colombeau constant $\widetilde{C}' = \left[(C'_{\varepsilon'})_{\varepsilon'}\right] > 0$ such that the inequalities (1)

$$\left[\left(\liminf_{\varepsilon\to 0} E\left[\left\|\mathbf{x}_{t'',\varepsilon'}^{\mathbf{x}_0,\varepsilon}(\omega)\right\|^2\right]\right)_{\varepsilon'}\right] \leq \widetilde{C}'\left[(\mathbf{I}_{\varepsilon'}^{\varepsilon})_{\varepsilon'}\right], \quad (102)$$

where

$$(\mathbf{I}^{\varepsilon}_{\varepsilon'})_{\varepsilon'} =$$

$$\int\limits_{-\infty}^{\infty} d\mathbf{q}'' \|\mathbf{q}''\|^2 \int\limits_{\mathbf{q}(t')=\mathbf{q}'}^{\mathbf{q}(t'')=\mathbf{q}''} [\mathbf{D}\mathbf{q}(t)] \exp\left[ -\frac{1}{2\varepsilon}\left( \int\limits_{t'}^{t''} \widehat{\mathcal{L}}_{\varepsilon'}(\dot{\mathbf{q}}(t), \mathbf{q}(t), t; \varepsilon) dt \right)_{\varepsilon'} \right] \tag{103}$$

and (2)

$$\left[\left( \liminf_{\varepsilon \to 0} \mathbf{E}\left[ \|\mathbf{x}^{\mathbf{x}_0,\varepsilon}_{t'',\varepsilon'}(\omega)\|^2 \right] \right)_{\varepsilon'} \right] \leq \widetilde{C}' \|\mathbf{U}(t)\| \tag{104}$$

is satisfies. Here a vector function $\mathbf{U}(t) = (U_1(t), \ldots, U_n(t))$ is the solution of the differential master equation

$$\dot{\mathbf{U}}(t) = \mathbf{J}_t \mathbf{U}(t) + \mathbf{b}(t), \mathbf{U}(0) = \mathbf{x}_0. \tag{105}$$

Here $\mathbf{b}(t) = \mathbf{b}(\mathbf{0}, t)$ and $\mathbf{J}_t = \mathbf{J}(t)$ is Jacobian i.e., $\mathbf{J}_t$ is $n \times n$-matrix:

$$\mathbf{J}(t) = \left[ \frac{\partial b_i(\mathbf{x}, t)}{x_j} \right]_{\mathbf{x}=0}. \tag{106}$$

**Proof**. For short, we will be considered the proof only for the case of the 1-dimensional Colombeau-Ito's SDE, without loss of generality.

Let us consider Feynman's path integral (76) corresponding to 1-dimensional Colombeau-Ito's SDE i.e.,

$$\mathbf{I}^{\varepsilon}_{\varepsilon'}(q',t',t'';L,m') = \left(\mathbf{E}_{L,m'}\left[x^{q',\varepsilon}_{t'',\varepsilon'}(\omega)\right]^2\right)_{\varepsilon'} =$$

$$\int_{-\infty}^{\infty} dq''\|q''\|^2 (p^{\varepsilon}_{\varepsilon'}(q',t'|q'',t'';L,m'))_{\varepsilon'} =$$

$$\int_{-\infty}^{\infty} dq''\|q''\|^2 \int_{\mathbf{q}(t')=q'}^{\mathbf{q}(t'')=q''} [\mathbf{D}q(t);L,m'] \exp\left[-\frac{1}{2\varepsilon}\left(\int_{t'}^{t''}\mathcal{L}_{\varepsilon'}(\dot{q}(t),q(t),t;\varepsilon)dt\right)_{\varepsilon'}\right] = \tag{107}$$

$$\int_{\mathbf{q}(t')=\mathbf{q}'} [\mathbf{D}q(t);L,m']\|q(t'')\|^2 \exp\left[-\frac{1}{2\varepsilon}\left(\int_{t'}^{t''}\mathcal{L}_{\varepsilon'}(\dot{q}(t),q(t),t;\varepsilon)dt\right)_{\varepsilon'}\right].$$

Let us rewrite now Lagrangian $\mathcal{L}_{\varepsilon'}(\dot{q}(t),q(t),t;\varepsilon)$ in the next equivalent form

$$\mathcal{L}_{\varepsilon'}(\dot{q}(t),q(t),t;\varepsilon) = \mathcal{L}'_{\varepsilon'}(\dot{q}(t),q(t),t;\varepsilon) - 2\varepsilon b_{11}(q_{\varepsilon'}(t),t)'. \tag{108}$$

Here

$$\begin{gathered}\mathcal{L}'_{\varepsilon'}(\dot{q}(t),q(t),t;\varepsilon) = \left[\dot{q}(t) - \widehat{b}(q_{\varepsilon'}(t),t) - b_2(q_{\varepsilon'}(t),t)\right]^2, \\ \widehat{b}(x,t) = b^0(t) + xb^1(t)\end{gathered} \tag{109}$$

and

$$b_2(x,t) = \sum_{\alpha,2\leq|\alpha|\leq r} b^{\alpha}(t)x^{\alpha} \tag{110}$$

Let us rewrite now Eq.(110) in the next form

$$b_2(x,t) = b_{2,2}(x,t) + b_{2,3}(x,t). \tag{111}$$

Here

$$b_{2,2}(x,t) = \sum_{\alpha,|\alpha|=2} b^{\alpha}(t)x^{\alpha}$$

$$b_{2,3}(x,t) = \sum_{\alpha,|\alpha|\geq 3} b^{\alpha}(t)x^{\alpha} \qquad (112)$$

From Eq.(109)-Eq.(110) we obtain

$$\mathcal{L}'_{\varepsilon'}(\dot{q}(t),q(t),t;\varepsilon) = \left[\left[\dot{q}(t) - \widehat{b}(q_{\varepsilon'}(t),t)\right] - b_2(q_{\varepsilon'}(t),t)\right]^2 =$$

$$\left[\dot{q}(t) - \widehat{b}(q_{\varepsilon'}(t),t)\right]^2 - 2b_2(q_{\varepsilon'}(t),t)\left[\dot{q}(t) - \widehat{b}(q_{\varepsilon'}(t),t)\right] +$$

$$+b_2^2(q_{\varepsilon'}(t),t) = [\dot{q}(t)]^2 - 2\dot{q}(t)\widehat{b}(q_{\varepsilon'}(t),t) + \widehat{b}^{\,2}(q_{\varepsilon'}(t),t) - \qquad (113)$$

$$-2\dot{q}(t)b_2(q_{\varepsilon'}(t),t) + 2b_2(q_{\varepsilon'}(t),t)\widehat{b}(q_{\varepsilon'}(t),t).$$

Substitution Eq.(112) into Eq.(113) gives

$$\mathcal{L}'_{\varepsilon'}(\dot{q}(t), q(t), t; \varepsilon) =$$

$$[\dot{q}(t)]^2 - 2\dot{q}(t)b^0(t) - 2\dot{q}(t)q_{\varepsilon'}(t)b^1(t) + [b^0(t)]^2 + 2q_{\varepsilon'}(t)b^0(t)b^1(t) +$$

$$+[q_{\varepsilon'}(t)b^1(t)]^2 - 2\dot{q}(t)b_2(q_{\varepsilon'}(t), t) + 2b_2(q_{\varepsilon'}(t), t)\widehat{b}(q_{\varepsilon'}(t), t) =$$

$$[\dot{q}(t)]^2 - 2\dot{q}(t)b^0(t) - 2\dot{q}(t)q_{\varepsilon'}(t)b^1(t) + [b^0(t)]^2 + 2q_{\varepsilon'}(t)b^0(t)b^1(t) +$$

$$+[q_{\varepsilon'}(t)b^1(t)]^2 - 2\dot{q}(t)b_2(q_{\varepsilon'}(t), t) + 2b_2(q_{\varepsilon'}(t), t)b^0(t) + \tag{114}$$

$$+2q_{\varepsilon'}(t)b_2(q_{\varepsilon'}(t), t)b^1(t) =$$

$$[\dot{q}(t)]^2 - 2\dot{q}(t)b^0(t) - 2\dot{q}(t)q_{\varepsilon'}(t)b^1(t) + [b^0(t)]^2 + 2q_{\varepsilon'}(t)b^0(t)b^1(t) -$$

$$-2\dot{q}(t)b_{2,2}(q_{\varepsilon'}(t), t) - 2\dot{q}(t)b_{2,3}(q_{\varepsilon'}(t), t) + 2b_{2,2}(q_{\varepsilon'}(t), t)b^0(t) +$$

$$+2b_{2,3}(q_{\varepsilon'}(t), t)b^0(t) + 2q_{\varepsilon'}(t)b_2(q_{\varepsilon'}(t), t)b^1(t).$$

Substitution Eq. (114) into Eq. (107) gives

$$\mathbf{I}^{\varepsilon}_{\varepsilon'}(q', t', t''; L, m') =$$

$$\exp\left[-\frac{1}{2\varepsilon}\int_{t'}^{t''}[b^0(t)]^2 dt\right]\int_{-\infty}^{\infty} dq'' \| q'' \|^2 \int_{\mathbf{q}(t')=q'}^{\mathbf{q}(t'')=q''} [\mathbf{D}q(t); L, m'] \times$$

$$\times\exp\left[-\frac{1}{2\varepsilon}\int_{t'}^{t''}\{[\dot{q}(t)]^2 - 2\dot{q}(t)b^0(t) - 2\dot{q}(t)q_{\varepsilon'}(t)b^1(t) + 2q_{\varepsilon'}(t)b^0(t)b^1(t) + dt\right. \tag{115}$$

$$+[q_{\varepsilon'}(t)b^1(t)]^2 - 2\dot{q}(t)b_{2,2}(q_{\varepsilon'}(t), t) - 2\dot{q}(t)b_{2,3}(q_{\varepsilon'}(t), t) + 2b_{2,2}(q_{\varepsilon'}(t), t)b^0(t) +$$

$$\left.+2b_{2,3}(q_{\varepsilon'}(t), t)b^0(t) + 2q_{\varepsilon'}(t)b_2(q_{\varepsilon'}(t), t)b^1(t)\}dt\right]$$

Using replacement $q(t) = p(t)\sqrt{2\varepsilon}$ into Feynman path integral (115), we obtain

$$\mathbf{I}_{\varepsilon'}^{\varepsilon}(q',t',t'';L,m') =$$

$$\exp\left[-\frac{1}{2\varepsilon}\int_{t'}^{t''}[b^0(t)]^2dt\right]\int_{-\infty}^{\infty}dq''\|q''\|^2\int_{\mathbf{q}(t')=\frac{q'}{\sqrt{2\varepsilon}}}^{\mathbf{q}(t'')=\frac{q''}{\sqrt{2\varepsilon}}}\left[\mathbf{D}p(t);\tfrac{L}{\sqrt{2\varepsilon}},m'\right]\times \tag{116}$$

$$\times\exp\left[-\int_{t'}^{t''}\left\{[\dot{p}(t)]^2-\tfrac{2}{\sqrt{2\varepsilon}}\dot{p}(t)b^0(t)-2\dot{p}(t)p_{\varepsilon'}(t)b^1(t)+\tfrac{2}{\sqrt{2\varepsilon}}p_{\varepsilon'}(t)b^0(t)b^1(t)+\right.\right.$$

$$\left.\left.+[p_{\varepsilon'}(t)b^1(t)]^2-2\sqrt{2\varepsilon}\,\dot{p}(t)\breve{b}_2(p_{\varepsilon'}(t),t)+2b_2(p_{\varepsilon'}(t),t)\widehat{b}\left(\sqrt{2\varepsilon}\,p_{\varepsilon'}(t),t\right)dt\right\}dt\right]$$

Here

$$\breve{b}_2(p_{\varepsilon'}(t),t) = \sum_{2\leq\alpha\leq r}\left(\sqrt{2\varepsilon}\right)^{|\alpha|-2}b^{\alpha}(t)(p_{\varepsilon'}(t))^{\alpha}, \tag{117}$$

and

$$p_{\varepsilon'}(t) = \frac{p(t)}{1+(\varepsilon')^l\varepsilon^l\left[p^2(t)+\int_{t'}^{t''}p^2(t)dt\right]^l},l\geq 3. \tag{118}$$

Let us rewrite now Feynman path integral (116) in the next equivalent form

$$\mathbf{I}^{\varepsilon}_{\varepsilon'}(q',t',t'';L,m') =$$

$$\exp\left[-\frac{1}{2\varepsilon}\int_{t'}^{t''}[b^0(t)]^2dt\right]\int_{-\infty}^{\infty}dq''\|q''\|^2\int_{\mathbf{q}(t')=\frac{q'}{\sqrt{2\varepsilon}}}^{\mathbf{q}(t'')=\frac{q''}{\sqrt{2\varepsilon}}}\left[\mathbf{D}p(t);\frac{L}{\sqrt{2\varepsilon}},m'\right]\times$$

$$\times\exp\left[-\int_{t'}^{t''}[\dot{p}(t)]^2dt\right]\times \tag{119}$$

$$\exp\left[-\int_{t'}^{t''}\left\{-\frac{2}{\sqrt{2\varepsilon}}\dot{p}(t)b^0(t)-2\dot{p}(t)p_{\varepsilon'}(t)b^1(t)+\frac{2}{\sqrt{2\varepsilon}}p_{\varepsilon'}(t)b^0(t)b^1(t)+[p_{\varepsilon'}(t)b^1(t)]^2\right\}dt\right]\times$$

$$\times\exp\left[\left\{2\int_{t'}^{t''}\sqrt{2\varepsilon}\,\dot{p}(t)\check{b}_2(p_{\varepsilon'}(t),t)-2b_2(p_{\varepsilon'}(t),t)\widehat{b}\left(\sqrt{2\varepsilon}\,p_{\varepsilon'}(t),t\right)\right\}dt.\right]$$

Assume that $1/p+1/q=1$ and $q=1/\varepsilon.$ Then

$$p=\frac{1}{1-\varepsilon}=1+\varepsilon+o(\varepsilon). \tag{120}$$

Using now Corollary 1, from Eq.(119) we obtain

$$\mathbf{I}_{\varepsilon'}^{\varepsilon}(q',t',t'';L,m') \leq$$

$$\exp\left[-\frac{1}{2\varepsilon}\int_{t'}^{t''}[b^0(t)]^2 dt\right]\left[\int_{-\infty}^{\infty} dq'' \| q'' \|^2 \int_{\mathbf{q}(t')=\frac{q'}{\sqrt{2\varepsilon}}}^{\mathbf{q}(t'')=\frac{q''}{\sqrt{2\varepsilon}}}\left[\mathbf{D}p(t);\frac{L}{\sqrt{2\varepsilon}},m'\right]\times\right.$$

$$\times\exp\left[-\int_{t'}^{t''}[\dot{p}(t)]^2 dt\right]\times$$

$$\exp\left[-(1+\varepsilon)\int_{t'}^{t''}\left\{-\frac{2}{\sqrt{2\varepsilon}}\dot{p}(t)b^0(t) - 2\dot{p}(t)p_{\varepsilon'}(t)b^1(t) + \right.\right. \tag{121}$$

$$\left.\left.\left.\frac{2}{\sqrt{2\varepsilon}}p_{\varepsilon'}(t)b^0(t)b^1(t) + [p_{\varepsilon'}(t)b^1(t)]^2\right\}dt\right]\right]^{1-\varepsilon}\times$$

$$\left[\int_{-\infty}^{\infty} dq'' \| q'' \|^2 \int_{\mathbf{q}(t')=\frac{q'}{\sqrt{2\varepsilon}}}^{\mathbf{q}(t'')=\frac{q''}{\sqrt{2\varepsilon}}}\left[\mathbf{D}p(t);\frac{L}{\sqrt{2\varepsilon}},m'\right]\exp\left[-\int_{t'}^{t''}[\dot{p}(t)]^2 dt\right]\times\right.$$

$$\left.\times\exp\left[\frac{1}{\varepsilon}\int_{t'}^{t''}\left\{\sqrt{2\varepsilon}\,\dot{p}(t)\breve{b}_2(p_{\varepsilon'}(t),t) - 2b_2(p_{\varepsilon'}(t),t)\widehat{b}\left(\sqrt{2\varepsilon}\,p_{\varepsilon'}(t),t\right)\right\}dt\right]\right]^{\varepsilon}.$$

Therefore

$$\left(\mathbf{I}_{\varepsilon'}^{\varepsilon}(q',t',t'';L,m')\right)_{\varepsilon'} \leq$$

$$\leq \left(\left(\left(\mathbf{I}_{\varepsilon'}^{\varepsilon;1}(q',t',t'';L,m')\right)^{1-\varepsilon}\right)_{\varepsilon'}\right)\left(\left(\left(\mathbf{I}_{\varepsilon'}^{\varepsilon;2}(q',t',t'';L,m')\right)^{\varepsilon}\right)_{\varepsilon'}\right) \tag{122}$$

Here

$$\left(\mathbf{I}_{\varepsilon'}^{\varepsilon,1}(q',t',t'';L,m')\right)_{\varepsilon'} =$$

$$\Bigg(\exp\left[-\frac{1}{2\varepsilon(1-\varepsilon)}\int_{t'}^{t''}[b^0(t)]^2dt\right]\left[\int_{-\infty}^{\infty}dq''\|q''\|^2\int_{\mathbf{q}(t')=\frac{q'}{\sqrt{2\varepsilon}}}^{\mathbf{q}(t'')=\frac{q''}{\sqrt{2\varepsilon}}}\left[\mathbf{D}p(t);\tfrac{L}{\sqrt{2\varepsilon}},m'\right]\times\right.$$

$$\times\exp\left[-\int_{t'}^{t''}[\dot{p}(t)]^2dt\right]\times \tag{123}$$

$$\exp\Bigg[-(1+\varepsilon)\int_{t'}^{t''}\Big\{-\tfrac{2}{\sqrt{2\varepsilon}}\dot{p}(t)b^0(t)-2\dot{p}(t)p_{\varepsilon'}(t)b^1(t)+$$

$$\left.\tfrac{2}{\sqrt{2\varepsilon}}p_{\varepsilon'}(t)b^0(t)b^1(t)+[p_{\varepsilon'}(t)b^1(t)]^2\Big\}dt\Bigg]\right]\Bigg)_{\varepsilon'}$$

and

$$\left(\left(\mathbf{I}_{\varepsilon'}^{\varepsilon,2}(q',t',t'';L,m')\right)^{\varepsilon}\right)_{\varepsilon'} = \int_{-\infty}^{\infty}dq''\int_{\mathbf{q}(t')=\frac{q'}{\sqrt{2\varepsilon}}}^{\mathbf{q}(t'')=\frac{q''}{\sqrt{2\varepsilon}}}\left[\mathbf{D}p(t);\tfrac{L}{\sqrt{2\varepsilon}},m'\right]\times$$

$$\times\exp\left[-\int_{t'}^{t''}[\dot{p}(t)]^2dt\right]\times \tag{124}$$

$$\exp\left[\frac{1}{\varepsilon}\int_{t'}^{t''}\left\{\sqrt{2\varepsilon}\,\dot{p}(t)\breve{b}_2(p_{\varepsilon'}(t),t)-2b_2(p_{\varepsilon'}(t),t)\widehat{b}\left(\sqrt{2\varepsilon}\,p_{\varepsilon'}(t),t\right)\right\}dt\right].$$

(**I**) Let us evaluate now path integral $\left(\mathbf{I}_{\varepsilon'}^{\varepsilon,1}(q',t',t'';L,m')\right)_{\varepsilon'}$. From Eq.(123) using replacement into Feynman path integral in the RHS of the Eq. (123), we obtain

$$\left(\mathbf{I}^{\varepsilon,1}_{\varepsilon'}(q',t',t'';L,m')\right)_{\varepsilon'} = \int\limits_{-\infty}^{\infty} dq'' \qquad (125)$$

Let us estimate now path integral in the RHS of the Eq. (125), using saddle point approximation [18].Denoting the global minimum of the action

$$\widehat{\mathbf{S}} = \int\limits_{t'}^{t''} \left[\dot{q}(t) - \widehat{b}(q_{\varepsilon'}(t),t)\right]^2 dt \qquad (126)$$

by $\breve{q}(t,\varepsilon')$, it follows that it satisfies the extremality conditions for the minimizing path $\breve{q}(t,\varepsilon')$ is

$$\frac{d\breve{q}(t,\varepsilon')}{dt} - \widehat{b}(\breve{q}(t,\varepsilon'),t) = 0 \qquad (127)$$

In the limit $\varepsilon' \to 0$ from Eq. (127) we obtain

$$\frac{d\breve{q}(t)}{dt} - \widehat{b}(\breve{q}(t),t) = 0. \qquad (128)$$

Here we let $\breve{q}(t) = \breve{q}(t,0)$. Therefore from Eq.(125) and Eq. (128) we obtain

$$\left(\mathbf{I}^{\varepsilon,1}_{\varepsilon'}(q',t',t'';L,m')\right)_{\varepsilon'} = (\|\breve{q}(t,\varepsilon')\|)_{\varepsilon'} + o(\varepsilon). \qquad (129)$$

In the limit $\varepsilon \to 0$ finally we obtain

$$\left(\lim_{\varepsilon\to 0} \mathbf{I}^{\varepsilon,1}_{\varepsilon'}(q',t',t'';L,m')\right)_{\varepsilon'} = (\|\breve{q}(t,\varepsilon')\|)_{\varepsilon'} \qquad (130)$$

Let us evaluate now path integral $\left(\mathbf{I}^{\varepsilon,2}_{\varepsilon'}(q',t',t'';L,m')\right)$. Let us rewrite Eq.(117) in the following form

$$\breve{b}_2(p_{\varepsilon'}(t),t) = \sum_{2\leq\alpha\leq r}\left(\sqrt{2\varepsilon}\right)^{|\alpha|-2} b^{\alpha}(t)(p_{\varepsilon'}(t))^{\alpha},$$

$$\breve{b}_2(p_{\varepsilon'}(t),t) = \breve{b}_{2,2}(p_{\varepsilon'}(t),t) + \breve{b}_{2,3}(p_{\varepsilon'}(t),t), \tag{131}$$

$$\breve{b}_{2,2}(p_{\varepsilon'}(t),t) =$$

$$\breve{b}_{2,3}(p_{\varepsilon'}(t),t) =$$

Substitution Eqs.(131) into Eq.( 124) gives

$$\left(\left(\mathbf{I}^{\varepsilon,2}_{\varepsilon'}(q',t',t'';L,m')\right)^{\varepsilon}\right)_{\varepsilon'} = \tag{132}$$

# Appendix III.

Let $\Re_i = \{\Omega_i,\Sigma_i,\mathbf{P}_i\}$ $i = 1,2$ be a probability spaces such that: $\Omega_1 \cap \Omega_2 = \varnothing$.Let $\mathbf{W}(t,\omega)$ be a Wiener process on $\Re_1$.

**Proposition 1**.Assume that (1) $\varpi \in \Omega_2,\ \varphi(t,\varpi),\alpha(t,\varpi) \in \mathcal{L}_1([0,T])$ $\mathbf{P}_2$-o.s., $\sup_{t\in[0,T]}|\varphi(t,\varpi)| < \infty$ $\mathbf{P}_2$-o.s., $\sup_{t\in[0,T]}|\alpha(t,\varpi)| < \infty$ $\mathbf{P}_2$-o.s., and (2) the inequality

$$\varphi(t,\varpi) \leq \alpha(t,\varpi) + L_{\varpi}\int_0^t \varphi(s,\varpi)ds \tag{1}$$

$\mathbf{P}_2$-o.s. is satisfied. Then the inequality

$$\varphi(t,\varpi) \leq \alpha(t,\varpi) + L_{\varpi}\int_0^t e^{L_{\varpi}(t-s)}\alpha(s,\varpi)ds. \tag{2}$$

$\mathbf{P}_2$-o.s. is satisfied.

**Proposition 2**.(**I**) Assume that: (1) let $\eta_n(t) = \eta_n(t,\omega,\varpi), n = 1,2,\ldots,\omega \in \Omega_1,\varpi \in \Omega_2$ be

the solutions of the Ito’s SDE’s

$$\boldsymbol{\eta}_n(t) = \boldsymbol{\phi}(t) + \int_0^t \mathbf{A}_n(\boldsymbol{\eta}_n(s), s, \varpi)ds + \int_0^t \mathbf{B}_n(\boldsymbol{\eta}_n(s), s, \varpi)d\mathbf{W}(s,\omega). \quad (3)$$

and let $\tilde{\eta}_n(t) = \tilde{\eta}_n(t,\omega,\varpi), n = 1,2,\ldots$ be the solutions of the Ito SDE's

$$\tilde{\boldsymbol{\eta}}_n(t) = \boldsymbol{\phi}(t) + \int_0^t \tilde{\mathbf{A}}_n(\tilde{\boldsymbol{\eta}}_n(s), s, \varpi)ds + \int_0^t \tilde{\mathbf{B}}_n(\tilde{\boldsymbol{\eta}}_n(s), s, \varpi)d\mathbf{W}(s,\omega). \quad (4)$$

Here $\mathbf{A}_n(x,t,\varpi) = (A_{n,1}(x,t,\varpi),\ldots,\mathbf{A}_{n,d}(x,t,\varpi)), x \in \mathbb{R}^d$,
$\tilde{\mathbf{A}}_n(x,t,\varpi) = \left(\tilde{A}_{n,1}(x,t,\varpi),\ldots,\tilde{\mathbf{A}}_{n,d}(x,t,\varpi)\right)$, $x \in \mathbb{R}^l$, $\mathbf{B}_n(x,t,\varpi) = \left\{B_n^{i,j}(x,t,\varpi)\right\}, i,j = 1,\ldots,l$
is $d \times d$ matrix, $\tilde{\mathbf{B}}_n(x,t,\varpi) = \left\{\tilde{B}_n^{i,j}(x,t,\varpi)\right\}, i,j = 1,\ldots,d$ is $l \times l$ matrix, $l \geq d$.

(2) The inequalities

$$(5.1)\ (\|\mathbf{A}_n(x,t,\varpi)\|^2 + \|\mathbf{B}_n(x,t)\|^2) \leq K_{n,\varpi}(1 + \|x\|^2),$$

$$(\|\tilde{\mathbf{A}}_n(x,t,\varpi)\|^2 + \|\tilde{\mathbf{B}}_n(x,t,\varpi)\|^2) \leq K_{n,\varpi}(1 + \|x\|^2),$$

$$(5.2)\|\mathbf{A}_n(x,t,\varpi) - \mathbf{A}_n(y,t,\varpi)\| + \|\mathbf{B}_n(x,t,\varpi) - \mathbf{B}_n(y,t,\varpi)\| \leq K_{n,\varpi}\|x-y\|,$$

$$\|\tilde{\mathbf{A}}_n(x,t,\varpi) - \tilde{\mathbf{A}}_n(y,t,\varpi)\| + \|\tilde{\mathbf{B}}_n(x,t,\varpi) - \tilde{\mathbf{B}}_n(y,t,\varpi)\| \leq K_{n,\varpi}\|x-y\|, \quad (5)$$

$$(5.3)\|\mathbf{A}_n(x,t,\varpi) - \tilde{\mathbf{A}}_n(x,t,\varpi)\| \leq \delta_{1,n,\varpi}\|x\|,$$

$$(5.4)\|\mathbf{B}_n(x,t,\varpi) - \tilde{\mathbf{B}}_n(x,t,\varpi)\| \leq \delta_{2,n,\varpi}\|x\|$$

is satisfied. Then the inequality

$$\sup_{0\leq t\leq T}\mathbf{E}_{\Omega_1}\left[\|\boldsymbol{\eta}_n(t,\varpi)-\tilde{\boldsymbol{\eta}}_n(t,\varpi)\|^2\right]\leq$$

$$\leq e^{L_{n,\varpi}}(T\delta^2_{1,n,\varpi}+\delta^2_{2,n,\varpi})\mathbf{E}_{\Omega_1}\left[\int_0^T\left(\|\tilde{\boldsymbol{\eta}}_n(t,\varpi)\|^2\right)dt\right], \tag{6}$$

$$L_{n,\varpi}=3(1+T)K^2_{n,\varpi}$$

$\mathbf{P}_2$-o.s. is satisfied.

**(II)** Let $\boldsymbol{\tau}_{U_n}=\tau_n(\omega,\varpi)$ be the random variable equal to the time at which the sample function of the process $\tilde{\boldsymbol{\eta}}_n(t,\varpi)$ first leaves the bounded neighborhood $U_n\ni 0$,and let $\tau_n(t,\varpi)=\min(\tau_n(\omega,\varpi),t)$. Assume that: (1) $\forall n\in\mathbb{N}:U_n\subset U_{n+1}$ and $\bigcup_{n\in\mathbb{N}}U_n=\mathbb{R}^d$, (2) $\sup_{n\in\mathbb{N}}\left(\mathbf{E}_{\Omega_1}\left[\int_0^T\left(\|\tilde{\boldsymbol{\eta}}_n(t,\varpi)\|^2\right)dt\right]\right)<\infty$ $\mathbf{P}_2$-o.s.

Then the inequality

$$\sup_{0\leq t\leq T}\mathbf{E}_{\Omega_1}\left[\|\boldsymbol{\eta}_n(\tau_n(t,\varpi))-\tilde{\boldsymbol{\eta}}_n(\tau_n(t,\varpi))\|^2\right]\leq$$

$$\leq e^{L_{n,\varpi}}(T\delta^2_{1,n,\varpi}+\delta^2_{2,n,\varpi})\left\{\sup_{n\in\mathbb{N}}\left(\mathbf{E}_{\Omega_1}\left[\int_0^T\left(\|\tilde{\boldsymbol{\eta}}_n(t,\varpi)\|^2\right)dt\right]\right)\right\}, \tag{6'}$$

$$L_{n,\varpi}=3(1+T)K^2_{n,\varpi}.$$

$\mathbf{P}_2$-o.s. is satisfied.

**Proof.**(**I**) From Eq.(3) and Eq.(4) one obtain

$$\boldsymbol{\eta}_n(t) - \tilde{\boldsymbol{\eta}}_n(t) =$$

$$\int_0^t \mathbf{A}_n(\boldsymbol{\eta}_n(s), s, \varpi)ds + \int_0^t \mathbf{B}_n(\boldsymbol{\eta}_n(s), s, \varpi)d\mathbf{W}(s,\omega) - \int_0^t \tilde{\mathbf{A}}_n(\tilde{\boldsymbol{\eta}}_n(s), s, \varpi)ds -$$

$$-\int_0^t \tilde{\mathbf{B}}_n(\tilde{\boldsymbol{\eta}}_n(s), s, \varpi)d\mathbf{W}(s,\omega) = \boldsymbol{\zeta}_n(t,\varpi) + \int_0^t [\mathbf{A}_n(\boldsymbol{\eta}_n(s), s, \varpi) - \mathbf{A}_n(\tilde{\boldsymbol{\eta}}_n(s), s, \varpi)]ds + \tag{7}$$

$$+\int_0^t [\mathbf{B}_n(\boldsymbol{\eta}_n(s), s, \varpi) - \mathbf{B}_n(\tilde{\boldsymbol{\eta}}_n(s), s, \varpi)]d\mathbf{W}(s,\omega).$$

Here

$$\boldsymbol{\zeta}_n(t,\varpi) = \int_0^t \left[ \mathbf{A}_n(\tilde{\boldsymbol{\eta}}_n(s), s, \varpi) - \tilde{\mathbf{A}}_n(\tilde{\boldsymbol{\eta}}_n(s), s, \varpi) \right] ds +$$

$$(8)$$

$$+\int_0^t \left[ \mathbf{B}_n(\tilde{\boldsymbol{\eta}}_n(s), s, \varpi) - \tilde{\mathbf{B}}_n(\tilde{\boldsymbol{\eta}}_n(s), s, \varpi) \right] d\mathbf{W}(s,\omega).$$

From Eq.(8) and inequalities (5.1)-(5.2) one obtain the inequality

$$\mathbf{E}_{\Omega_1}\left[ \|\boldsymbol{\eta}_n(t,\varpi) - \tilde{\boldsymbol{\eta}}_n(t,\varpi)\|^2 \right] \leq 3\mathbf{E}_{\Omega_1}\left[ \|\boldsymbol{\zeta}_n(t,\varpi)\|^2 \right] +$$

$$+L_{n,\varpi} \int_0^t \mathbf{E}_{\Omega_1}\left[ \|\boldsymbol{\eta}_n(s,\varpi) - \tilde{\boldsymbol{\eta}}_n(s,\varpi)\|^2 \right] ds, \tag{9}$$

$$L_{n,\varpi} = 3(1+T)K^2_{n,\varpi}.$$

Using Proposition 1, from inequality (9) one obtain the inequality

$$\mathbf{E}_{\Omega_1}\left[\|\boldsymbol{\eta}_n(t)-\tilde{\boldsymbol{\eta}}_n(t)\|^2\right] \leq 3\mathbf{E}_{\Omega_1}\left[\|\boldsymbol{\zeta}_n(t,\varpi)\|^2\right]+L_{n,\varpi}\int_0^t e^{L_{n,\varpi}(t-s)}\mathbf{E}_{\Omega_1}\left[\|\boldsymbol{\zeta}_n(s,\varpi)\|^2\right]ds. \quad (10)$$

From inequality (5.3) one obtain the inequality

$$\sup_{0\leq t\leq T}\left\|\mathbf{E}_{\Omega_1}\left[\int_0^t\left[\mathbf{A}_n(\tilde{\boldsymbol{\eta}}_n(s),s,\varpi)-\tilde{\mathbf{A}}_n(\tilde{\boldsymbol{\eta}}_n(s),s,\varpi)\right]ds\right]\right\|^2 \leq$$

$$\leq T\int_0^T\mathbf{E}_{\Omega_1}\left[\left\|\mathbf{A}_n(\tilde{\boldsymbol{\eta}}_n(s),s,\varpi)-\tilde{\mathbf{A}}_n(\tilde{\boldsymbol{\eta}}_n(s),s,\varpi)\right\|^2\right]ds \leq \quad (11)$$

$$\leq T\delta^2_{1,n,\varpi}\left[\int_0^T\mathbf{E}_{\Omega_1}\left[\|\tilde{\boldsymbol{\eta}}_n(s)\|^2\right]ds\right].$$

From inequality (5.4) one obtain the inequality

$$\mathbf{E}_{\Omega_1}\left[\sup_{0\leq t\leq T}\left\|\int_0^t\left[\mathbf{B}_n(\tilde{\boldsymbol{\eta}}_n(s),s,\varpi)-\tilde{\mathbf{B}}_n(\tilde{\boldsymbol{\eta}}_n(s),s,\varpi)\right]d\mathbf{W}(s,\omega)\right\|^2\right] \leq$$

$$\leq 4\mathbf{E}_{\Omega_1}\left[\int_0^T\left[\left\|\mathbf{B}_n(\tilde{\boldsymbol{\eta}}_n(s),s,\varpi)-\tilde{\mathbf{B}}_n(\tilde{\boldsymbol{\eta}}_n(s),s,\varpi)\right\|^2\right]ds\right] \leq \quad (12)$$

$$\leq \delta^2_{2,n,\varpi}\left[\int_0^T\mathbf{E}_{\Omega_1}\left[\|\tilde{\boldsymbol{\eta}}_n(s)\|^2\right]ds\right].$$

Let $\mathfrak{R}_i = \{\Omega_i,\Sigma_i,\mathbf{P}_i\}$ $i = 1,2$ be a probability spaces such that: $\Omega_1 \cap \Omega_2 = \varnothing$.
Let $\mathbf{W}(t,\omega)$ be a Wiener process on $\mathfrak{R}_1$ and let $\mathbf{W}(t,\varpi)$ be a Wiener process on $\mathfrak{R}_2$.Let us consider now a family (x ̇_(t,D,ε^')^(x_0,ε) (ω,ϖ))_(ε^' )of the Colombeau generalized stochastic processes which is a solution of the Colombeau-Ito's stochastic equation with stochastic coefficients:

.

# **Appendix IV**. The Pringsheim limit for double sequences

of random variables.

**Definition 1**.[55].Let $u = (u_{mn})$ be a double sequence of real numbers. A double sequence $(u_{mn})$ is said to be Pringsheim convergent (or $P$- convergent) to $s \in \mathbb{R}$ if for a given $\varepsilon > 0$ there exists a positive integer $N \in \mathbb{N}$ such that $|u_{mn} - s| < \varepsilon$ for all nonnegative integer $m, n \geq N$, and we write $P\text{-}\lim_{m,n\to\infty} u_{mn} = s$.

**Definition 2**.[55]. A double sequence $u = (u_{mn})$ is bounded if there exists a real number $C > 0$ such that $|u_{mn}| \leq C$ for all nonnegative $m, n$.

The $P$-limits superior and inferior can be defined using the concept of subsequential $P$-limits of the double sequence $(u_{m,n})$.

**Definition 3**.An element $\xi$ of the extended real numbers $\overline{\mathbb{R}}$ is a subsequential P-limit of $(u_{mn})$ if there exists a strictly increasing sequences of natural numbers $(m_l, n_k)$ such that $\xi = P\text{-}\lim_{l,k\to\infty} u_{m_l,n_k}$. If $E \subseteq \overline{\mathbb{R}}$ is the set of all subsequential $P$-limits of $(u_{m,n})$ , then

$$\begin{aligned} P\text{-}\overline{\lim}_{m,n\to\infty} u_{mn} &= \sup E, \\ P\text{-}\underline{\lim}_{m,n\to\infty} u_{mn} &= \inf E. \end{aligned} \tag{1}$$

**Remark1**. Whenever the $P$-limit exists, the $P$-limit inferior and limit superior are both equal to it; therefore, each can be considered a generalization of the ordinary P-limit which is primarily interesting in cases where the ordinary $P$-limit does not exist.

Whenever $P\text{-}\overline{\lim}_{m,n\to\infty} u_{mn}$ and $P\text{-}\underline{\lim}_{m,n\to\infty} u_{mn}$ both exist, we have

(i) $P\text{-}\underline{\lim}_{m,n\to\infty} u_{mn} \leq P\text{-}\overline{\lim}_{m,n\to\infty} u_{mn}$.

(ii) $P\text{-}\lim_{m,n\to\infty} u_{mn} = s$ if and only if $P\text{-}\underline{\lim}_{m,n\to\infty} u_{mn} = P\text{-}\overline{\lim}_{m,n\to\infty} u_{mn} = s$.

(iii) $P\text{-}\overline{\lim}_{m,n\to\infty}(-u_{mn}) = -(P\text{-}\underline{\lim}_{m,n\to\infty} u_{mn})$.

(iv) $P\text{-}\overline{\lim}_{m,n\to\infty}(u_{mn} + v_{mn}) \leq P\text{-}\overline{\lim}_{m,n\to\infty} u_{mn} + P\text{-}\overline{\lim}_{m,n\to\infty} v_{mn}$.

(v) $P\text{-}\underline{\lim}_{m,n\to\infty}(u_{mn} + v_{mn}) \geq P\text{-}\underline{\lim}_{m,n\to\infty} u_{mn} + P\text{-}\underline{\lim}_{m,n\to\infty} v_{mn}$.